\documentclass[12pt,a4paper, %flexing, 
titlepage, oneside, openright]{scrbook}
\usepackage{amsthm}
\usepackage{graphicx}
\usepackage{epsfig}
\def\d{\partial}
\def\be{\begin{eqnarray}}
\def\ee{\end{eqnarray}}

\def\d{\partial}

\def\slash#1{\, /\kern-0.6em{#1}}

\begin{document}
\begin{center}

%\pagenumbering{roman}
\thispagestyle{empty}
{\Huge \bf Black holes and  
\vskip 0.3cm
the positive cosmological constant}
\vskip 0.3cm

\vskip 2.0 true cm
Thesis Submitted for the degree of\\
Doctor of Philosophy (Science)\\
of \\
UNIVERSITY OF JADAVPUR
\vskip 3.0 true cm
2011 \\
{\large \bf SOURAV BHATTACHARYA}\\
S. N. Bose National Centre for Basic Sciences\\
Block-JD, Sector III, \\
Salt Lake \\
Kolkata  700098\\
India
\end{center}
\frontmatter
\newpage
%\thispagestyle{empty}
%\begin{figure}
%\centering%keepaspectratio
%\rotatebox{0}{
%\includegraphics[height=10cm,width=8cm]{sou_pre.pdf}}
%%\includegraphics[height=10.0cm,keepaspectratio]{sou_pre.ps}}
%\includegraphics{sig1.ps}}
%\caption{The Kruskal diagram for the de Sitter spacetime (\ref{anex9'}) with each point understood over a 2-sphere.}[height=7.5cm,width=9.0cm]
%\end{figure}
%\newpage

%\begin{center}
%{\bf {\large {\underline {CERTIFICATE FROM THE SUPERVISOR(S)}}}}
%\end{center}
%This is to certify that the thesis entitled ``Black holes and the positive cosmological constant" submitted by Sri / Smt. Sourav Bhattacharya, who got his/her name registered on 12/12/2008 for the award of Ph.D (Science) degree of Jadavpur University, is absolutely based upon his/her own work under the supervision of Dr. Amitabha Lahiri and that neither this thesis nor any part of it has been submitted for either any degree/diploma or any other academic award anywhere before.

 %              \vskip 4cm

  %             \qquad \qquad\qquad\qquad   \qquad   \qquad            (Signature of the Supervisor(s) date with official seal)
%%\pagebreak
%\frontmatter
\newpage
%%\frontmatter %%
\thispagestyle{empty}
\vskip 6.0cm
\begin{center}
{\Huge \bf  Dedicated}
\vskip 0.5 cm
{\Huge \bf to }
\vskip 0.5cm
{ \Huge \bf   My Parents}
%%%%{\Huge \bf Other\\ Family members}
\end{center}
\pagebreak
\newpage
%%\frontmatter %%

%\vskip 2.0cm
\begin{center}
{\large \bf ACKNOWLEDGMENTS}
\end{center}
 
As a sense of fulfillment at the completion of this phase of my academic career,
 I wish to express my gratitude to all those who made this thesis possible.

It has been my privilege to work under the able guidance of my thesis 
supervisor, Dr. Amitabha Lahiri. His insights into various problems and insistence
 on clarity have been most useful and inspiring. I express my deep sense of 
 gratitude to him for his nice behavior, all sorts of help and for showing me proper way through stimulating discussions throughout the time period. 

 I am grateful to Prof. Abhijit Mookherjee, ex. Director of Satyendra Nath Bose National Centre for Basic Sciences, for giving me the opportunity to work here.
 I am grateful to Prof. A. K. Raychaudhuri, the Director of Satyendra Nath Bose National Centre for Basic Sciences, for extending my fellowship to complete my thesis.

 I wish to acknowledge with sincere gratitude 
 Prof. Jayanta Kr. Bhattacharjee for all the academic and non-academic help I received from him.
I sincerely acknowledge 
Prof. Rabin Banerjee, Dr. Biswajit Chakraborty, Dr. Samir Kr. Paul, Dr. Partho Guha,
Dr. Debashish Gangopadhyay for academic and non academic help that I got from them during my working period.

 I sincerely acknowledge Prof. Amit Ghosh for his help and very useful suggestions in numerous occasions. I also thank
Prof. Parthasarathi Majumdar and Prof. Ghanashyam Date for many useful discussions.  

I thank all the  administrative staff of SNBNCBS 
for helping me in many ways.  In particular, I am thankful to  the office of Dean (AP), the Library staff and all account Section staff for providing me excellent assistance.    

 It is my pleasure to thank my senior colleagues and friends Saikat da, Chandrasekhar da, Saurabh da, Sunandan da, Anirban da, Prasad da and Sailesh for 
%\newpage
%\begin{figure}
%\centering%keepaspectratio
%\rotatebox{0}{
%\includegraphics[height=10cm,width=8cm]{sou_pre.pdf}}
%%\includegraphics[height=10.0cm,keepaspectratio]{sou_pre.ps}}
%\includegraphics{sig2.ps}}
%\caption{The Kruskal diagram for the de Sitter spacetime (\ref{anex9'}) with each point understood over a 2-sphere.}[height=7.5cm,width=9.0cm]
%\end{figure}
having excellent academic and non-academic interactions throughout. 

I also wish to thank my friends Debmalya, Subhashis, Arghya, Debraj, Sujoy, Rudra, Bibhas, Rudranil, Biswajit, Sudeep and Moutushi for their nice cooperation and help.
 
I have had a very nice time with my friends Mitalidi, Abhinab, Hena and Anirban over all these years.

I sincerely acknowledge Pradeep Chakrabarty and Dr. Shyamal Sanyal for their extreme kindness
and affection towards me.
Finally and most importantly, I express my whole hearted gratitude 
to my parents Mr. Uday Sankar Bhattacharya and Mrs. Chhabi Bhattacharya, my uncles Mr. Ashish Ganguly, Mr. Sujit Sen, Dr. Ajay Sankar Bhattacharya, Mr. Bijoy Sankar Bhattacharya, and my aunts and cousins. It is the love and unflinching support of my full family that enabled me to pursue this career. 

%\begin{center}
%%%%%%%%%%%%%%%%%%%%%%%%%%%%%%%%%%%% 
%%%%%%%%%%%%%% FIG   %%%%%%%%%%%%%%%%%%%%%%%%%%%%%%%%%%%
%\begin{figure}[h]
%\centering%keepaspectratio
%\rotatebox{0}{
%\includegraphics[height=10cm,width=8cm]{sou_pre.pdf}}
%%\includegraphics[height=10.0cm,keepaspectratio]{sou_pre.ps}}
%\includegraphics[height=2cm,width=12.0cm]{bhatta2.eps}}
%\caption{The Kruskal diagram for the de Sitter spacetime (\ref{anex9'}) with each point understood over a 2-sphere.}
%\end{figure}
%%%%%%%%%%%%%%%%%%%%%%%
%\end{center} 
%\vskip 1cm
%\vskip 1.0cm
%\qquad\qquad \qquad\qquad\qquad \qquad\qquad\qquad \qquad\qquad\qquad\qquad 
Sourav Bhattacharya\\
S.N. Bose National Centre for Basic Sciences,\\
Kolkata, India.

\noindent  
%\newpage
\tableofcontents
\mainmatter
%\pagenumbering{arabic}
\chapter{ Motivation and overview}
%\pagenumbering{arabic}
%\Section{Historical background}
%%%%%e%%%%%%%%%%%%%%%%%%%%%%%%%%%%%%%%%%%%%%%%
The cosmological constant $\Lambda$ appears in the Einstein equations as
\begin{eqnarray}
G_{ab}+\Lambda g_{ab}=8\pi G T_{ab}, 
\label{s11i}
\end{eqnarray}
where $g_{ab}$ is the metric and $T_{ab}$ is the energy-momentum tensor corresponding to any matter field present. $G$ is Newton's constant.
$G_{ab}$ is the Einstein tensor computed from the Ricci tensor $R_{ab}$ and the Ricci scalar $R$. The Christoffel symbols $\Gamma^{\mu}_{\nu\lambda}$, components of the Ricci tensor $R_{\mu \nu}$, the Ricci scalar $R$ and components of the Einstein tensor $G_{\mu\nu}$ are given by
\begin{eqnarray}
\Gamma^{\mu}_{\nu\lambda}&=&\frac12g^{\mu\beta}\left[\partial_{\nu}g_{\lambda \beta}+ \partial_{\lambda}g_{\nu \beta}-\partial_{\beta}g_{\nu\lambda}\right],\\
R_{\mu \nu}&=&\partial_{\sigma}\Gamma^{\sigma}_{\mu\nu}-\partial_{\nu}\Gamma^{\sigma}_{\sigma\mu}+\Gamma^{\sigma}_{\sigma\lambda}\Gamma^{\lambda}_{\nu\mu}-\Gamma^{\sigma}_{\nu\lambda}\Gamma^{\lambda}_{\mu\sigma},\\
R&=&g^{\mu\nu}R_{\mu\nu},\\
G_{\mu\nu}&=&R_{\mu\nu}-{\frac{1}{2}} Rg_{\mu\nu},
\label{adch42extra}
\end{eqnarray}
where $g^{\mu\nu}$ are the components of the inverse metric tensor $g^{ab}$ : $g^{\mu\nu}g_{\mu \lambda}=\delta^{\nu}{}_{\lambda}$.

We will set $c=1$ throughout the thesis. Our convention for the sign of the metric will be mostly positive, $(-,~+,~+,~+,\dots)$. If not otherwise mentioned we will always be working in (3+1)-dimensions.
Throughout the thesis we will adopt Einstein's summation convention, i.e. if not otherwise mentioned repeated indices will always be summed over. 
Throughout the thesis we will adopt the abstract index notation as described in \cite{Wald:1984rg}, i.e. we will use the lowercase Latin alphabet to denote tensors and dual tensors
whereas will use the Greek alphabet to denote their components. A `vector' sign  over any quantity (like $\vec{X}$) will always represent a spacelike vector. Any pair of tensorial indices appearing with parenthesis or square bracket will always denote symmetrization or anti-symmetrization respectively.
If not otherwise mentioned, $\Lambda$ will mean a positive $\Lambda$ throughout the thesis. 

The cosmological constant $\Lambda$ was first introduced by Einstein himself to achieve a stationary cosmological model of our universe. The Friedmann-Robertson-Walker (FRW) spacetime can be regarded as the first proposed model attempting to provide a dynamics of our universe. 
Below we very briefly review the FRW cosmology and the inclusion of $\Lambda$ referring the reader to \cite{Wald:1984rg, weinbergbook, Weinberg:2008zzc} for details.

In the FRW model it is assumed on the basis of the observed cosmological data such as the distribution of the distant galactic masses and the $X$-ray or the $\gamma$-ray spectra emitted from them that our universe is spatially isotropic in a large scale. It can be shown that spatial isotropy implies spatial homogeneity too. Then the spatial homogeneity and isotropy together imply that the spacetime can be foliated by a family of spacelike hypersurfaces $\Sigma$ of constant curvature. This means that the components of the Riemann tensor $\widetilde{R}_{abcd}$ over $\Sigma$ is a multiple of the identity operator, $\widetilde{R}_{\mu \nu}{}^{\alpha \beta}=k\delta_{[\mu}{}^{[\alpha}\delta_{\nu]}{}^{\beta]}$, where $k$ is a constant \cite{Wald:1984rg}.

Then it can be shown that there exist only three independent metrics over $\Sigma$ 
\begin{eqnarray}
d\Sigma^2=\frac{dR^2}{1-kR^2}+R^2\left(d\theta^2+\sin^2\theta d\phi^2\right), 
\label{s12i}
\end{eqnarray}
with $k=0,~\pm 1$ and $R$, $\theta$, $\phi$ are the usual spherical polar coordinates. 
$k=0$ represents flat spatial section, i.e. spatial section with zero curvature. $k=\pm1$ represent respectively constant positive-curvature (3-sphere) and constant negative-curvature (3-hyperboloid) spacelike surfaces. The cosmological redshift and luminosity data indicate that our universe is spatially flat $k=0$, although the reason behind this is not yet very well understood.   

A convenient ansatz for the full spacetime metric can be formed in the following way. Let us consider a family of timelike observers orthogonal to $\Sigma$ with tangent vector $\{u^a\}$ and proper time $\tau$. We choose this class of observers in such a way that $u_a\sim \nabla_a \tau$, so that each $\tau={\rm constant} $ hypersurface coincides with one and only one $\Sigma$. Then flowing the $\Sigma$'s along $u^a$ we `cover' the entire spacetime.
 A physically reasonable question then would be, how does such an observer see the entire spacetime evolve? To answer this, we take the following most general ansatz for the full spacetime metric preserving the spatial homogeneity and isotropy,   
\begin{eqnarray}
g_{ab}=u_au_b+a^2(\tau)\Sigma_{ab},\quad{\rm {or~ equivalently,}}\quad %\nonumber \\
ds^2=-d\tau^2+a^2(\tau)d\Sigma^2,
\label{s13i}
\end{eqnarray}
where $a(\tau)$ is smooth function known as the scale factor and $d\Sigma^2$ is given by 
Eq.~(\ref{s12i}). We have taken $u_au^b=-1$, because had we chosen instead the norm to be some $-f^2(\tau)$, we could have easily redefined a new `time' by $\tau^{\prime}=\int f(\tau)d\tau$ to get the form of Eq.~(\ref{s13i}).  

So all that we have to do now is to solve the Einstein equations with the ansatz (\ref{s13i}) and with some reasonable energy-momentum tensor $T_{ab}$. At the cosmological length scale we are interested in, $T_{ab}$ comes from stellar objects such as stars and galaxies. Since the cosmological length scale is very large compared to the dimensions of those stellar objects, we may treat them as grains of dust or perfect fluid. By the assumption of isotropy, the flow line of those stellar objects must coincide with the world lines $\{u^a\}$ of the observers. So $T_{ab}$ takes the form
\begin{eqnarray}
T_{ab}=\rho u_a u_b+P\left(g_{ab}+u_au_b\right),
\label{s14i}
\end{eqnarray}
where $\rho(x)$ and $P(x)$ are smooth functions regarded respectively as the energy density and pressure of the fluid. Using Eq.s~(1.2)-(\ref{s14i}) one then obtains the following two independent Einstein's equations $G_{\mu}{}^{\nu}=8\pi G T_{\mu}{}^{\nu}$ with $\Lambda=0$,
\begin{eqnarray}
\frac{3}{a^2}\left(\frac{d a}{d\tau}\right)^2=8\pi G\rho-\frac{3k}{a^2},\quad\frac{3}{a}\frac{d^2 a}{d\tau^2}=-4\pi G\left(\rho+3P\right).
\label{s15i}
\end{eqnarray}
Eq.s~(\ref{s15i}) give the general evolution for a spatially homogeneous and isotropic universe. The most astonishing thing of these equations are that, given $\rho>0$ and $P\geq 0$, the universe cannot be static, i.e. independent of $\tau$. To see this we first note that $a(\tau)$ cannot be negative because that will give negative proper distance $a(\tau)\sqrt{d\Sigma^2}$ between two spacelike separated points. Then the second of Eq.s~(\ref{s15i}) shows that we have always ${\displaystyle\frac{d^2a}{d\tau^2}<0}$. Also, the first of the above equations shows that the universe is either expanding : ${\displaystyle\frac{d a}{d\tau}\geq0}$, or contracting : ${\displaystyle\frac{d a}{d\tau}\leq0}$, where the equality holds only when expansion goes over to contraction and vice versa. This leads to many interesting features \cite{Wald:1984rg} of the FRW universes but we will not go into them here.   

Einstein himself was not happy with the FRW solutions which predict dynamic universes. His objective was to construct a static or at least a quasistatic universe to comply with the extremely slow motion of the stars surrounding us. In order to achieve this he introduced a positive fundamental constant $\Lambda$, called the cosmological constant, into the Einstein equations to get Eq.~(\ref{s11i}). With the inclusion of $\Lambda$ Eq.s~(\ref{s15i}) modify to 
\begin{eqnarray}
\frac{3}{a^2}\left(\frac{d a}{d\tau}\right)^2=8\pi G\left(\rho+\frac{\Lambda}{8\pi G}\right)-\frac{3k}{a^2},\quad\frac{3}{a}\frac{d^2 a}{d\tau^2}=-4\pi G\left[\rho+3\left(P-\frac{\Lambda}{8\pi G}\right)\right].
\label{s15i'}
\end{eqnarray}
The second of Eq.s~(\ref{s15i'}) shows that positive $\Lambda$ has a negative `pressure' and thus it may `balance' the positive pressure of other matter fields. In particular, Einstein was successful to obtain a static solution for $k=+1$, namely Einstein's static universe,   
\begin{eqnarray}
ds^2=-d\tau^2+d\Psi^2+\sin^2\Psi\left(d\theta^2+\sin^2\theta d\phi^2\right),
\label{s16i}
\end{eqnarray}
where we have used the usual 3-sphere coordinates. For many interesting geometrical properties of (\ref{s16i}) we refer our reader to \cite{Wald:1984rg, Hawking:1973uf}. 

After this the redshift observations of Hubble came in 1929~\cite{Hubble}. This proved that our universe is indeed expanding, which was predicted earlier by the $\Lambda=0$ FRW cosmology. Thus Einstein's motivation for introducing $\Lambda$ was ruled out. After this $\Lambda$ was included in general relativity in numerous occasions but any sufficient physical motivation was absent.

The story began to change from the end of the last century. The spectral and photometric observations of 10 type Ia supernovae (SNe Ia) revealed the striking possibility that our universe is not only expanding but doing so with an acceleration with $k=0$ \cite{Riess:1998cb, Perlmutter:1998np}. This means that both ${\displaystyle\frac{da}{d\tau}}$  and ${\displaystyle\frac{d^2 a}{d \tau^2}}$ must be positive. Since we have assumed $\rho>0$ and $P\geq0$ we see from Eq.s~(\ref{s15i'}) that the accelerated expansion is possible only for a positive $\Lambda$ due to its negative pressure. In those observations various cosmological data such as redshift factor, luminosity and the Hubble constant were measured. Then these observed data were matched with theoretical calculations made from an FRW universe. It was found that the observed data matches exceedingly well with a $k=0$ FRW universe undergoing accelerated expansion. This shows that there is a strong possibility that our universe is indeed endowed with a positive cosmological constant! So now we have a strong physical motivation to study $\Lambda>0$ gravity. There are a few models other than a positive $\Lambda$ using exotic matter fields which exert negative `pressure' and hence may also give rise to the accelerated expansion.  
All such matter fields are known as the dark energy. However in this thesis
we will not concern ourselves with forms of dark energy referring our reader to \cite{Weinberg:2008zzc, pilar, copeland} and references therein for exhaustive theoretical and phenomenological discussions on this.

The above was a very brief overview of the physical motivation to study gravity with a positive $\Lambda$. Since the observed value of $\Lambda$ is very small $\sim 10^{-52}\rm{m}^{-2}$, \footnote{This is in fact a few times larger than the observed density of matter other than $\Lambda$, i.e. those with $P\geq0$, which means that the present dynamics of our universe is dominated by $\Lambda $ in large scale \cite{Weinberg:2008zzc}.} it would be reasonable to ask why we should not neglect the effect of $\Lambda$ in local physics. Or more precisely, how strong are the perturbative effects due to $\Lambda$? Are there any non-perturbative effects too? We will review these topics in the remaining part of this Chapter and attempt to answer a few of them in this thesis. 

This Chapter is organized as follows. In the next Section we discuss various exact solutions with positive $\Lambda$ and their global properties. In Section 1.2 we discuss the no hair theorems and uniqueness problems. In Section 1.3 we discuss some perturbative calculations and geodesics in $\Lambda>0$ spacetimes. In Section 1.4 we discuss de Sitter black hole thermodynamics and Hawking radiation and the organization of the thesis. Each of this Sections is an introduction to the problems we address in the remaining part of this thesis.     

%%%%%%%%%%%%%%%%%%%%%%%%%%%%%%%%%%%%%%%%%%%%%%%%%%%%%%%%%%%%%%%%%%%%%
\section{Exact solutions with $\Lambda>0$ and causal properties of a cosmological event horizon}
%%%%%%%%%%%%%%%%%%%%%%%%%%%%%%%%%%%%%%%%%%%%%%%%%%%
\subsection{Exact solutions}
%%%%%%%%%%%%%%%%%%%%%%%%%%%%%%%%%%%%%%%%%%%%%%%%
In this Section we will discuss a few exact solutions with positive $\Lambda$ and introduce the cosmological event horizon. 

Let us start with the simplest $\Lambda$-vacuum, viz., the de Sitter spacetime. If one solves the Einstein equations (\ref{s11i}) for $\Lambda>0$ and $T_{ab}=0$ with the spatially homogeneous and isotropic FRW ansatz (\ref{s13i}) with flat spatial sections, $k=0$ in Eq.s~(\ref{s12i}), one obtains (see e.g. \cite{Weinberg:2008zzc}) in Cartesian coordinates
\begin{eqnarray}
ds^2=-d\tau^2+e^{2\sqrt{ \frac{\Lambda}{3}}\tau}\left(dx^2+dy^2+dz^2\right),
\label{dsin0}
\end{eqnarray}
or in the usual spherical polar coordinates 
\begin{eqnarray}
ds^2=-d\tau^2+e^{ 2\sqrt{ \frac{\Lambda}{3}} \tau}
\left(dR^2+R^2d\theta^2+R^2\sin^2\theta d\phi^2\right).
\label{dsin1}
\end{eqnarray}
 The spacetime (\ref{dsin0}) or (\ref{dsin1}) is known as the de Sitter spacetime. This can also be constructed by embedding a four-dimensional `surface' in five-dimensional Minkowski spacetime \cite{Hawking:1973uf}.

The de Sitter spacetime possesses a Killing vector field 
\begin{eqnarray}
\xi^a=(\partial_{\tau})^a\pm\sqrt{ \frac{\Lambda}{3}} R(\partial_{R})^a,\quad \xi^a\xi_a=-\left(1-\frac{\Lambda R^2 e^{ 2\sqrt{ \frac{\Lambda}{3}} \tau}}{3}\right). 
\label{dsin2'}
\end{eqnarray}
So $\xi^a$ is timelike as long as 
${\displaystyle R e^{\sqrt{ \frac{\Lambda}{3}} \tau}< \sqrt{\frac{3}{\Lambda}}}$.
Then by making the coordinate transformations \cite{Kastor:1992nn}
\begin{eqnarray}
 e^{ \sqrt{ \frac{\Lambda}{3}} \tau} R=r,~\quad \tau=t+\frac{1}{2}\sqrt{\frac{3}{\Lambda}}\ln\left \vert 1-\frac{\Lambda r^2}{3}\right\vert,
\label{dsin2}
\end{eqnarray}
 the metric (\ref{dsin1}) can be brought to a manifestly static form
\begin{eqnarray}
ds^2=-\left(1-\frac{\Lambda r^2}{3} \right)dt^2+ \left(1-\frac{\Lambda r^2}{3}\right)^{-1}dr^2+r^2\left(d\theta^2+\sin^2\theta d\phi^2\right).
\label{dsin3}
\end{eqnarray}
The timelike Killing field $\xi^a=(\partial_{t})^a$ becomes null at ${\displaystyle r_{\rm C}=\sqrt{\frac{3}{\Lambda}}}$. Outside $r_{\rm C}$ the timelike Killing field becomes spacelike and the metric functions (\ref{dsin3}) flip sign. Thus the chart in (\ref{dsin2}) covers only the region
${\displaystyle 0\leq r< \sqrt{\frac{3}{\Lambda}}}$ of the spacetime. The null surface at $r=r_{\rm C}$ is called the cosmological event horizon. It is a Killing horizon and hence is not an artifact of the coordinates. We note here that $r_{\rm C}$ is not a particle horizon. We recall that, if the Big Bang started at $\tau=0$, the particle horizon $R_{\rm {max}}(\tau)$ in the FRW spacetime defines the maximum radial distance from which an observer can receive light signals 
\cite{Weinberg:2008zzc}
\begin{eqnarray}
\int_{0}^{R_{\rm {max}}}\frac{dR}{1-kR^2}=\int_{0}^{\tau}\frac{d\tau^{\prime}}{a(\tau^{\prime})},
\label{parthor}
\end{eqnarray}
For $k=0$ we have the maximum proper distance $\displaystyle{d_{\rm {max}}(\tau)}$,
\begin{eqnarray}
 d_{\rm {max}}(\tau)= a(\tau)R_{\rm {max}}=a(\tau)\int_{0}^{\tau}\frac{d\tau^{\prime}}{a(\tau^{\prime})},
\label{parthor'}
\end{eqnarray}
where $a(\tau^{\prime})$ in the integrand corresponds to different cosmological era that the universe has passed through since the Big Bang and $a(\tau)$ outside the integral represent the present era where the observer is. Thus ${\displaystyle d_{\rm {max}}(\tau)}$ depends on $\tau$ and so does not equal ${\displaystyle \sqrt{\frac{3}{\Lambda}}}$ if we take the present metric to be de Sitter. In fact a particle horizon can be defined in any spacetime irrespective of whether it possesses a timelike Killing field or not and has nothing to do with any isometry of the spacetime.   
From now on in this thesis `the cosmological horizon' or `the cosmological event horizon' will always stand for the cosmological Killing horizon.

An interesting feature of the de Sitter spacetime is that the length scale ${\displaystyle \sqrt{\frac{3}{\Lambda}}}$ of the cosmological event horizon is observer independent as long as the observers are connected by spatial isometries and time translation. To see this let us explicitly consider the transformations on (\ref{dsin0})
\begin{eqnarray}
\tau^{\prime}=\tau+\tau_0,\quad\vec{x^i}^{\prime}=D^i{}_j\vec{x^j}+\vec{\epsilon^i},
\label{dsin4}
\end{eqnarray}
where $\tau_0$ and $\vec{\epsilon^i}$ are constants and $D^i{}_j$ is the usual SO(3) rotation matrix with constant components. 
Since any finite continuous transformation can be achieved by
successive infinitesimal transformations generated from the identity, we assume (\ref{dsin4}) to be infinitesimal, i.e. $D_i{}^j=\delta_{i}{}^{j}+\omega_{i}{}^{j}$, with $\omega$ infinitesimal. Then the invariance of the norm of a vector under rotation shows
that $\omega_{ij}$ is antisymmetric in its indices. 

Using the antisymmetry of $\omega$, we find from Eq.~(\ref{dsin4})
\begin{eqnarray}
\delta_{ij}\vec{x^i}^{\prime}\vec{x^j}^{\prime}&=&\delta_{ij}\left(\vec{x^i}+ \vec{\epsilon^i} \right)\left( \vec{x^j}+ \vec{\epsilon^j} \right)+{\cal{O}}(\epsilon\cdot\omega,~\omega^2),\\
\label{dsin4''}
\delta_{ij}d\vec{x^i}^{\prime}d\vec{x^j}^{\prime}&=&\delta_{ij}d\vec{x^i}d\vec{x^j}.
\label{dsin4'}
\end{eqnarray}
Eq.~(\ref{dsin4'}) shows that the de Sitter metric (\ref{dsin0}, \ref{dsin1}) remains formally invariant under the transformations (\ref{dsin4}):
\begin{eqnarray}
ds^2&=&-d\tau^2+e^{2\sqrt{ \frac{\Lambda}{3}}\tau}\left(dx^2+dy^2+dz^2\right),\nonumber\\
&=&-d\tau^{\prime 2}+e^{ 2\sqrt{ \frac{\Lambda}{3}}(\tau^{\prime}-\tau_0)}\left(dx^{\prime2}+dy^{\prime2}+dz^{\prime2}\right)\nonumber\\
&=&-d\tau^{\prime 2}+e^{ 2\sqrt{ \frac{\Lambda}{3}}\tau^{\prime}}\left(d\tilde{x}^{\prime2}+d\tilde{y}^{\prime2}+d\tilde{z}^{\prime2}\right)\nonumber\\
&=&-d\tau^{\prime 2}+e^{ 2\sqrt{ \frac{\Lambda}{3}}\tau^{\prime}}
\left(d\tilde{R}^{\prime2}+\tilde{R}^{\prime2} d\theta^{\prime2}+\tilde{R}^{\prime2}\sin^2\theta^{\prime} d\phi^{\prime2}\right),
\label{dsin5}
\end{eqnarray}
where in the second line we have used Eq.s~(\ref{dsin4}) and (\ref{dsin4'}), in the third line
 we have defined the scale transformations ${\displaystyle \vec{x}^{\prime i}\to
e^{-\sqrt{ \frac{\Lambda}{3}} \tau_0}\vec{x}^{i}}$, and in the
last line we have defined the new radial variable ${\displaystyle\tilde{R}^{\prime 2}= e^{-2\sqrt{ \frac{\Lambda}{3}} \tau_0}\delta_{ij}\vec{x^i}^{\prime}\vec{x^j}^{\prime}}$ and accordingly the new polar and azimuthal angles $\theta^{\prime}$ and $\phi^{\prime}$. Eq.s~(\ref{dsin5}) show that the Killing field $\xi^a$ in Eq.~(\ref{dsin2'}) also remains formally invariant under (\ref{dsin4}) --- we have only to replace $\tau$ and $R$ with $\tau^{\prime}$ and $\tilde{R}^{\prime}$ respectively. With the same replacement we may define the transformations (\ref{dsin2}) and arrive at Eq.~(\ref{dsin3}) but now $(t,~r,~\theta,~\phi)$ properly replaced with some $(t^{\prime},~\tilde{r}^{\prime},~\theta^{\prime},~\phi^{\prime})$. Thus under the transformations (\ref{dsin4}) the static chart (\ref{dsin3}) still shows a cosmological horizon at ${\displaystyle\tilde{r}_{\rm C}^{\prime}=\sqrt{\frac{3}{\Lambda}}}$. This shows that for observers connected by (\ref{dsin4}) in the de Sitter spacetime the cosmological horizon remains unchanged in the length scale. In other words each such observer will `see' the cosmological horizon at a spatial distance ${\displaystyle\sqrt{\frac{3}{\Lambda}} }$ from himself or herself.

For many other interesting geometrical properties of the de Sitter spacetime we refer our reader to \cite{Hawking:1973uf}.
 
How will the de Sitter spacetime change if a self-gravitating mass sits within it? Or what will be the black hole solution within the de Sitter universe? We will mention a few such solutions without giving any derivation, referring the reader to e.g. \cite{Carter:1968ks}. 

The simplest case will be to assume spherical symmetry and vacuum. A suitable ansatz for the metric is
\begin{eqnarray}
ds^2=-\lambda^2(r,~t)dt^2+f^2(r,~t)dr^2+r^2\left(d\theta^2+\sin^2\theta d\phi^2\right).
\label{s21i}
\end{eqnarray}
With this we solve Eq.~(\ref{s11i}) with $T_{ab}=0$. One finds that $R_{tr}=0$ implies $\partial_t\lambda^2=0=\partial_t f^2$ and obtains the static solution
\begin{eqnarray}
ds^2=-\left(1-\frac{2MG}{r}-\frac{\Lambda r^2}{3} \right)dt^2+ \left(1-\frac{2MG}{r}-\frac{\Lambda r^2}{3}\right)^{-1}dr^2+r^2\left(d\theta^2+\sin^2\theta d\phi^2\right), \nonumber \\
\label{s22i}
\end{eqnarray}
known as the Schwarzschild-de Sitter solution. $M$ is a constant which can be interpreted for $\Lambda=0$ as the ADM mass of the spacetime. Setting $M=0$ in Eq.~(\ref{s22i}) recovers the de Sitter universe (\ref{dsin3}).   

An electrically charged generalization of (\ref{s22i}) can easily be 
achieved by taking the Maxwell field as the source
\begin{eqnarray}
\mathcal{L}=-\frac{1}{4}F_{ab}F^{ab},
\label{s23i}
\end{eqnarray}
where $F_{ab}=\nabla_{[a} A_{b]}$, and $A_b$ is the gauge field. Due to the spherical symmetry we may take 
${\displaystyle A_a=\frac{Q}{r}(dt)_a}$, where the constant $Q$ is the electric charge. The energy-momentum tensor for the Maxwell field (\ref{s23i}) is
\begin{eqnarray}
T_{ab}=F_{ac}F_{b}{}^{c}+\mathcal{L}g_{ab}.
\label{s24}
\end{eqnarray}
With all these we may solve Eq.s~(\ref{s11i}) for (\ref{s21i}) to obtain
\begin{eqnarray}
ds^2=-\left(1-\frac{2MG}{r}-\frac{\Lambda r^2}{3}+\frac{Q^2}{r^2} \right)dt^2&+& \left(1-\frac{2MG}{r}-\frac{\Lambda r^2}{3}+\frac{Q^2}{r^2}\right)^{-1}dr^2\nonumber\\&+&r^2\left(d\theta^2+\sin^2\theta d\phi^2\right),
\label{s25i}
\end{eqnarray}
known as the Reissner-N\"{o}rdstrom-de Sitter solution.

The Reissner-N\"{o}rdstrom-de Sitter solution can be further generalized to the rotating spacetime known as the Kerr-Newman-de Sitter solution, 
\begin{eqnarray}
ds^2=\rho^2\left(\Delta_r^{-1}dr^2+\Delta_{\theta}^{-1}d\theta^2\right)+\frac{\Delta_{\theta}}{\rho^{2}\Sigma^{2}}\left[adt-\left(r^2+a^2\right)d\phi\right]^2-\frac{\Delta_{r}}{\rho^{2}\Sigma^{2}}\left[dt-a\sin^2\theta d\phi\right]^2,\nonumber\\
\label{s26i}
\end{eqnarray}
where 
\begin{eqnarray}
\rho^2&=&r^2+a^2\cos^2\theta,\quad \Delta_r=\left(r^2+a^2\right)\left(1-\frac{\Lambda r^2}{3}\right)-2MGr+Q^2,\nonumber\\
 \Delta_{\theta}&=&\left(1+\frac{\Lambda a^2}{3}\cos^2\theta\right),\quad {\rm and}~ \Sigma=\left(1+\frac{\Lambda a^2}{3}\right).
\label{knds1}
\end{eqnarray}
 The gauge field of this solution is given by
\begin{eqnarray}
A_a=\frac{Qr}{\rho^2 \Sigma}\left[(dt)_a-a\sin^2\theta (d\phi)_a\right]. 
\label{knds1}
\end{eqnarray}
The parameter $a$ is related to the rotation of the spacetime. For $Q=0$, Eq.~(\ref{s26i}) is known as the Kerr-de Sitter solution. There exist a few other exact solutions with positive $\Lambda$, one of which will be shown in Chapter 4 to describe a de Sitter cosmic string spacetime. 
%But now let us come to the most non-trivial
%and the most interesting feature of the $\Lambda>0$ spacetimes, namely the cosmological horizon.

Let us now consider the Schwarzschild-de Sitter spacetime (\ref{s22i}). The metric (\ref{s22i}) has singularities at $r=0$ and at points corresponding to ${\displaystyle\left(1-\frac{2MG}{r}-\frac{\Lambda r^2}{3}\right)=0}$, i.e. points where the timelike Killing field $(\partial_t)^a$ becomes null, defining the Killing horizons of the spacetime. It is easy to compute from the metric functions (\ref{s22i}) the invariant 
\begin{eqnarray}
 R_{abcd}R^{abcd}=\frac{48G^2M^2}{r^6}+\frac{\Lambda^2}{4}, 
\label{adintro1}
\end{eqnarray}
which shows that like the Schwarzschild spacetime, $r=0$ is a genuine or curvature singularity for (\ref{s22i}) and hence cannot be removed by any coordinate transformation.  
The points at which the timelike Killing vector field becomes null define the horizons of the spacetime. In order to find these points we solve ${\displaystyle\left(1-\frac{2MG}{r}-\frac{\Lambda r^2}{3}\right)=0}$, or equivalently the cubic equation 
\begin{eqnarray}
 r^3-\frac{3r}{\Lambda}+\frac{6MG}{\Lambda}=0.
\label{adintro2}
\end{eqnarray}
This can be solved by the usual Cardan-Tantaglia method. Let $r=m+n$, so that
\begin{eqnarray}
 r^3&=&m^3+n^3+3mn\left(m+n\right)=m^3+n^3+3mnr \nonumber\\ &\Rightarrow& r^3-3mnr-\left(m^3+n^3\right)=0. 
\label{adiin3}
\end{eqnarray}
Comparing Eq.s~(\ref{adintro2}) and (\ref{adiin3}) we have
\begin{eqnarray}
 m^3n^3=\frac{1}{\Lambda^3},\quad\left(m^3+n^3\right)=-\frac{6MG}{\Lambda},  
\label{adiin4}
\end{eqnarray}
which shows that $m^3$ and $n^3$ are the roots of the quadratic equation
\begin{eqnarray}
 x^2-\left(m^3+n^3\right)x+m^3n^3=x^2+\frac{6MG}{\Lambda}x+\frac{1}{\Lambda^3}=0.  
\label{adiin5}
\end{eqnarray}
We solve this to find
\begin{eqnarray}
 m=(-1)^{\frac13}\left[ \frac{3MG}{\Lambda}-\frac{1}{\Lambda^{\frac32}}\sqrt{\left(9M^2G^2\Lambda-1\right)}\right]^{\frac13},~n=(-1)^{\frac13}\left[ \frac{3MG}{\Lambda}+\frac{1}{\Lambda^{\frac32}}\sqrt{\left(9M^2G^2\Lambda-1\right)}\right]^{\frac13}.  \nonumber\\
\label{adiin6}
\end{eqnarray}
 Noting that 
\begin{eqnarray}
(-1)^{\frac13}\equiv\left\{-1,~\frac{1+\sqrt{3}i}{2},~ \frac{1-\sqrt{3}i}{2}\right\},
\label{adiin7}
\end{eqnarray}
the three roots $r=m+n$ of Eq.~(\ref{adintro2}) subject to Eq.s~(\ref{adiin4}) are the following
\begin{eqnarray}
 r_1&=&-\left[ \left(\frac{3MG}{\Lambda}-\frac{1}{\Lambda^{\frac32}}\sqrt{\left(9M^2G^2\Lambda-1\right)}\right)^{\frac13}+\left(\frac{3MG}{\Lambda}+\frac{1}{\Lambda^{\frac32}}\sqrt{\left(9M^2G^2\Lambda-1\right)}\right)^{\frac13}\right],\nonumber\\
r_2&=& \left(\frac{3MG}{\Lambda}-\frac{1}{\Lambda^{\frac32}}\sqrt{\left(9M^2G^2\Lambda-1\right)}\right)^{\frac13}\frac{1+\sqrt{3}i}{2}+\left(\frac{3MG}{\Lambda}+\frac{1}{\Lambda^{\frac32}}\sqrt{\left(9M^2G^2\Lambda-1\right)}\right)^{\frac13}\frac{1-\sqrt{3}i}{2},\nonumber\\
r_3&=&\left(\frac{3MG}{\Lambda}-\frac{1}{\Lambda^{\frac32}}\sqrt{\left(9M^2G^2\Lambda-1\right)}\right)^{\frac13}\frac{1-\sqrt{3}i}{2}+\left(\frac{3MG}{\Lambda}+\frac{1}{\Lambda^{\frac32}}\sqrt{\left(9M^2G^2\Lambda-1\right)}\right)^{\frac13}\frac{1+\sqrt{3}i}{2}.\nonumber\\
\label{adiin8}
\end{eqnarray}
There are three solutions depending upon the sign of the discriminant $\Delta=\left(9M^2G^2\Lambda\right.\\ \left.-1\right)$. For $\Delta>0$, $r_1$ in Eq.s~(\ref{adiin8}) is negative and the other two are complex conjugates of each other. So there is no actual horizon for this case and thus the curvature singularity at $r=0$ is naked. Also this situation seems unlikely for the observed tiny value of $\Lambda$. Thus we may ignore positive $\Delta$.  

For $\Delta=0$, we have
\begin{eqnarray}
 r_1=-\frac{2}{\sqrt{\Lambda}},~ r_2=\frac{1}{\sqrt{\Lambda}}=r_3,
\label{adiin9}
\end{eqnarray}
known as the Nariai class solution.
The most likely situation subject to the tiny value of $\Lambda$ is $\Delta<0$. Then the quantities within parenthesis in Eq.~(\ref{adiin8}) become complex. Writing $\sqrt{ 9M^2G^2\Lambda-1}=i\sqrt{1-9M^2G^2\Lambda}$, we find the following three real roots      
\begin{eqnarray}
r_3&=&r_{\rm{H}}=\frac{2}{\sqrt{\Lambda}}\cos\left[\frac{1}{3}
\cos^{-1}\left(3MG\sqrt{\Lambda}\right)+\frac{\pi}{3}\right],\nonumber\\~
r_2&=&r_{\rm{C}}=\frac{2}{\sqrt{\Lambda}}\cos\left[\frac{1}{3}
\cos^{-1}\left(3MG\sqrt{\Lambda}\right)-\frac{\pi}{3}\right], \nonumber \\ r_1&=&r_{\rm{U}}=-\left(r_{\rm{H}}+r_{\rm{C}}\right).
\label{s27i}
\end{eqnarray}
$r_{\rm{H}}$ and $ r_{\rm{C}}$ are positive, thereby defining two true horizons of the spacetime. The larger root $r_{\rm{C}}$ is known as the cosmological event horizon and the smaller root $r_{\rm{H}}$ is known as the black hole event horizon. The negative root $r_{\rm{U}}$ is unphysical. Thus for $ 3MG\sqrt{\Lambda}<1$, Eq.~(\ref{s22i}) represents a Schwarzschild black hole sitting inside the cosmological horizon. For $3MG\sqrt{\Lambda}=1$ the roots $r_{\rm{H}}$ and $r_{\rm{C}}$ merge and we recover the degenerate case of Eq.~(\ref{adiin9}).  

Since the observed value of $\Lambda$ is very small, let us now find the expressions for $r_{\rm H}$ and $r_{\rm C}$ in the limit $3MG\sqrt{\Lambda}\ll1$. We first note that if $\cos \phi=x$, we have $\left(\frac{\pi}{2}-\phi\right)=\sin^{-1}x\approx x$ for $x\ll1$. Then for $3MG\sqrt{\Lambda}\ll1 $, the quantity $\cos^{-1}\left(3MG\sqrt{\Lambda}\right)$ in Eq.s~(\ref{s27i}) can be approximated with ${\displaystyle\left(\frac{\pi}{2}-3MG\sqrt{\Lambda}\right)}$, giving
\begin{eqnarray}
r_{\rm{H}}\approx \frac{2}{\sqrt{\Lambda}}\cos\left(\frac{\pi}{2}-MG\sqrt{\Lambda} \right) =\frac{2}{\sqrt{\Lambda}}\sin\left(MG\sqrt{\Lambda}\right)
=2MG\left[1+{\cal{O}}\left(MG\sqrt{\Lambda}\right)^2\right],\nonumber\\
\label{s27i'}
\end{eqnarray}
i.e. the Schwarzschild radius in the leading order, and
\begin{eqnarray}
r_{\rm C}\approx \frac{2}{\sqrt{\Lambda}}\cos\left(\frac{\pi}{6}+MG\sqrt{\Lambda}\right)
&=&\frac{2}{\sqrt{\Lambda}}\left[\frac{\sqrt{3}}{2}\cos\left(MG\sqrt{\Lambda}\right) -\frac12\sin \left(MG\sqrt{\Lambda}\right)\right]\nonumber\\&=& \sqrt{\frac{3}{\Lambda}}\left[1-
{\cal{O}}\left(MG\sqrt{\Lambda}\right)\right], 
\label{s27i''}
\end{eqnarray}
i.e. the de Sitter horizon radius in the leading order. The observations show $\Lambda\sim 10^{-52}{\rm m}^{-2}$, so that $r_{\rm C}\sim10^{26}{\rm m}$.

Unlike the de Sitter spacetime, the Schwarzschild-de Sitter spacetime is neither spatially homogeneous nor isotropic, due the presence of the mass term $M$. This implies that unlike the de Sitter spacetime the length scale of the cosmological horizon will not be invariant for the Schwarzschild-de Sitter spacetime for observers connected by spacetime translations and spatial rotations. However for a black hole with $M$ of the order of a few solar mass $M_{\odot}$, we have $3MG\sqrt{\Lambda}\sim 10^{-22}$. Then for such a black hole the horizon lengths are given by Eq.s~(\ref{s27i'}) and (\ref{s27i''}), i.e.
 $r_{\rm H}\sim 10^4$m. Also for length scales $r\gg 2GM$, 
 for example $r\sim 10^{20}$m, the metric (\ref{s22i}) becomes de Sitter up to a very good approximation and so in this region the spatial homogeneity and isotropy are restored approximately. Thus in the region far away from a tiny black hole, the transformations of Eq.s~(\ref{dsin4}) can be regarded as isometries up to a very good approximation, and all observers connected by them will find the cosmological horizon at ${\displaystyle\sqrt{\frac{3}{\Lambda}}}$ from himself or herself. On the other hand, for a galactic centre black hole with $M\sim 10^{9}M_{\odot}$, we have $3MG\sqrt{\Lambda}\sim 10^{-14}$, so that $r_{\rm H}$ and $r_{\rm C}$ are still well approximated by Eq.s~(\ref{s27i'}), (\ref{s27i''}) and $r_{\rm H} \sim 10^{12}$m. Then for $r\gg 2GM$,
for example $r\sim 10^{23}$m, we recover spatial homogeneity and isotropy approximately and all observers connected by isometry transformations in this region will still find the cosmological horizon has length scale ${\displaystyle\sqrt{\frac{3}{\Lambda}}}$. In \cite{Wald:1984rgnew1} it was shown that the spatially anisotropic expanding cosmological models evolve to the de Sitter or the de Sitter cosmological multi-black hole spacetime of \cite{Kastor:1992nn}, asymptotically in time. Then from the two extreme and realistic examples considered above we may conclude that at sufficiently large distance from a gravitating object of compact mass distribution, an observer at asymptotic late time can find himself or herself in a universe surrounded by a cosmological horizon of size ${\displaystyle\sqrt{\frac{3}{\Lambda}}}$.   

Under some reasonable conditions on the parameters the Reissner-N\"{o}rdstrom-de Sitter and the Kerr-Newman-de Sitter solutions, Eq.s~(\ref{s25i}), (\ref{s26i}), also exhibit respectively a charged non-rotating and a charged rotating black hole sitting inside the de Sitter universe. In Chapter 4 we will construct a cylindrically symmetric de Sitter spacetime and see that this also exhibits a cosmological horizon. Like the $\Lambda=0$ spacetimes, the solutions (\ref{s25i}), (\ref{s26i}) also exhibit Cauchy horizons, i.e. Killing horizons located inside the black hole \cite{Gibbons:1977mu}. Like the Kerr or the Kerr-Newman spacetime the Kerr-de Sitter or the Kerr-Newman-de Sitter solutions also exhibit ergospheres, i.e. a `closed' surface over which the timelike Killing field which is not orthogonal to any spacelike hypersurface becomes null, within which it is spacelike and which intersects the black hole horizon at two diametrically opposite points $\theta=0,~\pi$.

%%%%%%%%%%%%%%%%%%%%%%%%%%%%%% Maximal extensions%%%%%%%%%%%%%%%%%%%%%%%
\subsection{Maximal analytic extension at the cosmological event horizon}
%%%%%%%%%%%%%%%%%%%%%%%%%%%%%%%%%%%%%%%%%%%%%%%%%%%%%%%%%%%%%%%%%
We have seen that for known solutions of the Einstein equations the addition of a positive $\Lambda$ gives an outer boundary or outer null surface, namely the cosmological event horizon. What are the causal properties of such a horizon? To answer this we recall that in order to understand the causal properties of a black hole event horizon one constructs a maximally extended coordinate system, namely the Kruskal coordinates, to remove the coordinate singularities at the black hole horizon, see e.g. \cite{Wald:1984rg, Birrell, Gourgoulhon:2005ng}. We will do the same for the cosmological horizon. Let us choose for simplicity the de Sitter spacetime and consider the static chart (\ref{dsin3}) which manifestly exhibits the cosmological horizon. Along a radial ($\theta,~\phi={\rm {constant}}$) and null ($ds^2=0$) geodesic in (\ref{dsin3}) we have ${\displaystyle\frac{dt}{dr}=\pm\left(1-\frac{\Lambda r^2}{3}\right)^{-1}\to\pm \infty}$ for ${\displaystyle r\to r_{\rm C}=\sqrt{\frac{3}{\Lambda}}}$. Thus in this 
chart the two branches of the light cone merge and becomes vertical as one moves towards $ r_{\rm C}$. So in order to understand the causal structure of the spacetime at or around $r_{\rm C}$, let us derive a maximally extended or Kruskal-like chart to remove the coordinate
singularity at $r_{\rm C}$ and construct a well behaved light cone structure there. Precisely, our objective will be to obtain a coordinate system $(T,~X)$ such that the $(t,~r)$ part of (\ref{dsin3}) becomes conformally flat with no singularity at least at or around  $r=r_{\rm C}$. We will see that nothing can come in from the cosmological event horizon along a causal curve and hence it acts as an outer causal boundary of our universe. 
The Kruskal extension for the Schwarzschild-de Sitter spacetime (\ref{s22i}) will also be derived, although for a different purpose, in Chapter 4.

For radial and null  geodesics in (\ref{dsin3}) we have
\begin{eqnarray}
\left(1-\frac{\Lambda r^2}{3}\right)dt^2=\left(1-\frac{\Lambda r^2}{3}\right)^{-1}dr^2, 
\label{anex1}
\end{eqnarray}
which means along such geodesics
\begin{eqnarray}
t=\pm r_{\star}+~{\rm {constant}}, 
\label{anex2}
\end{eqnarray}
where $r_{\star}$ is the tortoise coordinate defined by 
\begin{eqnarray}
 r_{\star}=\int\frac{dr}{1-\frac{r^2}{r_{\rm C}^2}}.  
\label{anex3}
\end{eqnarray}
Integrating, we find
\begin{eqnarray}
 r_{\star}=\frac{r_{\rm C}}{2}\ln\left\vert \frac{1+\frac{r}{r_{\rm C}} }  { 1-\frac{r}{r_{\rm C}} }\right\vert,  
\label{anex4}
\end{eqnarray}
which shows that $r_{\star}=0$ at the origin of the polar coordinates $r=0$ and $ r_{\star}\to +\infty $
as $r \to r_{\rm C}$. Using Eq.~(\ref{anex4}) we rewrite the $(t,~r)$ part of the metric (\ref{dsin3}) in a conformally flat form 
\begin{eqnarray}
 ds^2\vert_{{\rm{Radial}}}=\left(1+\frac{r(r_{\star})}{r_{\rm C}}\right)^2e^{-\frac{2r_{\star}}{ r_{\rm C}} }\left[-dt^2+dr_{\star}^2\right],  
\label{anex5}
\end{eqnarray}
where $r$ as a function of $r_{\star}$ can be found from Eq.~(\ref{anex4}). Now we define outgoing and incoming null
coordinates $u$ and $v$ as
\begin{eqnarray}
 u=t-r_{\star}, \quad v=t+r_{\star}. 
\label{anex6}
\end{eqnarray}
In terms of these null coordinates  Eq.~(\ref{anex5}) becomes 
\begin{eqnarray}
 ds^2\vert_{{\rm{Radial}}}=-\left(1+\frac{r(u,~v)}{r_{\rm C}}\right)^2e^{\frac{(u-v)}{ r_{\rm C}} }dudv,
\label{anex7m}
\end{eqnarray}
which after defining a set of new null coordinates $\overline{u}$ and $\overline{v}$,
\begin{eqnarray}
\overline{u}=r_{\rm{C}}e^{\frac{u}{ r_{\rm C}} },\quad \overline{v}=-r_{\rm{C}}e^{-\frac{v}{ r_{\rm C}} },
\label{anex8}
\end{eqnarray}
becomes
\begin{eqnarray}
 ds^2\vert_{{\rm{Radial}}}=-\left(1+\frac{r(\overline{u},~\overline{v} )}{r_{\rm C}}\right)^2 d \overline{u}d\overline{v}.
\label{anex7}
\end{eqnarray}
Let us now define new timelike and spacelike coordinates $T$ and $X$ such that
\begin{eqnarray} 
 T=\frac{ \overline{u}+\overline{v}}{2}=r_{\rm C}e^{-\frac {r_{\star}} {r_{\rm C}} }\sinh\left(\frac{t}{r_{\rm C}}\right),
\quad
 X=\frac{ \overline{u}-\overline{v}}{2}=r_{\rm C}e^{-\frac {r_{\star}} {r_{\rm C}} }\cosh\left(\frac{t}{r_{\rm C}}\right),
\label{anex8''}
\end{eqnarray}
where we have used Eq.s~(\ref{anex6}) and (\ref{anex8}). Eq.s~(\ref{anex8''}) show the following relationship between $T$, $X$, and $t$, $r$ :
\begin{eqnarray}
 X^2-T^2=r_{\rm C}^2 e^{ -\frac {2r_{\star}} {r_{\rm C}} }=r_{\rm C}^2\left\vert\frac{r_{\rm C}-r }{r_{\rm C}+r}\right\vert,\quad
%\label{youd12}
%\end{eqnarray}
%
%\begin{eqnarray}
 \frac{T}{X}=\tanh\left(\frac{t}{r_{\rm{C}}}\right).
\label{youd1}
\end{eqnarray}
With the new coordinates $T$ and $X$ the metric (\ref{anex7}) becomes
\begin{eqnarray}
 ds^2\vert_{{\rm{Radial}}}=\left(1+\frac{r(T,~X)}{r_{\rm C}}\right)^2 \left[-dT^2+dX^2\right],
\label{anex9}
\end{eqnarray}
where $r$ as a function of $(T,~X)$ can be found from Eq.s~(\ref{youd1}). 
Since this metric
does not contain any singularity, it can be regarded as the analytic continuation of the de Sitter metric for $r\geq r_{\rm C}$. In terms of $T$ and $X$ the analytically continued full de Sitter metric becomes  
\begin{eqnarray}
 ds^2=\left(1+\frac{r(T,~X)}{r_{\rm C}}\right)^2 \left[-dT^2+dX^2\right]+r^2(T,~X)\left(d\theta^2+\sin^2\theta d\phi^2\right).
\label{anex9'}
\end{eqnarray}
Eq.s~(\ref{youd1}) show that at $r=r_{\rm C}$ we have $X=\pm T$, so that $t\to \pm \infty$, i.e. the horizon
has two temporal components ${\cal{C^{+}}}$ and ${\cal{C^{-}}}$ known respectively as the future and past cosmological horizons. This means that an outgoing particle will take infinite Killing time $t$ to reach the future horizon, ${\cal{C^{+}}}$. On the other hand, if an incoming particle starting from the horizon is to be detected somewhere inside the horizon, it must have started infinite Killing time ago from ${\cal{C^{-}}}$. 
%\begin{center}
%%%%%%%%%%%%%%%%%%%%%%%%%%%%%%%%%%%% 
%%%%%%%%%%%%%% FIG   %%%%%%%%%%%%%%%%%%%%%%%%%%%%%%%%%%%
\begin{figure}
\centering%keepaspectratio
\rotatebox{0}{
\includegraphics[height=7.8cm,width=8.8cm]{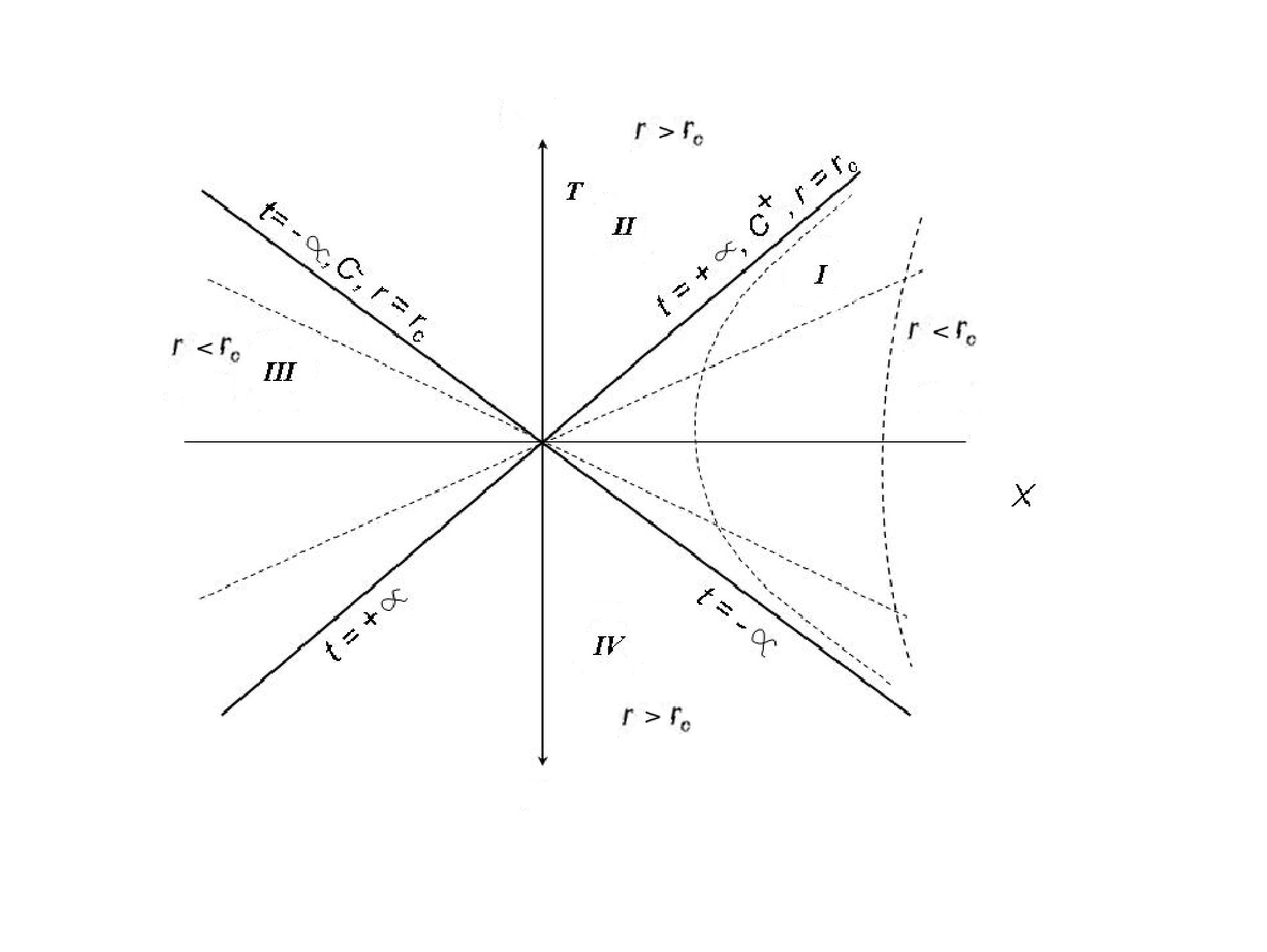}}
\caption{The Kruskal diagram for the de Sitter spacetime (\ref{anex9'}) with each point understood over a 2-sphere.}
\end{figure}
%%%%%%%%%%%%%%%%%%%%%%%
%\end{center} 
Fig. 1.1 shows the spacetime diagram of the analytically extended de Sitter spacetime.
In this diagram each point is understood as tangent to a 2-sphere which means that the null surfaces ${\cal{C^{\pm}}}$ have topology ${\cal{R}}^1\times S^2$. The two branches of the light cone 
 in the $(T,~X)$ coordinate are at $\pm 45 \,^{\circ}$  to the vertical meaning we have indeed constructed a good coordinate system. The dashed curves and straight lines refer respectively
to $r=$~constant and $t=$~constant hypersurfaces defined by Eq.s~(\ref{youd1}). At ${\cal{C^{\pm}}}$ those hypersurfaces merge. The null surfaces 
${\cal{C^{\pm}}}$ divide the spacetime into four regions. Using Eq.s~(\ref{youd1}) we find that
\begin{eqnarray}
 r<r_{{\rm C}}~{\rm in~I,~III};\quad r>r_{{\rm C}}~{\rm in~II,~IV}.
\label{anexreg1}
\end{eqnarray}
The timelike Killing field $(\partial_t)^a$ is future directed in region I, hence region I represents the world where we are located.
 In II and IV, the Killing field is spacelike. The analytically continued region III can be thought of as the Killing time reversal of region I, i.e. $(\partial_t)^a$ is past directed timelike there.  

Let us now consider a causal curve in region I. Since the spacetime (\ref{dsin3}) can be foliated by $r={\rm constant}$ timelike hypersurfaces, 
any causal curve will be the union of points over those hypersurfaces. In other words, we can construct any infinitesimal causal displacement by flowing a particle along a $r=$~constant curve, and then flowing the particle perpendicular to it, remembering in each such step we must have $dT >0$.
So we will take the variation of $r={\rm constant}$ curves defined by the first of Eq.s~(\ref{youd1}) along a timelike or null vector $\tau^a$ with parameter $\tau$. Thus taking the Lie derivative of the first of Eq.s~(\ref{youd1}) we have along any causal curve followed by a particle located at $r=r_{\rm C}-\epsilon$ with $\epsilon \to 0$,
\begin{eqnarray}
 X\frac{\Delta X}{\Delta \tau}-T\frac{\Delta T}{\Delta \tau}=\frac {r_{\rm C}} {4}\frac{\Delta\left\vert r_{\rm C}-r\right\vert}{\Delta \tau}\nonumber\\
\Rightarrow X \Delta X-T\Delta T= \frac {r_{\rm C}}{4} \Delta \left\vert r_{\rm C}-r\right\vert=\epsilon^{\prime}~(\rm say),
\label{youd12}
\end{eqnarray}
with the causality requirement $\Delta T>0$. 

For a particle outgoing (incoming) at the horizon, we have $\epsilon^{\prime}\leq0~(\geq0)$, so that 
\begin{eqnarray}
\left(X \Delta X-T\Delta T\right)\big\vert_{\rm {outgoing,~I}}\leq 0,
\label{youd12'}
\end{eqnarray}
and
\begin{eqnarray}
%%\displaystyle\lim_{ r \to r_{\rm C}} 
\left(X \Delta X-T\Delta T\right)\big\vert_{\rm {incoming,~I}}\geq 0.
\label{youd12ineq}
\end{eqnarray}
We note from the figure that
since $X^2-T^2=0$ at $r=r_{\rm C}$, the $ r={\rm constant}$ curves become asymptotic to ${\cal{C^{\pm}}}$. 
For a particle infinitesimally close to ${\cal{C^{+}}}$ in region I with $T,~X >0$, we consider a displacement 
orthogonal to a $r=\rm{constant}$ hypersurface. We see from the diagram that for such a displacement for a particle outgoing at ${\cal{C^{+}}}$, we have $\Delta X <0$ whereas for an incoming particle we have $ \Delta X >0$. But the latter is not possible at ${\cal{C^{+}}}$ since for this we have $\Delta T<0$. Thus Eq.~(\ref{youd12'}) is possible at ${\cal{C^{+}}}$ but Eq.~(\ref{youd12ineq}) is not.
  
Similar arguments show that nothing can be outgoing at ${\cal{C^{-}}}$ but can be incoming in region I. 
 
Thus we have seen that in region I, nothing can come out of the future horizon $ {\cal{C^{+}}}$ and nothing can go into the past horizon $ {\cal{C^{-}}}$. Can a particle then cross $ {\cal{C^{+}}}$ and reach region II? The answer is yes for a proper observer, in the following way. We consider a particle moving along a timelike/null geodesic in (\ref{dsin3}) and outgoing at $ {\cal{C^{+}}}$. The trajectory is represented by the effective 1-dimensional central force motion\footnote{See Appendix}
\begin{eqnarray}
\frac{1}{2}\dot{r}^2+ \frac{1}{2}\left(1-\frac{\Lambda r^2}{3}\right)\left(\frac{L^2}{r^2}-k\right)=\frac{1}{2}E^2,
\label{s42i}
\end{eqnarray}
where the `dot' denotes differentiation with respect to a parameter $\tau$ along the geodesic; $E$,
 $L$ correspond respectively to the conserved energy and the total orbital angular momentum of the particle and $k=0~(-1)$ for a null (timelike) geodesic and ${\displaystyle\frac{dt}{d \tau}=E\left(1-\frac{\Lambda r^2}{3}\right)^{-1}}$. Eq.~(\ref{s42i}) shows that ${\displaystyle\frac{dr}{d \tau}= +E>0}$ at the horizon for an outgoing particle. This means that for a proper observer with `time' $\tau$, the particle cannot be at rest at the horizon and hence it must eventually disappear to reach region II in a finite interval of the parameter $\tau$.

For maximal extensions of the spacetimes mentioned in Eq.s~(\ref{s22i}), (\ref{s25i}), (\ref{s26i}), we refer our reader to \cite{Gibbons:1977mu}. These spacetimes also possess cosmological horizons under some reasonable conditions and the maximal extensions at the cosmological horizon show the similar features discussed above. These spacetimes possess black holes also. The black hole horizons in these spacetimes show similar properties as those in asymptotically flat spacetimes. The  
 Reissner-N\"{o}rdstrom-de Sitter solution, the Kerr-de Sitter
or the Kerr-Newman-de Sitter solutions exhibit Cauchy horizons located inside the black hole horizons. For $\Lambda=0$ it has been shown that the Cauchy horizons exhibit instability under gravitational perturbations \cite{Chandrasekhar:1985kt}. It is likely that such kind of instabilities will also be present for the de Sitter black holes.

The Killing horizons which show past and future components such as ${\cal{C^{\pm}}}$ under maximal analytic extensions are called eternal horizons. 
Now the question is how much of the maximal extension discussed above should be taken seriously?
We recall that for a black hole formed from a gravitational collapse of a stellar object the past Killing horizon for $t\to -\infty$ does not exist---because at asymptotic past the collapse had only begun \cite{Wald:1984rg}. Such arguments can be applied for the de Sitter horizon also. Eq.~(\ref{dsin2}) gives $\tau \to -\infty$ at ${\cal{C^{-}}}$. But we recall that our universe is evolving to the de Sitter space asymptotically in $\tau$ \cite{Wald:1984rgnew1}, so
we may discard ${\cal{C^{-}}}$. Then regions III and IV are absent in Fig. 1.1 and the only regions are I and II separated by ${\cal{C^{+}}}$.

In any case, the above discussions show that the most non-trivial common feature of the known and physically reasonable $\Lambda>0$ spacetimes is the existence of a cosmological event horizon. This is a Killing horizon and nothing can come in from it, thereby acting as a natural outer causal boundary of the spacetime. 
So the infinities of such a spacetime is not very meaningful to a physical observer located in region I. Therefore, no precise notion of asymptotics exist in such spacetimes. There may also be non-trivial boundary effects due to this horizon. Due to this reason, in particular, throughout this thesis our motivation will be to study gravity in such spacetimes without referring to the region beyond the cosmological horizon.

When in general does a spacetime have a cosmological horizon?
We have seen from the known exact solutions that the addition of a positive $\Lambda$ into the Einstein equations gives rise to a cosmological horizon.  An interesting question at this point would be what happens if there is self gravitating matter without any particular spatial symmetry? In particular, is there still an outer
(cosmological) event horizon? More generally, what is the general criterion
for the existence of a cosmological event horizon? We will address these questions in the next Chapter.

%%%%%%%%%%%%%%%%%%%%%%%%%%%%%%%%%%%%%%%%%%%%%%%%%%%%%%%%%%%%%%%%%%%%%%%%%%
\section{No hair theorems and uniqueness}
%%%%%%%%%%%%%%%%%%%%%%%%%%%%%%%%%%%%%%%%%%%%%%%%%%%%%%%%%%%%%%%%%%%%
The black hole no hair conjecture states that any gravitational collapse reaches a final stationary state characterized by a small number of parameters. A part of this conjecture has been proven rigorously, known as the no hair theorem~\cite{Chrusciel:1994sn, Heusler:1998ua,Bekenstein:1998aw}, which deals with the uniqueness of stationary black
hole solutions characterized by mass, angular momentum, and charges corresponding to long-range gauge fields only.

These proofs usually involve constructing a suitable positive definite quadratic vanishing volume integral from the matter equation of motion which gives the fields to be zero identically everywhere in the black hole exterior. In particular, it has been shown that
static spherically symmetric black holes in asymptotically flat spacetime do not support external fields corresponding to scalars in convex potentials, Proca-massive vector
fields~\cite{Bekenstein:1971hc}, or even gauge field corresponding to the Abelian Higgs model~\cite{Adler:1978dp, Lahiri:1993vg}. Physically the no hair conjecture means that most of the matter constituting a stellar object either go inside the event horizon or escape to infinity during the collapse. The most noteworthy thing in all these proofs \cite{Bekenstein:1971hc, Adler:1978dp, Lahiri:1993vg} is that none of them need to solve the Einstein equations explicitly $-$ they only use the matter equations of motion derived from standard Lagrangians. This implies that the non-existence of matter fields outside the black hole is a consequence of the formation of the horizon by a collapse --- and does not depend upon the particular equation the spacetime itself obeys.

However, the above references assume that the spacetime to be asymptotically flat so that sufficiently rapid fall off conditions on the matter fields can be imposed at infinity. But we recall from the previous discussions that it is very likely that we have a cosmological event horizon as an outer boundary of our universe.
If we have a black hole, the black hole event horizon will be located inside the cosmological horizon and the spacetime is known as the de Sitter black hole spacetime. We argued in
 the previous Section that the cosmological horizon has a length scale of the order of $10^{26}$~m,
which is of course large but not infinite. We also demonstrated that the cosmological horizon acts as a natural boundary of the universe beyond which no causal communication is possible, and the infinities or the asymptotic region in such a spacetime 
are not very meaningful. So in the most general case neither can we impose any precise asymptotic behavior on the matter fields in such spacetimes, nor should we set $T_{ab}=0$ in the vicinity of the cosmological horizon since it is not located at infinity.
Keeping the non-existence of any precise asymptotic behavior of the matter fields in mind,  the extension of the no hair theorems for de Sitter black holes seems an interesting task.

In particular, Price's theorem, which can be regarded as a time-independent perturbative no hair theorem~\cite{Price:1971gc} was proved in~\cite{Chambers:1994sz} for the Schwarzschild-de Sitter black hole spacetime by taking massless perturbations in SL(2,~C) spinorial representations. So it would be highly interesting to generalize all known non-perturbative no hair theorems for a static de Sitter black hole spacetime.  

We will address this problem in details in Chapter 3. We will discuss also some counterexamples of the no hair theorems \cite{Torii:1998ir, Martinez:2002ru} and also show that due to the non-trivial boundary effect at the cosmological horizon the no hair theorem for the Abelian Higgs model can fail for a static spherically symmetric de Sitter black hole. This has no $\Lambda=0$ analogue. In fact this shows that the existence of the cosmological horizon can affect local physics.      

We will also generalize some no hair theorems for a stationary axisymmetric de Sitter black hole spacetime. For $\Lambda=0$ the no hair proofs for a rotating black hole for scalar and Proca-massive vector fields were first given in~\cite{Bekenstein:1972ky} assuming time reversal symmetry of the matter equations and a particular form of the metric. For a discussion on (2+1)-dimensional no hair theorems see~\cite{Skakala:2009ss}. See also~\cite{Sen:1998bj} for a scalar no hair theorem with a non-minimal coupling in stationary asymptotically flat black spacetimes. 

 For $\Lambda=0$ it can be rigorously shown that in (3+1)-dimensions the only spherically symmetric vacuum solution to the Einstein equations is the Schwarzschild solution. This statement is known as the Birkhoff theorem \cite{Hawking:1973uf}. Similarly it can be showed that the only electrically charged, spherically symmetric solution 
to the Einstein equations is the Reissner-N\"{o}rdstrom solution. It can also be shown that, for an asymptotically flat spacetime the Kerr or Kerr-Newman family is the unique solution of the vacuum or electrovac Einstein's equations, see e.g. \cite{Chandrasekhar:1985kt, Mazur:2000pn, Robinson}. This statement is known as the Robinson-Carter theorem.     
For the uniqueness proof of asymptotically anti-de Sitter vacuum black holes we refer our readers to \cite{Boucher:1983cv, AyonBeato:2004if}. The uniqueness proof of stationary de Sitter black holes however, remains elusive. While we will not address this problem in this thesis, we will discuss briefly the progress on this topic at the end of Chapter 3.

%%%%%%%%%%%%%%%%%%%%%%%%%%%%%%%%%%%%%%%%%%%%%%%%%%%%%%%%%%%%%%%%%%%%%
\section{Perturbative studies and geodesics in $\Lambda>0$ spacetimes}
%%%%%%%%%%%%%%%%%%%%%%%%%%%%%%%%%%%%%%%%%%%%%%%%%%%%%%%%%%%
In this Section we will discuss perturbative stability of the de Sitter black holes and the motion of geodesics in such spacetimes. Although stability issues are not discussed in this thesis, we make this digression because this can also be interpreted as a perturbative and time-dependent version of the no hair theorems.

 For $\Lambda=0$, a complete analysis on perturbative stability can be found in \cite{Chandrasekhar:1985kt}. 
The analysis usually involves writing down an effective Schr\"{o}dinger-like equation in a given background. For example, we take a free massless scalar field $\Psi$ with moving in the Schwarzschild-de Sitter background (\ref{s22i}). If we take usual separation of variables : $\Psi(t,~r,~\theta,~\phi)={\displaystyle\frac{u(r)}{r}e^{-i\omega t} Y_{lm}(\theta,~\phi)}$, the scalar equation of motion reduces to 
\begin{eqnarray}
-\frac{d^2 u}{d r_{\star}^2}+\left(1-\frac{2MG}{r}-\frac{\Lambda r^2}{3}\right)\left[\frac{l(l+1)}{r^2}+\frac{2MG}{r^3}-\frac{\Lambda}{3} \right]u(r)=\omega^2 u(r),
\label{s41}
\end{eqnarray}
where $r_{\star}$ is the tortoise coordinate defined by ${\displaystyle r_{\star}=\int \left(1-\frac{2MG}{r}-\frac{\Lambda r^2}{3} \right)^{-1}dr}$. The next task is then to solve Eq.~(\ref{s41}) with appropriate boundary conditions. We know that an incoming observer takes infinite Killing time
$t$ to reach the black hole horizon and an outgoing observer takes infinite Killing time $t$ to reach  the cosmological horizon and nothing can come out of them. Also, since ${\displaystyle\left(1-\frac{2MG}{r}-\frac{\Lambda r^2}{3}\right)=0}$ on the horizons, Eq.~(\ref{s41}) shows that $u(r)$ has the form of plane waves at the horizons. So for the Schwarzschild-de Sitter spacetime, the appropriate boundary condition would be $-$ ingoing plane wave at the black hole event horizon, and outgoing plane wave at the cosmological event horizon. 
The solution of Eq.~(\ref{s41}) with this boundary condition was found in \cite{Suneeta:2003bj}. The frequencies $\omega$ are complex numbers and are known as the quasinormal modes. It is clear from the time dependence $\sim e^{-i\omega t}$ that if the complex part of these frequencies are negative, the perturbation decays at late time and the spacetime is then said to be stable under such perturbation. One then says that the perturbation either moves into the black hole or flows out of the cosmological horizon. It was shown in \cite{Suneeta:2003bj} that there are two stable quasinormal modes corresponding to the two horizons of the Schwarzschild-de Sitter spacetime, the complex part of each of which is determined by the respective horizon's surface gravity.

For gravitational perturbation and the stability of the Cauchy horizon of the Kerr-de Sitter and Reissner-N\"{o}rdstrom-de Sitter spacetimes we refer our readers to \cite{Chambers:1994sz, Chambers:1994ap}. It remains as an interesting task to study the perturbation and quasinormal modes for Maxwell and Dirac fields in various de Sitter black hole backgrounds. 

Next, let us come to the effect of positive $\Lambda$ on geodesics.
This in particular, is related to possible observable effects like gravitational lensing. 
 
For $\Lambda=0$, a complete study of geodesic motion can be found in \cite{Chandrasekhar:1985kt} and references therein. To see the effect of positive $\Lambda$ on geodesics let us consider a particle following a timelike or null geodesic in the Schwarzschild-de Sitter spacetime. Let $E$ and $L$ be the conserved
energy and orbital angular momentum associated with that geodesic. 
As in the case for the Schwarzschild spacetime \cite{Wald:1984rg, Chandrasekhar:1985kt}, or the de Sitter spacetime (Eq.~(\ref{s42i})), we can map this motion to an effective 1-dimensional non-relativistic central force problem of a unit rest mass test particle with energy ${\displaystyle\frac{1}{2}E^2}$ and total orbital angular momentum $L$, 
\begin{eqnarray}
\frac{1}{2}\dot{r}^2+\psi (r,~L)=\frac{1}{2}E^2,\quad \left(\dot{\theta}^2+
 \dot{\phi}^2\sin^2\theta\right)^{\frac12}=\frac{L}{r^2},
\label{s42}
\end{eqnarray}
where the dot denotes differentiation with respect to a parameter along the geodesic, ${\displaystyle \psi (r,~L)=\frac{1}{2}\left(1-\frac{2MG}{r}-\frac{\Lambda r^2}{3}\right)\left(\frac{L^2}{r^2}-k\right)}$ is the effective potential barrier and $k=-1~(0)$ for a timelike (null) geodesic. 
 So the effect of positive $\Lambda$ can be estimated by observing bending of light and motion of massive particles. A realistic problem on gravitational lensing would then be to find out the effect of $\Lambda$ over light bending in the field of a spherical stellar object by solving Eq.s~(\ref{s42}). In fact some progress have already been made in this topic recently \cite{Rindler:2007zz}-\cite{Schucker:2007}, showing that positive $\Lambda$ has a repulsive effect over the geodesics. 

The study of light bending is particularly interesting in cosmic string spacetimes. A cosmic string is a cylindrically symmetric distribution of mass confined in a compact region of spacetime and zero outside. For $\Lambda=0$ such spacetimes have been studied in e.g. \cite{Garfinkle:1985}-\cite{Gregory:1995}, most of which are curved space generalizations of the Nielsen-Olesen string solution \cite{Nielsen:1973cs}. We refer our reader to \cite{Vilenkin2:2000} for an exhaustive study and list of references on this topic. These spacetimes show a conical singularity or a deficit in the azimuthal angle $\phi$ in the asymptotic region,
\begin{eqnarray}
ds^2\Big \vert_{\rho \to \infty}=-dt^2+d\rho^2+dz^2+\delta^2 \rho^2\phi^2,
\label{s43}
\end{eqnarray}
where the constant $\delta$ equals $(1-4G\mu)$, $\mu$ being the string mass per unit length. This is known as the Levi-Civita spacetime. Using Eq.s~(1.2) and (1.3) we can compute the
components of the Ricci tensor for (\ref{s43}) --- they all vanish identically. On the other hand
if one studies the motion of a null geodesic for (\ref{s43}), one finds that the geodesic bends towards the string due to the conical singularity $\delta$. In other words light gets attracted towards the string in the asymptotic region even though the curvature is zero there. This may be regarded as the gravitational analogue of the Aharanov-Bohm effect, see \cite{Vilenkin2:2000} and references therein. 

The study of cosmic string spacetimes
with a positive $\Lambda$ is in particular, interesting due to the expected repulsive effect due to $\Lambda$. In the exterior of a de Sitter cosmic string both the attractive effect due to the string and repulsive effect due to the ambient cosmological constant should be present.
While for (\ref{s43}) light bends towards the string in the asymptotic region, it seems that for $\Lambda >0$ such an effect may become weaker as we move away from the string. With this motivation we will construct explicit solutions for the cosmic Nielsen-Olesen strings with $\Lambda>0$ in Chapter 4 (for both free cosmic string and string with a black hole) and deriving analogues of Eq.s~(\ref{s42}), we will discuss the motion of null geodesics in the free cosmic string background.

 %%%%%%%%%%%%%%%%%%%%%%%%%%%%%%%%%%%%%%%%%%%%%%%%%%%%%%%%%%%%%%%%%%%%%%%%%%%%%
\section{Positive mass, thermodynamics and Hawking radiation}
%%%%%%%%%%%%%
The notion of a mass or mass function for a spacetime
 is an important thing in general relativity. The principal physical criterion of this mass function should be the following. Firstly, it must be defined with respect to a timelike Killing field, secondly, one should be able to relate the mass to the 
geodesic motion for a Newtonian interpretation and finally the mass function must be a positive definite quantity. It is the third criterion that makes the problem very severe because
one cannot define a satisfactory notion of the gravitational
 Hamiltonian unless one goes to the asymptotic region in an asymptotically flat spacetime \cite{Wald:1984rg}. Only an approximate notion of this can be defined perturbatively and locally but the positivity of this quantity is far from obvious. 

For asymptotically flat spacetimes a gravitational mass can be defined in several ways. One is the Komar mass. This is proportional to the surface integral of the derivative of the norm of the timelike Killing field and thus is related directly to geodesic motion. In general the Komar integral will be positive definite only if the matter energy-momentum tensor $T_{ab}$ satisfies the strong
energy condition (SEC) : $\left(T_{ab}-\frac{1}{2}T g_{ab}\right) \xi^a \xi^b \geq 0$, for any timelike $\xi^a$ \cite{Wald:1984rg}. We will see in the next Chapter that positive $\Lambda$
violets SEC, thus the notion of Komar mass is not very meaningful for spacetimes with $\Lambda>0$.
 
The second is the Arnowitt-Deser-Misner (ADM) \cite{misner1, misner2, misner3} formalism. In this approach a gravitational Hamiltonian density is defined with respect to the timelike Killing field in the asymptotic region and the integral of this Hamiltonian density is computed in the asymptotic region. This integral is interpreted as the gravitational mass.   

It is known from the Raychaudhuri equation that a matter field would converge geodesics only if it satisfies the SEC \cite{Wald:1984rg, Hawking:1973uf}. Also, it is known that the 
SEC implies the weak energy condition i.e., the positivity of the energy density. Using these two facts an approach to define gravitational mass and to prove its positivity was developed in \cite{geroch, penrose} for asymptotically flat spacetimes. Since positive $\Lambda$ repels geodesics even though it has a positive energy density, the above approach does not hold for any $\Lambda>0$ spacetime.

 The positivity conjecture of the ADM mass was first proved in \cite{shon1,
shon2}. Soon afterward, a remarkable proof of the positivity of the ADM mass was given in \cite{Witten:81} using spacelike spinors. 
This proof involved the assumption of a spinor over a spacelike non-singular Cauchy surface. This result was 
generalized for black holes in asymptotically flat or anti-de Sitter spacetimes in \cite{Hawking:83}. The $\Lambda \leq 0$ spacetimes usually have well defined asymptotic structure or infinities which are accessible to the
geodesic observers. The references mentioned above consider
explicit asymptotic structures of such spacetimes at spacelike infinities which are uniquely Minkowskian or anti-de Sitter. Thus
the positivity of the ADM mass for $\Lambda \leq 0$ spacetimes is quite well understood. 

Let us now take an account of the progress made so far on the definition of mass or mass functions for spacetimes with positive $\Lambda$. The very first approach can be found in \cite{Gibbons:1977mu}, where a mass function was defined
at the black hole and cosmological horizons using the integral of their respective surface gravities for stationary de Sitter black hole spacetimes. After that a perturbative approach was developed for asymptotically Schwarzschild-de Sitter 
spacetimes \cite{Abbott:82}. In this approach one considers metric perturbation in a region far away from the black hole, but inside the cosmological event horizon.
 The background spacetime in this region is de Sitter. A local gravitational energy momentum tensor was constructed and with respect to the background de Sitter timelike Killing field the mass of the perturbation was defined. This approach is similar to the usual Hamiltonian formulation of general
relativity. For asymptotically Schwarzschild-de Sitter spacetimes the mass
in this asymptotic region with respect to the de Sitter background was found to be $M$, i.e. the mass parameter of the Schwarzschild-de Sitter metric (\ref{s22i}). The spinorial proof of ADM mass was generalized later \cite{Shiromizu:94, kastor} to show that the mass defined in the sense of \cite{Abbott:82} with respect to the background de Sitter spacetime is indeed a positive definite quantity.  

How do quantities well defined on a black hole horizon change under infinitesimal variation of its mass? To answer this, we compute the variation of $M$ for a Schwarzschild black hole \cite{Smarr:1972kt}, 
\begin{eqnarray}
\delta M=\frac{\kappa_{\rm{H}}}{8\pi} \delta A_{\rm{H}},
\label{s51}
\end{eqnarray}
where ${\displaystyle\kappa_{\rm{H}}=\frac{1}{4 M}}$ is the horizon's surface gravity, and $A_{\rm{H}}$ is the horizon area. Similar variation can be made for charged and rotating black holes giving additional terms in the above equation \cite{Smarr:1972kt}. Equations like (\ref{s51}) are known as the Smarr formula.

Before we interpret formula (\ref{s51}), we mention here some interesting results. It was shown in \cite{Hawking:72} that by no classical physical process the area of black hole event horizon can be decreased $-$ it either increases or at least remains the same. Also it can be  shown that for a Killing horizon with a hypersurface orthogonal null Killing field, the surface gravity $\kappa_{\rm{H}}$ is a constant over the horizon (see e.g. \cite{Wald:1984rg}). It can also be demonstrated that the surface gravity of a black hole horizon
cannot be brought to zero by a finite number of physical processes (see e.g. \cite{Roman} 
 and references therein).

Now a question may be asked from the thermodynamical point of view. If we throw an object with some entropy into a black hole, is the entropy lost forever? If the answer is yes, clearly we violate the second law of thermodynamics. So the answer must be no. Keeping in mind that the black hole horizon's area can never be decreased  and in fact it increases when we throw objects within, it was proposed in \cite{Bekenstein:1973ur} that the horizon's area is proportional to the black hole entropy, ${\displaystyle S_{{\rm {BH}}}=\frac{A_{\rm H}}{4}}$. So the area theorem \cite{Hawking:72} is basically an analogue of the second law of thermodynamics. Also one may identify $\kappa_{\rm{H}}$ which is a constant over the horizon, to be the horizon's equilibrium temperature--- which is an analogue of the zeroth law. Moreover the impossibility of reaching zero surface gravity by a 
finite number of physical processes can be regarded as the analogue of the third law. 
So Eq.~(\ref{s51}) looks like the second law equation of thermodynamics with pressure $P=0$. A rigorous
formulation of the four laws of black hole mechanics in asymptotically flat spacetimes
can be found in \cite{Bardeen}.
We further refer our reader to \cite{Padmanabhan:2003gd} for a vast review and interesting issues on this topic.

If an object has a certain non-zero temperature and entropy, we know that it must emit thermal radiation. So in order to check whether black hole thermodynamics has any physical meaning, one has to see whether the black hole can make thermal emissions. While classically it cannot, it was shown in \cite{Hawk, Hartle:1976tp} that quantum mechanically a black hole can emit thermal radiation at temperature ${\displaystyle\frac{\kappa_{\rm{H}}}{2\pi}}$, in absolute agreement with Eq.~(\ref{s51}). This remarkable phenomenon is known as the Hawking radiation. Hawking's first calculations \cite{Hawk} was done in asymptotically flat spacetime and it was relied on the `in' and `out' particle states 
defined in the future and past null infinities ${\cal{I}^{\pm}}$. The incoming quantum fields  from ${\cal{I}^{-}}$ gets scattered by a collapsing object which forms a black hole in the asymptotic future. The outgoing waves which are not trapped by the future black hole horizon reaches ${\cal{I}^{+}}$. Using the geometric optics approximation it was very remarkably shown that the outgoing waves at ${\cal{I}^{+}}$ are thermal. Later the proof for the black hole radiance was rederived using path integrals \cite{Hartle:1976tp}.

The semiclassical tunneling method \cite{Kraus:1994}-\cite{Yale:2008kx} is an alternative way to model particle emission from a black hole using relativistic single particle quantum mechanics. The basic scheme of this method is to compute the imaginary part of the `particle' action by integrating the equation of motion across the horizon along an outgoing complex path. This integral gives the emission probability from the event horizon. From the expression of the emission probability one identifies the temperature of the radiation. This alternative approach of the Hawking radiation has received great attention during last few years. It is noteworthy that both of these methods deal only with the near horizon geometry, they can be very useful alternatives particularly when the spacetime has no well defined asymptotic structure or infinities like the de Sitter or de Sitter black hole spacetimes. We shall discuss particle creation via the semiclassical method in Chapter 5. 

In spacetimes with $\Lambda>0$ the issues of black hole thermodynamics and Hawking radiation are rather complicated and less understood. The very first attempt to study thermodynamics and particle creation in such spacetimes can be found in \cite{Gibbons:1977mu}. The generalization of Eq.~(\ref{s51})
for the Schwarzschild-de Sitter spacetime was found to be
\begin{eqnarray}
\kappa_{\rm{H}} \delta A_{\rm{H}}+\kappa_{\rm C}\delta A_{\rm{C}}=0,
\label{s52}
\end{eqnarray}
where $A_{\rm C}$ and $\kappa_{\rm C}$ are respectively the area and the surface gravity of the cosmological horizon. It can be shown that $\kappa_{\rm C}$ is a constant over the cosmological horizon. Then it turns out that it should also radiate at temperature ${\displaystyle\frac{\kappa_{\rm C}}{2\pi}}$. Thus even the de Sitter spacetime has a  Hawking-like temperature. In \cite{Gibbons:1977mu}
 the region between the two horizons were separated by an opaque membrane and the thermal radiation coming from both the horizons were studied separately using path integral quantization. It was shown that the cosmological event horizon indeed radiates thermal `particles' at temperature ${\displaystyle\frac{\kappa_{\rm{C}}}{2\pi}}$. However the problem arises when one does not isolate the two horizons. Then thermal radiation from both the horizons will mix and the resultant spectrum would be non-thermal. Then there exists no well defined notion of temperature in de Sitter black hole spacetimes. 
 Moreover, what will be the entropy of such spacetimes?
It has been argued in \cite{Goheer:2002vf}-\cite{Urano:2009xn} that for a multi-horizon spacetime like de Sitter black holes, the entropy area law may break down. So far it is only clear that both the black hole and the cosmological horizon radiate at temperatures proportional to their respective surface gravities. But can we treat these two radiations in equal footing? 

We will address some of the above issues in Chapter 5. We will rederive Eq.~(\ref{s52}) using the mass function of \cite{Abbott:82}. We will also prove the universality of Hawking or Hawking-like radiation from any stationary Killing horizon via the semiclassical complex path method and then we will discuss Hawking radiation in the Schwarzschild-de Sitter spacetime. However, it remains as an interesting problem to construct a meaningful quantum field theory of the Hawking radiation in de Sitter black hole spacetimes.

The thesis is organized as follows. In the next Chapter we will construct a general existence proof of cosmological event horizons for general static and stationary axisymmetric spacetimes.
Using the geometrical set up developed in Chapter 2, we will discuss various no hair theorems for static and stationary axisymmetric de Sitter black holes in Chapter 3. Chapter 4 will be devoted to the discussions of cosmic Nielsen-Olesen string solutions with positive $\Lambda$. Chapter 5 will concern with thermodynamics and the Hawking radiation for the Schwarzschild-de Sitter spacetime. A proof of the universality of the Hawking or Hawking-like radiation 
from Killing horizons of stationary spacetimes will be given via the semiclassical complex path method. Finally we summarize the thesis in Chapter 6 mentioning some future directions.

%%%%%%%%%%%%%%%%%%%%%%%%%%

%\begin{figure}[t]
%\makebox[\textwidth]{
%{\framebox[5cm]{\rule{0pt}{5cm}}
%\psfig{file=photo.ps}
%\psfig{file=photo.ps}}
%\caption{Regge trajectories \,(taken from \cite{Bali:2000gf})} \label{white}
%\end{figure}

%%%%%%%%%%%%%%%%%%%%%%%%%%%%%%%%%%%%%%%%%%%%%%%%%%%%%%%%%%%%%%%%%%%%%%%%%jhep1

%%%%%%%%%%%%%%fourth Chapter begins%%%%%%%%%%%%%%%%%%%%%%%%%%%%%%%%%%%%%%%%

%%%%%%%%%%%%%%%%%%%%%%%%%%%%%%%%%%%%%%%%%%%%%%%%%%%%%%%%%%%%%%%%%%%%%%%%%%%%%%%%%%%%%%%%%%%
\chapter{On the existence of cosmological event horizons}

%%%%%%%%%%%%%%%%%%%%%%%%%%%%%%%%%%%%%%%%%%%%55
In the previous Chapter we mentioned a few exact solutions
with $\Lambda>0$ and described the causal structure of a cosmological event horizon. 
 In general we expect that a positive cosmological constant
 implies the existence of a cosmological event horizon,
 i.e. an event horizon which acts as an outer causal boundary of the spacetime. As we have seen, if we add a positive cosmological constant $\Lambda$ into the Einstein equations,
 we find de Sitter space in the
absence of matter for a spatially homogeneous and isotropic universe.
We have seen in the previous Chapter that this solution exhibits an outer Killing horizon of
size $\sqrt {\frac {3} {\Lambda}}$.
 On the other hand, if we assume the spacetime to be
static and spherically symmetric, or stationary and axisymmetric, the solution to the vacuum
Einstein equations is the Schwarzschild-de Sitter or the Kerr-de Sitter~\cite{Carter:1968ks}. When $\Lambda$ and the
other parameters of these solutions (such as the mass parameter $M$ and the rotation parameter $a$) obey certain conditions
between them, we obtain static or stationary black hole spacetimes
embedded within a cosmological Killing horizon.  
There exist a few other exact solutions of the Einstein equations with a positive $\Lambda$, all exhibiting cosmological horizons under some reasonable conditions. 
  
It is interesting to note the following in this context.
All known solutions with or without matter and for $\Lambda \leq 0$ do not exhibit cosmological event horizons. We recall that the energy-momentum tensor $T_{ab}$ corresponding to 
any physical matter field obeys the weak energy condition (WEC), i.e. for any
timelike $n^a$ one has $T_{ab}n^an^b\geq 0$. We also recall two other energy conditions, namely the null energy condition (NEC) and the strong energy condition (SEC). The former one states that for any future directed null vector field $u^a$, $T_{ab}u^au^b\geq0$.
The strong energy condition states that for any future directed timelike $n^a$,
$\left(T_{ab}-\frac{1}{2}Tg_{ab}\right)n^an^b\geq 0$.
The cosmological constant term, appearing as $8\pi G T_{ab}\equiv-\Lambda g_{ab}$ on the right hand side of the Einstein equations obeys WEC for $\Lambda>0$, i.e. the vacuum energy density corresponding to $\Lambda>0$ is positive. Since a positive $\Lambda$ and any physical matter field both satisfy WEC, we ask here why a positive $\Lambda$ implies the existence of a cosmological or outer horizon. In other words, why is the global structure of spacetimes 
with $\Lambda>0$ are so different than those with $\Lambda\leq0$?

Secondly, what happens if we have matter fields in de Sitter or de Sitter black hole spacetimes? In particular, is there still an outer (cosmological) event horizon? More generally, what is the criterion for the existence of a cosmological event horizon? Or, can we find a class
of matter fields which also imply the existence of a cosmological event horizon? 
In this Chapter we focus on this question and establish a general criterion for the existence of an outer or cosmological horizon. We discuss two kinds of spacetime, one static, and the other stationary and axisymmetric. An inner or black hole event horizon is
not assumed, although one may be present. In fact we will see that the presence of the inner horizon does not affect our result anyway. Our region of interest of the spacetime will be the region inside the cosmological horizon, or if a black hole is present the region between the black hole and the cosmological horizon. 

Let us now come to our assumptions.
We assume that the spacetime is regular, i.e. there is no
naked curvature singularity anywhere in our region of interest.
Since the curvature is related to the energy-momentum tensor, this
assumption implies that the scalar invariants constructed out of the energy-momentum tensor $T_{ab}$ are bounded
everywhere in our region of interest.

We assume that the spacetime connection $\nabla$ is torsion free, i.e. for any smooth spacetime function $\varepsilon(x)$ we have identically
\begin{eqnarray}
\nabla_{[a}\nabla_{b]}\varepsilon(x)=0.
\label{s1''}
\end{eqnarray}
 We also assume that the energy-momentum tensor satisfies the
weak energy condition, i.e., the most reasonable energy condition expected from any physical matter field : $T_{ab}n^an^b\geq0$ for any timelike $n^a$. This is the only energy condition we assume any matter field we are concerned with must satisfy. We assume in the following that a `closed' and null outer horizon already exists and then find the condition that the
energy-momentum tensor must fulfill for the Einstein equations to
hold.  We find that the strong energy condition must be violated by
the energy-momentum tensor, at least over some portion of the spacelike
hypersurface inside the outer horizon. While the simplest example of such a  matter field is a positive cosmological
constant, we also find conditions on the energy-momentum
tensor due to ordinary matter satisfying the strong energy condition so that $\Lambda>0$ implies an outer horizon.

%%%%%%%%%%%%%%%%%%%%%%%%%%%%%%%%%%%%%%%%%%%%%%%%%%%%%%%%%%%%%%%%%%%%
\section{Static spacetimes}
%%%%%%%%%%%%%%%%%%%%%%%%%%%%%%%%%%%%%
In the previous Chapter we discussed the features of a cosmological event horizon by considering the de Sitter spacetime. For known static or stationary solutions, a cosmological horizon, like a black hole horizon is a Killing
horizon with future and past components. Let us now generalize the notion of Killing horizons for an arbitrary static spacetime in a coordinate independent way. 

We start with a spacetime which is static in our region of interest.
So in this region the spacetime is endowed with a timelike Killing
vector field $\xi^a$,
\begin{eqnarray}
\nabla_{a}\xi_b+\nabla_b \xi_{a}=0,
\label{s1'}
\end{eqnarray}
with norm $\xi_{a}\xi^a=-\lambda^2$. Since the spacetime is
static, $\xi^a$ is by definition orthogonal to a family of spacelike
hypersurfaces $\Sigma$, and we have the Frobenius condition of hypersurface orthogonality
\begin{eqnarray}
\xi_{[a}\nabla_{b}\xi_{c]}=0.
\label{s2'}
\end{eqnarray}
%Since a null vector field is both parallel and orthogonal to itself,  $\xi^a$ is also orthogonal to ${\cal{H}}$.
A Killing horizon ${\cal{H}}$ of the spacetime is defined to be a 3-dimensional surface to which $\xi^a$ is normal and becomes null, i.e. $\lambda^2=0$ over ${\cal{H}}$~\cite{ Wald:1984rg, Gibbons:1977mu}. 
A normal to the $\lambda^2=0$ surface ${\cal{H}}$ is $R_a=\zeta(x)\nabla_a \lambda^2$ also, where $\zeta(x)$ is a smooth spacetime function. Using the torsion-free condition (\ref{s1''}) it is easy to see that $R_a$ satisfies the Frobenius condition of hypersurface orthogonality, meaning ${\cal{H}}$ is a null hypersurface with normal $R^a$ or $\xi^a$. 
The region of our interest of the spacetime is given by ${\cal{H}}\cup \Sigma\cup \gamma_{\xi}$, where $\gamma_{\xi}$ denotes the orbits of the timelike Killing field $\xi^a$. 
 
Let us consider the Killing identity for $\xi^a$ 
\begin{eqnarray}
\nabla_a\nabla^a \xi_b=-R_{ab}\xi^a\,,
\label{s3'}
\end{eqnarray}
and contract both sides of Eq.~(\ref{s3'}) by $\xi^b$ to obtain
\begin{eqnarray}
\nabla_a\nabla^a \lambda^2=2R_{ab}\xi^a\xi^b-
2\left(\nabla_a\xi_b\right)
\left(\nabla^a\xi^b\right).
\label{s4'}
\end{eqnarray}
Also we use the Killing equation~(\ref{s1'}) and the Frobenius condition~(\ref{s2'}) to get
\begin{eqnarray}
\nabla_a \xi_b=\frac{1}{\lambda}\left(\xi_b\nabla_a \lambda-
\xi_a \nabla_b \lambda \right).
\label{s5'}
\end{eqnarray}
Substituting this into Eq.~(\ref{s4'}) we obtain
\begin{eqnarray}
\nabla_a\nabla^a \lambda^2=2R_{ab}\xi^a\xi^b+
4\left(\nabla_a\lambda\right)
 \left(\nabla^a\lambda\right).
\label{s6'}
\end{eqnarray}

Now we wish to project Eq.~(\ref{s6'}) onto the spacelike hypersurfaces $\Sigma$.
In order to do this, we consider the usual projector $h_{a}{}^{b}$ or the induced metric 
$h_{ab}$ over $\Sigma$
\begin{eqnarray}
h_{a}{}^{b}=\delta_{a}{}^{b}+\lambda^{-2}\xi_a\xi^b \Rightarrow h_{ab}=g_{ab}+\lambda^{-2}\xi_a\xi_b.
\label{s7'}
\end{eqnarray}
The $\Sigma$ projection $\omega$ of any spacetime tensor $\Omega$ is given via the projector by 
\begin{eqnarray}
\omega_{a_1a_2\dots}{}^{b_1b_2\dots}=h_{a_1}{}^{c_1}\dots h^{b_1}{}_{d_1}\dots 
\Omega_{c_1c_2\dots}{}^{d_1d_2\dots}.
\label{adch21}
\end{eqnarray}
We also denote the induced connection over $\Sigma$ by $D_a$ defined via the projector \\ $h_{a}{}^{b}$ : $D_a\equiv h_{a}{}^{b}\nabla_b$. We define the action of the induced connection $D_a$ over $\Sigma$ as \cite{Wald:1984rg}, 
\begin{eqnarray}
D_a\omega_{a_1a_2\dots}{}^{b_1b_2\dots}=h_{a_1}{}^{c_1}\dots h^{b_1}{}_{d_1}\dots 
h_{a}{}^{b}\nabla_b\Omega_{c_1c_2\dots}{}^{d_1d_2\dots}.
\label{adch22}
\end{eqnarray}
It is easy to see that the above definition satisfies the Leibniz rule and $D_a$
is compatible with the induced metric $h_{ab}$. Then we have using Eq.~(\ref{adch22})
\begin{eqnarray}
D_a \lambda=h_{a}{}^{b}\nabla_b\lambda=\nabla_a \lambda+\lambda^{-2}\xi_a\left(\pounds_{\xi}\lambda\right).
\label{adt1}
\end{eqnarray}
But the Killing equation (\ref{s1'}) implies $\pounds_{\xi}\lambda=0$ identically. So we have $\nabla_a \lambda=D_a \lambda$. Also, using Eq.~(\ref{s7'}) and $\pounds_{\xi}\lambda^2=0$, we see that   
\begin{eqnarray}
\nabla_a\nabla^a \lambda^2= \frac{1}{\sqrt{-g}}\d_a\left[\sqrt{-g}g^{ab}\d_b \lambda^2\right]
&=&
\frac{1}{\lambda \sqrt{h}}\d_a\left[\lambda \sqrt{h}\left\{-\lambda^{-2}\xi^a\xi^b+h^{ab} \right\}\d_b \lambda^2\right] \nonumber\\&=&
\frac{1}{\lambda \sqrt{h}}\partial_a\left[\lambda\sqrt{h}h^{ab}\partial_b \lambda^2\right]=\frac{1}{\lambda}D_a\left(\lambda D^a \lambda^2 \right),\nonumber\\
\label{adt2}
\end{eqnarray}
where $g$ is determinant of the spacetime metric $g_{ab}$, $h$ is the determinant of the induced metric $h_{ab}$ defined by Eq.~(\ref{s7'}) and $h^{ab}$ is the inverse of $h_{ab}$. 
With all these, we have the $\Sigma$-projection of Eq.~(\ref{s6'})  
\begin{eqnarray}
D_a\left(\lambda D^a \lambda^2\right) 
 =  2\lambda\left[R_{ab}\xi^a\xi^b+2 \left(D_a \lambda\right)
 \left(D^a \lambda\right)\right].
\label{s14}
\end{eqnarray}
Now we wish to integrate this equation over the spacelike 
hypersurfaces $\Sigma$ with the horizon or horizons acting as a boundary. But before
we go into that, it is worthwhile to
spend a few words about how we may perform the integration at the horizon
 where $\lambda= 0$ and the induced metric $h_{ab}$ defined by Eq.~(\ref{s7'}) becomes singular. It seems that it is difficult
to say without using any particular coordinate system, whether the invariant
$\lambda\left(D_a \lambda\right)\left(D^a \lambda\right)$ appearing in Eq.~(\ref{s14})   
remains bounded over the horizon. In order to bypass this 
difficulty we multiply both sides of Eq.~(\ref{s14}) by
$\lambda^{n+1}$, where $n$ is an arbitrary positive integer. 
Now we integrate Eq.~(\ref{s14}) by parts over 
the spacelike hypersurfaces $\Sigma$ to obtain
\begin{eqnarray}
\oint_{\partial\Sigma}\lambda^{n+1} D_a \lambda^2d\gamma^{(2)a} =
2\int_{\Sigma}\left[\lambda^{n+1} R_{ab}\xi^a\xi^b+ 
\left(n+2\lambda\right)\lambda^n
\left(D_a \lambda\right)\left(D^a \lambda\right)\right],
\label{s15}
\end{eqnarray}
where the surface integral on the left hand side is calculated over
the boundary of $\Sigma$, i.e., over the horizon or horizons. The volume element $d\gamma^{(2)a}$ corresponds to the `closed' and regular spacelike 2-surfaces located at the horizons.

As we have assumed, the spacetime has a closed outer
boundary or cosmological horizon, so that $\lambda=0$ there. 
Hence, by choosing the integer $n$ in Eq.~(\ref{s15}) to be sufficiently large and positive it may be guaranteed that each of the invariant terms appearing in the right hand side of
Eq.~(\ref{s15}), including $R_{ab}\xi^a\xi^b$ remains bounded as $\lambda \to 0$.
  
 If we have a black hole present in the spacetime, the inner boundary
is the black hole event horizon, and we must also have $\lambda=0$
there. Then the surface integrals over the horizons in Eq.~(\ref{s15}) vanish, and we finally get
\begin{eqnarray}
\int_{\Sigma}
\left[\lambda^{n+1}
R_{ab}\xi^a\xi^b+ (n+2\lambda)\lambda^n
 \left(D_a\lambda\right)
 \left(D^a\lambda\right)\right]=0.
\label{s16}
\end{eqnarray}
On the other hand, since we have assumed that there is no naked curvature singularity anywhere in our region of interest,
when any inner or black hole horizon is absent, we are free to
shrink the inner boundary or surface of Eq.~(\ref{s15})
to a non-singular point or `centre', where the inner surface integral gives zero. Thus
Eq.~(\ref{s16}) also holds for non-singular spacetimes without
a black hole.  

The second term in Eq.~(\ref{s16}) is a spacelike inner product and
hence positive definite over $\Sigma$, so we must have a negative
contribution from the first term $R_{ab}\xi^a\xi^b$. In other
words, the existence of an outer or cosmological Killing horizon implies
\begin{eqnarray}
R_{ab}\xi^a\xi^b<0,
\label{s17}
\end{eqnarray}
at least over some portion of $\Sigma$, so that the integral
in Eq.~(\ref{s16}) vanishes. Using the Einstein equations 
\begin{eqnarray}
R_{ab}-\frac{1}{2}R g_{ab}= 8\pi GT_{ab},
\label{s18}
\end{eqnarray}
we see that the condition~(\ref{s17}) implies that the strong
energy condition (SEC) must be violated by the energy-momentum tensor
\begin{eqnarray}
\left(T_{ab}-\frac{1}{2}Tg_{ab}\right)\xi^a\xi^b<0,
\label{s19}
\end{eqnarray}
at least over some portion of $\Sigma$. A positive
cosmological constant $\Lambda$, appearing on the right hand side
of the Einstein equations (\ref{s18}) as $8\pi G T_{ab}\equiv-\Lambda g_{ab}$, violates the SEC, because in that case
\begin{eqnarray}
\left(T_{ab}-\frac{1}{2}Tg_{ab}\right)\xi^a\xi^b=-\frac{\Lambda \lambda^2}{8\pi G}\leq 0,
\label{s20f'}
\end{eqnarray}
 where the equality holds only on the horizons. We now split the total energy-momentum tensor $T_{ab}$ as
\begin{eqnarray}
8\pi G T_{ab}=-\Lambda g_{ab}+8\pi GT^{\rm{N}}_{ab},
\label{s20}
\end{eqnarray}
where the superscript `N' denotes `normal' matter fields satisfying the SEC. Then Eq.~(\ref{s16}) becomes 
\begin{eqnarray}
  \int_{\Sigma}\lambda^{n+1}
  \left[X^{\rm{N}}+ \frac{(n+2\lambda)}{\lambda}
(D_a \lambda)( D^a \lambda) -
    \Lambda\lambda^2\right]=0, 
\label{s21}
\end{eqnarray}
where $X^{\rm{N}}$ corresponds to $ T^{\rm{N}}_{ab}$, i.e. the normal
matter fields satisfying the SEC and thus is a positive definite quantity. So for the cosmological horizon to exist, we must have
\begin{eqnarray}
\int_{\Sigma}\lambda^{n+1}
\left[X^{\rm{N}}-\Lambda\lambda^2\right]<0.
\label{s22}
\end{eqnarray}
In other words, if there is to be an outer horizon, the positive cosmological constant term has to dominate the integral.
It is interesting to note that the observed values of $\Lambda$ and
matter densities satisfying the SEC in the universe do satisfy this requirement \cite{Weinberg:2008zzc}. On the other hand,
a universe endowed with a positive $\Lambda$ and in which all matter is restricted to a
finite region in spacetime also satisfies this requirement. This has relevance in discussions of late time behavior of de Sitter black holes formed by collapse.

We provide here an example of a matter field other than positive $\Lambda$ violating SEC and hence may give rise to a cosmological event horizon. We consider a real scalar field $\psi$ in a double well potential
\begin{eqnarray}
{\cal{L}}=-\frac{1}{2}(\nabla_a\psi)(\nabla^a\psi)-\frac{k}{4}\left(\psi^2-v^2\right)^2,
\label{cral21}
\end{eqnarray}
where $k$ and $v$ are constants and $k>0$. The energy-momentum tensor corresponding to the scalar field is given by
\begin{eqnarray}
T_{ab}=\nabla_a \psi\nabla_b\psi +g_{ab}{\cal{L}}.
\label{cral22'}
\end{eqnarray}
 The potential $V(\psi)=\frac{k}{4}\left(\psi^2-v^2\right)^2$ has a maximum at $\psi=0$, and two minima at $\psi=\pm v$. Now let us suppose a stationary configuration where $\psi$ assumes a constant value at the maximum of $V(\psi)$. Then the energy-momentum tensor (\ref{cral22'}) becomes $T_{ab}=-\frac{kv^4}{4}g_{ab}$, which violates SEC since $k$ is positive. In fact the Einstein-Hilbert Lagrangian in that case is ${\cal{L}}_{\rm{EH}}=\left(R-\frac{kv^4}{4}\right)$, which gives the Schwarzschild-de Sitter solution (\ref{s22i}) with $\Lambda=\frac{kv^4}{8}$.

%%%%%%%%%%%%%%%%%%%%%%%%%%%%%%%%%%%%%%%%%%%%%%%%%%
\section{Stationary axisymmetric spacetimes}
%%%%%%%%%%%%%%%%%%%%%%%%%%%%%%%%%%%%%%%%%%%%%%%%

We now generalize the above result for the static spacetimes to stationary axisymmetric
spacetimes, in general rotating, which satisfy some additional
geometric constraints. The basic scheme will be the same as before, i.e. to use
the Killing identities and to construct quadratic vanishing integrals over the spacelike hypersurface inside the outer horizon. 

We assume that the 
spacetime is endowed with two commuting Killing fields $\xi^a$, and $\phi^a$,
\begin{eqnarray}
\nabla_{(a}\xi_{b)} &=& 0 =
 \nabla_{(a}\phi_{b)} \,,\\
\left[\xi,~\phi\right]^a &=& \pounds_{\xi}\phi^a=-\pounds_{\phi}\xi^a=\xi^b\nabla_b \phi^a-\phi^b\nabla_b \xi^a=0\,.
\label{2k}
\end{eqnarray}
$\xi^a$ is locally timelike with norm $-\lambda^2$ and generates the stationarity. 
$\phi^a$ is a locally spacelike Killing field with closed orbits
and norm $f^2$ and hence generates the axisymmetry. We also assume that the vectors orthogonal to
$\xi^a$ and $\phi^a$ span integral 2-submanifolds. The existence of the two commuting Killing fields $(\xi^a,~\phi^a)$ and the integral 2-submanifolds orthogonal to them are the additional constraints mentioned. We note that
all known stationary axisymmetric spacetimes obey these restrictions. Let us denote the basis vectors of this spacelike 2-submanifolds by $\{\mu^a,~\nu^a\}$, with $\mu^a\nu_a=0$. 

For a stationary spacetime, $\xi^a$ is not orthogonal to $\phi^a$, so
in particular there is no spacelike hypersurface both tangent to
$\phi^a$ and orthogonal to $\xi^a$. Let us first construct a family
of spacelike hypersurfaces. We first define $\chi_{a}$ as
\begin{eqnarray}
\chi_a=\xi_a-\frac{1}{f^2}\left(\xi_b\phi^b\right)
\phi_a \equiv \xi_a+\alpha \phi_a,
\label{sa2}
\end{eqnarray}
so that $\chi_a\phi^a=0$ everywhere. We note that
\begin{eqnarray}
\chi_a\chi^a=-\beta^2\equiv
-\left(\lambda^2+\alpha^2 f^2\right),
\label{sa3}
\end{eqnarray}
i.e., $\chi_a$ is timelike when $\beta^2>0$. We also note that by construction $\chi^a\mu_a=0=\chi^a\nu_a$.
Therefore we may now choose an orthogonal basis for the spacetime as $\left\{\chi^a,~\phi^a,~\mu^a,~\nu^a\right\}$. We also note that $\chi^a$ is not a Killing field,
\begin{eqnarray}
\pounds_{\chi}g_{ab}=\nabla_{(a}\chi_{b)} = \phi_a\nabla_b \alpha
+\phi_b\nabla_a \alpha.
\label{sa5}
\end{eqnarray}
Next we recall that the necessary and sufficient condition that an
arbitrary subspace of a manifold forms an integral submanifold or a
hypersurface is the existence of a Lie algebra of the basis vectors
of that subspace (see e.g.~\cite{Wald:1984rg} and references
therein). So our assumption that $\{\mu^a,~\nu^a\}$ span integral 2-submanifolds implies 
\begin{eqnarray}
\left[\mu,~\nu \right]^a=\mu^b\nabla_b \nu^a -\nu^b\nabla_b \mu^a=g_1(x) \mu^a+g_2(x)\nu^a,
\label{adt3}
\end{eqnarray}
where $g_1(x)$ and $g_2(x)$ are arbitrary smooth functions.
We contract Eq.~(\ref{adt3}) separately by $\xi_a$ and $\phi_a$, and use the fact that both  these 1-forms are orthogonal to $\mu^a$ and $\nu^a$ to find
\begin{eqnarray}
\mu^{[a}\nu^{b]}\nabla_a \xi_b=0=\mu^{[a}\nu^{b]}\nabla_a \phi_b.
\label{adt4}
\end{eqnarray}
This shows that $\nabla_{[a} \xi_{b]}$ or $\nabla_{[a} \phi_{b]}$ can be expanded as
\begin{eqnarray}
\nabla_{[a} \xi_{b]}=\omega_{1[a}\xi_{b]}+\omega_{2[a}\phi_{b]},\quad\nabla_{[a} \phi_{b]}=\omega_{3[a}\xi_{b]}+\omega_{4[a}\phi_{b]},
\label{adt4l}
\end{eqnarray}
where the $\omega$'s are arbitrary 1-forms. Clearly, Eq.s~(\ref{adt4l}) guarantee that Eq.s~(\ref{adt4}) hold identically. Multiplying by $\xi_a\phi_b$ and antisymmetrizing, we see that Eq.s~(\ref{adt4l}) are equivalent to
\begin{eqnarray}
\xi_{[a}\phi_b \nabla_c\phi_{d]} = 0
=\phi_{[a}\xi_b \nabla_c\xi_{d]}.
\label{cral22}
\end{eqnarray}
Using now the definition of $\chi_a$
given in Eq.~(\ref{sa2}) we obtain from Eq.s~(\ref{cral22}) the following conditions for the existence of the 2-submanifolds
\begin{eqnarray}
\chi_{[a}\phi_b \nabla_c\phi_{d]} &=& 0\,,\\
\label{frob1}
\phi_{[a}\chi_b \nabla_c\chi_{d]} &=& 0\,.
\label{frobenius}
\end{eqnarray}
We now expand Eq.~(\ref{frobenius}), use Eq.~(\ref{sa5}), and contract by $\chi^b\phi^a$ to obtain
\begin{eqnarray}
f^2\left[\beta^2\nabla_{[d}\chi_{c]}+2\beta\chi_{[d}\nabla_{c]}\beta\right]+\beta^2\left[\phi_c(\pounds_{\phi}\chi_d)-\phi_d(\pounds_{\phi}\chi_c)\right]=0.
\label{adt5}
\end{eqnarray}
But the commutativity of the two Killing fields $[\xi,~\phi]^a=0$ gives
\begin{eqnarray}
\pounds_{\phi}\chi^a=\pounds_{\phi}\left(\xi^a+\alpha\phi^a\right)=\left(\pounds_{\phi}\alpha\right) \phi^a=-\left\{\pounds_{\phi}\left(\frac{\xi\cdot\phi}{\phi^2}\right)\right\}\phi^a=0.
\label{cral23}
\end{eqnarray}
So, $\pounds_{\phi}\chi_a=(\pounds_{\phi}\chi^b)g_{ab}+\chi^b(\pounds_{\phi}g_{ab})=0$. Thus we obtain from Eq.~(\ref{adt5}) the following
\begin{eqnarray}
\nabla_{[a}\chi_{b]}=
2\beta^{-1}\left(\chi_b\nabla_a \beta-
\chi_a\nabla_b \beta\right),
\label{sa8}
\end{eqnarray}
which shows that $\chi^a$ satisfies the Frobenius condition,
\begin{eqnarray}
\chi_{[a}\nabla_b\chi_{c]}=0.
\label{sa4}
\end{eqnarray}
So there exists a family of spacelike hypersurfaces $\Sigma$ orthogonal
to $\chi^a$, spanned by $\left\{\phi^a,~\mu^a,~\nu^a\right\}$. However we should note that unlike in the case of the static spacetime, $\chi^a$ is not a Killing
vector field here, Eq.~(\ref{sa5}). 

Since the timelike Killing vector field $\xi^a$ is not hypersurface orthogonal for the present case, $\xi^a\xi_a=-\lambda^2=0$ does not define the horizon of the spacetime. In fact in all the known cases of the stationary axisymmetric black hole spacetimes the horizon is located inside a $\lambda^2=0$ surface, i.e. the ergosphere. Within the ergosphere $\xi^a$ becomes spacelike and the region between the ergosphere and the horizon is known as the ergoregion. Since there is no timelike Killing vector field within the ergoregion, no observer can be stationary in this region. For many interesting effects due to the ergosphere, including superradiant scattering, we refer our reader to \cite{Wald:1984rg, Chandrasekhar:1985kt} and references therein.  So, let us now define the horizons of such spacetimes.

We will show below that any `closed' surface ${\cal{H}}$ orthogonal to $\chi^a$, on which $\chi^a$ is null (i.e., $\beta^2=0$), is a Killing
horizon of a stationary axisymmetric spacetime. We note here that since $\chi^a$
is null over ${\cal{H}}$, it is evident that ${\cal{H}}$ is a null surface of dimension three.
In order to show that ${\cal{H}}$ is a Killing horizon, we will construct a congruence of null geodesics over ${\cal{H}}$ and consider the Raychaudhuri equation. But
before we do that we need to construct null geodesics over ${\cal{H}}$ explicitly.

 To construct this, we first note that the normal to the $\beta^2=0$ surface
is $\nabla_a \beta^2$. Also since the vector field $\chi^a$
is hypersurface orthogonal, Eq.~(\ref{sa4}), we may take the
 following ansatz for it~\cite{Gourgoulhon:2005ng} on the
$\beta^2=0$ surface
\begin{eqnarray}
\chi_a=e^{\rho}\nabla_a u,
\label{k}
\end{eqnarray}
where $\rho$ and $u$ are differentiable functions on that surface. This ansatz satisfies the Frobenius condition, Eq.~(\ref{sa4}), identically. 

Using Eq.~({\ref{k}}) and the torsion free condition $\nabla_{[a}\nabla_{b]}u=0$, we now compute
\begin{eqnarray}
\nabla_{[a}\chi_{b]}=\chi_{[b}\nabla_{a]}\rho,
\label{cral24}
\end{eqnarray}
which can be rewritten as 
\begin{eqnarray}
2\nabla_a\chi_b= \chi_{[b}\nabla_{a]}\rho+\phi_{(a}\nabla_{b)}\alpha,
\label{cral25f'}
\end{eqnarray}
using Eq.~(\ref{sa5}). We contract this equation by $\chi^b$.
We note that by the commutativity of the two Killing fields
\begin{eqnarray}
\pounds_{\chi}\alpha=\pounds_{\xi}\alpha+\alpha\pounds_{\phi}\alpha= -\pounds_{\xi}\left\{\frac{\xi\cdot\phi}{\phi\cdot\phi}\right\}-\alpha\pounds_{\phi}\left\{\frac{\xi\cdot\phi}{\phi\cdot\phi}\right\}=0,
\label{dalpha}
\end{eqnarray}
identically. This, along with the orthogonality $\chi_a\phi^a=0$ give over any $\beta^2=0$ surface ${\cal{H}}$,    
\begin{eqnarray}
2\chi^b\nabla_a \chi_b=
\nabla_a\left({\chi_b\chi^b}\right)=-\nabla_a\beta^2=-2\kappa \chi_a,
\label{k1}
\end{eqnarray}
where $\kappa:=-\frac12\pounds_{\chi}\rho$ is a function over ${\cal{H}}$. Since $\chi^a$ is null over ${\cal{H}}$, Eq.~(\ref{k1}) shows that the 1-form 
$\nabla_a \beta^2$, which is normal to the $\beta^2=0$ surface ${\cal{H}}$, is also null on that surface
and linearly dependent with $\chi_a$. We also note that since $\nabla_a\beta^2$ satisfies the Frobenius condition of hypersurface orthogonality, the 3-dimensional null surface ${\cal{H}}$ is a hypersurface. 

Now we take the Lie derivative of Eq.~(\ref{k1}) with respect to $\chi^a$,
\begin{eqnarray}
\chi^a\nabla_a\nabla_b\beta^2+\nabla_a\beta^2(\nabla_b\chi^a)=-2(\pounds_{\chi}\kappa) \chi_b-2\kappa(\pounds_{\chi}g_{ab})\chi^a.
\label{cral26}
\end{eqnarray}
Let us first consider the left hand side of the above equation. We rewrite this as
\begin{eqnarray}
\chi^a\nabla_a\nabla_b\beta^2+\nabla_a\beta^2(\nabla_b\chi^a)=\nabla_{[a}\nabla_{b]}\beta^2+\nabla_{b}\left(\pounds_{\chi} \beta^2\right).
\label{cral27}
\end{eqnarray}
The first term on the right hand side of Eq.~(\ref{cral27}) is zero by the torsion free condition. On the other hand, using Eq.~(\ref{sa5}) and the orthogonality of $\chi_a$ and $\phi_a$, we have 
\begin{eqnarray}
\pounds_{\chi} \beta^2=-\chi^a\chi^b\pounds_{\chi}g_{ab}=-\chi^a\chi^b\left(\phi_{(a}\nabla_{b)}\alpha\right)=0 
\label{cral27'}
\end{eqnarray}
identically, so that the left hand side of Eq.~(\ref{cral26}) vanishes. Let us now consider the right hand side of Eq.~(\ref{cral26}). We have using Eq.~(\ref{sa5}) 
\begin{eqnarray}
(\pounds_{\chi}g_{ab})\chi^a=\left(\phi_a\nabla_b\alpha+\phi_b\nabla_a\alpha\right) \chi^a=0,
\label{cral28}
\end{eqnarray}
where we have used the orthogonality of $\chi_a$ and $\phi_a$ and the fact that
$\pounds_{\chi}\alpha=0$. So Eq.~(\ref{cral26}) gives
\begin{eqnarray}
\pounds_{\chi}\kappa=0,
\label{cral29}
\end{eqnarray}
over ${\cal{H}}$.

Let us now define a null geodesic $k^a$ over the $\beta^2=0$ surface ${\cal{H}}$
in the following way~\cite{Wald:1984rg}. Let $k^a:=e^{-\kappa \tau}\chi^a$,
so that $k_ak^a=0$, where $\tau$ is the parameter along $\chi^a$, defined so that $\chi^a\nabla_a \tau=1$. Then we have 
\begin{eqnarray}
k^a\nabla_ak_b&=&e^{-2\kappa\tau}\left[\chi^a\nabla_a\chi_b-\chi_b\pounds_{\chi}(\kappa \tau)\right]\nonumber\\
&=&e^{-2\kappa\tau}\left[\chi^a(-\nabla_b\chi_a+\phi_{(a}\nabla_{b)}\alpha)-\kappa\chi_b(\pounds_{\chi} \tau)\right]=0,
\label{cral210'}
\end{eqnarray}
using Eq.s~(\ref{sa5}), (\ref{cral29}), (\ref{k1}), (\ref{dalpha}) along with the orthogonality $\chi^a\phi_a=0$. We know that a null vector field can be thought of as orthogonal to itself since it has vanishing norm. Thus $\chi^a$ is tangent to ${\cal{H}}$ as well. Then since $k^a$
is proportional to $\chi^a$, Eq.~(\ref{cral210'}) says that it is a null geodesic over ${\cal{H}}$.   

Now we are ready to consider the Raychaudhuri equation for the null geodesics $\{k^a\}$ \cite{Wald:1984rg, Hawking:1973uf} over ${\cal{H}}$, 
\begin{eqnarray}
\frac{d\theta}{ds}=-\frac{1}{2}\theta^2
-\hat{\sigma}_{ab}\hat{\sigma}^{ab}+\hat{\omega}_{ab}
\hat{\omega}^{ab}-R_{ab}k^ak^b,
\label{k3}
\end{eqnarray}
where $s$ is an affine parameter along a geodesic,
$\theta$, $\hat{\sigma}_{ab}$ and $\hat{\omega}_{ab}$
are respectively the expansion, shear and rotation of the geodesics over ${\cal{H}}$ 
\begin{eqnarray}
\theta=\hat{h}^{ab}\widehat{\nabla_a k_b},\quad
\hat{\sigma}_{ab}=\widehat{\nabla_{(a} k_{b)}}-\frac{1}{2}\theta
\hat{h}_{ab},\quad
\hat{\omega}_{ab}=\widehat{\nabla_{[a} k_{b]}}.
\label{expressions}
\end{eqnarray}
 The `hat' over the tensors denotes that they are defined on a spacelike 2-plane orthogonal to 
 $k^a$ and $\hat{h}_{ab}$ is the induced metric on this plane. Since ${\cal{H}}$ is a 3-dimensional surface, it is clear that the spacelike 2-plane is a subspace of ${\cal{H}}$.

In order to solve the Raychaudhuri equation (\ref{k3}), we first have to determine $\theta$, $\hat{\sigma}_{ab}$ and $\hat{\omega}_{ab}$ for our spacetime. Using the definition of the null geodesic $k^a$ and using
%
%\begin{eqnarray}
%k_{[a}\nabla_{b]}k_c=e^{-2\kappa \tau}
%\left[\chi_{[a}\nabla_{b]}\chi_{c}- 
%\chi_{c}\chi_{[a}\nabla_{b]}\left(\kappa \tau\right)
%\right],
%\label{k2}
%\end{eqnarray}
%
Eq.s~(\ref{sa5}) and (\ref{sa4}) we compute over ${\cal{H}}$,
\begin{eqnarray}
k_{[a}\nabla_{b]}k_c=e^{-2\kappa \tau}
\left[\frac{1}{2}\chi_{(a}\phi_{b}\nabla_{c)}\alpha-
\chi_c\nabla_a\chi_b-\chi_b\phi_a\nabla_c \alpha
-\chi_b\phi_c\nabla_a \alpha-
\chi_{c}\chi_{[a}\nabla_{b]}\left(\kappa \tau\right)
\right]. \nonumber \\
\label{k2}
\end{eqnarray}
 Let us choose the basis of the spacelike 2-plane tangent to ${\cal{H}}$ as $\phi^a$ and some $X^a$, with $\phi_aX^a=0$.
The appearance of $\phi^a$ as a basis vector of these 2-planes is guaranteed due to fact that ${\cal{H}}$ is by definition a `closed' surface. The induced metric $\hat{h}_{ab}$ over this spacelike 2-plane is given by
\begin{eqnarray}
\hat{h}_{ab}=f^{-2}\phi_a\phi_b+X^{-2}X_aX_b,
\label{cral210}
\end{eqnarray}
where $X^2$ is the norm of $X^a$.

We are now ready to compute $(\theta,~\hat{\omega}_{ab},~
\hat{\sigma}_{ab})$ given in Eq.~(\ref{expressions}). Let us first contract 
Eq.~(\ref{k2}) by the inverse induced metric $\hat{h}^{ab}=f^{-2}\phi^a\phi^b+X^{-2}X^aX^b$. We recall that the commutativity of the two Killing fields implies that $\pounds_{\phi}\alpha=0$ everywhere, Eq.~(\ref{cral23}). We also use the orthogonality of $(\phi^a,~\chi^a)$ and $(X^a,~\chi^a)$ to obtain
\begin{eqnarray}
k_a(f^{-2}\phi^b\phi^c+X^{-2}X^bX^c)\nabla_b k_c=0,
\label{adt6}
\end{eqnarray}
which shows that the expansion $\theta$ given in Eq.~(\ref{expressions}) vanishes. Similarly, by contracting Eq.~(\ref{k2}) by $\phi^{[b}X^{c]}$, we find 
\begin{eqnarray}
k_a\phi^{[b}X^{c]}\nabla_b k_c=0,
\label{adt7}
\end{eqnarray}
which shows that the components of the rotation $\hat{\omega}_{ab}$ also vanish. However if we 
contract Eq.~(\ref{k2}) by $\phi^{(b}X^{c)}$, we see that the components of the
 shear $\hat{\sigma}_{ab}$ does not vanish
\begin{eqnarray}
k_a \phi^{(b}X^{c)}\nabla_{b}k_c=
\frac{1}{2}e^{-\kappa \tau}\phi^{(b}X^{c)}
\phi_c\left(\nabla_b \alpha\right) k_a.
\label{shear}
\end{eqnarray}
Eq.~(\ref{shear}) and Eq.~(\ref{expressions}) give
\begin{eqnarray}
\hat{\sigma}_{ab}=\widehat{\nabla_{(a} k_{b)}}=\frac{1}{2}e^{-\kappa \tau}
\phi_{(a} \widehat{\nabla}_{b)}\alpha,
\label{cral211}
\end{eqnarray}
where we have used $\theta=0$. With these and
using Einstein's equations, Eq.~(\ref{k3}) over ${\cal{H}}$ becomes
\begin{eqnarray}
8\pi G\left[T_{ab}-\frac{1}{2}Tg_{ab}\right]k^ak^b=-\frac{1}{2}e^{-2\kappa \tau}
f^2\left(\widehat{\nabla}_a\alpha\right)
\left(\widehat{\nabla}^a\alpha\right).
\label{k4}
\end{eqnarray}
According to our assumption, there is no naked curvature singularity anywhere
in our region of interest. So the invariants constructed from the energy-momentum tensor must be bounded everywhere. Also since $k^a$ is null over ${\cal{H}}$, this implies that the second term on the left hand side of Eq.~(\ref{k4}) is zero. Since the inner product on the right hand side of Eq.~(\ref{k4}) is spacelike, we finally find that the null energy condition is violated over ${\cal{H}}$ 
\begin{eqnarray}
T_{ab}k^ak^b\Big\vert_{\cal{H}}\leq0\Rightarrow T_{ab}\chi^a\chi^b\Big\vert_{\cal{H}}\leq0.
 \label{k4'}
\end{eqnarray}
Since $\chi^a$ is timelike outside ${\cal{H}}$, this also implies by
continuity, the violation of the weak energy condition (WEC) outside ${\cal{H}}$. But by our assumption we are not violating WEC, so we have a contradiction unless
\begin{eqnarray}
 \left(\widehat{\nabla}_a\alpha\right)
\left(\widehat{\nabla}^a\alpha\right)
=0,
\label{adt8}
\end{eqnarray}
 over the spacelike section of ${\cal{H}}$. On the other hand, we recall that 
since $\chi_a$ is a null normal to ${\cal{H}}$, it is also 
tangent to it. This, along with Eq.s~(\ref{adt8}), (\ref{dalpha}) imply that $\alpha$
is indeed a constant over any 3-dimensional $\beta^2=0$ surface ${\cal{H}}$. Thus when $\beta^2=0$, the vector field $\chi^a=\xi^a+\alpha \phi^a$ coincides
with a Killing field and hence the horizon or horizons we have defined are Killing horizons. This is actually an old result \cite{Carter:69}, which we have rederived using a different method.

After this necessary digression, we are now ready to find the existence criterion for the cosmological horizon. Using the Killing identities $\nabla_{a}\nabla^a\xi_b=
-R_{b}{}^{a}\xi_a$, and $\nabla_{a}\nabla^a\phi_b=-R_{b}{}^{a}\phi_a$, and also the
orthogonality $\chi_a\phi^a=0$, we obtain
\begin{eqnarray}
\chi^b\nabla_a\nabla^a\chi_b
=-R_{ab}\chi^a\chi^b
+2\chi^a\left(\nabla_c\phi_a\right)\left(\nabla^c \alpha\right)\,,
\label{sa6}
\end{eqnarray}
which is equivalent to
\begin{eqnarray}
\nabla_a\nabla^a\beta^2
=2R_{ab}\chi^a\chi^b-2\left(\nabla^c\chi^a\right)\left(\nabla_c\chi_a\right)
-4\chi^a\left(\nabla_c\phi_a\right)\left(\nabla^c \alpha\right).
\label{sa7}
\end{eqnarray}
In order to simplify Eq.~(\ref{sa7}) in terms of the norms and derivatives of the norms of various vector fields, we have to find the expressions for $\nabla_a\chi_b$ and $\nabla_a \phi_b$. We find easily from Eq.~(\ref{sa5}) and the Frobenius condition (\ref{sa4}) that,
\begin{eqnarray}
\nabla_a\chi_b=
\beta^{-1}\chi_{[b}\nabla_{a]} \beta+\frac{1}{2}\phi_{(a}\nabla_{b)} \alpha.
\label{adt9}
\end{eqnarray}
Next we note that the subspace spanned by
$\left\{\chi^a,~\mu^a,~\nu^a \right\}$ do not form a hypersurface.
This is because, as we have mentioned earlier, the necessary and sufficient condition that an
arbitrary subspace of a manifold forms an integral submanifold or a
hypersurface is the existence of a Lie algebra of the basis vectors
of that subspace. The condition in Eq.~(\ref{sa4}) followed from this. On the
other hand, Lie brackets among $\left\{\chi^a,~\mu^a,~\nu^a
\right\}$ do not close. For example,
\begin{eqnarray}
[\chi,~\mu]^a=[\xi,~\mu]^a+\alpha[\phi,~\mu]^a
+\phi^a \left(\mu^b\nabla_b \alpha\right).
\label{sa10}
\end{eqnarray}
Since $\mu^a$ is not a Killing field, the last term on the right
hand side of Eq.~(\ref{sa10}) is not zero. A similar argument holds
for $\nu^a\,$ also.  Therefore the vectors spanned by
$\left\{\chi^a,~\mu^a,~\nu^a\right\}$ do not form a Lie
algebra. This implies that we cannot write a Frobenius condition like
$\phi_{[a}\nabla_b\phi_{c]}=0$.

However, according to our assumptions, there are integral spacelike
2-submanifolds orthogonal to both $\chi^a$ and $\phi^a$. These are
spanned by $\left\{\mu^a,~\nu^a\right\}$ which form a Lie algebra, Eq.~(\ref{adt3}). We project Eq.~(\ref{adt3}) over $\Sigma$ via the projector $h_{a}{}^{b}=\delta_{a}{}^{b}+\beta^{-2}\chi_a\chi^b$, use the orthogonalities $\chi^a\mu_a=0=\chi^a\nu_a$ and the fact that
for the spacelike vector fields $\mu^a$ and $\nu^a$,
\begin{eqnarray}
h^{a}{}_b\mu^b=\mu^a,\quad h^{a}{}_b\nu^b=\nu^a,
\label{adt10f}
\end{eqnarray}
to obtain
\begin{eqnarray}
\left(\mu^b\nabla_b\nu^a-\nu^b\nabla_b \mu^a\right)=\left(\mu^b D_b\nu^a-\nu^bD_b \mu^a\right)=g_{1}(x)\mu^a+g_2(x)\nu^a,
\label{adt10}
\end{eqnarray}
where $D_a$ is the induced connection over $\Sigma$ : $D_a\equiv h_{a}{}^{b}\nabla_b$.
The action of $h_{a}{}^b$ and $D_a$ can be defined exactly in the same way as what we did for the static spacetime.
We contract Eq.~(\ref{adt10}) with $\phi_a$. Then the arguments similar to which gave Eq.s~(\ref{adt4}), (\ref{adt4l}) now yield over $\Sigma$
\begin{eqnarray}
\mu^a\nu^bD_{[a}\phi_{b]}=0\Rightarrow D_{[a}\phi_{b]}=\rho_{[a}\phi_{b]},
\label{adt10f}
\end{eqnarray}
where $\rho_{a}$ is an arbitrary 1-form over $\Sigma$. Antisymmetrizing with $\phi_c$, we obtain a Frobenius-like condition over $\Sigma$ :
\begin{eqnarray}
\phi_{[a}D_{b}\phi_{c]}=0.
\label{sa11}
\end{eqnarray} 
Precisely, the above equation shows a foliation of $\Sigma$ into the family of spacelike
2-submanifolds spanned by $\{\mu^a,~\nu^a\}$, orthogonal to $\phi^a$. We now compute
\begin{eqnarray}
D_a\phi_b=h_{a}{}^{c}h_{b}{}^{d}
\nabla_c\phi_d
=\nabla_a \phi_b +\beta^{-2}
\left(\chi_a \phi^c\nabla_b \chi_c
-\chi_b \phi^c\nabla_a \chi_c \right).
\label{sa12}
\end{eqnarray}
Using the expression of $\nabla_a\chi_b$ given in Eq.~(\ref{adt9}), we can rewrite this as
\begin{eqnarray}
D_a\phi_b=\nabla_a\phi_b+\frac{f^2}{2\beta^2}
\chi_{[a} \nabla_{b]} \alpha.
\label{sa13}
\end{eqnarray}
It follows from Eq.~(\ref{sa13}) that we can write the Killing
equation for $\phi_a$ over $\Sigma$ as
\begin{eqnarray}
D_a\phi_b + D_b\phi_a=0.
\label{sa14}
\end{eqnarray}
Using this equation and the Frobenius-like condition of Eq.~(\ref{sa11}) we find that
\begin{eqnarray}
D_a\phi_b=f^{-1}\left(\phi_b D_af-\phi_a D_bf\right).
\label{sa14f}
\end{eqnarray}
 Finally, substituting this expression into Eq.~(\ref{sa13}) we arrive at the following
\begin{eqnarray}
\nabla_a\phi_b=\frac{1}{f}\phi_{[b} D_{a]} f+
\frac{f^2}{2\beta^2}\chi_{[b} \nabla_{a]} \alpha.
\label{sa16}
\end{eqnarray}
These are all that is needed to simplify Eq.~(\ref{sa7}). We
Substitute Eq.s~(\ref{adt9}), (\ref{sa16}) into Eq.~(\ref{sa7}), note that $\pounds_{\phi}\alpha=0=\pounds_{\chi}\alpha$ (Eq.s (\ref{cral23}), (\ref{dalpha})), $\chi^aD_af=0$ since $D_af$ is spacelike, and also the orthogonality $\chi^a\phi_a=0$, to find
\begin{eqnarray}
\nabla_a\nabla^a \beta^2=2R_{ab}\chi^a\chi^b+
4\left(\nabla_a\beta\right)
\left(\nabla^a\beta\right)
+f^2\left(\nabla_a 
\alpha\right)\left(\nabla^a \alpha\right).
\label{sa17}
\end{eqnarray}
 Eq.s (\ref{cral27'}) and (\ref{dalpha}) also imply that $\nabla_a\beta=D_a\beta$ and $\nabla_a\alpha=D_a\alpha$. With this, using the same line of arguments as for Eq.~(\ref{s21}), we get
\begin{eqnarray}
\int_{\Sigma}\beta^{n+1}
\left[X^{\rm{N}}+ \frac{(n+2\beta)}{\beta}
\left(D_a \beta\right)
\left( D^a \beta\right)
+\frac{f^2}{2}\left(D_a \alpha\right)
\left(D^a \alpha\right) 
 -\Lambda\beta^2\right]=0,
\label{sa18}
\end{eqnarray}
if the spacetime has an outer or cosmological Killing horizon. If we set $\alpha=0$, we recover the static case of Eq.~(\ref{s21}).

$X^{\rm N}=0$ in Eq.~(\ref{sa18}), corresponds to the Kerr-de Sitter solution
\cite{Carter:1968ks}. If $X^{\rm N}$ corresponds to the Maxwell field, Eq.~(\ref{sa18}) corresponds to the Kerr-Newman-de Sitter solution \cite{Carter:1968ks}. We note that the assumption of the existence of integral 2-submanifolds orthogonal to both the Killing fields $\xi^a$ and $\phi^a$ was crucial to this proof. For a completely general
stationary axisymmetric spacetime such submanifolds may not exist, and thus the existence of an outer horizon is not guaranteed in such cases, even for $\Lambda >0$.

%%%%%%%%%%%%%%%%%%%%%%%%%%%%%%%%%%%%%%%

%%%%%%%%%%%%%%%%%%%%%%%%%%%%%%%
We now summarize the results obtained in this Chapter as follows. For general static or stationary axisymmetric
spacetimes, an outer or cosmological Killing horizon exists only if
$R_{ab}n^{a}n^{b}< 0$ for a hypersurface orthogonal timelike
$n^{a}$, at least over some portion of the region of interest of
the manifold. This implies the violation of the strong energy
condition by the matter fields. The violation of the SEC can be achieved
either through a positive $\Lambda$, for which there is strong observational evidence~\cite{Riess:1998cb, Perlmutter:1998np}, or through some exotic matter.

Now we consider the Raychaudhuri equation for the timelike geodesics $\{u^a:~u^a\nabla_au^b=0, ~u^au_a=-1\}$ \cite{Wald:1984rg, Hawking:1973uf},  
\begin{eqnarray}
\frac{d\theta}{ds}=-\frac{1}{3}\theta^2
-\sigma_{ab}\sigma^{ab}+\omega_{ab}
\omega^{ab}-R_{ab}u^au^b.
\label{adt11}
\end{eqnarray}
As before, $\theta$, $\sigma_{ab}$ and $\omega_{ab}$ are respectively the expansion, shear and rotation for the timelike geodesics given by 
\begin{eqnarray}
\theta=h^{ab}\nabla_a u_b,\quad
\sigma_{ab}=\nabla_{(a}u_{b)}-\frac13\theta h_{ab},\quad
\omega_{ab}=\nabla_{[a}u_{b]},
\label{adt11'}
\end{eqnarray}
where $h^{ab}$ is the inverse of the induced metric $h_{ab}=g_{ab}+u_au_b$ over a spacelike 3-plane orthogonal to $\{u^a\}$. If in addition we assume that those 3-planes are hypersurfaces, we have the Frobenius condition $u_{[a}\nabla_b u_{c]}=0$, contracting which by $u^a$ we find 
\begin{eqnarray}
\nabla_{[b}u_{c]}=\omega_{bc}=0.
\label{adt11''}
\end{eqnarray}
We also note from Eq.~(\ref{adt11'}) that $\sigma_{ab}$ is spacelike,
\begin{eqnarray}
u^a\sigma_{ab}=0=u^b\sigma_{ab},
\label{adt11'''}
\end{eqnarray}
so that the second term on the right hand side of Eq.~(\ref{adt11}) is non-negative.
By Einstein's equations (\ref{s19})
we have $R_{ab}u^au^b=8\pi G\left(T_{ab}-\frac{1}{2}Tg_{ab}\right)u^au^b$. We can as before split $T_{ab}$ into two parts, one satisfying SEC and the other violating it, having contributions of opposite signs in Eq.~(\ref{adt11}). We denote them by $X^{\rm N}$ and $-X^{\rm AN }$ respectively with $ X^{\rm N},~X^{\rm AN }\geq 0$. Putting in these all together we find that Eq.~(\ref{adt11}) implies
\begin{eqnarray}
\frac{d\theta}{ds}+\frac{1}{3}\theta^2+X^{\rm N}-X^{\rm AN }\leq0.
\label{adt11f}
\end{eqnarray}
We consider first $X^{\rm AN }=0$, then we have always ${\displaystyle\frac{d\theta}{ds}\leq0}$. In other words the geodesics will either remain parallel or converge with increasing $s$. On the other hand Eq.~(\ref{adt11f}) shows that the inclusion of $X^{\rm AN }$ (due to a positive $\Lambda$ for example) would decrease the convergence rate, even ${\displaystyle\frac{d\theta}{d s}}$ may be positive.
Thus a positive $\Lambda$ repels geodesics. We will demonstrate this repulsive effect explicitly in Chapter 4.  

An interesting question in this context which we have not tried to answer is : how does an outer or cosmological horizon form? More precisely, we know that attractive gravity or collapse can form the black hole event horizon; similarly, can the repelled outgoing geodesics form an outer horizon? Can we suggest a clear mechanism for forming such outer horizons? This may have relevance in the discussion of non-stationary spacetimes with outer horizons or particle creation in such spacetimes.

Finally we emphasize that the existence of an outer horizon implies that positive $\Lambda$ must dominate the integrals in Eq.s~(\ref{s21}), (\ref{sa18}). Since observations suggest that $\Lambda$ is significantly larger than the normal
matter density of our universe \cite{Weinberg:2008zzc}, it is very likely that our universe is indeed endowed with an outer boundary. This motivates us to further study spacetimes with outer horizons.

%%%%%%%%%%%%%%%%%%%%%%%%%%%%%%%%%%%%%%%%%%%%%%%%%%%%%%%%%%%%%%%%%%%%%%%%%%%%
\chapter{Black hole no hair theorems}

%%%%%%%%%%%%%%%%%%%%%%%%%%%%%%%%%%%%%%%%%%%%%%%%%%%
%\Section{Attachment of monopoles to the flux tube for confinement}
%\label{monopole_attachment}
In the following we will use the geometrical setup and results derived in the previous Chapter to prove no hair theorems for black holes in a de Sitter universe. We have already reviewed
this topic in Section 1.3. As we have mentioned earlier, a lot of effort to prove these theorems corresponding to various matter fields have been given for asymptotically flat spacetimes~\cite{Bekenstein:1971hc, Adler:1978dp, Lahiri:1993vg}. On the other hand, 
Price's theorem~\cite{Price:1971gc}, a perturbative no hair theorem stating that only static solutions to massless wave equations with spin ${\displaystyle s=0,~\frac{1}{2},~1,~\frac{3}{2},~2 }$~for a spherically symmetric black hole background have an angular momentum less than $s$, was proved for $\Lambda> 0$ some years ago~\cite{Chambers:1994sz}.
But no non-perturbative version of this theorem about the existence of static or stationary matter fields has been established for spacetimes with a positive $\Lambda.$ In this Chapter we will establish
classical no hair theorems for various matter fields in such spacetimes. 

We will consider two kind of spacetimes $-$ one static and spherically symmetric and the other is stationary and axisymmetric. However, we will see later that the assumption of spherical symmetry is not required for most of the proofs for the static spacetime. 

 We will use the geometrical setup developed in the previous Chapter to carry out these proofs. We consider only the region between the black hole horizon
and the cosmological horizon, and hence ignore the asymptotic behavior of both
the metric and matter fields. We will see that except for the stationary axisymmetric spacetimes, we do not have to explicitly use any
equations for the metric like Einstein's equations at all, beyond assuming the existence of a
cosmological horizon. We find that it is possible to extend most of the well
known no hair theorems to black holes in a universe with
$\Lambda>0\,.$ We also find one clear exception, that of the
Abelian Higgs model, which indeed shows that the existence of the outer boundary, i.e. the cosmological horizon may even change the local physics considerably.

%%%%%%%%%%%%%%%%%%%%%%%%
\section{No hair theorems for static spacetimes}
%%%%%%%%%%%%%%%%%%%%%%%%%%%%%  
Let us start with a static and spherically symmetric spacetime.
 By a static black hole spacetime with $\Lambda> 0$ we will mean a spacetime with two
Killing horizons, between which there is a timelike Killing vector field
$\xi^a$ orthogonal to a family of spacelike hypersurfaces
$\Sigma$. So $\xi^a$ satisfies the Frobenius condition of hypersurface orthogonality stated in Eq.~(\ref{s2'}). By spherical symmetry we mean that the spacetime can be foliated by 2-spheres.
The norm $\lambda^2(r) = - \xi^{a}\xi_{a}$ vanishes at two
values $r_{\rm{H}}< r_{\rm{C}}$ of $r\,,$ which is a suitable radial coordinate.
We will call $r_{\rm{H}}$ and $r_{\rm{C}}$ the black hole and the cosmological horizon respectively. The spacetime manifold is divided into three regions. The region $r< r_{\rm{H}}$ is the black hole region and may contain a
spacetime curvature singularity. The points of this region do not lie to the
past of $\Sigma$, for which $r_{\rm{H}} < r < r_{
\rm{C}}$, while the points of $\Sigma$ do not lie to the future of the region $r > r_{
\rm{C}}\,$ (see the discussions of Chapter 1). We will not be concerned with the world beyond the cosmological horizon $r>r_{\rm C}$, or the world inside the black hole horizon $r<r_{\rm H}$. So the
asymptotic behavior of the metric or matter fields will not be relevant to our
computations. In particular, apart from the assumption of the existence of the outer or the cosmological horizon, we do not assume the metric to be asymptotically de Sitter or Schwarzschild-de Sitter. The region of our interest is $r_{\rm H}\leq r\leq r_{\rm C}$ and we assume that there is no naked curvature singularity anywhere in this region.

The various no hair theorems will be taken to be statements about
the corresponding classical fields on the spacelike hypersurfaces
$\Sigma$ between the two horizons. We will not not look for explicit solutions
of matter or Einstein's equations, but will only prove general statements about their existence. 

Since we have assumed that there is no curvature singularity in the region $r_{\rm H}\leq r\leq r_{\rm C}$, and since curvature is related to the energy-momentum tensor, the scalar invariants constructed out of the energy-momentum tensor are bounded everywhere in this region. 

Let $X$ be a Killing field of the spacetime, then $\pounds_{X}g_{ab}=0$. Then, since the curvature tensors are computed from the metric functions, we have 
$\pounds_{X}R_{ab}=0=\pounds_{X}R$. So the Einstein equations show $\pounds_{X}T_{ab}=0$. 
A matter field which appears in the energy-momentum tensor is a physical matter field.
Since we will not neglect backreaction, we assume that any physical matter field also obeys the symmetry of the spacetime itself, because otherwise the energy-momentum tensor may itself destroy that symmetry. So if $Y$ is any physical matter field or a component of it, we will impose
\begin{eqnarray}
\pounds_{X}Y=0.
\label{adch31}
\end{eqnarray}
We note that the above arguments does not hold if $Y$ is a gauge field.

Our main goal in order to prove the no hair theorems will be to project any matter equation of motion onto the spacelike hypersurfaces $\Sigma$ and to construct vanishing positive definite quadratic volume integrals. In order to do this we will use the induced metric or the projector $h_{a}{}^{b}$ described in the previous Chapter.

%Although the calculations are on a spacelike hypersurface $\Sigma$
%orthogonal everywhere to a timelike Killing vector, it is
%convenient to use covariant notation without resorting to explicit
%coordinates. Let $h^{a}{}_{b}=\delta^{a}{}_{b}+\lambda^{-2}\xi^{a}\xi_{b}$ denote the projection
%tensor which projects vectors to $\Sigma$ and let
%$D_a$ denote the induced connection on
%$\Sigma$. Then for a rank $p$ antisymmetric tensor $\Omega$ whose
%Lie derivative with respect to $\xi^a$ vanishes,
%
%\begin{eqnarray}
%D_{a}\left(\lambda \omega^{a b \dots
%c}\right)=\lambda 
%\left(\nabla_{a}\Omega^{a b' \dots c'  
%}\right)h^{b}{}_{b'}\dots h^{c}{}_{c'}\,, 
%\label{formlemma}
%\end{eqnarray}
%
%where $\omega$ is the $\Sigma$-projection of $\Omega\,.$
%, and $\pounds_{\xi}\Omega=0$. 
%This is essentially the
%statement that the 4-divergence of $\Omega$ is the same as its
%3-divergence when both $\Omega$ and the metric are time
%independent. All our proofs will be based on this result.
%%%%%%%%%%%%%%%%%%%%
\subsection{Scalar field}
%%%%%%%%%%%%%%%%%%%%% 
We start with a real scalar field $\psi$ moving in a
potential $V(\psi)$,
\begin{eqnarray}
{\cal{L}}=-\frac 1 2 \nabla_{a}\psi\nabla^{a}\psi-V(\psi),
\label{adch32}
\end{eqnarray}
where any mass term is included in $V(\psi)$. The equation of motion for $\psi$ is
\begin{eqnarray}
\nabla_{a}\nabla^{a}\psi=V'(\psi),
\label{adch33}
\end{eqnarray}
where the `prime' denotes differentiation with respect to $\psi$. We will now project Eq.~(\ref{adch33}) onto the spacelike hypersurfaces $\Sigma$. We note that
since a non-vanishing $V(\psi)$ enters into the energy-momentum tensor which has a backreaction, we have $\pounds_{\xi} \psi=0$ everywhere (Eq.~(\ref{adch31})). Then exactly the same procedure which led to Eq.~(\ref{s14}) now yields the $\Sigma$-projection of Eq.~(\ref{adch33}), 
\begin{eqnarray}
D_{a}(\lambda D^{a}\psi) =
\lambda  V'(\psi),
\label{eqscal}
\end{eqnarray}
where $D_a$ is the induced connection over $\Sigma$ as defined in the previous Chapter.
We now 
multiply both sides of Eq.~(\ref{eqscal}) by $V'(\psi)$ and integrate by parts
over $\Sigma$ to obtain
\begin{eqnarray}
&& \int_{\partial \Sigma}\lambda
V'(\psi)n^aD_a\psi 
 + \int_{\Sigma}\lambda\left[V''(\psi) 
\left(D^{a}\psi\right) \left(D_{a}\psi\right)
+ V'^2(\psi)\right]=0,
\label{integrals}
\end{eqnarray}
where $\partial\Sigma$ corresponds to the boundary of $\Sigma$, which is the union of the black hole and the cosmological horizon where $\lambda(r)=0$. Since we have assumed the spacetime to be spherically symmetric, it is clear that $\partial \Sigma$ are two 2-spheres of radius $r_{\rm H}$ and $r_{\rm C}$ located at the respective horizons. $n^{a}$ is the $\Sigma$-ward pointing spacelike unit normal to these 2-spheres. Since $(D^{a}\psi)( D_a\psi$)
appears in the energy-momentum tensor $T_{ab}$, it must be bounded over the two horizons. We then use $\left(D_a\psi-n_a n^bD_b\psi\right)^2\geq0$ to have the Schwarz inequality
\begin{eqnarray}
|n^a D_a\psi|^{2}\leq
 \left(D^{a}\psi\right)
\left(D_{a}\psi\right).
\label{schtz}
\end{eqnarray}
For generic $V(\psi)$, the boundedness of invariants of $T_{ab}$ over $\partial\Sigma$ or the horizons implies that $\psi$ must also be bounded there. 
Also Eq.~(\ref{schtz}) shows that $n^a D_a\psi$ is bounded over the horizons. Thus it follows that the integral on $\partial\Sigma$ in Eq.~(\ref{integrals}) vanishes leaving us only with the vanishing volume integral
\begin{eqnarray}
\int_{\Sigma}\lambda\left[V''(\psi) 
\left(D^{a}\psi\right) \left(D_{a}\psi\right)
+ V'^2(\psi)\right]=0\,.\quad
\label{integrals2}
\end{eqnarray}
 Since $\Sigma$ is spacelike and $D_a$ is the induced connection over it,
$\left(D_a\psi\right)\left(D^a\psi\right)$ is a spacelike inner product i.e.,
 non-negative. So if $V(\psi)$ is convex i.e., if $V''(\psi)\geq 0$ for all values of $\psi$, 
Eq.~(\ref{integrals2}) shows that $\psi$ is a constant at its minimum
everywhere on $\Sigma\,,$ which is the no hair result. Since we have $\pounds_{\xi}\psi=0$ everywhere, $\psi$ remains constant throughout the spacetime. 
For $V(\psi)=0$, we can multiply the field equation over $\Sigma$ by $\psi$ and
insist that $\psi$ be measurable at the horizons, and the no hair
result follows. 

The proof in general does not apply to a non-convex $V(\psi)$. For example, a
real scalar field in the double well potential
${\displaystyle V(\psi)=\frac{\alpha}{4}(\psi^2-v^2)^2}$ can have a non-trivial
static solution in $\Sigma$ which may be unstable~\cite{Torii:1998ir}. Another interesting and not so obvious case is that of the conformal scalar with ${\displaystyle V(\psi)
= \frac{1}{12}R \psi^2}$. The part of the action containing
$\psi$ is invariant under local conformal transformations $\psi\to \omega^2(x)\psi$. Then it
turns out that the conservation equation $\nabla_a T^{ab}=0$ is also conformally invariant
with $T=T_{ab}g^{ab}=0$ \cite{Wald:1984rg}. 
 So by appropriately choosing the conformal factor of the transformation we can make $\psi$ or $n^aD_a\psi$ diverge at $\partial\Sigma$ in Eq (\ref{integrals}) without causing a curvature singularity. Then the $\partial\Sigma$ integral can be non-zero, which allows a non-trivial
configuration of $\psi$ on $\Sigma$. In fact a static spherically symmetric solution with conformal scalar hair with $\Lambda>0$ is known~\cite{Martinez:2002ru}. 
 The proof also will not apply to scalars with a kinetic term of the wrong sign, as in
phantom models of dark energy~\cite{Izquierdo:2005ku}. Of course,
in such models a static black hole may not form in the first place, and a statement of no hair theorems may not be possible. 

%%%%%%%%%%%%%%%%%%%
\subsection{Massive vector field}
%%%%%%%%%%%%%%%%%%%%%%%%%%
Let us now consider the Proca-massive vector field $A^b$ with the Lagrangian 
\begin{eqnarray}
\mathcal{L} = -\frac{1}{4} F_{ab}F^{ab} - \frac{1}{2} m^2
A_{b}A^{b}\,,
\label{adch34} 
\end{eqnarray}
where $F_{ab}=\nabla_{[a}A_{b]}$. The equation of motion for $A_b$ is
\begin{eqnarray}
\nabla_{a}F^{ab}=m^2A^b.
\label{eompr} 
\end{eqnarray}
Let us define the potential $\psi$  and the electric field $e^{a}$ as
\begin{eqnarray}
\psi:=\lambda^{-1}\xi_{a} A^{a}\quad
e^a:=\lambda^{-1}\xi_{b}F^{ab}.
\label{adch35} 
\end{eqnarray}
Using these definitions we compute the following
\begin{eqnarray}
\lambda e_a =\xi^bF_{ab}=\nabla_{a}\left(\xi^bA_b\right)-\left\{(\nabla_a\xi^b)A_b+\xi^b\nabla_b A_a\right\}=\nabla_a(\lambda \psi)-\pounds_\xi A_a.
\label{elecdef1}
\end{eqnarray}
We now project Eq.~(\ref{elecdef1}) onto $\Sigma$ via the projector $h_{a}{}^{b}=\delta_{a}{}^{b}+\lambda^{-2}\xi_a\xi^b$. First we note from the definition (\ref{adch35}) that  $\xi^ae_a=0$ identically, i.e. the electric field $e^a$ is spacelike. Also $A_b$ is a physical matter field which appears in the energy-momentum tensor, so by Eq.~(\ref{adch31}) we have $\pounds_{\xi}A_b=0$. Thus the first of the definitions (\ref{adch35}) gives us $\pounds_{\xi}(\lambda \psi)=0$ and we have    
\begin{eqnarray}
D_{a}(\lambda \psi)=h_{a}{}^{b}\nabla_{b}(\lambda \psi)=\nabla_{a}(\lambda \psi)+\lambda^{-2}\xi_a(\pounds_{\xi}(\lambda \psi))=\nabla_{a}(\lambda \psi).
\label{elecdef'}
\end{eqnarray}
Thus the $\Sigma$ projection of Eq.~(\ref{elecdef1}) reads
\begin{eqnarray}
D_a(\lambda \psi)=\lambda e_a.
\label{elecdef2}
\end{eqnarray}
Also from the definition of the electric field $e^a$ given in (\ref{adch35}) we compute
\begin{eqnarray}
 \nabla_a\left(\lambda e^a\right) =\left(\nabla_a\xi_{b}\right)F^{ab}+\xi_b\left(\nabla_a F^{ab}\right)=\left(\nabla_a\xi_{b}\right)F^{ab}+m^2\lambda\psi,
\label{elecdef3}
\end{eqnarray}
using Eq.~(\ref{eompr}). Substituting the expression for $\nabla_a\xi_b$ from Eq.~(\ref{s5'}) into Eq.~(\ref{elecdef3}), we obtain
\begin{eqnarray}
 \nabla_a \left(\lambda e^a\right) =\lambda^{-1}\left(\xi_b\nabla_a\lambda- \xi_a\nabla_b\lambda\right)F^{ab}+m^2\lambda\psi=2e^a\nabla_a\lambda+m^2\lambda\psi,
\label{elecdef4'}
\end{eqnarray}
which we rewrite as 
\begin{eqnarray}
 \nabla_ae^a =\lambda^{-1}e^a\nabla_a\lambda+m^2\psi.
\label{elecdef4}
\end{eqnarray}
Now we project this equation onto $\Sigma$ using the projector $h_{a}{}^{b}$. We find, using Eq.~(\ref{adch22}), 
\begin{eqnarray}
 D_ae^a =h^{a}{}_{b}\nabla_ae^b=\nabla_a e^a+\lambda^{-2}\xi_b \xi^a\nabla_a e^b.
\label{elecdef5}
\end{eqnarray}
Let us look at the second term of Eq.~(\ref{elecdef5}). Using the orthogonality $\xi_ae^a=0$, we rewrite this term as  
\begin{eqnarray}
\lambda^{-2}\xi_b \xi^a\nabla_a e^b=-\lambda^{-2}\xi^ae^b\nabla_a\xi_b=-\lambda^{-1}e^a\nabla_a\lambda,
\label{elecdef6}
\end{eqnarray}
where we have once again used the expression for $\nabla_a\xi_b$ from Eq.~(\ref{s5'}). Combining Eq.s~(\ref{elecdef4}), (\ref{elecdef5}) and (\ref{elecdef6}), we have the equation for $e^a$ over $\Sigma$,
\begin{eqnarray}
D_{a}e^{a}-m^2\psi=0\,.
\label{adch36}
\end{eqnarray}
Multiplying both sides of this equation by $\lambda\psi$, using Eq.~(\ref{elecdef2}) and integrating by parts over $\Sigma$, we find
\begin{eqnarray}
\int_{\partial \Sigma}\lambda \psi e^{a}n_{a} + \int_\Sigma
\lambda \left[e_a e^a + m^2\psi^2\right] =0,
\label{adch37}
\end{eqnarray}
where $\partial\Sigma$ denote two 2-spheres located at the two horizons and $n^a$ is the $\Sigma$-ward unit normal to $\partial\Sigma$ as before. Since both $\psi^2$ and $e_a e^a$ appear in the energy-momentum tensor, $\psi$ must be finite and by the Schwarz inequality
$e^a n_a$ is finite making the $\partial\Sigma$ integrals vanishing as before.
On the other hand, since $e^a$ is spacelike, the integrand in Eq.~(\ref{adch37}) is positive definite. This implies that $\psi = 0 = e_a\,$ on $\Sigma$. Also the vanishing Lie derivatives $\pounds_{\xi}\psi=0=\pounds_{\xi}e_a$ of the physical fields imply that they are zero throughout the spacetime. So this shows that the black hole does not carry any electric charge corresponding to the Proca-massive vector field.

We note that the above arguments for vanishing of the electric charge does not hold for the Maxwell field, $m=0$ in Eq.~(\ref{adch34}), for the following reason. For $m=0$, the Lagrangian (\ref{adch34}) has a gauge symmetry under the local gauge transformation $A\to A+d\chi(x)$, where $\chi(x)$ is any arbitrary differentiable function. This means that for $m=0$, the vector potential $A_b$ is not a physical quantity. In that case we cannot take $\pounds_{\xi}A_b=0$ in Eq.~(\ref{elecdef1}), because we can always make a local gauge transformation to make $A_b$ non-static. Thus the integral (\ref{adch37}) will then carry a term containing $\pounds_{\xi}A_b$ which may be positive or negative and the above arguments cannot be used there.

Let us now investigate the behaviour of the spacelike components of the vector field $A_b$.
We multiply Eq.~(\ref{eompr}) by the projector $h_{a}{}^{b}$ to write
\begin{eqnarray}
\lambda h^{b}{}_{c}\nabla_a F^{ac}=m^2\lambda a^b,
\label{adch38}
\end{eqnarray}
where $a^b$ is the $\Sigma$-projection of $A^b$ : $a^b=h^{b}{}_{a}A^a$.
We wish to relate Eq.~(\ref{adch38}) to the induced connection $D_a$ and the
projected or the `magnetic' field tensor $f_{ab}$ on $\Sigma$, 
\begin{eqnarray}
f_{ab}:=h^{c}{}_{a}h^{d}{}_{b} F_{cd}=D_{[a} a_{b]}.
\label{adch39}
\end{eqnarray}
In order to do this, we consider the 3-divergence
$D_{a}\left(\lambda f^{ab}\right)$. Using Eq.~(\ref{adch22}) we have
\begin{eqnarray}
D_{a}\left(\lambda f^{ab}\right)&=& h^{b}{}_{e}h^{f}{}_{a}\nabla_f
\left(\lambda F^{ae}\right) \nonumber \\
&=& h^{b}{}_{e} \nabla_a\left(\lambda
  F^{ae}\right)+\lambda^{-2}h^{b}{}_{e}\xi_a\xi^f\nabla_f
\left(\lambda F^{ae}\right). 
\label{adch310}
\end{eqnarray}
The Killing equation for $\xi_a$ imply $\pounds_{\xi}\lambda=0$. Also since $F_{ab}$ appears 
in the energy-momentum tensor we have by Eq.~(\ref{adch31}), 
\begin{eqnarray}
\pounds_{\xi}F^{ab}=0=\xi^c\nabla_cF^{ab}-F^{ac}\nabla_c\xi^b-F^{cb}\nabla_c\xi^a.  
\label{adch312}
\end{eqnarray}
Then Eq.~(\ref{adch310}) becomes 
\begin{eqnarray}
D_{a}\left(\lambda f^{ab}\right)=\lambda h^{b}{}_{e} \nabla_a F^{ae}+h^{b}{}_{e}F^{ae}\nabla_a \lambda
+\lambda^{-1}\xi_a h^{b}{}_{e}\left[F^{ce}\nabla_c
  \xi^a+F^{ac}\nabla_c\xi^e\right].
 %\nonumber\\
\label{adch311}
\end{eqnarray}
Substituting the expression for $\nabla_a\xi_b$ from Eq.~(\ref{s5'}) into Eq.~(\ref{adch311}), and using Eq.~(\ref{adch35}), we find
\begin{eqnarray}
D_{a}\left(\lambda f^{ab}\right)=\lambda h^{b}{}_{e} \nabla_a F^{ae}
+\lambda^{-1}\xi_a h^{b}{}_{e}\left[e^{[e}\nabla^{a]}\lambda+\lambda^{-1}F^{ac}\left(\nabla_c\lambda \right)\xi^e\right]=\lambda h^{b}{}_{e} \nabla_a F^{ae},
 \nonumber\\
\label{adch313}
\end{eqnarray}
where we have used $\pounds_{\xi}\lambda=0$, $\xi^ae_a=0$ and $\xi_ah^{a}{}_{b}=0$. Comparing Eq.s~(\ref{adch313}) and (\ref{adch38}) we finally arrive at the
equation of motion for the magnetic field, 
\begin{eqnarray}
D_{a}\left(\lambda f^{ab}\right) - m^2\lambda a^{b} = 0.
\label{procamag}
\end{eqnarray}
In a coordinate dependent language, Eq.~(\ref{procamag}) simply says that a 4-divergence should become a 3-divergence for time independent field. Multiplying both sides of this equation by $a_{b}$, using Eq.~(\ref{adch39}) for the projected tensor $f_{ab}$ and
integrating by parts over $\Sigma$, we obtain
\begin{eqnarray}
\int_{\partial \Sigma} \lambda
a_{b}f^{ab}n_{a}+\int_{\Sigma}\lambda\left(\frac{1}{2}
(f^{ab})^2+m^2 (a^{b})^2\right) = 0\,. 
\end{eqnarray}
Since both $a^{b}$ and $f_{ab}$ appear in $ T_{ab}$, these
must be regular, which ensures that the $\partial\Sigma$ integrals
vanish as before. On the other hand both $a_b$ and $f_{ab}$ are spacelike, which means that the second integral is over a sum of squares. So $a_b =
0 = f_{ab}$ on $\Sigma$. This along with the fact that the Lie derivatives of $a_b$ and $f_{ab}$ vanish along $\xi_a$ is the desired no hair result.

As we have mentioned earlier, the no hair result does not hold for the massless case due to the local gauge symmetry of the Lagrangian. In fact the generalization of the Reissner-N\"{o}rdstrom solutions with a positive cosmological constant is known and is given by Eq.~(\ref{s25i}).

%\vskip .6cm

There are two gauge-invariant Lagrangians which describe a massive Abelian gauge field. The no hair conjecture fails for both of these cases in the presence of a positive $\Lambda$, as we will see below.
%%%%%%%%%%%%%%%%%%%%%%%%%%%%%%%
\subsection{The $B\wedge F$ theory}
%%%%%%%%%%%%%%%%%%%%%%%%%%%%%%%%%%

The first mechanism we consider is described by the Lagrangian
\begin{eqnarray}
\mathcal{L}= -\frac14 F_{ab}F^{ab} -
\frac{1}{12}H_{abc}H^{abc}+\frac m4
\epsilon^{abcd}B_{ab}F_{cd},
\label{bflag}
\end{eqnarray}
where $B_{ab}$ is an antisymmetric tensor potential and
$H_{abc}= (\nabla_a B_{bc} + {\rm {cyclic}})$ is its field
strength, and $F_{ab}=\nabla_{[a}A_{b]}$ is the Maxwell field tensor. 
In addition to the local gauge symmetry of the Maxwell field, the above Lagrangian is also
invariant under the gauge transformation $B\to B+d\omega(x)$, where $\omega(x)$ is an arbitrary differentiable 1-form.

This system describes equally well either a massive
vector or a massive antisymmetric tensor field. A static, spherically symmetric,
asymptotically flat black hole can carry a topological charge corresponding to the $B$ field,
with both $F_{ab}$ and $H_{abc}$ vanishing everywhere
outside the black hole horizon~\cite{Allen:90, Bowick:1988xh}. We wish to show below that a similar solution exists in presence of a cosmological horizon as well.

Let us first derive the equations of motion for $A_b$ and $B_{ab}$
 \begin{eqnarray}
 \nabla_{a}F^{ab}  &=& -\frac m6
\epsilon^{bcde}  
H_{cde}\,,\label{Feom}\\
 \nabla_{c}H^{abc}  &=& - \frac m2
 \epsilon^{abcd} F_{cd}\,. \label{Heom} 
\end{eqnarray}
We define for our convenience the Hodge dual $H_a$ of the 3-form $H_{abc}$ by
 \begin{eqnarray}
H^{a} :=\frac16
\epsilon^{abcd} H_{bcd}.
\label{adch314}
\end{eqnarray}
In terms of the dual field $H_a$, the equations of motion (\ref{Feom}) and (\ref{Heom}) become 
 \begin{eqnarray}
 \nabla_{a}F^{ab}  &=& -mH^b\,,\label{Feom'}\\
 \nabla_{[a}H_{b]}  &=& -  m F_{ab}\,. \label{Heom'} 
\end{eqnarray}
Let $f_{ab}$ and $h_{a}$ be the $\Sigma$-projections of $F_{ab}$ and $H_a$ respectively defined via the projector $h_{a}{}^{b}$
 \begin{eqnarray}
f_{ab}=h_{a}{}^{c}h_{b}{}^{d}F_{cd}=D_{[a}a_{b]},\quad h_a=h_{a}{}^{b}H_b,
 \label{adch315} 
\end{eqnarray}
where $a_b$ is the $\Sigma$ projection of the gauge field $A_b$ : $a_b=h_{b}{}^{a}A_a$.
Then following exactly the same method which led to Eq.~(\ref{procamag}), we now obtain the following  `magnetic equations' : 
\begin{eqnarray}
D_b(\lambda f^{ab})= \lambda m
h^{a}\,, \qquad %\\\star
D_{[a} h_{b]} =
- m f_{ab}\,. 
\label{dualheom1}
\end{eqnarray}
Let us also define  $e^{a} = \lambda^{-1}\xi_{b}F^{ab}$
and $\psi = \lambda^{-1}\xi_a H^a\,,$ to find the `electric
equations' as before
\begin{eqnarray}
D_a e^a = - m\psi\,, \qquad % \\
D_a (\lambda\psi) = - \lambda m e_a+\pounds_\xi H_a=- \lambda m e_a,
\label{dualeheom2} 
\end{eqnarray}
where since $H_a$ is a physical matter field, we have set $\pounds_{\xi}H_a=0$.

Multiplying the first of Eq.s~(\ref{dualheom1}) by $h_a$, integrating by parts over $\Sigma$ and using the second of the Eq.s~(\ref{dualheom1}), we obtain
\begin{eqnarray}
\int_{\partial\Sigma} \lambda f^{ab} h_{a} n_{b} 
+\int_{\Sigma} m\lambda \left(\frac12 f^{ab}f_{ab} + 
h^{a}h_{a}  \right) = 0.
\label{bfmagnew}
\end{eqnarray}
%t
Similarly, multiplying the first of Eq.s~(\ref{dualeheom2}) by $\lambda\psi$, integrating by parts over $\Sigma$ and using the second of Eq.s~(\ref{dualeheom2}), we obtain
\begin{eqnarray}
 \int_{\partial\Sigma} \lambda \psi e_a n^a +\int_{\Sigma} 
m\lambda \Big( e_a e^a + \psi^2 \Big) = 0.
\label{bfelecnew}
\end{eqnarray}
Since $f^{ab}$, $h_a$, $\psi$ and $e^a$ appearing in Eq.s~(\ref{bfmagnew}) and (\ref{bfelecnew}) are physical matter fields, the surface integrals contribute nothing, by the same arguments presented earlier. So it follows that all
components of the field strengths $H_{\mu\nu\rho}$ and
$F_{\mu\nu}$ vanish on $\Sigma$ and hence over the entire region of our interest of the spacetime by staticity. The solution for the Einstein equations is then the Schwarzschild-de Sitter spacetime, with an arbitrary topological charge $q$
corresponding to the gauge symmetry of the $B$-field, whose non-vanishing component due to the spherical symmetry is
\begin{eqnarray}
B_{\theta\phi} = \frac{q}{4\pi r^2}\,.
%\label{}
\end{eqnarray}
This charge should be measurable via a stringy Bohm-Aharanov
effect because the 2-form potential $B$ should couple to the world sheet field $X_{ab}$ of a moving string. So by performing an interference experiment one should be able to find out the topological charge $q$. For asymptotically flat
spacetimes this effect has been described in~\cite{Bowick:1988xh}. We note here that the
free Abelian 2-form, i.e. $m=0$ in Eq.~(\ref{bflag}), will leave the same kind of charge on the
black hole, the proof of $H = 0$ on $\Sigma$ proceeds in a similar fashion for that theory.

%%%%%%%%%%%%%%%%%%%%%%%%%%%
\subsection{The Abelian Higgs model}
%%%%%%%%%%%%%%%%%%%%%%%%%%%%%%%%%%%%
The other case for which the no hair conjecture fails with a positive $\Lambda$ is the Abelian Higgs model. In the absence
of cosmological constant, a static spherically symmetric black hole
does not carry electric or magnetic charge if the gauge field
becomes massive via the Higgs mechanism \cite{Adler:1978dp, Lahiri:1993vg}. However, as we
shall see below, the presence of a positive cosmological constant or the cosmological horizon
allows a charged black hole solution sitting in the false vacuum of the Higgs field. 

The Lagrangian for the Abelian Higgs model is
\begin{eqnarray}
\mathcal{L}= - \frac14 F_{ab}F^{ab} - \frac12 \left(\widetilde{\nabla}_a\Phi\right)^{\dagger} \left(\widetilde{\nabla}_a\Phi\right)-\frac{\alpha}{4}(\vert \Phi\vert^2 -v^2)^2, 
\label{lagabh1}
\end{eqnarray}
where $F_{ab}=\nabla_{[a}A_{b]}$ is the Maxwell field strength, $\Phi$ is a complex scalar namely the Higgs field and the gauge covariant derivative $\widetilde{\nabla}$ is defined as
 $\widetilde{\nabla}_a\Phi:=\left(\nabla_a+iqA_a\right)\Phi$. The parameters $q$, $v$ and $\alpha$ are real and $\alpha$ is positive. 

We write the Higgs field $\Phi$ as $\Phi=\rho e^{\frac{i\eta}{v}}$, where $\rho$ and $\eta$ are real fields. In terms of these fields the Lagrangian (\ref{lagabh1}) can be rewritten as
\begin{eqnarray}
\mathcal{L}= - \frac14 F_{ab}F^{ab} - \frac12
q^2\rho^2\left(A_a  + \frac{1}{qv}\nabla_a\eta\right)\left(A^a +
\frac{1}{qv}\nabla^a\eta\right)  
 - \frac12 \nabla_{a}\rho
\nabla^{a}\rho- \frac{\alpha}{4}\left(\rho^2 -v^2\right)^2\,. \nonumber \\
\label{lagabh2}
\end{eqnarray}
Then it is easily seen that the Lagrangian is invariant under the local gauge transformations
\begin{eqnarray}
A_a\to A_a+\nabla_a\chi(x),~\eta\to \eta-vq\chi(x),
\label{gaugetr} 
\end{eqnarray}
where $\chi(x)$ is an arbitrary differentiable function.

The equations of motion corresponding to the two degrees of freedom $A_b$ and $\rho$ are given by
\begin{eqnarray}
\nabla_aF^{ab} - q^2 \rho^2
\left(A^b + \frac{1}{qv}\nabla^b\eta\right) &=& 0\,, \label{maxeq1}\\
\nabla_{a} \nabla^{a}\rho - q^2\rho\left(A_a + 
\frac{1}{qv}\nabla_a \eta\right)^2 - \alpha\rho
(\rho^2-v^2)
&=& 0.  
\label{scaleq1}
\end{eqnarray}
Let us first concentrate on the electromagnetic equation (\ref{maxeq1}). We project this equation onto $\Sigma$, and following exactly the same route which led to Eq.~(\ref{procamag}), obtain the magnetic equation,
\begin{eqnarray}
D_a(\lambda f^{ab}) - \lambda q^2 \rho^2
\left(a^b + \frac{1}{qv}D^b\eta\right) &=& 0\,, \label{mageq}
\end{eqnarray}
where $f^{ab}$ and $a^b$ are respectively the $\Sigma$ projections of $F^{ab}$ and $A^b$, defined via the projector as before, and $D_b\eta=h_{b}{}^{a}\nabla_a\eta$. We also define the potential $\psi$ and the electric field $e^a$ as in Eq.s~(\ref{adch35}) and do the following computations :
\begin{eqnarray}
\lambda e_a =\xi^bF_{ab}=\xi^b\left[\nabla_{[a}\left(A_{b]}+\frac{1}{qv}\nabla_{b]}\eta\right)\right]&=&\xi^b\nabla_{[a}A_{b]}+\frac{1}{qv}\xi^b\nabla_{[a}\nabla_{b]}\eta\nonumber\\
&=&\nabla_{a}\left(\xi^bA_b+\frac{1}{qv}\pounds_{\xi}\eta\right)-\pounds_{\xi}\left(A_a+\frac{1}{qv}\nabla_a\eta\right) \nonumber\\ &=&\nabla_a\left(\lambda \left(\psi+\frac{1}{qv\lambda}\dot{\eta}\right)\right)-\pounds_\xi\left(A_a+\frac{1}{qv}\nabla_{b}\eta\right),\nonumber\\
\label{elecdef1ah}
\end{eqnarray}
where we have used the torsion-free condition $\nabla_{[a}\nabla_{b]}\eta=0$ and we have written $\dot{\eta}=\pounds_{\xi}\eta$. But the quantity ${\displaystyle \left(A_a+\frac{1}{qv}\nabla_{b}\eta\right)}$ is a physical field which appears in the energy-momentum tensor derived from (\ref{lagabh2}), so by Eq.~(\ref{adch31}) the Lie derivative in the last line of Eq.~(\ref{elecdef1ah}) vanishes giving 
\begin{eqnarray}
\lambda e_a =\nabla_a\left(\lambda \left(\psi+\frac{1}{qv\lambda}\dot{\eta}\right)\right).
\label{elecdef1ah2}
\end{eqnarray}
But the definition of $e^a$ in Eq.~(\ref{adch35}) shows that it is spacelike. So Eq.~(\ref{elecdef1ah2}) is basically a spacelike equation over $\Sigma$ and we may replace $\nabla_a$
by $D_a$ to have
\begin{eqnarray}
\lambda e_a =D_a\left(\lambda \left(\psi+\frac{1}{qv\lambda}\dot{\eta}\right)\right).
\label{elecdef1ah3}
\end{eqnarray}
Next, let us compute the divergence $\nabla_a e^a$ and determine this over $\Sigma$. Following exactly the same procedure which led to Eq.~(\ref{adch36}), we now have
\begin{eqnarray}
D_a e^a = q^2 \rho^2 \left(\psi + \frac1{qv\lambda}\dot{\eta}\right). 
\label{eleceq}
\end{eqnarray}
Let us now multiply Eq.~(\ref{mageq}) by ${\displaystyle\left(a^b + \frac{1}{qv}D^b\eta\right)}$ and integrate by parts over $\Sigma$ to find
\begin{eqnarray}
\int_{\partial \Sigma}\lambda \left(a_b +
\frac{1}{qv}D_b\eta\right)f^{ab}n_a 
+ \int_\Sigma \lambda\left[\frac12 f^{ab}f_{ab}
+ q^2\rho^2\left(a_b + \frac{1}{qv}D_b\eta\right)\left(a^b +
\frac{1}{qv}D^b\eta\right)\right]=0, \nonumber \\  
\label{monopole}
\end{eqnarray}
where we have used the fact that $D_a$ is torsion-free : $D_{[a}D_{b]}\eta=0$, which can be derived from $\nabla_{[a}\nabla_{b]}\eta=0$ using the projector. 

The $\Sigma$ integral in Eq.~(\ref{monopole}) can be non-vanishing only if the
$\partial\Sigma$ integral is also non-vanishing, which means that
the norm of either $f_{ab}$ or $\left(a_b + D_b\eta\right)$ must
diverge on one or both of the horizons. However, since we have assumed spherical
symmetry of the spacetime, a non-vanishing magnetic field strength $f_{ab}$ essentially corresponds to the
magnetic monopole. This implies that then $\left(a_b + D_b \eta\right)$ must be both spherically symmetric and divergent over the horizon. But we know that this is impossible. This rules out the possibility of non-vanishing of the $\partial\Sigma$ integral in Eq.~(\ref{monopole}). Then, since all the inner products in the second integral in Eq.~(\ref{monopole}) are spacelike, we must have $f_{ab} =0=\left(a_b + D_b\eta\right)$~throughout $\Sigma$. This shows that a spherically symmetric static black hole spacetime with $\Lambda>0$ cannot have any magnetic charge corresponding to the Abelian Higgs model. 

To investigate the electric charge, we multiply the electric field equation (\ref{eleceq}) by ${\displaystyle \lambda \left(\psi+\frac{1}{qv\lambda}\dot{\eta}\right)}$, integrate by parts over $\Sigma$ and use Eq.~(\ref{elecdef1ah3}) to find
\begin{eqnarray}
\int_{\partial\Sigma}\lambda\left(\psi +
\frac1{ qv\lambda}\dot{\eta}\right)e^{a}n_{a}
+ \int_{\Sigma}\lambda\left[ e^{a}e_{a} +
q^2\rho^2\left(\psi +
\frac1{ qv\lambda}\dot{\eta}\right)^2\right]=0. 
\label{eleceom}
\end{eqnarray}
 Since $e_a e^a$ appears in $T_{ab}\,,$ we may
use Schwarz inequality to say that $e^{a}n_{a}$ is finite on
$\partial\Sigma\,.$ So the $\Sigma$ integral can be non-zero only
if ${\displaystyle \left(\psi+\frac{1}{\lambda qv}\dot{\eta}\right)}$ diverges over at least
one of the horizons. On the other hand we have already proved that ${\displaystyle \left(a_b+\frac{1}{qv}D_b\eta\right)=0}$ throughout $\Sigma$. So the Lagrangian (\ref{lagabh2}) shows that only the timelike part ${\displaystyle-\rho^2\left(\psi +
\frac{1}{ qv\lambda}\dot{\eta}\right)^2}$ of the quantity ${\displaystyle\rho^2\left(A_b+\frac{1}{qv}\nabla_b\eta\right)^2} $ appears in the energy-momentum tensor and hence must be bounded on the horizons by our assumption of regularity. Thus, in order to make ${\displaystyle\left(\psi+\frac{1}{qv\lambda}\dot{\eta}\right)}$ divergent on any of the horizons we must have $\rho=0$ over that horizon.

For asymptotically flat black hole spacetimes the energy-momentum tensor corresponding to (\ref{lagabh2}) vanishes at the spatial infinity. This implies that the magnitude $\rho$ of the Higgs field must reach $\pm v$ in the asymptotic region.
In particular, it has been 
shown for $\Lambda=0$ that $\rho$ cannot vanish on the horizon, and so the black
hole cannot have any electric charge~\cite{Lahiri:1993vg}. Let us now see
what happens for our present case of $\Lambda>0$. Since the cosmological horizon $r_{\rm C}$ is not located at spacelike infinity we cannot set $T_{ab}=0$
at $r_{\rm{C}}$. This means that we cannot impose $\rho \to \pm v$ as $r\to r_{\rm C}$. Since $\rho$ is a physical field, we may only assume that $\rho$ remains bounded on the horizons.

Let us now project the equation of motion (\ref{scaleq1}) for $\rho$ onto $\Sigma$. We recall that ${\displaystyle\left(a_b+\frac{1}{qv}D_b\eta\right)=0}$ throughout $\Sigma$ and since $\rho$ is a physical field we must have $\pounds_{\xi}\rho=0$. Then the procedure which led to Eq.~(\ref{eqscal}) now yields 
\begin{equation}
D_{a}\left(\lambda D^{a}\rho\right) = - \lambda q^2\rho\left(\phi + 
\frac{1}{qv\lambda}\dot{\eta}\right)^2 +  \lambda\alpha\rho
(\rho^2-v^2).   
\label{rhoeq}
\end{equation}  
Let us assume for the moment that $\rho$ vanishes on the black hole
horizon at $r = r_{
\rm{H}}$, and starts increasing with increasing $r$.
Then $\rho$ must increase monotonically from $\rho=0$ at $r=r_{
\rm{H}}$ to one of: 
\begin{eqnarray}
{\rm{(1)}}~ \rho &=& \rho_{\rm C} < v~{\rm{ at}}~r = r_{
\rm{C}};\nonumber\\
{\rm{(2)}}~ \rho &=& v~{\rm {at}}~r = r_{\rm V} \leq r_{\rm{C}};\nonumber\\
{\rm{(3)}}~ \rho &=& \rho_{\rm{max}} < v~{\rm{at~the
~turning~point}}~ r = r_{\rm{max}} < r_{
\rm{C}}.
\label{scalassump}
\end{eqnarray}  
For all the above three cases, we multiply Eq.~(\ref{rhoeq}) by $(\rho - v)$
and integrate over a region $\Omega$ to obtain 
\begin{eqnarray}
\int_{\partial\Omega}\lambda(\rho - v)n^a D_a\rho -
\int_\Omega\lambda \left[\left(D_a\rho\right)\left( D^a\rho\right) 
 - \rho(\rho-v)\left(\phi + \frac{1}{\lambda
qv} \dot\eta\right)^2  \right.\nonumber\\\left. +\alpha(\rho-v)^2\rho(\rho+v)\right]=0
\,.%\nonumber \\
\label{rhoeq2}
\end{eqnarray}
The region $\Omega$ and its boundary $\partial\Omega$ for the three
cases stated in Eq.~(\ref{scalassump}) are taken respectively to be ${\rm {(1)}}~\Omega =
\Sigma,~\partial\Omega = \partial\Sigma$;~${\rm {(2)}}~\Omega
=\Sigma\vert_{r<r_{\rm V}}~{\rm {and}}~\partial\Omega = {\rm {spheres~ at}}~ r_{
\rm{H}},~r_{\rm {V}}$,\\and~${\rm {(3)}}~\Omega = \Sigma\vert_{r < r_{\rm{max}}},~ 
\partial\Omega =~{\rm{ spheres~ at}}~ r_{
\rm{H}},~ r_{\rm{max}}$.

In all three cases, 
 the integral over $\partial\Omega$ in Eq.~(\ref{rhoeq2}) vanishes on the respective 2-spheres leaving us only with the vanishing volume integral over $\Omega$. On the other hand, since $0\leq \rho \leq v$ everywhere in $\Omega$, 
 all terms in the $\Omega$ integral of Eq.~(\ref{rhoeq2}) are positive definite. So we have a contradiction and $\rho$ cannot
increase from zero as $r$ increases from $r_{
\rm{H}}$. In particular, Eq.~(\ref{rhoeq2}) shows that either $\rho=v$ or $\rho=0$ throughout our region of interest.

Next, let us consider the reverse case of Eq.~(\ref{scalassump}), i.e. we assume that $\rho=0$ over $r_{\rm{H}}$ and decreases monotonically  to one of: 
\begin{eqnarray}
{\rm{(1)}}~ \rho &=& \vert \rho_{\rm C}\vert < \vert v\vert~{\rm{ at}}~r = r_{
\rm{C}};\nonumber\\
{\rm{(2)}}~ \rho &=& -v~{\rm {at}}~r = r_{\rm V} \leq r_{\rm{C}};\nonumber\\
{\rm{(3)}}~ \rho &=& \vert \rho_{\rm{min}}\vert < \vert v\vert~{\rm{at~the
~turning~point}}~ r = r_{\rm{min}} < r_{
\rm{C}}.
\label{scalassump2}
\end{eqnarray}  
In all three cases we multiply Eq.~(\ref{rhoeq}) by $(\rho+v)$ and integrate by parts to get 
\begin{eqnarray}
\int_{\partial\Omega}\lambda(\rho+v)n^a D_a\rho -
\int_\Omega\lambda \left[\left(D_a\rho\right)\left( D^a\rho\right) 
 - \rho(\rho+v)\left(\phi + \frac{1}{\lambda
qv} \dot\eta\right)^2 \right.\nonumber \\ \left.+  \alpha(\rho-v)\rho(\rho+v)^2\right]=0
\,.
\label{rhoeq3}
\end{eqnarray}
The surface integral vanishes as before. Also, since $-v\leq\rho\leq0 $ for this case, the volume integral comprises of positive definite quantities. So again we reach a contradiction and we must have either $\rho=0$ or $\rho=-v$ throughout our region of interest.

Thus we have seen that for a static and spherically symmetric $\Lambda>0$ black hole spacetime $\rho$ cannot vary in the region between the black hole and the cosmological horizon. In particular, $\rho$ can assume only three discrete values in this region : $\rho=0$ or $\rho=\pm v$.

Let us now look at the consequences of these. 
We consider first $\rho=\pm v$. Previously we argued that the surface integral of
Eq.~(\ref{eleceom}) can only be non-zero if $\rho$ vanishes on at least one of the horizons. So for $\rho=\pm v$, the surface integral of Eq.~(\ref{eleceom}) vanishes giving us a vanishing volume integral which comprises of positive definite quantities. Thus for non-zero $\rho$ we must have $e^a=0=\psi$
throughout $\Sigma$. Also the vanishing Lie derivatives $\pounds_{\xi}e^a=0=\pounds_{\xi}\psi$ imply that $e^a$ and $\psi$ must vanish throughout the spacetime. The black hole in that case will have no electric charge. Also we have proved earlier that it will have no magnetic charge as well. This is the usual no hair result for the Abelian Higgs model. The same kind of result
holds also for asymptotically flat spacetimes~\cite{Lahiri:1993vg}.

 There is however one exception $-$ we have found another solution $\rho=0$ for which the 
Lagrangian (\ref{lagabh2}) becomes
\begin{eqnarray}
\mathcal{L}= - \frac14 F_{ab}F^{ab} - \frac{\alpha v^4}{4}. 
\label{lagabh3}
\end{eqnarray}
The static, spherically symmetric solution to this Lagrangian is clearly the
Reissner-N\"ordstrom-de Sitter solution with a modified cosmological constant
${\displaystyle\Lambda^{\prime}=\Lambda+\frac{\alpha v^4}{8}}$. In other words, the solution $\rho=0$
represents an electrically charged static black hole sitting in the false vacuum of the Higgs field.
This has no $\Lambda=0$ analogue and this is contradictory to what one expects from the no hair
conjectures. This charged solution for the Abelian Higgs model clearly comes from the non-trivial boundary condition at the cosmological horizon. So we have seen that the existence of a cosmological horizon can change the local physics considerably.  

We also note here that the assumption of spherical symmetry is not
crucial for the proofs, except for the Abelian Higgs model. For all the other matter fields 
we have discussed, it is sufficient to assume that $\partial{\Sigma}$ comprises of closed and non-singular 2-surfaces located at the horizons, as we did in the previous Chapter. Then the assumption of regularity leads to the usual no hair results.
So for example, a static-axisymmetric black hole will be hairless for most field theories we considered, while dipole or
other axisymmetric hair cannot be ruled out for the Abelian Higgs model. In particular, we will discuss the cylindrically symmetric cosmic string solutions for this model in the next Chapter.

%%%%%%%%%%%%%%%%%%%%%%%%%%%%%%%%%%%%%%%%%%%%%%%%%%%%%%%%%%%%%%%%%%%%%
\section{No hair theorems for stationary axisymmetric spacetimes}
%%%%%%%%%%%%%%%%%%%%%%%%%%%%%%%%%%%%%%%%%%%%%%%%%%%%%%%%%%%%%%%%
In the following we will generalize some of the above no hair results for a stationary axisymmetric de Sitter black hole spacetime. We will use the geometrical set up developed in the previous Chapter for this purpose. Let us first summarize the results of Section 2.2 for convenience. 

The spacetime is stationary and axisymmetric endowed with two commuting Killing vector fields $\xi^a$ and $\phi^a$, generating respectively stationarity and axisymmetry of the spacetime. $\xi^a$ and $\phi^a$ have norms $-\lambda^2$ and $f^2$ respectively.
 
 The spacetime is assumed to be regular, i.e. there is no naked curvature singularity anywhere in our region of interest. 

 Since the spacetime is stationary$-$not static, $\xi^a\phi_a \neq 0$ and hence $\xi^a$ is not orthogonal to any family of spacelike hypersurfaces containing $\phi^a$. As in Section 2.2, we define a vector field ${\displaystyle\chi^a:= \xi^a-\frac{\xi \cdot \phi}{\phi \cdot \phi}\phi^a\equiv \xi^a+\alpha \phi^a}$, with norm $\chi^a\chi_a=-\beta^2$ such that $\chi^a\phi_a=0$ everywhere and $\chi^a$ is locally timelike. Since $\alpha$ is a spacetime function, $\chi^a$ is not a Killing vector field. Next we choose an orthogonal basis $\left\{\chi^a,~\phi^a,~\mu^a,~\nu^a\right\}$
for our spacetime with $\left\{\phi^a,~\mu^a,~\nu^a\right\}$ orthogonal spacelike basis vectors.
 We also assume that the spacelike 2-planes orthogonal to both $(\xi^a,~\phi^a)$ or $(\chi^a,~\phi^a)$ are integral 2-submanifolds. Then the vector field $\chi^a$ satisfies the Frobenius condition for hypersurface orthogonality and is orthogonal to the family of spacelike hypersurfaces $\Sigma$ spanned by $\left\{\phi^a,~\mu^a,~\nu^a\right\}$.     

The hypersurface orthogonal timelike vector field $\chi^a$ becomes null when $\beta^2=0$. It was shown in the previous Chapter that the vector field $\chi^a$ coincides with a Killing vector field on any closed surface ${\cal{H}}$ with $\beta^2=0$. So ${\cal{H}}$ defines the true or Killing horizon of the spacetime. The inner (outer) $\beta^2=0$ surface is the black hole (cosmological) Killing horizon.    
 
The assumptions on the matter fields are the same as those for the static spacetime discussed earlier. Since the spacetime is regular, the invariants constructed from the energy-momentum tensor are bounded everywhere in our region of interest including the horizons.
Also as before, we assume that any physical matter field obeys the symmetry of the spacetime, Eq.~(\ref{adch31}). 

We define the projector $h^a{}_{b}$ which projects spacetime tensors over $\Sigma$ as 
\begin{eqnarray}
h^{a}{}_{b}:=\delta^{a}{}_{b}+\beta^{-2}\chi^a\chi_b. 
\label{pro1}
\end{eqnarray}
The operation of the projector and the induced connection $D_a$ over $\Sigma$ can be defined in the similar manner as described in Chapter 2. Also since by our assumption the spacelike 2-planes spanned by $(\mu^a,~\nu^a)$ orthogonal to $\chi^a$ and $\phi^a$ are integral submanifolds, $\overline{\Sigma}$, we may define another projector to project spacetime tensors onto $\overline{\Sigma}$,
\begin{eqnarray}
\Pi^{a}{}_{b}:=\delta^{a}{}_{b}+\beta^{-2}\chi^a\chi_b-f^{-2}\phi^a\phi_b=h^{a}{}_{b}-f^{-2}\phi^a\phi_b. 
\label{pro2}
\end{eqnarray} 
The operation of $\Pi^{a}{}_{b}$ and the induced connection $\overline{D}_a\equiv\Pi_{a}{}^{b} \nabla_b \equiv\Pi_{a}{}^{b} D_b$ over
the 2-submanifolds $\overline{\Sigma}$ can be defined similarly as what was done for $h^{a}{}_{b}$.

With all this equipment, we are now ready to go into the no hair proofs.
%%%%%%%%%%%%%%%%%%%%%%%%%%%%%%%%%%%%%%%%%%%%%%%%%%%%%%%%%%%%%
\subsection{Scalar field}
%%%%%%%%%%%%%%%%%%%%%%%%%%
Let us start with the simplest case, that of a scalar field $\psi$ in a potential $V(\psi)$ with the Lagrangian of Eq.~(\ref{adch32}) and the equation of motion Eq.~(\ref{adch33}).

Since we are assuming stationarity and axisymmetry of the spacetime, we have
$\pounds_{\xi}\psi=0=\pounds_{\phi}\psi$, by Eq.~(\ref{adch31}). Then we have for the hypersurface orthogonal vector field $\chi^a$,
\begin{eqnarray}
\pounds_{\chi}\psi=\pounds_{(\xi+\alpha \phi)}\psi =\pounds_{\xi}\psi+\alpha (\pounds_{\phi}\psi)=0.
\label{saad31}
\end{eqnarray}
 Then following the same procedure which led to Eq.~(\ref{eqscal}) now yields the 
projection of Eq.~(\ref{adch33}) onto the $\chi$-orthogonal family of spacelike 
hypersurfaces $\Sigma$,
\begin{eqnarray}
\nabla_a\nabla^a\psi= \frac{1}{\beta h}\partial_{a}\left[\beta h g^{ab}\partial_b \psi\right]=\frac{1}{\beta h}\partial_{a}\left[\beta h \left(h^{ab}-\beta^{-2}\chi^a\chi^b\right)
\partial_b \psi\right]&=&\frac{1}{\beta h}\partial_{a}\left[\beta h h^{ab}\partial_b \psi\right]
\nonumber\\\Rightarrow D_a\left(\beta D^a\psi\right)&=&\beta V'(\psi),
\label{nh2}
\end{eqnarray}
where $h$ is the determinant of the induced metric $h_{ab}$ over $\Sigma$. We multiply Eq.~(\ref{nh2}) by $V'(\psi)$ and integrate by parts over $\Sigma$ to get 
\begin{eqnarray}
\int_{\partial \Sigma}\beta
V'(\psi)n^aD_a\psi 
+\int_{\Sigma}\beta\left[V''(\psi) 
\left(D^a\psi\right)\left(D_a\psi\right)
+ V'^2(\psi)\right]=0,
\label{nh3}
\end{eqnarray}
where $\partial\Sigma$ are the boundaries of $\Sigma$, i.e. spacelike closed 2-surfaces located at the horizons and $n^a$ is a unit spacelike vector normal to these 2-surfaces.

According to our assumption, there is no naked curvature singularity
anywhere between the horizons, including the horizons. This implies that
the invariants of the energy-momentum tensor must be bounded over the
horizons. Since $\left(\nabla_a\psi\right)\left(\nabla^a\psi\right)$ appears in the invariants constructed from the
energy-momentum tensor, this must be bounded on the horizons. Also $\pounds_{\chi}\psi=0$ implies that $\nabla_a\psi=D_a\psi$. As before we then use the Schwarz inequality,
$\big\vert n^aD_a\psi\big\vert^2\leq \left(D_a\psi\right)\left(D^a
\psi\right)$. Therefore the quantity $n^aD_a\psi$ remains bounded
on the horizons. Then since $\beta=0$ on the horizons, the surface
integrals in Eq.~(\ref{nh3}) vanish as before.

Since the inner product in the $\Sigma$ integral of Eq.~(\ref{nh3}) is
spacelike, it immediately follows that no non-trivial solution exists
for $\psi$ over $\Sigma$ for a convex potential. So like the static case, here we also find that for a convex $V(\psi)$ the scalar field $\psi$ is a constant located at the minimum of the potential $V(\psi)$. Then Eq.~(\ref{saad31}) ensures that we have the same trivial
solution throughout the spacetime, which is the standard no hair result for a scalar field.

Clearly the above no hair result will not hold for a non-convex $V(\psi)$. The arguments are similar to that presented for the static spacetime.

%%%%%%%%%%%%%%%%%%%%%%%%%%%%%%%%%%%%%%%%%%%%%%%%%%%%%%%%
\subsection{The Proca field}
%%%%%%%%%%%%%%%%%%%%%%%%%%%%%%
Next we consider the Proca-massive vector field with the Lagrangian (\ref{adch34}) satisfying the equation of motion Eq.~(\ref{eompr}). Although our objective will be to construct a
positive definite quadratic with a vanishing integral on $\Sigma$ as before, we will see below that
proving a no hair statement in this case is quite a bit more
complicated than in the case of the static spacetime. In particular, there will be effects of the spacetime rotation which will bring in some more technicalities. 

Let us start as before by defining the potential $\psi$ and the `electric'
field $e^a$ 
\begin{eqnarray}
\psi :=\beta^{-1}\chi_a A^a,\quad e^a := \beta^{-1}\chi_bF^{ab}.
\label{nh6}
\end{eqnarray}
We note from this definition that $e_a\chi^a=0$, i.e. $e^a$ is spacelike. Also we note that\newpage
\begin{eqnarray}
\beta e^a\phi_a=\chi_b\phi_aF^{ab}&=&\chi^b\phi^a\left(\nabla_aA_b-\nabla_b A_a\right)\nonumber\\
&=&\phi^a\left[\nabla_a\left(A_b\chi^b\right)-\chi^b\nabla_bA_a-A_b\nabla_a\chi^b \right]\nonumber\\
&=&\pounds_{\phi}\left(\beta\psi\right)-\phi^a\left[\pounds_{\xi}A_a+\alpha(\pounds_{\phi}A_a)+ A_b\phi^b(\nabla_a \alpha)\right],%\nonumber\\
\label{ephi1}
\end{eqnarray}
using $\chi_a=\xi_a+\alpha \phi_a$. Since $A_b$ is a physical matter field for the Proca theory, we have by Eq.~(\ref{adch31}) $\pounds_{\xi}A_a=0=\pounds_{\phi}A_a$ identically. Eq. (\ref{cral23}) shows $\pounds_{\phi}\alpha=0$. Also, the first of  Eq.s~(\ref{nh6}) gives
\begin{eqnarray}
\pounds_{\phi}\left(\beta\psi\right)=\pounds_{\phi}(g_{ab}A^a\chi^b)=A_b(\pounds_{\phi}\chi^b)
=0,
\label{ephi2}
\end{eqnarray}
using Eq.~(\ref{cral23}). Thus the right hand side of Eq.~(\ref{ephi1}) vanishes and we see that $e^a$ is orthogonal to $\phi_a$.

Since both $\psi$ and $e_a$ are physical matter fields appearing in the energy-momentum tensor,
we have $\pounds_{\xi}\psi=0= \pounds_{\xi}e_a$, and $\pounds_{\phi}\psi=0= \pounds_{\phi}e_a$. Then we compute
\begin{eqnarray}
\pounds_{\chi}\psi&=&\pounds_{\xi}\psi+\alpha(\pounds_{\phi}\psi)=0.
\label{saad32}
\end{eqnarray}
\begin{eqnarray}
\pounds_{\chi}e_a&=&\pounds_{\xi}e_a+\alpha (\pounds_{\phi}e_a)+e^b\phi_b\left(\nabla_a \alpha\right)=0,
\label{saad33}
\end{eqnarray}
where the orthogonality of $e^a$ and $\phi^a$ has been used. It also follows that
\begin{eqnarray}
\pounds_{\chi}e_a=0=\pounds_{\chi}\left(e^b g_{ab}\right)&=&\left(\pounds_{\chi}e^b\right)g_{ab}+e^b\left(\pounds_{\chi}g_{ab}\right)=\left(\pounds_{\chi}e^b\right)g_{ab}+e^b\left(\nabla_{(a}\chi_{b)}\right)\nonumber\\
&=&\left(\pounds_{\chi}e^b\right)g_{ab}+e^b\left(\phi_{(a}\nabla_{b)}\alpha\right)
=\left(\pounds_{\chi}e^b\right)g_{ab}+\phi_a\left(e^b\nabla_b\alpha\right),\nonumber\\
\label{saad34}
\end{eqnarray}
 where we have used Eq.~(\ref{sa5}) and that $e^a\phi_a=0$. Thus we have
\begin{eqnarray}
\pounds_{\chi}e^a=-\phi^a\left(e^b\nabla_b\alpha\right).
\label{nh7}
\end{eqnarray}
Let us now derive the analogues of Eq.s~(\ref{elecdef2}) and (\ref{adch36}) for the present case. Using the definitions (\ref{nh6}) we compute the following:
\begin{eqnarray}
\beta e_a=\chi^bF_{ab}= \chi^b\left(\nabla_a A_b-\nabla_b A_a\right)=\nabla_{a}\left(\beta\psi\right)-\pounds_{\chi}A_a.
\label{nh7int1}
\end{eqnarray}
Using $\pounds_{\xi}A_a=0=\pounds_{\phi}A_a$, we have $\pounds_{\chi}A_a=(A_b\phi^b)\nabla_a\alpha$. Thus Eq.~(\ref{nh7int1}) becomes
\begin{eqnarray}
\beta e_a=\nabla_{a}\left(\beta\psi\right)-(A_b\phi^b)\nabla_a\alpha.
\label{nh7int2}
\end{eqnarray}
Since $e^a$ is spacelike, and $\nabla_a\alpha=D_a\alpha$ by Eq.~(\ref{dalpha}), the $\Sigma$ projection of Eq.~(\ref{nh7int2}) is obtained simply by replacing the spacetime connection $\nabla_a$ with the induced connection $D_a$ on $\Sigma$,
\begin{eqnarray}
D_a(\beta \psi)=\beta e_a+(A_b\phi^b)D_a\alpha.
\label{nh7int3}
\end{eqnarray}
After that, using Eq.~(\ref{nh6}) we compute the following divergence,
\begin{eqnarray}
\nabla_a\left(\beta e^a\right) =\left(\nabla_a\chi_{b}\right)F^{ab}+\chi_b\left(\nabla_a F^{ab}\right)=\left(\nabla_a\chi_{b}\right)F^{ab}+m^2\beta\psi,
\label{dive1}
\end{eqnarray}
using the equation of motion (\ref{eompr}). We substitute the expression for $\nabla_a\chi_b$ from Eq.~(\ref{adt9}) into Eq.~(\ref{dive1}), the symmetric part of $\nabla_a\chi_b$ does not contribute and using Eq.~(\ref{nh6}) we obtain
\begin{eqnarray}
\nabla_a\left(\beta e^a\right) =\beta^{-1}\left(\chi_b\nabla_a\beta-\chi_a\nabla_a\beta \right)F^{ab}+m^2\beta\psi=2e^a\nabla_a\beta+m^2\beta\psi,
\label{dive2}
\end{eqnarray}
which we rewrite as
\begin{eqnarray}
\nabla_a e^a=\beta^{-1}e^a\nabla_a\beta+m^2\psi.
\label{dive3}
\end{eqnarray}
Let us now project this onto $\Sigma$ using the projector $h_{a}{}^{b}$ defined in Eq.~(\ref{pro1}). We write 
\begin{eqnarray}
 D_ae^a =h^{a}{}_{b}\nabla_ae^b=\nabla_a e^a+\beta^{-2}\chi_b \chi^a\nabla_a e^b,
\label{dive4}
\end{eqnarray}
and look at the last term. Using the orthogonality $\chi_a e^a=0$ we write
\begin{eqnarray}
\beta^{-2}\chi_b \chi^a\nabla_a e^b=-\beta^{-2}\chi^ae^b\nabla_a\chi_b&=&
-\beta^{-2}\chi^ae^b\left[\beta^{-1}(\chi_{[b}\nabla_{a]}\beta)+\frac12\phi_{(a}\nabla_{b)}\alpha\right]\nonumber\\
&=&-\beta^{-1}e^a\nabla_a\beta,
\label{dive5}
\end{eqnarray}
where we have substituted the expression for $\nabla_a\chi_b$ from Eq.~(\ref{adt9}) and used also the orthogonality of $e^a$ and $\phi^a$. Combining Eq.s~(\ref{dive3}), (\ref{dive4}) and (\ref{dive5}), we obtain the equation for $e^a$ over $\Sigma$,
\begin{eqnarray}
D_ae^a=m^2\psi,
\label{nh7.1}
\end{eqnarray}
which has the same form as the static equation (\ref{adch36}).
We now multiply Eq.~(\ref{nh7.1}) by $\beta\psi$ and use Eq.~(\ref{nh7int3}) and integrate by parts over $\Sigma$ to find
\begin{eqnarray}
\int_{\partial \Sigma}\beta \psi n^a e_a 
+\int_{\Sigma}\left[\beta \left(e_ae^a+m^2\psi^2\right) +\left(A_b\phi^b\right)e^aD_a \alpha\right]=0. 
\label{nh8}
\end{eqnarray}
 The terms $\psi^2$ and $e_a^2$ appear in the invariants of the
energy-momentum tensor, so are bounded on the horizons. This
implies as before that the surface integrals in Eq.~(\ref{nh8}) vanish, giving us the following
vanishing $\Sigma$ integral
\begin{eqnarray}
  \int_{\Sigma}\left[\beta \left(e_ae^a+ m^2\psi^2\right)
    + \left(A_b\phi^b\right)e^aD_a \alpha \right]=0. 
\label{nh9}
\end{eqnarray}
We recall that $e^a$ is a spacelike vector field and $\beta >0$ between the two horizons and vanishes on the horizons. So all but the last term in Eq.~(\ref{nh9}) are
positive definite. The last term is $e^a\pounds_{\chi}A_a$, so we cannot set this to zero, since $\chi^a$ is not a Killing field. Thus the non-existence of the electric charge for
the Proca field cannot be proven from Eq.~(\ref{nh9}) alone, and we 
need to make a more careful analysis of the rest of the equations of
motion. We note that if we set $\alpha=0$ in Eq.~(\ref{nh9}), we recover the static case.

Let us now project Eq.~(\ref{eompr}) onto
$\Sigma$. Let $a_b$ and $f_{ab}$ be the $\Sigma$ projections of $A_b$ and $F_{ab}$,
\begin{eqnarray}
a_b&=&h^{a}{}_{b}A_a,\\
f_{ab}&=&h_{a}{}^{c}h_{b}{}^{d}F_{cd}=D_{[a} a_{b]}.
\label{nh10}
\end{eqnarray}
We now multiply Eq.~(\ref{eompr}) by the projector to write
\begin{eqnarray}
\beta h^{b}{}_{c}\nabla_a F^{ac}=m^2\beta a^b.
\label{nh11}
\end{eqnarray}
In order to get an equation for $f_{ab}$ we consider the expression $D_{a}\left(\beta f^{ab}\right)$. Using the projector $h_a{}^{b}$ and its action discussed in Chapter 2 we have
\begin{eqnarray}
D_{a}\left(\beta f^{ab}\right)&=& h^{b}{}_{e}h^{f}{}_{a}\nabla_f
\left(\beta F^{ae}\right) \nonumber \\
&=& h^{b}{}_{e} \nabla_a\left(\beta
  F^{ae}\right)+\beta^{-2}h^{b}{}_{e}\chi_a\chi^f\nabla_f
\left(\beta F^{ae}\right). 
\label{add2}
\end{eqnarray}
In order to simplify this, we first recall from the previous Chapter that $\pounds_{\chi}\beta=0$.
 %
%\begin{eqnarray}
%\pounds_{\chi}\beta\equiv\frac12\beta^{-1}\chi^a\nabla_a\beta^2=-\frac12 \beta^{-1}\chi^a\chi^b\nabla_{(a}\chi_{b)}=-\frac12 \beta^{-1}\chi^a\chi^b\phi_{(a}\nabla_{b)}\alpha=0.
%\label{add2'}
%\end{eqnarray}
%
Also, since $\xi^a$ and $\phi^a$ are
Killing fields we have by Eq.~(\ref{adch31}), $\pounds_{\xi}F^{ab}=0=\pounds_{\phi}F^{ab}$. Then we find
\begin{eqnarray}
\pounds_{\chi}F^{ab}&=&\chi^c\nabla_cF^{ab}-F^{ac}\nabla_c\chi^b-F^{cb}\nabla_c\chi^a\nonumber\\
&=&\pounds_{\xi}F^{ab}+\alpha \left(\pounds_{\phi}F^{ab}\right)+\phi^{[a}F^{b]c}\nabla_c\alpha
\nonumber\\
&=&\phi^{[a}F^{b]c}\nabla_c\alpha.
\label{add2"}
\end{eqnarray}
Let us now look at Eq.~(\ref{add2}). Using $\nabla_a\beta=D_a\beta$, the orthogonality $\chi_a\phi^a=0$, and substituting the expression for $\chi^f\nabla_f F^{ae}$ from Eq.~(\ref{add2"}) we obtain
\begin{eqnarray}
D_{a}\left(\beta f^{ab}\right)&=&\beta h^{b}{}_{e} \nabla_a F^{ae}+h^{b}{}_{e}F^{ae}D_a \beta
 \nonumber\\ &+&\beta^{-1}\chi_a h^{b}{}_{e}\left[F^{ce}\nabla_c
  \chi^a+F^{ac}\nabla_c\chi^e-\left(F^{ce}\nabla_c\alpha\right)
  \phi^a -\left(F^{ac}\nabla_c \alpha\right) \phi^e\right]\nonumber\\
&=&\beta h^{b}{}_{e} \nabla_a F^{ae}+f^{ab}D_a \beta\nonumber\\
&+&\beta^{-1}\chi_a h^{b}{}_{e}\left[F^{ce}\nabla_c
  \chi^a+F^{ac}\nabla_c\chi^e-\left(F^{ac}\nabla_c \alpha\right) \phi^e\right].
\label{nh12}
\end{eqnarray}
Let us consider the first two terms within the square bracket of the above equation. Substituting the expression for $\nabla_a\chi_b$ from Eq.~(\ref{adt9}) we find
\begin{eqnarray}
\chi_a h^{b}{}_{e}\left[F^{ce}\nabla_c\chi^a+F^{ac}\nabla_c\chi^e\right]
&=& \chi_a h^{b}{}_{e}F^{ce}\left[\beta^{-1}\left(\chi^a \nabla_c\beta-\chi_c\nabla^a \beta\right)
+\frac{1}{2}\left(\phi^a\nabla_c\alpha+\phi_c\nabla^a\alpha\right)\right]
\nonumber\\&+&\chi_a h^{b}{}_{e}F^{ac}\left[\beta^{-1}\left(\chi^e \nabla_c\beta-\chi_c\nabla^e \beta\right)
+\frac{1}{2}\left(\phi^e\nabla_c\alpha+\phi_c\nabla^e\alpha\right)\right].\nonumber\\
\label{nh13}
\end{eqnarray}
Using Eq.s~(\ref{dalpha}), $\nabla_a\beta=D_a\beta$, the orthogonality of $\chi^a$ and $\phi^a$ and the fact that $h_{a}{}^b\chi^a=0$, we may simplify Eq.~(\ref{nh13}) to
\begin{eqnarray}
\chi_a h^{b}{}_{e}\left[F^{ce}\nabla_c\chi^a+F^{ac}\nabla_c\chi^e\right]
&=&-\beta f^{cb}D_c\beta+\frac12 \chi_aF^{ac}\left(\phi^b D_c\alpha+\phi_cD^b\alpha\right),\nonumber\\
\label{nh14'}
\end{eqnarray}
which, using the definition (\ref{nh6}) of the electric field $e^a$ and the previously derived
orthogonality $e_a\phi^a=0$, may be further simplified as
\begin{eqnarray}
\chi_a h^{b}{}_{e}\left[F^{ce}_c\chi^a+F^{ac}\nabla_c\chi^e\right]
&=&-\beta f^{cb}D_c\beta-\frac12 \beta e^c\left(\phi^bD_c\alpha+\phi_cD^b\alpha\right)\nonumber\\
&=&-\beta f^{cb}D_c\beta-\frac12\beta \phi^b e^c\nabla_c\alpha,
\label{nh14}
\end{eqnarray}
We substitute this expression into Eq.~(\ref{nh12}) to find, 
\begin{eqnarray}
D_{a}\left(\beta f^{ab}\right)&=&\beta h^{b}{}_{e} \nabla_aF^{ae}+f^{ab}D_a\beta \nonumber\\&+&\beta^{-1}\left[-\beta f^{cb}D_c\beta -\frac12 \beta \phi^b e^c\nabla_c \alpha \right]-\phi^b\left(\beta^{-1}\chi_aF^{ac}\right)\nabla_c\alpha\nonumber\\
&=&\beta h^{b}{}_{e} \nabla_aF^{ae}+\frac{1}{2}
\left(e^c\nabla_c\alpha\right)\phi^b,
\label{nh14''}
\end{eqnarray}
using once again the definition of $e^a$ and the fact that $h_{a}{}^{b}\phi^a=\phi^b$.
Combining Eq.~(\ref{nh14''}) with Eq.~(\ref{nh11}), we finally obtain the $\Sigma$ projection of the equation of motion (\ref{eompr}) for a stationary axisymmetric spacetime, 
\begin{eqnarray}
D_{a}\left(\beta f^{ab}\right)=m^2\beta a^b+\frac{1}{2}
\left(e^c\nabla_c\alpha\right)\phi^b.
\label{nhn}
\end{eqnarray}
If we multiply both sides of Eq.~(\ref{nhn}) by $a_b$ and integrate
it over $\Sigma$, we again end up with an integral which, like
Eq.~(\ref{nh9}), is not guaranteed to be positive definite.

In order to simplify the situation, we recall that by our assumption the spacelike 2-planes
orthogonal to both $\chi^a$ and $\phi^a$ are integral 2-submanifolds, $\overline{\Sigma}$. So let us further project Eq.~(\ref{nhn}) onto $\overline{\Sigma}$ using the projector $\Pi_{a}{}^{b}$ in Eq.~(\ref{pro2}). The $\overline{\Sigma}$ projections $\overline{a}_b$ and $\overline{f}_{ab}$ of $a_b$ and $f_{ab}$, or $A_b$ and $F_{ab}$ are given by
\begin{eqnarray}
\overline{a}_b&=&\Pi_b{}^{c}a_c=\Pi_b{}^{c}A_c,\\
\overline{f}_{ab}&=&\Pi_a{}^{c}\Pi_b{}^{d}f_{cd}=\Pi_a{}^{c}\Pi_b{}^{d}F_{cd}=\overline{D}_{[a}\overline{a}_{b]},
\label{2mani1}
\end{eqnarray}
where $\overline{D}$ is the induced connection over $\overline{\Sigma}$. We multiply Eq.~(\ref{nhn}) by $ f\Pi^{c}{}_{b}$, where $f^2=\phi^a\phi_a$ as before and we get
\begin{eqnarray}
f\Pi^{c}{}_{b}D_{a}\left(\beta f^{ab}\right)=f\Pi^{c}{}_{b}\left[m^2\beta a^b+\frac{1}{2}
\left(e^c\nabla_c\alpha\right)\phi^b\right]=f\beta m^2\overline{a}^c.
\label{2mani2}
\end{eqnarray}
Let us consider the left hand side of this equation. Using Eq.~(\ref{pro2}) we write this as
\begin{eqnarray}
f\Pi^{c}{}_{b}D_{a}\left(\beta f^{ab}\right)=f\left[\delta^{c}{}_{b}-f^{-2}\phi^c\phi_b\right]D_{a}\left(\beta f^{ab}\right)+\frac{f}{\beta^2}\chi^c\chi_bD_{a}\left(\beta f^{ab}\right).
\label{2mani3}
\end{eqnarray}
Since $f^{ab}$ is spacelike, $f^{ab}\chi_b=0$ and we may then rewrite the last term of the above equation as
\begin{eqnarray}
\frac{f}{\beta^2}\chi^c\chi_bD_{a}\left(\beta f^{ab}\right)=-\frac{f}{2\beta}\chi^cf^{ab}D_{[a}\chi_{b]}=\frac{f}{2\beta^2}\chi^cf^{ab}\chi_bD_a\beta=0,
\label{2mani4}
\end{eqnarray}
where we have substituted Eq.~(\ref{adt9}) with the index $a$ projected onto $\Sigma$. So Eq.~(\ref{2mani2}) can now be written as
\begin{eqnarray}
f\left[\delta^{c}{}_{b}-f^{-2}\phi^c\phi_b\right]D_a\left(\beta f^{ab}\right)=f\beta m^2\overline{a}^c.
\label{2mani5}
\end{eqnarray}
We note that since $D_a$ is spacelike, we always have $\chi_aD^a\equiv0$. Also, Eq.~(\ref{2mani4}) gives $\chi_bD_{a}\left(f^{ab}\right)=0$. These show that
\begin{eqnarray}
\overline{D}_{a}\left(f\beta \overline{f}^{ab}\right)=\Pi^{b}{}_{e}\Pi^{f}{}_{a}D_f
\left(f \beta f^{ae}\right)=\left[ \delta^{b}{}_{e}-f^{-2}\phi^b\phi_e\right]\left[\delta^{f}{}_{a}-f^{-2}\phi^f\phi_a\right]D_f\left(f \beta f^{ae}\right).\nonumber\\
\label{2mani6}
\end{eqnarray}
Next, using the projector (\ref{pro1}) and the definition (\ref{nh6}), we write the induced magnetic tensor $f^{ab}$ as
\begin{eqnarray}
f^{a b}=h^{a}{}_{c}h^{b}{}_{d}F^{cd}=F^{ab}+\beta^{-1}\left(\chi^a e^b-\chi^b e^a\right). 
\label{nh14'}
\end{eqnarray}
We take the Lie derivative of this equation with respect to the Killing field $\phi^a$. The first term, $\pounds_{\phi}F^{ab}$ vanishes by Eq.~(\ref{adch31}). By Eq.~(\ref{cral23}), we have $\pounds_{\phi}\chi_a=0=\pounds_{\phi}\chi^a$. This implies that $\pounds_{\phi}\beta^2=-\pounds_{\phi}\left(\chi^a\chi_a\right)=0$. Hence we further have $\pounds_{\phi}e^a=\pounds_{\phi}\left(\beta^{-1}\chi_bF^{ab}\right)=0$. Thus we find $\pounds_{\phi}f^{ab}=0$ identically, which also means $\pounds_{\phi}\left( \beta f^{ab}\right)=0$. Then starting from Eq.~(\ref{2mani6}) we follow exactly the same procedure that led Eq.~(\ref{adch38}) to Eq.~(\ref{procamag}) to now yield
\begin{eqnarray}
\overline{D}_{a}\left(f\beta \overline{f}^{ab}\right)=m^2f\beta
\overline{a}^b, 
\label{nh14}
\end{eqnarray}
Contracting both sides by $\overline{a}_b$ and integrating by parts over $\overline{\Sigma}$ we get
\begin{eqnarray}
 \int_{\partial\overline{\Sigma}}f\beta \overline{a}_b \overline{f}^{ab}m_a+\int_{\overline{\Sigma}}\beta f\left(\overline{f}_{ab}
  \overline{f}^{ab}+m^2\overline{a}^b\overline{a}_b\right)=0, 
\label{nh15}
\end{eqnarray}
where $\partial\overline{\Sigma}$ denotes the boundary of $\overline{\Sigma}$ and $m^a$ is a unit spacelike normal directed towards $\overline{\Sigma}$. This 1-dimensional boundary comprises of two spacelike curves located at the two horizons, $\beta^2=0$.
 Since $ \overline{a}_b$ and $\overline{f}^{ab}$ are both physical fields, the boundedness arguments over the horizons can be given for them as before and thus the integral over $ \partial\overline{\Sigma}$ in Eq.~(\ref{nh15}) vanishes leaving us with the vanishing spacelike
integral over $\overline{\Sigma}$. This shows us that $\overline{f}_{ab}=0=\overline{a}_b$ throughout the 2-submanifolds, $\overline{\Sigma}$. Next we write $\overline{a}_b$ as
\begin{eqnarray}
 \overline{a}_b=\Pi_{b}{}^{a}A_a=A_b+\beta^{-2}\chi_b\left(A_a\chi^a\right)-f^{-2}\phi_b \left(A_a\phi^a\right).
\label{nh14"}
\end{eqnarray}
Then we use $\pounds_{\phi}\chi^a=0$ and $\pounds_{\phi}A^a=0$ to find $\pounds_{\phi}\overline{a}_b=0$. The commutativity of the two Killing fields gives $\pounds_{\xi}\chi^a=0=\pounds_{\xi}\chi_a$. Also using $\pounds_{\xi}A_b=0$, we see from Eq.~(\ref{nh14"}) that
 $\pounds_{\xi}\overline{a}_b=0$. Using the vanishing of these two Lie derivatives we get
\begin{eqnarray}
 \pounds_{\chi}\overline{a}_b=\pounds_{\xi}\overline{a}_b+\alpha\left(\pounds_{\phi} \overline{a}_b\right)+\left(\overline{a}_c\phi^c\right)\nabla_b \alpha=0,
\label{nh14"'}
\end{eqnarray}
where we have used the orthogonality of $\phi_c$ and $\overline{a}_c$. The vanishing Lie derivatives along $\chi^a$ and $\phi^a$ show us that $\overline{a}_b=0$ throughout the spacetime. Similarly we can show 
$\overline{f}_{ab}$ vanishes throughout the spacetime. Thus we may take the following form
for the vector field $A_b$,
\begin{eqnarray}
A_b=\Psi_1(x) \chi_b+\Psi_2(x)\phi_b,
\label{nh16}
\end{eqnarray}
where $\Psi_1$ and $\Psi_2$ are some differentiable functions. Using $\pounds_{\phi}A_b=0=\pounds_{\phi}\chi_b$ and the orthogonality $\chi^a\phi_a=0$, we have
\begin{eqnarray}
\pounds_{\phi}\Psi_1=0=\pounds_{\phi}\Psi_2. 
\label{vanlee1}
\end{eqnarray}
On the other hand,
$\pounds_{\chi}A_b= \left(A_a\phi^a\right)\nabla_b \alpha$, $\pounds_{\chi}\phi_b= f^2\nabla_b \alpha$, along with the vanishing Lie derivatives $\pounds_{\chi}\alpha=0=\pounds_{\phi}\alpha$, and the orthogonality $\chi_a\phi^a=0$ imply 
\begin{eqnarray}
\pounds_{\chi}\Psi_1=0=\pounds_{\chi}\Psi_2.
\label{vanlee2}
\end{eqnarray}
In terms of $\Psi_1(x)$ and $\Psi_2(x)$, the field tensor $F_{ab}$ becomes
\begin{eqnarray}
F_{ab}=\left(\nabla_{[a}\Psi_1(x)\right)\chi_{b]}+ \left(\nabla_{[a}\Psi_2(x)\right)\phi_{b]} 
+\Psi_1(x)\nabla_{[a}\chi_{b]}+\Psi_2(x)\nabla_{[a}\phi_{b]}.
\label{nh17'}
\end{eqnarray}
Substituting Eq.s~(\ref{adt9}), (\ref{sa16}) into this, we compute the
Proca Lagrangian (\ref{adch34}) in terms of $\Psi_1(x)$ and $\Psi_2(x)$,
\begin{eqnarray}
  \mathcal{L}&=&\frac{1}{2}\left(\beta \nabla_a\Psi_1+2\Psi_1\nabla_a
    \beta\right)^2-\frac{1}{2}\left(f\nabla_a\Psi_2+2\Psi_2\nabla_a
    f\right)^2+f^2\Psi_2\left(\nabla_a
    \Psi_1\right)\left(\nabla^a\alpha\right) \nonumber \\ 
&& \qquad\qquad +\frac{f^4\Psi_2^2}{2\beta^2}\left(\nabla_a \alpha\right)
  \left(\nabla^a\alpha\right) 
  +\frac{2f^2}{\beta}\Psi_1\Psi_2\left(\nabla_a\beta\right)
  \left(\nabla^a\alpha\right)+\frac{m^2}{2}\left(\beta^2  
    \Psi_1^2-f^2\Psi_2^2\right). \nonumber\\
\label{nh17}
\end{eqnarray}
The equations of motion for the two degrees of freedom $\Psi_1$ and $\Psi_2$ are then
\begin{eqnarray}
 \nabla_a\left(\beta^2
  \nabla^a\Psi_1\right)-2\beta\left(\nabla_a\beta\right)\left(\nabla^a
  \Psi_1\right)+\nabla_a\left(2\beta \Psi_1\nabla^a
  \beta\right)-4\Psi_1\left(\nabla_a\beta\right)
\left(\nabla^a\beta\right) && \nonumber\\    +
\nabla_a\left(f^2\Psi_2\nabla^a\alpha\right)-
\frac{2f^2}{\beta}\Psi_2\left(\nabla_a\beta\right)
\left(\nabla^a\alpha\right)-m^2\beta^2\Psi_1&=&0, \nonumber\\
\label{nh18}
\end{eqnarray}
and
\begin{eqnarray}
\nabla_a\left(f^2 \nabla^a\Psi_2\right)-2f\left(\nabla_af\right)
\left(\nabla^a \Psi_2\right) + \nabla_a\left(2f
  \Psi_2\nabla^af\right) -  
4\Psi_2\left(\nabla_af\right)\left(\nabla^af\right) 
\qquad\qquad &&\nonumber\\
+ \frac{f^4\Psi_2}{\beta^2}\left(\nabla_a\alpha
\right)\left(\nabla^a\alpha\right)
+ \frac{2f^2}{\beta}\Psi_1\left(\nabla_a\beta \right)
\left(\nabla^a\alpha\right)+f^2\left(\nabla_a\Psi_1\right)
\left(\nabla^a\alpha\right)-m^2f^2\Psi_2&=&0. \nonumber\\
\label{nh19}
\end{eqnarray}
Let us now project Eq.s~(\ref{nh18}) and (\ref{nh19}) onto $\Sigma$ and form quadratic integrals following the same techniques described before. 

We have already shown that $\pounds_{\chi}\beta=0=\pounds_{\chi}\alpha$, which mean that $\nabla_a\beta=D_a\beta$ and $\nabla_a \alpha=D_a \alpha$. Using the commutativity of the Killing fields, Eq.~(\ref{2k}), 
we have $\pounds_{\chi}f^2=2\phi^a\pounds_{\chi}\phi_a =2f^2\pounds_{\phi}\alpha=0$. This means
that we also have $\nabla_af=D_af$. Then using Eq.s~(\ref{vanlee2}), and following exactly the same procedure that, starting from the respective equations of motion, led to Eq.s~(\ref{eqscal}), (\ref{nh2}), we now have of Eq.~(\ref{nh18}) written on $\Sigma$,
\begin{eqnarray}
  D_a\left(\beta^3 D^a\Psi_1\right)-2\beta^2 \left(D_a\beta\right)
  \left(D^a \Psi_1\right)+D_a\left(2\beta^2 \Psi_1 D^a
    \beta\right)-4\beta\Psi_1 \left(D_a\beta
  \right)\left(D^a\beta\right)+\qquad \nonumber\\ 
  D_a\left(\beta f^2\Psi_2
    D^a\alpha\right)-
  2f^2\Psi_2\left(D_a\beta\right)
  \left(D^a\alpha\right)-m^2\beta^3\Psi_1=0, \nonumber\\
\label{nh20}
\end{eqnarray}
and also Eq.~(\ref{nh19}) written on $\Sigma$,
\begin{eqnarray}
&&D_a\left(f^2\beta D^a\Psi_2\right)-2\beta f\left(D_a f
\right)\left(D^a \Psi_2\right)+D_a\left(2\beta f
  \Psi_2D^af\right)-4\beta \Psi_2\left(D_af\right)
\left(D^af\right)+\qquad\quad \nonumber\\ 
&&\frac{f^4\Psi_2}{\beta}\left(D_a\alpha \right)
\left(D^a\alpha\right)+
2f^2\Psi_1\left(D_a\beta\right)\left( D^a\alpha\right)+\beta
f^2\left(D_a\Psi_1\right)\left(D^a\alpha\right)-m^2\beta
f^2\Psi_2=0. \nonumber\\
\label{nh21}
\end{eqnarray}
We now multiply Eq.~(\ref{nh20}) by $\Psi_1$ and Eq.~(\ref{nh21})
by $\Psi_2$, add them and integrate by parts over $\Sigma$. The surface integrals
do not survive as we can see from the boundedness arguments over $\partial \Sigma$ presented earlier and we have
\begin{eqnarray}
\int_{\Sigma}\beta\left[ \left(\beta
    D_a\Psi_1+2\Psi_1D_a\beta\right)^2 +  
\left(fD_a\Psi_2+2\Psi_2D_a f\right)^2
-\frac{f^4\Psi_2^2}{\beta^2}\left(D_a\alpha\right)
\left(D^a\alpha\right) \right. \qquad\qquad \nonumber \\
\left. +m^2\left(\beta^2\Psi_1^2 +
  f^2\Psi_2^2\right)\right]=0. \nonumber\\
\label{nh22}
\end{eqnarray}
This is clearly not positive definite due to the presence of the 
third term which is negative. We can naively interpret that term as the centrifugal 
effect on the field due to the rotation of the spacetime. Let us now investigate
whether the rotation can actually be so large that the integrand in
Eq.~(\ref{nh22}) becomes negative and the matter field can really remain outside the black hole horizon.

In order to do this, let us consider the Killing identity for $\phi_b$
\begin{eqnarray}
\nabla_b\nabla^b\phi_a=-R_{a}{}^{b}\phi_b.
\label{nh23}
\end{eqnarray}
Contracting this by $\phi^a$ and substituting Eq.~(\ref{sa16}) into it
we get 
\begin{eqnarray}
\nabla_b\nabla^b f^2=\left[ 4\left(\nabla_a
    f\right)\left(\nabla^af\right)-\frac{f^4}{\beta^2}
  \left(\nabla_a\alpha  \right)\left(\nabla^a\alpha\right)-2R_{a
    b}\phi^a\phi^b\right].  
\label{nh24'} 
\end{eqnarray}
Since $\pounds_{\chi}f=0$, the above equation can be written on $\Sigma$ as
\begin{eqnarray}
D_b \left( \beta D^b f^2\right)=\beta\left[ 4\left(D_a
    f\right)\left(D^af\right)-\frac{f^4}{\beta^2}
  \left(D_a\alpha \right)\left(D^a\alpha\right)-2R_{a
    b}\phi^a\phi^b\right].  
\label{nh24}
\end{eqnarray}
Multiplying with
$\Psi_2^2$ and integrating by parts over $\Sigma$, we see that the $\partial\Sigma$ integral i.e., the integral over the horizons does not survive from the boundedness arguments and we obtain 
\begin{eqnarray}
\int_{\Sigma}\beta \left[4f \Psi_2 \left(D_a\Psi_2\right)
  \left(D^af\right)+ 4\Psi_2^2 \left(D_a f\right)
  \left(D^af\right)-\frac{\Psi_2^2f^4}{\beta^2}\left(D_a\alpha
  \right)\left(D^a\alpha\right)-2\Psi^2_2R_{a
    b}\phi^a\phi^b\right]=0. \nonumber\\
\label{nh25}
\end{eqnarray}
Subtracting Eq.~(\ref{nh25}) from Eq.~(\ref{nh22}) we get
\begin{eqnarray}
\int_{\Sigma}\beta\left[ \left(\beta D_a
    \Psi_1+2\Psi_1D_a\beta\right)^2+ f^2 \left(D_a\Psi_2\right)
  \left(D^a\Psi_2\right)+2\Psi^2_2R_{a b}\phi^a\phi^b 
+m^2\left(\beta^2\Psi_1^2+f^2\Psi_2^2\right)\right]=0.\nonumber\\
\label{nh26}
\end{eqnarray}
So the no hair result $\Psi_1=0=\Psi_2$ will follow from
Eq.~(\ref{nh26}) if $R_{ab}\phi^a\phi^b\geq 0$. In particular, using Einstein's equations
\begin{eqnarray}
R_{ab}\phi^a\phi^b=8\pi G\left(T_{ab}-\frac{1}{2}T
  g_{ab}\right)\phi^a\phi^b+\Lambda f^2. 
\label{nh27a}
\end{eqnarray}
We compute the energy-momentum tensor for the Proca Lagrangian (\ref{adch34}),
\begin{eqnarray}
T_{ab}=-\frac{2}{\sqrt{-g}}\frac{\delta S_{\rm{P}}}{\delta g^{ab}}= F_{ac}F_{b}{}^{c}+m^2A_aA_b +\mathcal{L}g_{ab},
\label{nh27}
\end{eqnarray}
where ${\displaystyle S_{\rm{P}}=\int d^4 x \sqrt{-g}{\mathcal {L}}}$ is the action corresponding to the Proca Lagrangian $ {\mathcal {L}}$. Eq.~(\ref{nh27}) yields
\begin{eqnarray}
\left(T_{ab}-\frac{1}{2}T
  g_{ab}\right)\phi^a\phi^b=\left(\frac12 b_a^2 + \frac12 f^2e_a^2 +
m^2  f^4 \Psi_2^2 \right),  
\label{nh28}
\end{eqnarray}
where $b_a=F_{ab}\phi^b$ and $e_a$ is the electric field defined in
Eq.~(\ref{nh6}). We have already proved that $e_a\phi^a=0$,  which means $b_a$
is spacelike.  The electric field $e^a$ is also spacelike as
mentioned earlier. So Eq.~(\ref{nh28}) consists of spacelike inner products and hence ${\displaystyle\left(T_{ab} -
  \frac{1}{2}T g_{ab}\right)\phi^a\phi^b\geq0}$ for the Proca
field. Putting in all this, we can rewrite Eq.~(\ref{nh26}) as
\begin{eqnarray}
\int_{\Sigma}\beta\left[ \left(\beta D_a \Psi_1+
    2\Psi_1D_a\beta\right)^2+ 
f^2 \left(D_a\Psi_2 \right)\left(D^a\Psi_2\right) 
+m^2\beta^2\Psi_1^2 \right. \qquad \qquad && \nonumber \\ 
 \left. + \left(m^2+2\Lambda\right)f^2\Psi_2^2 +
 16\pi G\Psi_2^2\left(\frac12 b_a^2 + \frac12 f^2e_a^2 +
m^2  f^4 \Psi_2^2 \right)
%%% \left(T_{ab}-\frac{1}{2}T g_{ab}\right)\phi^a\phi^b
\right] =0,
\label{nh29}
\end{eqnarray}
which gives
$\Psi_1=0=\Psi_2$ over $\Sigma$. Since
$\pounds_{\chi}\Psi_1=0=\pounds_{\chi}\Psi_2$, Eq.~(\ref{vanlee2}), we have
$\Psi_1=0=\Psi_2$ throughout the spacetime. This, combined with the
previous proof that $\overline{a}_b=0$, is the desired no hair result
for a de Sitter black hole for the Proca-massive vector field. 

Clearly, our proof is also valid for asymptotically flat
stationary axisymmetric spacetimes, $\Lambda=0$. We have only to
replace the outer boundary or the cosmological horizon by a
2-sphere at spacelike infinity with sufficiently rapid fall off
conditions imposed upon the fields. Our proof also applies to asymptotically
anti-de Sitter spacetimes provided we assume $m^2\geq 2|\Lambda|$
in Eq.~(\ref{nh29}). We note that this is not a
strong assumption --- it only means that the Compton wavelength of
the vector field is less than the cosmological length scale or the
AdS radius.

As in the static case, the no hair proof fails for the Maxwell field. The local gauge symmetry of the Lagrangian gives rise to a charged solution, namely the Kerr-Newman-de Sitter solution~\cite{Carter:1968ks}, given in Eq.(\ref{s26i}). 
%%%%%%%%%%%%%%%%%%%%%%%%%%%
%\vskip 1cm
%%%%%%%%%%%%%%%%%%%%%%%%%%%

Let us now summarize the discussions.
In this Chapter we have studied various static and stationary de Sitter black hole no hair theorems by restricting our attention to the region between the two horizons. Unlike usual investigations of black hole spacetimes, we have managed to completely bypass bothering about the asymptotic behavior, only we needed to assume that the cosmological horizon exists and there is no naked curvature singularity anywhere in our region of interest. 

Interestingly, we have seen in Section 3.1.4 that the Abelian Higgs model allows a static and spherically symmetric solution with electric charge which has no counterpart in the asymptotically flat case. This suggests the intriguing possibility that, even for
the $\Lambda=0$ black holes with hair, there may be additional classes of solutions for $\Lambda>0$, coming from non-trivial boundary conditions at the two horizons. For example, black holes pierced by a cosmic Nielsen-Olesen string~\cite{Gregory:1995}, black holes
with non-trivial external Yang-Mills and Higgs fields, or Skyrme black holes~\cite{Volkov:1998cc, Brihaye:2006kn}, may have more varied counterparts for $\Lambda>0$. Black holes with discrete gauge hair (see~\cite{Coleman:1991ku} for a review), because of the
underlying Higgs model, may be dressed differently for $\Lambda>0$. There may also be new axisymmetric solutions in a Higgs background. Other kinds of quantum hair such as the non-Abelian quantum hair~\cite{Coleman:1991ku, Lahiri:1992yz} or the spin-2 hair~\cite{Dvali:2006az}, whose existence are related to the topology of the spacetime, are likely to be present also for $\Lambda>0$.

Since the static, spherically symmetric charged solution corresponding to the Abelian Higgs model sits over the false vacuum $\rho=0$ of the Higgs field, it is likely that this solution will be unstable under perturbations. On the other hand the uncharged solution located at the true vacuum $\rho=\pm v$ of the Higgs field, should be stable under perturbations. Therefore if a charged solution forms initially, it should decay to the uncharged solution. It would be very interesting to study this decay mechanism.

We have also proven the no hair theorems for scalar and Proca-massive vector fields for a stationary axisymmetric de Sitter black hole spacetime. We note that in comparison to
the proof for a static spacetime, this proof contains some additional geometric constraints such as the commutativity of the two Killing fields $\xi^a$ and $\phi^a$ and the existence of spacelike 2-submanifolds orthogonal to them. Also to prove the theorem for the vector field we had to use explicitly in Eq.~(\ref{nh27a}) the Einstein equations. For a static spacetime
we did not need to do that.

For the static spacetime it is necessary to assume spherical symmetry in
order to prove the no hair theorem for the Abelian Higgs
model. In fact if we have a cylindrically symmetric matter distribution, we may have a cosmic string
piercing the horizons, as will be discussed in the next Chapter. It seems likely that we will have a string-like solution for a rotating axisymmetric de Sitter black hole also.

We wish to mention here that the no hair results proved here are not black
hole uniqueness theorems. It is well known that if one assumes spherical symmetry, the only solution to the vacuum Einstein equations in (3+1)-dimensions is the Schwarzschild spacetime, known as Birkhoff's theorem (see e.g. \cite{Hawking:1973uf}). Following \cite{Hawking:1973uf}, one can similarly generalize this result for $\Lambda>0$. For a discussion on this and for some subtle issues regarding the beyond horizon properties of the Schwarzschild-de Sitter spacetime see \cite{Cruz:1999gd}-\cite{Schleich:2009ix}.

The situation is however very different for stationary axisymmetric $\Lambda>0$ spacetimes.
 It has been proven that for $\Lambda=0$, the
Kerr spacetime is the only asymptotically flat black hole solution
of the vacuum Einstein equations in (3+1)-dimensions~\cite{Chandrasekhar:1985kt, Mazur:2000pn, Robinson}. The uniqueness of
 asymptotically anti-de Sitter black hole spacetimes was given in \cite{Boucher:1983cv} by a remarkable use of the Lindblom identity and the positivity of the gravitational mass. In (2+1)-dimensions, a result analogous
to Birkhoff's theorem was proven for the BTZ black hole in~\cite{AyonBeato:2004if}. However for $\Lambda > 0$, no proof of uniqueness of stationary axisymmetric black hole solutions is known~\cite{Robinson, Boucher:1983cv}. Although we note that our results reduce the Einstein-scalar (in convex potential) and Einstein-massive vector (with no gauge symmetry) systems to vacuum Einstein equations in the presence of a stationary axisymmetric black hole. So any proof of uniqueness of the Kerr-de Sitter spacetime, if it exists, will apply to these
systems as well. This remains as an interesting problem.

%%%%%%%%%%%%%%%%%%%%%%%%%%%%%%%%%%%%%%%%%%%%%%%%%%%%%%%%%%%%%%%%%%%%%%%%%%%%%%%%%%%%%%%%%%%%%%%%%%%%%%%%%
\chapter{Cosmic strings and positive $\Lambda$}
%\label{t'hooft_monopole_Chapter}

 In the previous Chapter we discussed static black hole solutions with the Abelian Higgs model for spherically symmetric mass distribution. In this Chapter we will also discuss some exact solutions with the Abelian Higgs model but for cylindrically symmetric mass distribution, namely cosmic string solutions with $\Lambda>0$.  Precisely, by cosmic string we mean a
vortex line (a cylindrically symmetric or axisymmetric mass distribution which is zero outside a compact region of space) in the Abelian Higgs model. It is well known that in flat spacetime the Abelian Higgs model shows vortex solutions~\cite{Nielsen:1973cs}, known as the Nielsen-Olesen string.

Let us come to our motivation for making this study with $\Lambda>0$.
The first motivation comes from the black hole no hair theorem with positive $\Lambda$ for the Abelian Higgs model discussed in Section 3.1.4. We found a charged solution which has no $\Lambda=0$ analogue. The black hole looks like the Reissner-N\"{o}rdstrom-de Sitter solution
with the Higgs field in the false vacuum. This of course disagrees with the usual no hair statement.

%% This new solution is purely an outcome of the existence of the cosmological event horizon. Precisely, in an asymptotically flat spacetime the boundary conditions are
%%imposed at infinity, while for a positive cosmological constant it
%%is both convenient and sufficient to impose them at the cosmological
%%horizon. In fact this new solution shows that the non-trivial boundary conditions at the cosmological event horizon may change the local physics considerably. 

In general, given some asymptotically flat solution
(corresponding to $\Lambda=0$) of some matter fields coupled to gravity, we may
find additional solutions, or at least qualitatively different ones, when
there is an outer or cosmological horizon (corresponding to $\Lambda > 0$).

We are motivated by these arguments to look at cylindrically symmetric cosmic strings in spacetimes with $\Lambda>0\,.$ While
the role of such cosmic strings in cosmological perturbations and
structure formation is ruled out and the contribution of these
strings to the primordial perturbation spectrum must be less than
9\% (see~\cite{Perivolaropoulos:2005wa} for a review and references), such strings could exist in small numbers. How does a positive cosmological constant or a
cosmological horizon affect the physics of the string? We discussed in the first Chapter that in asymptotically flat spacetimes a self gravitating cosmic string produces a conical singularity, or a deficit
angle~(see e.g. \cite{Vilenkin2:2000} and references therein). Due to this conical singularity light bends towards the string in the asymptotic region where the curvature is zero.
On the other hand, it is also known that a cosmological constant
affects the bending of light~\cite{Rindler:2007zz}-\cite{Schucker:2007} by a repulsive effect. So both the attractive and repulsive effects on the geodesics should be present in a string spacetime with $\Lambda> 0\,$.

In this Chapter we will present analytical results for a cosmic string in two kinds of
spacetime with a positive cosmological constant. The first one
is static and cylindrically symmetric, with an infinite string placed along the axis. We calculate the
angle deficit and the bending of light for this spacetime. The
other spacetime we consider is the Schwarzschild-de Sitter
spacetime, and a cosmic string stretched between the inner and
outer horizons. We consider both non-gravitating and gravitating strings and show that they can exist between the two horizons of this spacetime.

%%%%%%%%%%%%%%%%%%%%%%%%%%%%%%%%%%%%%%%%%%%%%%%%%%%%%%%%%%%
\section{Free cosmic string and angle deficit}
%%%%%%%%%%%%%%%%%%%%%%%%%%%%%%%%%%%%%%%%%%%%%%%%%%%%%%%%%%%
Let us start by constructing a suitable ansatz for a static and cylindrically symmetric  spacetime with the usual coordinatization $(t,~z,~\rho,~\phi)$. The coordinate vector fields $\{(\partial_{t})^a,~(\partial_{z})^a,~(\partial_{\phi})^a\}$ are Killing fields of this spacetime generating respectively staticity, space translation symmetry along the axis and rotational symmetry around the axis.  So none of the metric components are dependent on $(t,~z,~\phi)$. Since any 2-dimensional metric may be written in a conformally flat form \cite{Chandrasekhar:1985kt}, we may take the $(t,~z)$ part of the metric to be conformally flat. With this, we make the following ansatz
\begin{eqnarray}
ds^2=e^{A(\rho)}\left[-dt^2+ dz^2\right]+\rho^2 e^{B(\rho)}d\phi^2+ e^{C(\rho)}d\rho^2,
\label{adch41}
\end{eqnarray}
where $A(\rho)$, $B(\rho)$ and $C(\rho)$ are smooth functions. We can further simplify (\ref{adch41}) by redefining the radial variable as ${\displaystyle\rho^{\prime}:=\int e^{\frac{C(\rho)}{2}}d\rho}$. Then dropping the primes we arrive at the following simplified form
\begin{eqnarray}
ds^2=e^{A(\rho)}\left[-dt^2+ dz^2\right]+\rho^2 e^{B(\rho)}d\phi^2+ d\rho^2.
\label{metric'}
\end{eqnarray}
The orbits of the azimuthal spacelike Killing field
$(\partial_{\phi})^a$ are closed spacelike curves which shrink to a
point as $\rho \rightarrow 0$. We regard the set of points $\rho=0$ as the axis of the spacetime, then a convenient coordinatization will be to set the
metric to be locally flat on the axis, i.e.
\begin{eqnarray}
ds^2\stackrel{\rho\rightarrow 0}\longrightarrow
-dt^2+dz^2+\rho^2d\phi^2+d\rho^2.  
\label{bcmetric}
\end{eqnarray}
We can always do this as long as there is no curvature singularity
on the axis. With this coordinatization let us first solve the cosmological constant vacuum
equations, ${\displaystyle R_{ab}-\frac{1}{2}R g_{ab}+ \Lambda g_{ab}=0}$ or equivalently, $R_{ab}-\Lambda g_{ab}=0$ with $\Lambda>0$. Eq.s~(1.2)-(1.5) yield that the 
 cross components of $R_{\mu\nu}$~($\mu\neq\nu$) vanish identically for (\ref{metric'}). Since the $(t,~z)$ part of the metric (\ref{metric'}) is conformally flat and none of the metric functions depend upon these coordinates, we have $R_{tt}=-R_{zz}$. Then we arrive at the following three independent $\Lambda$-vacuum Einstein equations $R_{\mu}{}^{\nu}-\Lambda\delta_{\mu}{}^{\nu}=0$,
\begin{eqnarray}
R_{t}{}^{t}-\Lambda\delta_{t}{}^{t}=0\Rightarrow \frac{A^{\prime \prime}}{2}+ \frac{A^{\prime 2}}{2}+\frac{A^{\prime}}{2\rho}+\frac{A^{\prime} B^{\prime}}{4}+\Lambda=0,
\label{adch42}
\end{eqnarray}
\begin{eqnarray}
R_{\rho}{}^{\rho}-\Lambda\delta_{\rho}{}^{\rho}=0\Rightarrow A^{\prime \prime}+ \frac{B^{\prime \prime}}{2}+ \frac {B^{\prime 2}}{4}+\frac{A^{\prime 2}}{2}
+\frac {B^{\prime}}{\rho}+\Lambda=0,
\label{adch43}
\end{eqnarray}
and
\begin{eqnarray}
R_{\phi}{}^{\phi}-\Lambda\delta_{\phi}{}^{\phi}=0\Rightarrow \frac{B^{\prime \prime}}{2}+ \frac {B^{\prime 2}}{4}+\frac{A^{\prime} B^{\prime}}{2}+\frac{A^{\prime} +B^{\prime}}{\rho}+\Lambda=0,
\label{adch44}
\end{eqnarray}
where a `prime' denotes differentiation once with respect to $\rho$. Eq.s~(\ref{adch42})-(\ref{adch44}) can be solved for $A(\rho)$ and $B(\rho)$ in the following way. We add Eq.~(\ref{adch44}) with twice of Eq.~(\ref{adch42}) and subtract Eq.~(\ref{adch43}) from the result to get
\begin{eqnarray}
A^{\prime} B^{\prime}=-\left(\frac{2A^{\prime}}{\rho} +\frac{A^{\prime 2}}{2}+2\Lambda\right).
\label{adal41}
\end{eqnarray}
Next, we rewrite Eq.~(\ref{adch43}) as
\begin{eqnarray}
\left(A^{\prime}+\frac{B^{\prime}}{2}\right)^{\prime}+\left(A^{\prime}+\frac{B^{\prime}}{2}\right)^2-\frac{A^{\prime 2}}{2}-A^{\prime}B^{\prime}+\frac{B^{\prime}}{\rho}+\Lambda=0.
\label{adal42}
\end{eqnarray}
Substituting the expression for $A^{\prime} B^{\prime}$ from Eq.~(\ref{adal41}) into Eq.~(\ref{adal42}) we obtain 
\begin{eqnarray}
\left(A^{\prime}+\frac{B^{\prime}}{2}\right)^{\prime}+\left(A^{\prime}+\frac{B^{\prime}}{2}\right)^2+\frac{2}{\rho}\left(A^{\prime}+\frac{B^{\prime}}{2} \right)+3\Lambda=0,
\label{adch45}
\end{eqnarray}
which can be rewritten as
\begin{eqnarray}
\left(A^{\prime}+\frac{B^{\prime }}{2}+\frac{1}{\rho}\right)^{\prime}+\left(A^{\prime}+\frac{B^{\prime}}{2}+\frac{1}{\rho}\right)^2+3\Lambda=0.
\label{adch45}
\end{eqnarray}
We integrate Eq.~(\ref{adch45}) once to find 
\begin{eqnarray}
\left(A^{\prime}+\frac{B^{\prime }}{2}\right)=-\sqrt{3\Lambda}\tan\sqrt{3\Lambda}\left(\rho-k_1\right)-\frac{1}{\rho},
\label{adch46}
\end{eqnarray}
where $k_1$ is an integration constant. Substituting Eq.~(\ref{adch46}) into Eq.~(\ref{adch42}) we have
\begin{eqnarray}
A^{\prime\prime}-\sqrt{3\Lambda} A^{\prime}\tan\sqrt{3\Lambda}\left(\rho-k_1\right)+2\Lambda=0,\label{adch47}
\end{eqnarray}
which we integrate twice to obtain
\begin{eqnarray}
A(\rho)=\frac{2}{3} \ln \left \vert \cos \sqrt{3\Lambda}(\rho-k_1)\right\vert  - \frac{k_2}{\sqrt{3\Lambda}}\ln \left \vert \sec \sqrt{3\Lambda}(\rho-k_1)+ \tan \sqrt{3\Lambda}\left(\rho-k_1\right)\right\vert+k_3,\nonumber \\
\label{adch48}
\end{eqnarray}
where $k_2$ and $k_3$ are integration constants. It is clear that when Eq.~(\ref{adch48}) is substituted into (\ref{metric'}), we may rescale the coordinates $t$ and $z$ as $t\to e^{\frac{k_3}{2}}t$ and $z\to e^{\frac{k_3}{2}}z$. So without any loss of generality we set $k_3=0$. Now let us determine the other constants $k_1$ and $k_2$ subject to the boundary condition (\ref{bcmetric}), i.e. $A(0)=0=B(0)$, and also such that the limit $\Lambda \to 0$ recovers the flat spacetime. To do this we write Eq.~(\ref{adch48}) as
\begin{eqnarray}
A(\rho)=\frac{2}{3} \ln \left \vert \sin \left(\sqrt{3\Lambda}(\rho-k_1)+\frac{\pi}{2}\right)\right\vert  - \frac{k_2}{\sqrt{3\Lambda}}\ln \left \vert \tan \left( \frac{\sqrt{3\Lambda}}{2}(\rho-k_1)+\frac{\pi}{4}\right)\right\vert.
\label{adch49}
\end{eqnarray}
It is clear that a convenient choice which satisfies our requirements would be ${\displaystyle k_1=\frac{\pi}{2 \sqrt{3\Lambda}}}$ and $k_2=\sqrt{3\Lambda}$. With these choices Eq.~(\ref{adch49}) becomes ${\displaystyle A(\rho)=\ln \left \vert 2^{\frac{2}{3}}\cos^{\frac{4}{3}}\frac{\rho\sqrt{3\Lambda}}{2}\right \vert }$. The numerical factor $ 2^{\frac{2}{3}}$ can be absorbed by coordinate rescaling, so without any loss of generality we may take
\begin{eqnarray}
A(\rho)= \ln \left \vert \cos^{\frac{4}{3}}\frac{\rho\sqrt{3\Lambda}}{2}\right \vert.
\label{adch50}
\end{eqnarray}
 Substituting the expression for $A(\rho)$ into Eq.~(\ref{adch46}), and integrating we obtain
\begin{eqnarray}
B(\rho)= \ln \left \vert \sin^2\frac{\rho\sqrt{3\Lambda}}{2} \cos^{-\frac{2}{3}}\frac{\rho\sqrt{3\Lambda}}{2}\right \vert-\ln\left\vert \rho^2 k_4\right\vert,
\label{adch50}
\end{eqnarray}
where $k_4$ is a constant. The choice of $k_4$ which satisfies the boundary condition (\ref{bcmetric}) is ${\displaystyle k_4=\frac{3\Lambda}{4}}$. With all these, we arrive at   
a $\Lambda$-vacuum solution of the Einstein equations (\ref{adch42})-(\ref{adch44}) subject to the boundary condition (\ref{bcmetric})~\cite{Tian:1986, Linet:1986sr, BezerradeMello:2003ei},
\begin{eqnarray}
ds^2=\cos^{\frac{4}{3}}\frac{\rho \sqrt{3\Lambda}}{2}
\left(-dt^2+dz^2 \right) +\frac{4}{3\Lambda}\sin^2 \frac{\rho
 \sqrt{3\Lambda}}{2}\cos^{-\frac{2}{3}}\frac{\rho
 \sqrt{3\Lambda}}{2}d\phi^2+d\rho^2.
\label{vacuum'}
\end{eqnarray}
We note that the limit $\Lambda \to 0$ in the metric (\ref{vacuum'}) recovers the usual cylindrically symmetric flat spacetime. 

Now let us look at the singularities of the metric (\ref{vacuum'}). Clearly,
the metric (\ref{vacuum'}) is singular at ${\displaystyle\rho=\frac{n\pi}{\sqrt{3\Lambda}}}$,
where $n$ are integers. Of these points, those corresponding to
even $n$ look flat, with $n=0$ being the axis. On the other hand, the points
corresponding to odd $n$ are curvature singularities. The quadratic
invariant of the Riemann tensor shows a quartic divergence there : 
\begin{eqnarray}
R_{abcd}R^{abcd} \approx \frac{\Lambda^2}{\left(\frac{n\pi}{2}-
 \frac{\rho \sqrt{3\Lambda}}{2}\right)^4},\quad n~{\rm{odd}}.  
\label{curvature}
\end{eqnarray}
The timelike Killing vector field $(\partial_{t})^a$ becomes null at these odd $n$ singularities of (\ref{vacuum'}). So these points are Killing horizons of the spacetime. However Eq.~(\ref{curvature}) shows that these horizons are naked curvature singularities. 
 The singularities for $n>1$ appear to be unphysical or irrelevant, and will not concern us further. Our region of interest will be near the axis and far from the $n=1$ naked
singularity located at ${\displaystyle\rho=\frac{\pi}{\sqrt{3\Lambda}}}$. 

In this region, let us construct a string-like solution of the Einstein equations.
 We consider Einstein's
equations with a non-vanishing energy-momentum tensor : ${\displaystyle R_{ab}-\frac{1}{2}R
g_{ab}+\Lambda g_{ab}=8 \pi G T_{ab}} $. The energy-momentum tensor
$T_{ab}$ corresponds to the Abelian Higgs model with the Lagrangian
\begin{eqnarray}
{\cal{L}} = -\left(\widetilde{\nabla}_{a}\Phi\right)^{\dagger}\left(\widetilde{\nabla}^{a}\Phi\right)
- \frac{1}{4}\tilde{F}_{ab}\tilde{F}^{ab} -
\frac{\lambda}{4}\left(\Phi^{\dagger}\Phi - \eta^2\right)^2,
\label{lagrangian}
\end{eqnarray}
where~${\displaystyle\widetilde{\nabla}_{a}\equiv\nabla_{a}+ieA_{a}}$~is the usual gauge covariant
derivative,~$\tilde{F}_{ab}=\nabla_{a}A_{b}-\nabla_{b}A_{a}$ is the
electromagnetic field strength tensor and $\Phi$ is a complex
scalar. We mentioned that (\ref{lagrangian}) has string like solutions in flat
spacetime~\cite{Nielsen:1973cs}. Let us now briefly see what is meant by that. 
The equation of motion for the gauge field $A_b$ is 
\begin{eqnarray}
\nabla_a\widetilde{F}^{a}{}_{b}\equiv j_b=-\frac{ie}{2}\left(\Phi^{\dagger}\nabla_b \Phi- 
\Phi\nabla_b\Phi^{\dagger}\right)+e^2A_b\Phi^{\dagger}\Phi.
\label{gaugeeom1}
\end{eqnarray}
We also have for any $(2,~0)$ tensor $h^{ab}$, $\nabla_{[a}\nabla_{b]}h^{cd}=-R_{abe}{}^ch^{ed}-R_{abe}{}^dh^{ce}$, which implies
\begin{eqnarray}
\nabla_b\nabla_a\widetilde{F}^{ab}=\frac{1}{2}\nabla_{[b}\nabla_{a]}
\widetilde{F}^{ab}=\frac12\left[-R_{be}\widetilde{F}^{be}+R_{eb}\widetilde{F}^{eb}\right]=0,
 \label{gaugeeom2}
\end{eqnarray}
which means $j_b$ in Eq.~(\ref{gaugeeom1}) is conserved : $\nabla_{b}j^b=0$.
By a string solution corresponding to (\ref{lagrangian}) we mean a 
cylindrically symmetric and static vortex solution in which the field lines are confined
within a compact region of space. For this requirement it is necessary that the flux $S$ 
corresponding to $\widetilde{F}_{ab}$ is quantized. To see this we compute
\begin{eqnarray}
S=\int \widetilde{F}_{ab}d\sigma^{ab} =\oint A_b dx^b,
 \label{gaugeeom3}
\end{eqnarray}
where $\sigma_{ab}$ is a spacelike 2-surface and $x^b$ denotes the boundary of that surface.
Letting $\Phi=\eta X e^{i\chi}$ we get from Eq.~(\ref{gaugeeom1})
\begin{eqnarray}
A_b=\frac{j_b}{e^2\eta^2 X^2}-\frac{1}{e}\nabla_b \chi.
 \label{gaugeeom4}
\end{eqnarray}
We substitute Eq.~(\ref{gaugeeom4}) into Eq.~(\ref{gaugeeom3}) and perform the surface integral where there is no current, i.e. $j_b=0$,
\begin{eqnarray}
S=\oint A_b dx^b=-\frac{1}{e}\oint \nabla_b \chi dx^b.
 \label{gaugeeom5}
\end{eqnarray}
The phase $\chi$ of $\Phi$ need not be single valued. The only physical requirement is
that $\Phi$ is single valued. So we can take $\chi=2\pi m$ with $m$ integer to have the flux quantization relation
\begin{eqnarray}
S=-\frac{2\pi}{e}m.
 \label{gaugeeom6}
\end{eqnarray}
The integer $ m$ is called the winding number.
Thus (\ref{lagrangian}) allows a vortex solution when the flux is quantized. It can be 
further
shown by solving the equations of motion that (\ref{lagrangian}) allows in flat spacetimes a 
cylindrically symmetric, infinitely long field configuration which is only non-zero
within a compact region of space, i.e. a string solution.%~\cite{Nielsen:1973cs}. 

 For convenience of calculations we parametrize $\Phi$ and $A_{a}$ as~\cite{Nielsen:1973cs},
\begin{eqnarray}
\Phi=\eta X e^{i\chi}, \qquad
A_{a}=\frac{1}{e}\left[P_{a}-\nabla_{a}\chi\right].
\label{gaugecov}
\end{eqnarray}
 For vortex or string-like solutions we are looking for, we have 
seen that the phase
$\chi$ of $\Phi$ is multiple valued outside the string.
 On the other hand, $\chi$ is single valued inside
the string core, so the Lagrangian (\ref{lagrangian}) inside the core 
with (\ref{gaugecov}) becomes
\begin{eqnarray}
{\cal{L}}=-\eta^2 \nabla_{a}X\nabla^a X -\eta^2 X^2 P_a P^a
-\frac{1}{4 e^2} F_{ab}F^{ab}-\frac{\lambda \eta^4}{4}
\left(X^2-1\right)^2,
\label{lagrangian2}
\end{eqnarray}
where $F_{ab}=\nabla_{a} P_b -\nabla_{b}P_a$. We will denote
the core radius by $\rho_0$. Due to the staticity and the cylindrical symmetry of
the spacetime, the matter fields $X$ and $P_a$ depend on $\rho$ only. Also 
for the vortex solution the magnetic flux must be directed along $z$. This
 means that the gauge field $P_a$ is azimuthal. So we can take the following ansatz for $X$ and $P_a$
\begin{eqnarray}
X=X(\rho),\qquad P_a=P(\rho)\nabla_a \phi\,.
\label{fieldansatz}
\end{eqnarray}
Since we are looking for a string-like solution, the energy-momentum tensor is taken to be non-zero only inside the
string core ($0\leq \rho < \rho_0$), and zero outside. Let us first compute the components of the energy-momentum tensor corresponding to the Lagrangian (\ref{lagrangian2}). The energy-momentum tensor $T_{ab}$ of any matter field with action $S_{\rm M}$  is defined with respect to the variation of the inverse metric $g^{ab}$ by ${\displaystyle T_{ab}:=-\frac{2}{\sqrt{-g}}\frac{\delta S_{\rm{M}}}{\delta g^{ab}}}$. Since the metric (\ref{metric'}) with the boundary condition (\ref{bcmetric}) describes a general non-singular static cylindrically symmetric spacetime with or without matter fields, we may use (\ref{metric'}) to compute $T_{ab}$ for the Lagrangian (\ref{lagrangian2}). Then the various non-vanishing
components of energy momentum tensor $T_{ab}$ for the configuration of (\ref{fieldansatz}) in cylindrical coordinates are
\begin{eqnarray}
&T_{tt}&=\left[\eta^2 {X^{\prime}}^2 + \frac {\eta^2 X^2 P^2
     e^{-B}}{\rho^2}+\frac{{P^{\prime}}^2 e^{-B}}{2 e^2 \rho^2}
   +\frac{\lambda \eta^4}{4} \left(X^2-1\right)^2 \right]
 e^A. \nonumber \\  
&T_{\rho\rho}&= \left[\eta^2 {X^{\prime}}^2-\frac {\eta^2 X^2 P^2s
     e^{-B}}{\rho^2}+\frac{{P^{\prime}}^2 e^{-B}}{2 e^2 \rho^2}
     -\frac{\lambda \eta^4}{4} \left(X^2-1\right)^2
     \right]. \nonumber \\  
&T_{\phi \phi}&=\left[-\eta^2 {X^{\prime}}^2 + \frac {\eta^2 X^2
     P^2 e^{-B}}{\rho^2}+\frac{{P^{\prime}}^2 e^{-B}}{2 e^2
     \rho^2} -\frac{\lambda \eta^4}{4} \left(X^2-1\right)^2
     \right] \rho^2 e^B. \nonumber \\  
&T_{zz}&= -\left[\eta^2 {X^{\prime}}^2 + \frac {\eta^2 X^2 P^2
     e^{-B}}{\rho^2}+\frac{ {P^{\prime}}^2 e^{-B}}{2 e^2 \rho^2}
     +\frac{\lambda \eta^4}{4} \left(X^2-1\right)^2 \right]
     e^A. %\nonumber \\  
\label{emtensor}
\end{eqnarray}
Let us now fix the boundary conditions for $X(\rho)$ and $P(\rho)$ following~\cite{Nielsen:1973cs}. 
 For the string solution the Higgs field $X(\rho)$
should vanish as we approach the axis $\rho=0 $, and should approach its
vacuum expectation value outside the string $\rho\geq \rho_0$. The gauge field
$A_{\phi}$ should accordingly approach
${\displaystyle-\frac{1}{e}\partial_{\phi}\chi}$ away from the string and a
constant on the axis. We set this constant to be unity.
 In other words the boundary conditions on the fields for the string-like solution would be 
\begin{eqnarray}
X\rightarrow 0,~ P\rightarrow 1~ {\rm {as}}~ \rho \to 0,~{\rm{and}}~
 X\rightarrow 1,~ P\rightarrow 0~ {\rm{for}}~\rho > \rho_0.
\label{adch51}
\end{eqnarray}
We now return to our main goal of solving Einstein's equations $R_{ab}-\frac{1}{2}R
g_{ab}+\Lambda g_{ab}=8\pi G T_{ab}$ with $\Lambda>0$. The variation of the scalar and gauge field amplitudes $X$ and
$P\,,$ and hence of the energy-momentum tensor $T_{ab}$, Eq.~(\ref{emtensor}), across the `string surface' at $\rho = \rho_0$ is a problem of considerable interest
and has been studied numerically by various authors (see e.g.~\cite{Vilenkin2:2000, BezerradeMello:2003ei}).
However, here we are concerned about the existence of the cosmic string and
its effect on the geodesic motion. Accordingly, instead of trying to solve the Einstein equations with the full expression of $T_{ab}$ given in Eq.~(\ref{emtensor}), we will simplify the situation by assuming $X=0$, $P=1$ inside the string core and
$X=1$, $P=0$ outside. This means that the string core is assumed to be entirely in the false vacuum of the Higgs field. Note that this guarantees that the energy-momentum
tensor (\ref{emtensor}) is identically zero outside the string core, on the other hand inside the core now takes the form
\begin{eqnarray}
T_{ab}\approx-\frac{\lambda \eta^4}{4}g_{ab}.
\label{adch52}
\end{eqnarray}
The fields $X(\rho)$ and $P(\rho)$ are assumed to be smoothed out sufficiently rapidly at the string
surface at $\rho=\rho_0$ so that the local conservation
law for the energy-momentum tensor $\nabla_{a}T^{ab}=0$ remains valid.
 
 Now we solve Einstein's
equations $G_{ab}+\Lambda g_{ab}=8\pi G T_{ab}$ or equivalently, $R_{ab}-\Lambda g_{ab}=8 \pi G\left(T_{ab}-\frac12Tg_{ab}\right)$ with the general ansatz (\ref{metric'}), the boundary condition (\ref{bcmetric}) and $T_{ab}$ given in Eq.~(\ref{adch52}). Inside the core $0\leq \rho <\rho_0$, Einstein's equations are then 
\begin{eqnarray}
R_{\mu \nu}-\left(\Lambda+2\pi G \lambda \eta^4\right)g_{\mu \nu}&=&0,~%\nonumber \\ 
{\rm {or~ equivalently,}} \nonumber \\
R_{\mu}{}^{\nu}-\Lambda^{\prime}\delta_{\mu}{}^{\nu}&=&0,
\label{adch52'}
\end{eqnarray}
where 
\begin{eqnarray}
\Lambda^{\prime}=\Lambda+2\pi G \lambda \eta^{4} 
\label{lambdaprime}
\end{eqnarray}
 can be regarded as the `effective cosmological constant' inside the core. Thus with the general ansatz (\ref{metric'}), Eq.s~(\ref{adch52'}) will look the same as that of the vacuum equations (\ref{adch42})-(\ref{adch44}), except that $\Lambda$ is now replaced by $\Lambda^{\prime}$. Hence the solution in this region subject to the boundary condition (\ref{bcmetric}) is given by 
\begin{eqnarray}
ds^2\approx\cos^{\frac{4}{3}}\frac{\rho
\sqrt{3\Lambda^{\prime}}}{2}\left(-dt^2+dz^2\right)
+\frac{4}{3\Lambda^{\prime}}\sin^2 \frac{\rho
\sqrt{3\Lambda^{\prime}}}{2}\cos^{-\frac{2}{3}}\frac{\rho
\sqrt{3\Lambda^{\prime}}}{2}d\phi^2+d\rho^2,  
\label{metric3}
\end{eqnarray}
i.e. the same as that of (\ref{vacuum'}) with $\Lambda$ replaced by $\Lambda^{\prime}$.
Let us now solve for the vacuum region outside the string $(\rho\geq\rho_0)$. The Einstein equations in this region are given by (\ref{adch42})-(\ref{adch44}). We note that here we cannot impose the boundary condition (\ref{bcmetric}) since the vacuum region $\rho\geq\rho_0$ for the present case does not include the axis $\rho=0$. However we note that for $T_{ab}=0$,  the required solution must coincide with (\ref{vacuum'}). Keeping this in mind we see that the constant $k_4$ appearing in Eq.~(\ref{adch50}) does not equal ${\displaystyle\frac{3\Lambda}{4}}$ for the present case. Instead, we take ${\displaystyle k_4=\delta^{-2}\frac{3\Lambda}{4}}$, where $\delta$ is another constant with the requirement that for $T_{ab}=0$ we have $\delta=1$. Thus the vacuum solution for $\rho\geq \rho_0$ becomes 
\begin{eqnarray}
ds^2=\cos^{\frac{4}{3}}\frac{\rho
 \sqrt{3\Lambda}}{2}\left(-dt^2+dz^2\right)
+\delta^2\frac{4}{3\Lambda}\sin^2 \frac{\rho
 \sqrt{3\Lambda}}{2}\cos^{-\frac{2}{3}}\frac{\rho
 \sqrt{3\Lambda}}{2} d\phi^2+d\rho^2.  
\label{metricfinal}
\end{eqnarray}
The constant $\delta$ is related to the deficit in the azimuthal angle $\phi$. We have now to determine $\delta$. 

In~\cite{Ghezelbash:2002cc} the vortex solutions in the de Sitter space were
studied perturbatively and the authors proved the existence of
this $\delta$, but did not estimate it. Here we evaluate $\delta$
in the following way. Let us first compute
\begin{eqnarray}
\frac{1}{2 \pi}
\int \int\sqrt{g^{(2)}} d\rho d\phi 
\left(G_{t}\,^{t}+\Lambda\right)
\label{gbonet}
\end{eqnarray}
on $(\rho,~\phi)$ planes orthogonal to $(\partial_t,~\partial_{\phi})$.    
 $g^{(2)}$ is the determinant of the spacelike metric induced on these 2-planes. We compute $G_{t}\,^{t}$ from the general ansatz (\ref{metric'}) and Eq.~(\ref{adch42extra}) to have
\begin{eqnarray}
&& \frac{1}{2 \pi}
\int \int\sqrt{g^{(2)}} d\rho d\phi 8\pi G T_{t}{}^{t}=\frac{1}{2\pi}\int_{\rho=0}^{\rho_0}\int\sqrt{g^{(2)}} d\rho d\phi  
\left(G_{t}\,^{t}+\Lambda\right)\nonumber\\
&=&\frac{1}{2\pi}\int_{\rho=0}^{\rho_0}\int d\rho d\phi \rho e^{\frac{B}{2}}\left[\frac{A^{\prime\prime}}{2}+\frac{B^{\prime\prime}}{2}+ \frac{A^{\prime 2}}{4}+\frac{A^{\prime}B^{\prime}}{4}+\frac{B^{\prime}}{\rho}+\frac{A^{\prime}}{2\rho}+\frac{B^{\prime 2}}{4}+\Lambda\right]
\nonumber\\&=&\frac{1}{2\pi}\int_{\rho=0}^{\rho_0}\int d\rho d\phi\left[
\rho e^{\frac{B}{2}}\left(\frac{{A^{\prime}}^2}{4} + \Lambda\right) 
+ \left(\rho e^{\frac{B}{2}}\frac{{A^{\prime}}}{2}\right)' +
\left(\rho e^{\frac{B}{2}}\right)'' \right].%\nonumber\\
\label{mucalc}
\end{eqnarray}
We note that since $\delta=1$ when $\rho<\rho_0$, we may take the azimuthal angle $\phi$ to vary from $0$ to $2\pi$ in this region. Thus we get from Eq.~(\ref{mucalc})
\begin{eqnarray}
 \frac{d}{d\rho}\left(\rho e^{\frac{B}{2}}\right)\Bigg\vert^{\rho_0}_{0}+
 \left(\rho e^{\frac{B}{2}}\frac{A'}{2}
 \right)\Bigg\vert^{\rho_0}_{0}&=&4G \int_{\rho=0}^{\rho_0}\int_{0}^{2\pi}\sqrt{g^{(2)}} d\rho d\phi T_{t}{}^{t} -\int_{0}^{\rho_{0}}d\rho
 \rho e^{\frac{B}{2}}\left(\frac{{A^{\prime}}^2}{4}+\Lambda\right), \nonumber\\
&=&-4G\mu-\int_{0}^{\rho_{0}}d\rho
 \rho e^{\frac{B}{2}}\left(\frac{{A^{\prime}}^2}{4}+\Lambda\right),
\label{mu2}
\end{eqnarray}
where
\begin{eqnarray}
  \mu:=-\int_0^{2\pi}\int_0^{\rho_0}  d\phi  d\rho
\rho e^{\frac{B}{2}}T_{t}\,^{t}&\approx& \frac{\pi\lambda \eta^4}{\sqrt{3 \Lambda^{\prime}} } \int_0^{\rho_0} d\rho \sin \frac{\rho\sqrt{3\Lambda^{\prime}}  }{2}\cos^{-\frac{1}{3}}\frac{\rho\sqrt{3\Lambda^{\prime}}  }{2}\nonumber\\      &=&\frac{\pi \lambda
  \eta^4}{\Lambda^{\prime}}\left[1-\cos^{\frac{2}{3}} \frac{\rho_0
    \sqrt{3 \Lambda^{\prime}}}{2}\right] 
\label{mu} 
\end{eqnarray}
is the string mass per unit length. To get the approximate expression
for $\mu$ in Eq.~(\ref{mu}) we have used ${\displaystyle T_{t}\,^{t}=
-\frac{\lambda \eta^4}{4}}$ (Eq.~(\ref{adch52})) inside the core which is due to our approximation $X=0$ and $P=1$ there. On the other hand, outside the core $T_{t}\,^{t}=0$
identically, so we have used the metric functions $B(\rho)$ or $e^{B(\rho)}$ from 
Eq.~(\ref{metric3}) to evaluate the inside core integral. 

Now let us evaluate the total derivative terms on the left hand side of Eq.~(\ref{mu2}).
In order to do this, we will 
use the interior metric of Eq.~(\ref{metric3}) at $\rho=0$, but the
vacuum metric of Eq.~(\ref{metricfinal}) at the string surface
$\rho=\rho_0$. The reason for doing this is the following. Since we have assumed the energy-momentum tensor to be
non-vanishing only within the string core, the right hand side of
Eq.~(\ref{mucalc}) will have non-zero contributions only from $0\leq\rho\leq
\rho_0\,,$ as we have written. The integrand on the left hand side
Eq.~(\ref{mucalc}) also vanishes outside the string core according to vacuum
Einstein equations $G_{ab}+\Lambda g_{ab}=0$. Thus when we evaluate the left hand side of Eq.~(\ref{mu2}), we
must do so only up to the surface of the string $\rho_0$, i.e., where the
energy-momentum tensor vanishes. But at that point we have the
vacuum solution of Eq.~(\ref{metricfinal}), so that is what we should
use at the upper limit of the integration. Thus we find from Eq.~(\ref{mu2})
\begin{eqnarray}
 1-\delta \left(\cos^{\frac{2}{3}}\frac{\rho_0 \sqrt{3
 \Lambda}}{2}-\frac{1}{3}\cos^{-\frac{4}{3}}\frac{\rho_0 \sqrt{3
 \Lambda}}{2} \sin^2 \frac{\rho_0 \sqrt{3 \Lambda}}{2}\right)=
 4G\mu + \int_{0}^{\rho_{0}}d\rho \rho e^{\frac{B}{2}}\left(\Lambda
 +\frac{{A^{\prime}}^2}{4}\right)\,.\, \nonumber\\
\label{delta}
\end{eqnarray}
The integrals on the right hand side of Eq.~(\ref{delta}) cannot be
evaluated explicitly, since neither the integrand can be written as a 
total derivative, nor do we know the detailed behavior of the fields or the
metric near the string surface at $\rho=\rho_0\,.$ However, we may
make an estimate of these integrals using the expressions of the metric
coefficients inside the core. This means that we ignore the details
of the fall off of the energy-momentum tensor near $\rho =
\rho_0$ and we take the metric functions (\ref{metric3}) up to $\rho_0$. Then using $A$ 
and $B$ from the metric of Eq.~(\ref{metric3}) we obtain from Eq.~(\ref{delta}) an 
 approximate expression for $\delta$,
\begin{eqnarray}
\delta= \frac{1-4G\mu-\frac{2\Lambda}{\Lambda^{\prime}}
 \left(1-\cos^{\frac{2}{3}}\frac{\rho_0 
\sqrt{3\Lambda^{\prime}}}{2} \right)
+\frac{1}{3}\left(1-\cos^{-\frac{4}{3}}\frac{\rho_0 \sqrt{3\Lambda^{\prime}}}{2} \right)
+\frac{2}{3}\left(1-\cos^{\frac{2}{3}}\frac{\rho_0
 \sqrt{3\Lambda^{\prime}}}{2} \right)}
{\left(\cos^{\frac{2}{3}}\frac{\rho_0 \sqrt{3
\Lambda}}{2}-\frac{1}{3}\cos^{-\frac{4}{3}}\frac{\rho_0 \sqrt{3
\Lambda}}{2} \sin^2 \frac{\rho_0 \sqrt{3 \Lambda}}{2}\right)}\,.\nonumber\\
\label{deltafinal}
\end{eqnarray}
This result may be compared with one obtained in~\cite{Deser:1983dr}
where the authors considered point particles of equal masses $m$ as source and solved 
Einstein's equations in (2+1)-dimensional de Sitter space. The particles 
may be considered to be punctures created in spacelike planes by an infinitely thin 
long string, i.e., a $\delta$-function string. 
A conical singularity away from the string was found, with an angle deficit
$\delta=(1-4 G m)$. Our result, Eq.~(\ref{deltafinal}), includes corrections dependent on $\Lambda$, 
which we may think of as coming from the finite thickness of the string.

Now let us try to simplify Eq.~(\ref{deltafinal}) for a realistic situation as the following.
First we observe that the size of the core $\rho_0$ for a thin string is of the order of $(\sqrt\lambda
\eta)^{-1}\,,$ at least when the winding number is
small~\cite{Brihaye:2008uy}. This is essentially because the metric
is flat on the axis $\rho=0$ and hence we may approximate $\rho_0$ for a thin string by its flat spacetime value. Also the scale of symmetry breaking $\eta$ is small
compared to the Planck scale in theories of particle physics in
which cosmic strings appear. For example, the grand unified scale
is about $10^{16}$ GeV, so that $G\eta^2 \sim 10^{-3}\,.$ Further, the observed value of $\Lambda$ is of the order of $10^{-52}~{\rm m}^{-2}$ and the cosmological horizon has size ${\displaystyle \sim {\cal{O}}(\Lambda^{-\frac{1}{2}})}$ which is of course, very large. Therefore we also have for a thin string, $\rho_0^2\Lambda \ll1\,.$ 
%Thus we find that $\rho_0^2\Lambda' \ll 1\,$ as well, where
%$\Lambda'$ is given by Eq.~(\ref{lambdaprime})\,.
Next we expand $G\mu$ using the expression given in Eq.~(\ref{mu}), we
find ${\displaystyle \mu = \frac\pi4 \lambda\eta^4\rho_0^2}$ approximately, and thus
$G\mu \sim G\eta^2 \ll 1\,$ for the GUT scale strings. We also find under these assumptions an approximate expression for $\delta$ from Eq.~(\ref{deltafinal}), 
\begin{eqnarray}
\delta \approx 1 - 4G\mu\left(1 + \frac34\rho_0^2\Lambda +
G\mu\right)\,.   
\label{deltacorr}
\end{eqnarray}
Since the string radius is much smaller than the cosmological horizon size ($\rho_0\sqrt{\Lambda}\ll1$), Eq.~(\ref{deltacorr}) shows that the leading correction to $\delta$ due to the cosmological constant $\Lambda$
is of a higher order of smallness. The
meaning of $\delta$ is obvious in spacetimes with vanishing
cosmological constant, for which Eq.~(\ref{delta}) was worked out
in e.g.~\cite{Garfinkle:1985} (see also \cite{Vilenkin2:2000}). It was found that $\delta\approx (1 - 4G\mu),$
where $G\mu\ll 1$ as before, and ${\cal O}(G^2\mu^2)$ terms
were neglected. Then asymptotically one gets the cylindrically symmetric flat spacetime with a conical singularity called the Levi-Civita spacetime given by Eq.~(\ref{s43}).
%
%\begin{eqnarray}
%ds^2=-dt^2+d\rho^2+dz^2+\rho^2\delta^2 d\phi^2.
%\label{Levi-Civita}
%\end{eqnarray}
%
In this spacetime the azimuthal angle $\phi$ runs from $0$ to
$2\pi\delta$, which is less than $2\pi$. So Eq.~(\ref{s43}) is the Minkowski spacetime
minus a wedge which corresponds to a deficit $2\pi (1-\delta)$ in
the azimuthal angle. The difference of initial and final azimuthal
angles of a null geodesic i.e., light ray in the geometrical
optics approximation, at $\rho \rightarrow \infty$ is
${\displaystyle\frac{\pi}{\delta}}$~\cite{Vilenkin2:2000}. Therefore light bends towards the string even
 though the curvature of spacetime is
zero away from the axis. Thus one may regard the bending of light in the asymptotic
region $\rho \to \infty$ as the gravitational analogue of the Aharanov-Bohm effect.

Thus we have seen that for a positive cosmological constant, the metric in the exterior of
the string is given by Eq.~(\ref{metricfinal}) and approximate expressions for the defect term $\delta$ is given in Eq.~(\ref{deltafinal}) or
Eq.~(\ref{deltacorr}). We compare Eq.~(\ref{metricfinal}) with the string-free vacuum
solution of Eq.~(\ref{vacuum'}) to see that, similar to the
asymptotically flat spacetime, the deficit in the azimuthal angle
in spacetime with a positive cosmological constant is also
$2\pi(1-\delta)\,,$ but now with $\delta$ given by
Eq.s~(\ref{deltafinal}) or (\ref{deltacorr}). Also we have already argued that the $\Lambda$ correction to $\delta$ is very tiny for realistic cases like GUT strings. 

However the bending of null geodesics will
be quite different in (\ref{metricfinal}) from that in an asymptotically flat cosmic string
spacetime. The difference comes from the background curvature produced by $\Lambda$.
Let us now look into this effect.

Since our spacetime (\ref{metricfinal}) has a translational
isometry along $(\partial_z)^a$, for the sake of simplicity we can consider
null geodesics on the $z=0$ plane. 
On this plane we consider the two other Killing fields $(\partial_t)^a$ and
$(\partial_{\phi})^a$. We recall that if
$\xi^a$ is a Killing field, then for any geodesic with tangent vector $u^a$, the quantity $g_{ab}u^a \xi^b$ is conserved along the geodesic.\footnote{See Appendix}

 We will refer to the conserved quantities associated with these two Killing fields as the energy $E$ and the angular momentum $L$ respectively. 
We have for the spacetime (\ref{metricfinal}),
\begin{eqnarray}
 E=-g_{ab}u^a(\partial_t)^b
=\cos^{\frac{4}{3}}\frac{\rho \sqrt{3\Lambda}}{2} \dot{t},
\label{energy}
\end{eqnarray}
and
\begin{eqnarray} 
L= g_{ab}u^a(\partial_\phi)^b 
=\delta^2
\frac{4}{3\Lambda}\sin^2 \frac{\rho
\sqrt{3\Lambda}}{2}\cos^{-\frac{2}{3}}\frac{\rho
\sqrt{3\Lambda}}{2} \dot{\phi},
\label{angularmom}
\end{eqnarray}
where the `dot' denotes differentiation with respect to an affine parameter $s$ along the geodesic. Also using the expression for the metric (\ref{metricfinal}) for the null geodesics on the $z=0$ plane we have 
\begin{eqnarray} 
0= g_{ab}u^a u^b&=&-\cos^{\frac{4}{3}}\frac{\rho\sqrt{3\Lambda}}{2}{\dot{t}}^2+{\dot{\rho}}^2+\delta^2
\frac{4}{3\Lambda}\sin^2 \frac{\rho
\sqrt{3\Lambda}}{2}\cos^{-\frac{2}{3}}\frac{\rho
\sqrt{3\Lambda}}{2} {\dot{\phi}}^2\nonumber\\
&=&-\frac{E^2}{\cos^{\frac{4}{3}}\frac{\rho\sqrt{3\Lambda}}{2}}+{\dot{\rho}}^2+\frac{3\Lambda L^2}{4\delta^2 \sin^2 \frac{\rho
\sqrt{3\Lambda}}{2}\cos^{-\frac{2}{3}}\frac{\rho
\sqrt{3\Lambda}}{2}},
\label{geoadd1}
\end{eqnarray}
where Eq.s~(\ref{energy}) and (\ref{angularmom}) have been used to eliminate $\dot{t}$ and $\dot{\phi}$. From Eq.s~(\ref{angularmom}) and (\ref{geoadd1}) we now obtain
\begin{eqnarray}
{\frac{d\phi}{d\rho}}={\frac{3\Lambda L}{4  E \delta^2}
 \frac{\cos^{\frac{4}{3}}\frac{\rho\sqrt{3\Lambda}}{2}}
      {\sin^2\frac{\rho \sqrt{3\Lambda}}{2} \left[1-\frac{3
         \Lambda L^2}{4 E^2 \delta ^2 } \cot^2 \frac{\rho
         \sqrt{3\Lambda}}{2}\right]^{\frac{1}{2}}} }~.  
\label{bending1}
\end{eqnarray}
Since both $(\rho,~\phi)$ are smooth functions of the affine parameter $s$,
the derivative on the left hand side of Eq.~(\ref{bending1}) is well
defined. By setting $\dot{\rho}=0$ in Eq.~(\ref{geoadd1}) we find the
distance of closest approach to the string, also known as the impact parameter, 
\begin{eqnarray}
\rho_{\rm{c}}=\frac{\, 2}{\sqrt{3 \Lambda}}\tan^{-1}
\frac{\sqrt{3\Lambda }L}{2E \delta}~.
\label{impact}
\end{eqnarray}
Let us now consider a null geodesic in the region between the string 
surface $\rho_0$ and the singularity at ${\displaystyle \rho= \frac{\pi}{\sqrt{3 \Lambda}}}$. We 
look at it when it is traveling between two spacetime points 
$(t_1,~\rho_{\rm{m}},~ \phi_1)$ and $(t_2,~\rho_{\rm{m}},~\phi_2)$
in the same region. We have kept the initial
and final radial distances equal $(=\rho_{\rm{m}})$ for simplicity of interpretation only. 

The spacetime we are considering has a rotational isometry along $(\partial_{\phi})^a$. This means that we can always rotate, without any loss of generality, a radial line going through $\rho=0$  and joining $\phi=0$ 
and $\phi=\pi$, to make it perpendicular to the radial line joining $\rho=0$ and $\rho=\rho_{\rm c}$.
Then, since we have chosen the observed initial and final radial points to be equal ($=\rho_{\rm m}$), the radial line joining $\rho=0$ and $\rho=\rho_{\rm c}$ divides a $(\rho,~\phi)$  plane into two symmetric halves. 
Thus the spacelike part of the trajectory of
the geodesic is symmetric about the line joining $\rho=0$ and $\rho_{\rm{c}}$ and
 the change in the azimuthal angle due to this trajectory obtained from Eq.~(\ref{bending1}) is
\begin{eqnarray}
\Delta \phi=\phi_2-\phi_1&=&\frac{3\Lambda L}{4  E \delta^2}
\int_{(\rho_{\rm m }~\phi_1)}^{(\rho_{\rm m }~\phi_2)}
 \frac{\cos^{\frac43 }\frac{\rho\sqrt{3\Lambda}}{2}}
      {\sin^2\frac{\rho \sqrt{3\Lambda}}{2} 
\left[1-\frac{3\Lambda L^2}{4E^2\delta^2}\cot^2\frac{ \rho\sqrt{3\Lambda}}{2}                  \right]^{\frac12}  }d\rho\nonumber\\
&=&\frac{3\Lambda L}{2  E \delta^2}\int_{\rho_{\rm c}}^{\rho_{\rm m }}
 \frac{\cos^{\frac43 }\frac{\rho\sqrt{3\Lambda}}{2}}
      {\sin^2\frac{\rho \sqrt{3\Lambda}}{2} 
\left[1-\frac{3\Lambda L^2}{4E^2\delta^2}\cot^2\frac{ \rho\sqrt{3\Lambda}}{2}                  \right]^{\frac12}  }d\rho.
\label{bending2}
\end{eqnarray}
Eq.~(\ref{bending2}) along with the expression for $\rho_c$, Eq.~(\ref{impact}),
determines the change of $\phi$ with $\rho$. The full expression
for the integral in Eq.~(\ref{bending2}) is rather messy and we
will look at two special cases only, to have some insight. First, we consider $\rho$ to be much
smaller than the radius of the cosmological singularity ${\displaystyle\left(\rho
\ll \frac{\pi}{\sqrt{3\Lambda}} \right)}$. Keeping only up to quadratic terms of the trigonometric functions in Eq.~(\ref{bending2}), we then have approximately  
\begin{eqnarray}
\Delta \phi\approx \frac{2}{\delta}
\sec^{-1}\left(\sqrt{1 + k^2}    
\frac{\rho E\delta}{L} \right)
\Bigg\vert_{\rho_ {{\rm {c}}}  }^{\rho_{\rm{m}}}
 - \frac{4 k}  {3 \delta\sqrt{1 + k^2}} 
\left(\frac{\rho^2 3 \Lambda}{4}-\frac{k^2}{1 + k^2}
\right)^{\frac{1}{2} }
\Bigg\vert_{\rho_c}^{\rho_{\rm{m}}},
\label{bending3}
\end{eqnarray}
where ${\displaystyle k = \frac {\sqrt{3\Lambda} L} {2 E \delta}}$.  The second
term in Eq.~(\ref{bending3}) is negative and hence the repulsive effect
of positive $\Lambda$ is manifest in this term. In the $\Lambda
\rightarrow 0$ limit only the first term survives. In that case ${\displaystyle \rho_{\rm c} =\frac{L}{E\delta}}$ and
in the limit $\rho_{\rm{m}} \rightarrow \infty$, we recover the well
known formula ${\displaystyle\Delta \phi= \frac{\pi}{\delta}}$ \cite{Vilenkin2:2000}. 

 Next, near the naked
singularity located at ${\displaystyle\rho =\frac{\pi}{\sqrt{3 \Lambda}}}$, we 
approximate $\cos \frac{\rho\sqrt{3\Lambda} } {2}\approx
\left( \frac{\pi}{2}-\frac{\rho\sqrt{3\Lambda}}{2} \right)$ and
integrate Eq.~(\ref{bending2}) to get
\begin{eqnarray}
\Delta \phi\approx -\frac {6 k}{\delta}
\left[\frac{1}{7}\left(\frac{\pi}{2}- \frac{\rho
 \sqrt{3
     \Lambda}}{2}\right)^{\frac{7}{3}
 }+\frac{k^2}{26}\left(\frac{\pi}{2}- \frac{\rho
 \sqrt{3
     \Lambda}}{2}\right)^{\frac{13}{3} }+\dots\right]
_{\rho_ {{\rm {c} }} }^{\rho_{\rm{m}}}.
\label{bending4}
\end{eqnarray}
%

%%%%%%%%%%%%%%%%%%%%%%%%%%%%%%%%%%%%%%%%%%%%%%%%%%%%%%%%%%%%%%%%%%
\section{Black hole pierced by a string}
In the previous Section we have discussed a static cylindrically symmetric free cosmic string spacetime. In this Section we will discuss cosmic strings stretching between the horizons of a spherically symmetric de Sitter black hole, namely the Schwarzschild-de Sitter black hole.  
%%%%%%%%%%%%%%%%%%%%%%%%%%%%%%%%%%%%%%%%%%%%%%%%%%%%%%%%%%%%%%%%%%%
\subsection{Case 1. Non-self gravitating string}
%%%%%%%%%%%%%%%%%%%%%%%%%%%%%
Let us first consider a static and non-gravitating cylindrical distribution of energy-momentum corresponding to the Abelian Higgs model (\ref{lagrangian2}). It was shown in \cite{Vilenkin4:1981} that
if a cosmic string pierces the horizon of a Schwarzschild black hole, the
resulting spacetime has a conical singularity as well. In~\cite{Gregory:1995} it was shown by
considering the equations of motion of the matter fields that an
Abelian Higgs string (for both self gravitating and non-self gravitating energy-momentum) can pierce a Schwarzschild black hole. In the following we will adopt the method described in ~\cite{Gregory:1995} to establish that both the horizons of a Schwarzschild-de Sitter black hole can be similarly pierced by a non-gravitating Nielsen-Olesen string.

First we derive the equations of motion
for the fields $X$ and $P$ from (\ref{lagrangian2}),
\begin{eqnarray}
\nabla_{a}\nabla^{a}X- X P_{a}P^{a}-\frac{\lambda
 \eta^2}{2}X\left(X^2-1\right)=0,
\label{eom1}
\end{eqnarray}
\begin{eqnarray}
\nabla_{a}F^{ab}-2 e^2 \eta^2 X^2 P^{b}=0.
\label{eom2}
\end{eqnarray}
We consider for a moment the flat spacetime metric written in cylindrical coordinates 
\begin{eqnarray}
ds^2=-dt^2+d\rho^2+dz^2+\rho^2d\phi^2,
\label{flatmetric}
\end{eqnarray}
 and take the scalar field $X$ to be cylindrically
symmetric, $X=X(\rho)$. Also, we take the gauge field $P_a$ to be azimuthal and cylindrically symmetric as well : $P_{a}=P(\rho)\nabla_{a}\phi$. Then with this ansatz the equations of
motion (\ref{eom1}) and (\ref{eom2}) in the flat background (\ref{flatmetric}) become
\begin{eqnarray}
\frac{d^2X}{d \rho^2}+\frac{1}{\rho}\frac{dX}{d\rho}-                                              
\frac{X P^2}{\rho^2} -\frac{X}{2}(X^2-1)=0,
\label{noeq1}
\end{eqnarray}
\begin{eqnarray}
\frac{d^2 P}{d \rho^2}-\frac{1}{\rho}\frac{dP}{d\rho}-                                             
\frac{2 e^2}{\lambda} X^2 P=0.
\label{noeq2}
\end{eqnarray}
In Eq.s~(\ref{noeq1}) and (\ref{noeq2}) we have scaled $\rho$ by $\left(\sqrt{\lambda}\eta
\right)^{-1}$ to convert it to a dimensionless radial
coordinate. These are the equations which were shown
in~\cite{Nielsen:1973cs} to have string-like solutions. We wish to
show that these equations hold also in the Schwarzschild-de Sitter
background spacetime up to a very good approximation if the string thickness is small compared to
the black hole horizon size, and if we neglect the backreaction of
the string on the metric.

We consider the Schwarzschild-de Sitter metric written in the spherical polar
coordinates 
\begin{eqnarray}
ds^2= -\left(1-\frac{2MG}{r}-\frac{\Lambda r^2}{3}\right)dt^2+
\left(1-\frac{2MG}{r}-\frac{\Lambda
r^2}{3}\right)^{-1}dr^2+r^2d\Omega^2.
\label{sdsmetric}
\end{eqnarray}
As we discussed in Section 1.2, for $3MG\sqrt{\Lambda}<1$, solutions to $g_{tt}=0$ give three horizons in this spacetime $-$ the black hole
event horizon at $r=r_{\rm {H} }$, the cosmological horizon at $r=r_{\rm{C}}$
and an `unphysical horizon' at $r=r_{\rm{U}}$ with $r_{\rm{U}}<0$. We discussed that since the observed value of $\Lambda$ is very small we can take  $3MG\sqrt{\Lambda}\ll 1$ for a realistic situation and we then have
%%s27i',s27i''
 %
\begin{eqnarray}
r_{\rm{H}}\approx2GM,\quad r_{\rm{C}}\approx \sqrt{\frac{3}{\Lambda}}, 
\label{horsize}
\end{eqnarray}
with $r_{\rm{U}}=-\left( r_{\rm{H}}+r_{\rm{C}}\right)$. Thus Eq.s~(\ref{horsize}) show that under the condition $3MG\sqrt{\Lambda}\ll1$, we have $r_{\rm{H}}\ll r_{\rm{C}}$ and hence $r_{\rm {U}} \approx -r_{\rm{C}}$. The string we are looking for is thin compared to the horizon size $r_{\rm{H}}$, i.e. we assume further that
\begin{eqnarray}
\frac{1}{\sqrt{\lambda}\eta} \ll 2MG \ll \sqrt{\frac{3}{\Lambda}}~.
\label{approx}
\end{eqnarray}
We now expand the field equations in the Schwarzschild-de Sitter
background (\ref{sdsmetric}). In other words, we neglect the backreaction on the
metric due to the string. Then Eq.~(\ref{eom1}) becomes
\begin{eqnarray}
\frac{1}{r^2} \partial_{r} \left[r^2
 \left(1-\frac{2MG}{r}-\frac{\Lambda r^2}{3}\right)
 \partial_{r}X\right] &+&\frac{1}{r^2 \sin^2
 \theta}\partial_{\theta}\left(\sin \theta \partial_{\theta}X
\right) \nonumber \\
&-&\frac{X P^2}{r^2 \sin^2 \theta} -\frac{\lambda \eta^2}{2}
X\left(X-1\right)=0.
\label{scalareq}
\end{eqnarray}
For the string solution the matter distribution is 
cylindrically symmetric. For convenience of calculations we consider a string
along the axis $\theta=0$, although our arguments will be valid for
$\theta=\pi$ as well. Let us define
as before a dimensionless cylindrical radial coordinate $\rho =
r\sqrt{\lambda}{\eta}\sin\theta$. For a cylindrically symmetric
matter distribution both $(X,~P)$ will be functions of $\rho$
only. With this we now write the $r$ and $\theta$ derivatives of Eq.~(\ref{scalareq}) in terms of $\rho$ derivatives to have
\begin{eqnarray}
\left(\sin^2\theta -\frac{2MG \sqrt{\lambda}\eta
 \sin^3\theta}{\rho}-\frac{\overline{\Lambda} \rho^2}{3}\right)
\left[\frac{d^2 X}{d\rho^2}+\frac{2}{\rho}\frac{d X}{d\rho}\right]+
\left(\frac{2MG \sqrt{\lambda}\eta \sin^3\theta}{\rho^2}-\frac{2
 \overline{\Lambda} \rho}{3}\right) \frac{dX}{d\rho}&&\nonumber
\\  
+\left[ \frac{1}{\rho}\frac{dX}{d\rho}\cos^2 \theta
 -\frac{1}{\rho}\frac{dX}{d\rho} \sin^2\theta+
 \frac{d^2X}{d\rho^2} \cos^2\theta \right]-\frac{X
 P^2}{\rho^2}-\frac{1}{2} X\left(X-1\right)=0,&&\nonumber\\
\label{scalareq2}
\end{eqnarray}
where $\overline {\Lambda}=\frac{\Lambda}{\lambda \eta^2}$ is a
dimensionless number which by Eq.~(\ref{approx}) is much less than unity. 
We note that inside the core, $\sin \theta \ll 1 $ and the dimensionless string radius $\rho$ is less than unity. We also have then,
\begin{eqnarray}
\frac{2MG \sqrt{\lambda}\eta\sin^3\theta}{\rho}=\frac{2MG}{r}\sin^2\theta\ll1,\quad 
\frac{2MG \sqrt{\lambda}\eta \sin^3\theta}{\rho^2}=\frac{2MG}{\rho r}\sin^2\theta\ll1.
\label{extension1f}
\end{eqnarray}
Putting these in all together, Eq.~(\ref{scalareq2}) reduces to Eq.~(\ref{noeq1}),
i.e. the flat space equation of motion for the Abelian Higgs
model in the leading order. Outside the string core $(\rho>1)$, we may as before set
$X=1$. Thus we may conclude that Eq.~(\ref{scalareq2}), and hence
Eq.~(\ref{scalareq}) gives rise to a configuration of the scalar field
$X(\rho)$ similar to that of the Nielsen-Olesen string under the reasonable assumptions we have made. A similar
calculation for the gauge field  equation (\ref {eom2}) shows that it reduces to Eq.~(\ref{noeq2}). These are sufficient to show that the Schwarzschild-de
Sitter spacetime allows a thin and uniform Nielsen-Olesen string along
the axis $\theta=0$ in the region $r_{\rm{H}}< r < r_{\rm{C}}$.

However from the calculations done above we cannot conclude how the string behaves
at or near the horizons. The two horizons at $r_{\rm{H}}$ and $r_{\rm{C}}$
are two coordinate singularities in the metric in Eq.~(\ref{sdsmetric}). Clearly we cannot expand
the field equations in this singular coordinate system at or around the 
horizons. In order to perform this expansion we need to use maximally
extended charts which will be free from coordinate 
singularities, and 
has only the curvature singularity at $r=0$. So following the procedure described in Chapter 1 for the de Sitter spacetime, let us first construct Kruskal-like patches at the two horizons to remove the two coordinate singularities.

We first construct a Kruskal-like patch for the black hole horizon $r_{\rm H}$.
We consider radial ($\theta,~\phi={\rm{constant}}$), null geodesics $(ds^2=0)$ in the Schwarzschild-de Sitter spacetime,
\begin{eqnarray}
&&\left(1-\frac{2MG}{r}-\frac{\Lambda r^2}{3}\right)dt^2=\left(1-\frac{2MG}{r}-\frac{\Lambda r^2}{3}\right)^{-1}dr^2 \nonumber\\
&&\Rightarrow\frac{dt}{dr}=\pm \frac{1}{\left(1-\frac{2MG}{r}-\frac{\Lambda r^2}{3}\right)}~,
\nonumber\\
\label{extension1}
\end{eqnarray}
which means along such geodesics 
\begin{eqnarray}
t=\pm r_{\star}+{\rm{constant}},
\label{extension2}
\end{eqnarray}
where $r_{\star}$ is the tortoise coordinate defined by
\begin{eqnarray}
 r_{\star}=\int\frac{dr}{\left(1-\frac{2MG}{r}-\frac{\Lambda
   r^2}{3}\right)}.
\label{extension2'}
\end{eqnarray}
In order to integrate Eq.~(\ref{extension2'}), we break the integrand into partial fractions
\begin{eqnarray}
\frac{1}{\left(1-\frac{2MG}{r}-\frac{\Lambda
   r^2}{3}\right)}=-\frac{3r}{\Lambda
 \left(r-r_{\rm{H} } \right)\left(r-r_{\rm{U}}\right)\left(r-r_{\rm{C}}\right)}=\left[\frac{\alpha}{r-r_{\rm H}}+ \frac{\beta}{r-r_{\rm C}}+\frac{\gamma}{r-r_{\rm U}}\right],\nonumber\\
\label{extension1n}
\end{eqnarray}
where $\alpha$, $\beta$, $\gamma$ are three constants. Solving Eq.~(\ref{extension1n})
we find them to be
\begin{eqnarray}
\alpha=\frac{3
r_{\rm{H}} }{\Lambda\left(r_{\rm{C}}-r_{\rm{H}}\right)\left(r_{\rm{H}}-r_{\rm{U}}\right)},\quad
\beta=-\frac{3
r_{\rm{C}}}{\Lambda\left(r_{\rm{C}}-r_{\rm{H}}\right)\left(r_{\rm{C}}-r_{\rm{U}}\right)},\quad
\gamma=-\frac{3r_{\rm{U}} }{\Lambda\left(r_{\rm{C}}-r_{\rm{U}}\right)\left(r_{\rm{H}}-r_{\rm{U}}\right)}~.
\nonumber\\  
\label{kruskal2}
\end{eqnarray}
Substituting Eq.s~(\ref{extension1n}) and (\ref{kruskal2}) into Eq.~(\ref{extension2'}) and integrating we find
\begin{eqnarray}
r_{\star}=\frac{3}{\Lambda}\left[\frac{
r_{\rm{H}} }{\left(r_{\rm{C}}-r_{\rm{H}}\right)\left(r_{\rm{H}}-r_{\rm{U}}\right)}   \ln \left \vert \frac{r}{r_{\rm{H}} }-1\right\vert  -\frac{
r_{\rm{C}}}{\left(r_{\rm{C}}-r_{\rm{H}}\right)\left(r_{\rm{C}}-r_{\rm{U}}\right)}
\ln \left \vert\frac{r}{r_{\rm{C}}}-1\right\vert \right.  \nonumber\\
\left.-\frac{r_{\rm{U}} }{\left(r_{\rm{C}}-r_{\rm{U}}\right)\left(r_{\rm{H}}-r_{\rm{U}}\right)}\ln \left \vert
\frac{r}{r_{\rm{U}}  }-1\right\vert\right].  
\label{extension3}
\end{eqnarray}
We note that $r_{\star}\to -\infty(+\infty)$ as one reaches $r_{\rm H}(r_{\rm C})$. In the ($t$, $r_{\star}$) coordinates the radial part of (\ref{sdsmetric}) becomes
\begin{eqnarray}
ds^2_{\rm{radial}}=\left(1-\frac{2MG}{r}-\frac{\Lambda
 r^2}{3}\right)\left(-dt^2+dr_{\star}^{2} \right),
\label{extension4}
\end{eqnarray}
where $r$ is understood as a function of the new coordinate $r_{\star}$ and can be found from Eq.~(\ref{extension3}). 
We also note from Eq.~(\ref{adintro2}) that we have always $r_{\rm H}r_{\rm C}r_{\rm U}=-\frac{6MG}{\Lambda} $. Then we have
\begin{eqnarray}
\left(1-\frac{2MG}{r}-\frac{\Lambda r^2}{3}\right)&=&-\frac{\Lambda r_{\rm H}r_{\rm C}r_{\rm U} }
{3r}\left[\frac{r}{r_{\rm{H}}}-1\right]
\left[\frac{r}{ r_{\rm C} }-1\right] \left[\frac{r}{r_{\rm{U}} }-1\right]\nonumber\\
&=&\frac{2MG}{r} \left[\frac{r}{r_{\rm{H}} }-1\right]\left[\frac{r}{r_{\rm{C}} }-1\right] \left[\frac{r}{r_{\rm{U}} }-1\right].
\label{exal41}
\end{eqnarray}
Now let us define the outgoing and incoming null coordinates ($u$, $v$) as
\begin{eqnarray}
u=t-r_{\star}, \quad{\rm{and}}\quad v=t+r_{\star}.
\label{extension5}
\end{eqnarray}
With these null coordinates and Eq.~(\ref{exal41}) the radial metric (\ref{extension4}) becomes
\begin{eqnarray}
ds^2_{\rm{radial}}=-\frac{2MG}{r} \left[\frac{r}{r_{\rm{H}} }-1\right]\left[\frac{r}{r_{\rm{C}} }-1\right] \left[\frac{r}{r_{\rm{U}} }-1\right] du dv.
\label{extension6}
\end{eqnarray}
Using Eq.~(\ref{extension3}) we eliminate $\left(\frac{r}{r_{\rm H}}-1\right)$ from Eq.~(\ref{extension6}) to get
\begin{eqnarray}
ds^2_{\rm{radial}}=-\frac {2MG}{r}e^{\frac {v-u} {2\alpha} }\left\vert\frac{r}{r_{\rm{U}} }-1 \right\vert^
{1-\frac{\gamma}{\alpha}}\left\vert\frac{r}{r_{\rm{C}}}-1 \right\vert^
{1-\frac{\beta}{\alpha}} du dv.
\label{extensionad}
\end{eqnarray}
Now we define timelike and spacelike Kruskal coordinates ($T$, $Y$) by
\begin{eqnarray}
T:=\frac{e^{\frac{v}{2\alpha}} -e^{-\frac{u}{2\alpha}} }{2};\qquad
Y:= \frac{e^{\frac{v}{2\alpha}} +e^{-\frac{u}{2\alpha}} }{2}~.
\label{extension7}
\end{eqnarray}
From Eq.s~(\ref{extension3}), (\ref{extension5}) we find that $T$ and $Y$ satisfy the following relations
\begin{eqnarray}
Y^2-T^2&=&\left[\frac{r}{r_{\rm{H}} }-1 \right]
e^{\frac{\beta}{\alpha}\ln\left\vert\frac{r}{r_{\rm{C}}} - 1\right\vert +
 \frac{\gamma}{\alpha}\ln\left\vert\frac{r}{r_{\rm{U}}}-1\right\vert},  
\label{kruskal3}
\end {eqnarray}
\begin{eqnarray}
\frac{T}{Y}&=&\tanh \left(\frac{t}{2\alpha}\right),
\label{extension8}
\end {eqnarray}
which show that at $r=r_{\rm {H}}$, $t\to\pm \infty$, i.e. we have future and past horizons.
In terms of ($T$, $Y$), the full spacetime metric of Eq.~(\ref{sdsmetric}) finally becomes
\begin{eqnarray}
ds^2=\frac{8M G\alpha^2}{r} \left\vert\frac{r}{r_{\rm{U}} }-1 \right\vert^
{1-\frac{\gamma}{\alpha}}\left\vert\frac{r}{r_{\rm{C}}}-1 \right\vert^
{1-\frac{\beta}{\alpha}}\left(-dT^2+dY^2\right)+r^2 d\Omega^2,
\label{kruskal1}
\end{eqnarray}
where $r$ as a function of $(T,~Y)$ is understood and can be found from Eq.~(\ref{kruskal3}).
The metric (\ref{kruskal1}) is manifestly nonsingular at $r=r_{\rm H}$. Thus ($T$,
$Y$) indeed define a well behaved coordinate system around the black hole event
horizon. The Kruskal diagram at the black hole event horizon of the Schwarzschild-de Sitter
black hole shows features similar to the black hole horizons of asymptotically flat spacetimes.
When $r\rightarrow r_{\rm H} (\approx 2MG,~ {\rm under~ our~ approximation})$, from
Eq.~(\ref{kruskal3}) we have after scaling $r\rightarrow \sqrt{\lambda} \eta r$ to get a dimensionless variable,
\begin{eqnarray}
r\approx 2MG\sqrt{\lambda}\eta+ 2MG\sqrt{\lambda}\eta
\left[e^{-\frac{\beta}{\alpha}\ln\left\vert\frac{2MG}{r_{\rm C}}-1\right\vert
   - \frac{\gamma}{\alpha}\ln\left\vert
   \frac{2MG}{r_{\rm U}}-1\right\vert}\right]\left(Y^2-T^2\right).    
\label{kruskal4}
\end{eqnarray}
We now expand the Higgs field equation of motion (\ref{eom1}) in the vicinity of the black hole
event horizon $r_{\rm{H}}\approx 2GM$ using the analytically extended chart (\ref{kruskal1}). Denoting the conformal factor of the $(T,~Y)$ 
part of (\ref{kruskal1}) by $f(T,~Y)$, Eq.~(\ref{eom1}) becomes 
\begin{eqnarray}
\frac{1}{f\lambda \eta^2}\left[-\partial_{T}^2X+\partial_{Y}^2X-
\frac{2}{r}\left(
\partial_{T}X\right)\left(\partial_{T}r\right) +
\frac{2}{r}\left(\partial_{Y}X\right)\left(\partial_{Y}r\right)\right]\nonumber\\
+\frac{1}{r^2\sin^2\theta}
\partial_{\theta}\left(\sin\theta \partial_{\theta}X\right)
-\frac{XP^2}{r^2\sin^2\theta}-\frac{X}{2}(X-1)=0,
\label{eombhk1}
\end{eqnarray}
where any $r$ appearing above is understood (and also will
be understood below) as dimensionless
$(r\to \sqrt{\lambda}\eta r)$, whereas $T$ and $Y$ are dimensionless according to
our definition, Eq.s~(\ref{extension7}). 
Using Eq.~(\ref{kruskal4}) we have the following derivatives of $r(T,~Y)$ in the vicinity
of the black hole horizon,
\begin{eqnarray}
\partial_{T}r=-AT,\quad \partial_{Y}r=AY,
\label{eombhk2}
\end{eqnarray}
where 
$A=4MG\sqrt{\lambda}\eta
e^{-\frac{\beta}{\alpha}\ln\left\vert\frac{2MG}{r_{\rm C}} -
1\right\vert-\frac{\gamma}{\alpha}\ln\left\vert\frac{2MG}{r_{\rm U}} -
1\right\vert}$. We also compute the following derivatives of
 the scalar field $X(\rho)$
\begin{eqnarray}
\partial_{T}X&=&\frac{\partial X}{\partial\rho }\frac{\partial \rho}{\partial r}
\frac{\partial r}{\partial T}= -AT\sin\theta
\frac{\partial X}{\partial \rho},\\
\partial_{Y}X&=& AY \sin\theta \frac{\partial X}{\partial \rho},\\
\partial_{T}^2X&=&-A\sin\theta\frac{\partial X}{\partial \rho}+A^2T^2
\sin^2\theta \frac{\partial^2 X}{\partial \rho^2},\\
\partial_{Y}^2X&=&A\sin\theta\frac{\partial X}{\partial \rho}+A^2Y^2
\sin^2\theta \frac{\partial^2 X}{\partial \rho^2}.
\label{derivatives}
\end{eqnarray}
where $\rho=r\sin \theta$, with $r$ dimensionless as mentioned before, is the dimensionless transverse radial coordinate. Substituting Eq.s~(\ref{eombhk2})-(\ref{derivatives}) into Eq.~(\ref{eombhk1}),
and converting as before the $\theta$-derivatives into $\rho$-derivatives 
we have
\begin{eqnarray}
\frac{1}{f\lambda \eta^2}\left[A^2\left(Y^2-T^2\right)\sin^2\theta \frac{d^2 X}{d \rho^2} +2A\sin\theta \frac{d X}{d \rho}+
\frac{2}{\rho}\left(Y^2-T^2\right)A^2\sin^2\theta \frac{d X}{d \rho} \right]   \nonumber \\
+\left(\frac{1}{\rho}\frac{dX}{d\rho}\cos^2 \theta
 -\frac{1}{\rho}\frac{dX}{d\rho} \sin^2\theta+
 \frac{d^2X}{d\rho^2} \cos^2\theta \right)
-\frac{XP^2}{\rho^2}-\frac{X}{2}\left(X^2-1\right)=0.
\label{horizoneq1}
\end{eqnarray}
Let us now compare the various terms in Eq.~(\ref{horizoneq1}) using 
Eq.s~(\ref{horsize}), (\ref{approx}). Using Eq.~(\ref{kruskal2}), ${\displaystyle\frac{1}{f\lambda \eta^2}\approx 
\frac{1}{16 G^2 M^2\lambda \eta^2 }}$, hence 
${\displaystyle\frac{A^2}{f\lambda \eta^2 }\approx 1}$. Hence ${\displaystyle\frac{A}{f\lambda \eta^2 }\sim {\cal{O}}(A^{-1})}$ which is much less than unity. For a thin string we have as before $\sin\theta \ll1$
inside the core.
Also, Eq.~(\ref{kruskal3}) or 
(\ref{kruskal4}) shows that the quantity
$(Y^2-T^2)$ becomes infinitesimal as $r\to r_{\rm H} \approx 2GM$. We further
have as $r\to r_{\rm H}$, ${\displaystyle\frac{2A\sin\theta} {f\lambda \eta^2}=\frac{\sin\theta}
{2GM\sqrt{\lambda} \eta}\approx \frac{\sin^2\theta}{\rho}}$. Putting these in 
all together, we see that Eq.~(\ref{horizoneq1}) reduces to the Nielsen-Olesen
equation (\ref{noeq1}) in the leading order.

A similar procedure can be applied to Eq.~(\ref{eom2}), which reduces to the gauge field equation (\ref{noeq2}) up to a very good approximation.

The chart defined in Eq.~(\ref{kruskal1}) is however manifestly singular at the cosmological horizon  $r=r_{\rm C}$.
So for calculations at the cosmological horizon, we have to use another Kruskal-like chart nonsingular there. We derive the following
\begin{eqnarray}
ds^2= \frac{8MG\beta^2}{r} \left\vert\frac{r}{r_{\rm U}}-1
\right\vert^{1-\frac{\gamma}{\beta}}\left\vert \frac{r}{r_{\rm H}}-1
\right\vert^{1-\frac{\alpha}{\beta}}\left(-{dT^{\prime}}^2 +
       {dY^{\prime}}^2\right)+r^2 d\Omega^2,
\label{kruskal5}
\end{eqnarray}
where $T^{\prime}$ and $Y^{\prime}$ are respectively the Kruskal timelike and spacelike
coordinates at the cosmological horizon : 
\begin{eqnarray}
T^{\prime}:&=&\frac{e^{\frac{v}{2\beta}} -e^{-\frac{u}{2\beta}} }{2},\qquad
Y^{\prime}:= \frac{e^{\frac{v}{2\beta}} +e^{-\frac{u}{2\beta}} }{2}~,\nonumber\\
Y^{\prime 2}-T^{\prime 2}&=&\left[1-\frac{r}{r_{\rm{C}} } \right]
e^{\frac{\alpha}{\beta}\ln\left\vert\frac{r}{r_{\rm{H}}} - 1\right\vert +
 \frac{\gamma}{\beta}\ln\left\vert\frac{r}{r_{\rm{U}}}-1\right\vert}, \quad 
\frac{T^{\prime}}{Y^{\prime}}=\tanh \left(\frac{t}{2\beta}\right),
\label{extension8ch}
\end {eqnarray}
where $(\alpha,~\beta,~\gamma ) $ are given by Eq.~(\ref{kruskal2}) and the null 
coordinates $(u,~v)$ are given by Eq.s~(\ref{extension5}), (\ref{extension3}). 
The chart (\ref{kruskal5}) is well defined at or around $r=r_{\rm C}$. This can be
derived exactly in the same manner as (\ref{kruskal1}).

Following exactly the same procedure as before we can show that Eq.s
(\ref{eom1}), (\ref{eom2}) reduce to flat space Eq.s~(\ref{noeq1}),
(\ref{noeq2}) respectively, in the leading order. Thus the flat space equations of motion
hold on both black hole and cosmological horizons. We also note that, replacing ${\displaystyle\left(\frac{r}{r_{\rm H}}-1\right)}$ and ${\displaystyle\left(1-\frac{r}{r_{\rm C}}\right)}$ in Eq.s (\ref{kruskal3}) and (\ref{extension8ch}) by their respective modulus, we can make the coordinate systems described in
Eq.s~(\ref{kruskal1}) and (\ref{kruskal5}) well behaved beyond the horizons also (i.e. regions with $r<r_{\rm H}$, and $r>r_{\rm C}$).
Then we may also use these charts to expand the field equations in regions infinitesimally beyond the horizons. For
$r\rightarrow (r_{\rm H}-0)$ the scalar field equation (\ref{horizoneq1})
still holds and the quantity $\left(Y^2-T^2\right)$ is still infinitesimal which can be
neglected anyway. Similar arguments using the chart of Eq.~(\ref{kruskal5}) show that for the region
$r\rightarrow (r_{\rm C}+0)$ the desired string equations exist. 

Thus we have seen that with the string-like boundary conditions on $X$ and $P$ and the
approximations of Eq.~(\ref{approx}), the configuration of cylindrically symmetric non-gravitating matter
fields are like the Nielsen-Olesen string within, at or even slightly
beyond the horizons of a Schwarzschild-de Sitter black hole. Hence we conclude that a Schwarzschild-de
Sitter black hole can be pierced by a thin Nielsen-Olesen string if
the back reaction of the matter distribution to the background
spacetime can be ignored.

%%%%%%%%%%%%%%%%%%%%%%%%%%%%%%%%%%%%%%%%%%%%%
\subsection{Case 2. Self gravitating string}
%%%%%%%%%%%%%%%%%%%%%%%%%%%%%%%555555555555555
Finally we come to the topic of the backreaction of the string on
 the Schwarzschild-de Sitter spacetime. If we place a string
along the $z$-axis, in the most general case the metric functions will be $z$-dependent. If
we set the cosmological constant to be zero in Eq.~(\ref{sdsmetric}), the 
resulting (Schwarzschild) spacetime
would be asymptotically flat. Then we could use Weyl
coordinates~\cite{J.L.Synge:1960zz} to write the metric in an explicitly static and axisymmetric form,
\begin{eqnarray}
ds^2= -B^2 dt^2+\rho^2 B^{-2}d\phi^2+
A^2  \left[d\rho^2+dz^2\right],
\label{harmonic2}
\end{eqnarray}
where the functions $A$ and $B$ depend on $(\rho,~ z)$ only.
It would be relatively easy to determine the existence of cosmic
strings from the equations of motion of the gauge and Higgs fields
written in these coordinates. In particular, using these coordinates
it was shown in \cite{Gregory:1995} by iteratively solving the Einstein equations that
if a thin and self gravitating Abelian Higgs string pierces the horizon of a Schwarzschild black hole, the
resulting spacetime has a conical singularity at the exterior of the string
\begin{eqnarray}
ds^2= -\left(1-\frac{2MG}{r}\right)dt^2+\left(1-\frac{2MG}{r}\right)^{-1}dr^2+r^2\left(d\theta^2+\left(1-4G\mu\right)^2d\phi^2\right),\nonumber\\
\label{finalad1}
\end{eqnarray}
where $\mu$ is the string mass per unit length.

On the other hand, when the
cosmological constant is non-vanishing, it is no longer possible to
write down the metric in the form of Eq.~(\ref{harmonic2}). In fact if one tries
to solve the $\Lambda$-vacuum Einstein equations $R_{ab}-\frac{1}{2}Rg_{ab}+\Lambda g_{ab}=0$ with the ansatz (\ref{harmonic2}), one would get $\Lambda=0$ identically.
 It turns out that
 with a $\Lambda$, positive or negative, we must take $g_{\rho\rho}\neq g_{zz}$ in 
Eq.~(\ref{harmonic2}). But with this even the vacuum Einstein equations become extremely difficult to handle. We were unable to find a
suitable generalization of the Weyl coordinates, which are
needed to solve Einstein's equations coupled to the gauge and Higgs
fields of the Abelian Higgs model.

However, we can bypass this problem and still find an approximate solution for the
exterior of a thin string in the following way. We first note that
inside the string core and near the axis $(\theta=0,~\pi)$, we can set $X\approx0$ 
and $P\approx1$. 
Then the Lagrangian (\ref{lagrangian2}) in that region 
becomes ${\displaystyle{\cal{L}}\approx -\frac{\lambda \eta^4}{4}}$. This gets added to the cosmological constant as before to give 
$\Lambda^{\prime}$ given in Eq.~(\ref{lambdaprime}). With this 
`modified cosmological constant' $\Lambda^{\prime}$ if we solve Einstein's equations
inside the string core and near the axis with a spherically symmetric ansatz, we get
\begin{eqnarray}
ds^2= -\left(1-\frac{2MG}{r}-\frac{\Lambda^{\prime} r^2}{3}\right)dt^2+
\left(1-\frac{2MG}{r}-\frac{\Lambda^{\prime}
r^2}{3}\right)^{-1}dr^2+r^2d\theta^2+ r^2\sin^2\theta d\phi^2, \nonumber\\
\label{adfinal5}
\end{eqnarray}
i.e. the Schwarzschild-de Sitter spacetime with a modified cosmological constant.
Outside the string core we have
$\Lambda$-vacuum and we choose the following ansatz for this region :%\newpage
\begin{eqnarray}
ds^2= -\left(1-\frac{2MG}{r}-\frac{\Lambda r^2}{3}\right)dt^2+
\left(1-\frac{2MG}{r}-\frac{\Lambda
r^2}{3}\right)^{-1}dr^2+r^2d\theta^2+ \delta^2 r^2\sin^2\theta d\phi^2,\nonumber\\
\label{adfinal6}
\end{eqnarray}
where $\delta$ is a constant to be determined. It can be checked that (\ref{adfinal6}) indeed 
satisfies $G_{ab}-\Lambda g_{ab}=0$. 

In order to determine $\delta$ we first note that
for the Schwarzschild-de Sitter spacetime in spherical coordinates $(t,~r,~\theta,~\phi)$,
 we can define a transverse radial 
coordinate $R=r\sin\theta$ for the string core. If the string is very thin compared to the  black hole (and hence the cosmological horizon) 
we have $R\ll r$ inside the core for any $r_{\rm H} \leq r\leq r_{\rm C}$.
 Then inside the core we may define
 new coordinates $(t,~r,~R,~\phi)$ to replace the polar angle $\theta$ by $R$ to have
\begin{eqnarray}
dR=dr\sin\theta-r\cos\theta d\theta\approx -rd\theta \Rightarrow dR^2\approx r^2d\theta^2,
\label{adfin2}
\end{eqnarray}
for a thin string  placed along $\theta=0$ or $\theta=\pi$.

So we make a general ansatz for the metric inside the core for a thin string
\begin{eqnarray}
ds^2= -A^2(r,~R)dt^2+B^2(r,~R)dr^2+dR^2+C^2(r,~R)d\phi^2.
\label{ans}
\end{eqnarray}
We note that Eq.~(\ref{adfinal5}) is only a special case of (\ref{ans}) with
 $ R=r\sin\theta,~
 A^2(r,~R)=B^{-2}(r,~R)=\left(1-\frac{2MG}{r}-\frac{\Lambda^{\prime} r^2}{3}\right)$,
$C(r,~R)=R$. On the other hand since the string is very `thin', Eq.~(\ref{adfinal6})
also describes a special case of (\ref{ans}) just outside the string core with
the same $R$ and $A^2(r,~R)=B^{-2}(r,~R)=\left(1-\frac{2MG}{r}-\frac{\Lambda r^2}{3}\right)$,
$C(r,~R)=\delta R$. 

Next, we use the Killing identity for the azimuthal Killing field $\phi^a=(\partial_{\phi})^a$ for (\ref{ans}),
 $\nabla_a\nabla^a\phi_b=-R_{ab}\phi^a$, and contract by $\phi^b$. We also note that the Killing vector field $\phi^a$ in the spacetime (\ref{ans}) is orthogonal to the $(t,~r,~R)$ hypersurfaces. Then following exactly the same way which led to Eq.~(\ref{s6'}), we now obtain
\begin{eqnarray}
\nabla_a\nabla^a C^2=4(\nabla_a C)(\nabla^a C)-2R_{ab}\phi^a\phi^b.
\label{adfinal3}
\end{eqnarray}
Next we project Eq.~(\ref{adfinal3}) over the $(t,~r,~R)$ hypersurfaces exactly in the same manner as discussed in Chapter 2. We denote the induced  connection over those hypersurfaces by $\widetilde{D}_a$ and we have
\begin{eqnarray}
&&\widetilde{D}_a\left(C\widetilde{D}^a C^2\right)= 2C\left[2 \left(\widetilde{D}^a C\right)
\left(\widetilde{D}_a C\right)-R_{ab}\phi^a\phi^b\right]  \nonumber\\
&&\Rightarrow\widetilde{D}_a\widetilde{D}^a C=-C^{-1} R_{ab}\phi^a\phi^b.
\label{adfinal4'}
\end{eqnarray}
Since $(\partial_t)^a$ is also a Killing field for (\ref{ans}), we may similarly further project this equation onto the $(r,~R)$ surfaces to find
\begin{eqnarray}
\overline{D}_a\left(A\overline{D}^a C\right)=-C^{-1}A R_{ab}\phi^a\phi^b,
\label{adfinal4'f}
\end{eqnarray}
where $\overline{D}_a$ denotes the induced connection over the $(r,~R)$ surfaces.
In order to determine $\delta$ in Eq.~(\ref{adfinal6}),
 we will integrate Eq.~(\ref{adfinal4'f}) up to the string surface $R_0$.
The situation greatly simplifies if we assume as before that within the string core $(0\leq R<R_0)$, we have $X\approx 0$, $P\approx 1$, i.e. ${\displaystyle T_{ab}\approx-\frac{\lambda \eta^4}{4}g_{ab}}$,
and $X=1$, $P=0$ for $R\geq R_{0}$, i.e. $ {\displaystyle T_{ab}=0}$, so that using Einstein's equations we find inside the core,
\begin{eqnarray}
R_{ab}\phi^a\phi^b=8\pi G\left(T_{ab}-\frac{1}{2}T g_{ab}\right)\phi^a\phi^b+\Lambda C^2\approx- 8\pi GT_{t}{}^{t}g_{\phi\phi}+\Lambda C^2\nonumber\\=\left(- 8\pi GT_{t}{}^{t}+\Lambda\right) C^2. 
\label{falsevac}
\end{eqnarray}
Also under this assumption, the inside core metric is entirely given by (\ref{adfinal5}), so that we ignore the $r$ dependence of $C(r,~R)$, and the $R$ dependence of $A(r,~R)$ and then Eq.~(\ref{adfinal4'f}) simplifies to,
\begin{eqnarray}
\overline{D}_a\left(\overline{D}^a C\right)=-C^{-1} R_{ab}\phi^a\phi^b= C\left[8\pi GT_{t}{}^{t}-\Lambda\right],
\label{adfinal4'f2}
\end{eqnarray}
using Eq. (\ref{falsevac}). We also note that under the same approximation, we can ignore the $R$ dependence of $B$. Then the left hand side of Eq. (\ref{adfinal4'f2}) equals ${\displaystyle\frac{\d^2 C}{\d R^2}}$. With this, let us now integrate Eq.~(\ref{adfinal4'f2}) in the following way,
\begin{eqnarray}
\int_{R=0}^{R_0} dR \frac{\d^2 C}{\d R^2}
=\frac {1}{ 2 \pi} \int_{R=0}^{R_0}\oint dR  d\phi C\left[8\pi GT_{t}{}^{t}-\Lambda\right].
\label{adfinal8d}
\end{eqnarray}
%
%%where $r_1$ and $r_2$ are such that the proper radial distance equals unity, i.e.
%
%%\begin{eqnarray}
%%\int_{r=r_1}^{r_2}B dr=1,
%%\label{adfinal8dff}
%%\end{eqnarray}
%
%%ignoring $R$ dependence of $B$.
We define the string mass per unit length $\mu$ by 
\begin{eqnarray}
\mu:=-\int_{R=0}^{R_0} dR \oint d\phi C T_{t}{}^{t}, 
\label{mueq}
\end{eqnarray}
so that Eq.~(\ref{adfinal8d}) becomes 
\begin{eqnarray}
\frac{\d C}{\d R}\Bigg\vert_{R=0}^{R_0}
=-4G\mu- \int_{R=0}^{R_0} dR C\Lambda.
\label{adfinal8d'}
\end{eqnarray}
 Then using $C(R\to0)=R$ and $C(R)=\delta R$ for the inside core and outside core metric functions we have
\begin{eqnarray}
\delta=\left(1-4G \mu\right)-\int_{R=0}^{R_0} dR  C\Lambda. 
\label{adfinal9}
\end{eqnarray}
To evaluate the integral of Eq. (\ref{adfinal9}), we have to know the details of how $C$ varies across the string surface at $R=R_0$.  However, as before we can make an estimate of that term by taking entirely the inside core value $C=R$. Then the above further simplifies to
\begin{eqnarray}
\delta=\left(1-4G \mu-\frac{\Lambda R_{0}^2}{2}\right). 
\label{adfinal9ff'}
\end{eqnarray}
Thus we have shown that under our approximations, the exterior of a thin self-gravitating Abelian Higgs string in the Schwarzschild-de Sitter spacetime also exhibits a conical singularity 
\begin{eqnarray}
ds^2= -\left(1-\frac{2MG}{r}-\frac{\Lambda r^2}{3}\right)dt^2+
\left(1-\frac{2MG}{r}-\frac{\Lambda
r^2}{3}\right)^{-1}dr^2+r^2d\theta^2\nonumber\\+
\left(1-4G\mu-\frac{\Lambda {R_0}^2}{2}\right)^2 r^2\sin^2\theta d\phi^2.
\label{p4}
\end{eqnarray}
 This generalizes the result of~\cite{Deser:1983dr} for the 3-dimensional de Sitter space without black hole. The limit $R_0 \to 0$ recovers the result for a string of vanishing thickness.

It remains as an interesting task to study the motion of null geodesics for (\ref{p4})
since this would exhibit both the attractive effect due to the string
 and repulsive effect due to ambient positive $\Lambda$, as the free cosmic string spacetime we studied earlier.
 Generalization of the spacetime (\ref{p4}) for rotating case would also be interesting since in such spacetimes an additional repulsive effect due to the rotation 
should be present.
 
%%%%%%%%%%%%%%%%%%%%%%%%%%%%%%%%%%%%%%%%%%%%%%%%%%%%%%%%%%%%%%%%%%%%%%%%%%%%%%%%%%%%%%
%%%%%%%%%%%%%%%%%%%%%%%%%%%%%%%                  %%%%%%%%%%%%%%%%%%%%%%%%%
%%%%%%%%%%%%%%%%%%%%%%%%%%%%%%%%%%%%%%%%%%%%%%%%%%%%%%%%%%%%%%%%%%%%%%%%%%%%%%%%%%%%%%%%%%%%%%%%%%%%%
%%\chapter{Mass function for a Schwarzschild-de Sitter spacetime}
%%%%%%%%%%%%%%%%%%%%%%%%%%%%%%%%%%%%%%%%%%%%%%%%%%%%%%%%%%%%%%%%%%%%%%%%%%%%%%%%%%%%%

%%We can easily relate our mass function $U(r,~M)$ to geodesic motion and provide a Newtonian interpretation. 

%%%%%%%%%%%%%%%%%%%%%%%%%%%%%%%%%%%%%%%%%%%%%%%%%%%%%%%%%%%%%%%%%%%%%%%%%%%%%%%%%%%%%%%%%%%%%%%%%%%%%%%%%%%%%%%%%%%%%%%%%%%%%%%%%%%%%%%%%%%%%%%%%%%%%%%%%%%%%%%%%%%%%%%%%%%%%%%%%%%%%%%%%%%%%%%%
%%%\section{Variation of the mass function and the Smarr formula }

%%%%%%%%%%%%%%%%%%%%%%%%%%%%%%%%%%%%%%%%%%%%%%%%%%%

%%%%%%%%%%%%%%%%%%%%%%%%%%%%%%%%%%%%%%%%%%%%%%%%%%%%%%%%%%%%%%%%%%%%%%%%%%%%%%%%%%%%%%%%%%%%%%%

%%%%%%%%%%%%%%%%%%%%%%%%%%%%%%%%%%%%%%%%%%%%%%%%%%%%%%%%%%%%%%%%%%%%%%%%%%%
\chapter{Thermodynamics and Hawking radiation in the Schwarzschild-de Sitter spacetime}
%%%%%%%%%%%%%%%%%%%%%%%%%%%%%%%%%%%%%%%%%%%%%
In this Chapter we will discuss thermodynamics and particle creation or the Hawking radiation in the Schwarzschild-de Sitter spacetime. 

We reviewed in the first Chapter the problem of defining a positive definite mass function and 
thermodynamics in the Schwarzschild-de Sitter spacetime \cite{Gibbons:1977mu, Abbott:82, Shiromizu:94, kastor}. In Section 5.1, we will give a simple and alternative derivation of 
Eq.~(\ref{s52}) using the mass function derived in \cite{Abbott:82}. This will motivate us to study particle creation in the Schwarzschild-de Sitter spacetime.  

We mentioned in Chapter 1 that the very first approach to explain and compute particle creation
by the cosmological event horizon appeared in \cite{Gibbons:1977mu}, using the path integral formalism developed in \cite{Hartle:1976tp}. The arguments are the following. 
In the maximally extended spacetime diagram at the cosmological event horizon (Fig. 1.1),
region III is endowed with a past directed timelike Killing field. So in this region a `particle' can have negative energy. If a particle-antiparticle pair is produced in this region, the particle with negative energy or the antiparticle resides within it whereas the positive energy particle propagates through region IV and finally emerges through ${\cal{C}^{-}} $ in region I. Hence an observer located in region I will register an incoming particle at asymptotic late time. The ratio of probabilities for a particle to emerge from ${\cal{C}^{-}}$ and
to disappear through ${\cal{C}^{+}}$ was shown to be of the type ${\displaystyle\sim e^{-\frac{2\pi E}{\kappa_{\rm C}}}}$, where $E$ is the positive energy of a particle and $\kappa_{\rm C}$
is the surface gravity of the cosmological event horizon. This shows that the 
incoming particle flux from the cosmological horizon is thermal with a temperature ${\displaystyle\frac{\kappa_{\rm C}}{2\pi}}$.    
For a de Sitter spacetime with a black hole, for example the Schwarzschild-de Sitter spacetime,
the region between the cosmological and the black hole horizons were separated by a thermally opaque membrane and particle creation by each of the horizons were studied independently.
To explain particle creation by black hole a particle-antiparticle pair was considered just outside the black hole event horizon. The antiparticle with a negative energy is swallowed by the hole whereas the particle with a positive energy moves away. The ratio of probabilities for a particle to emerge from the black hole horizon and to move into it was shown to be like 
${\displaystyle\sim e^{-\frac{2\pi E}{\kappa_{\rm H}}}}$, where $ \kappa_{\rm H}$ is the surface gravity of the black hole event horizon. This shows that the black hole radiates with a temperature ${\displaystyle\frac{\kappa_{\rm H}}{2\pi}}$. On the other hand, the arguments same as that of the de Sitter space were used to show that the cosmological horizon also emits thermal radiation with temperature ${\displaystyle\frac{\kappa_{\rm C} }{2\pi}}$.   

A quantum field theoretic approach for particle creation near the horizons of a de Sitter
black hole background was developed in~\cite{Traschen:1999zr}. This approach does not
consider division of the region between the two horizons into two thermally disconnected part. 
The set of two different Kruskal-like coordinates were used to make mode expansions at the two horizons. The Bogoliubov
coefficients between these modes were computed. It was shown that the particle spectra at the two horizons are non-thermal since the surface gravities of the two horizons are in general  different. The exceptions to this are the $3MG\sqrt{\Lambda}=1$ limit of the Schwarzschild-de Sitter spacetime and also the $Q\to MG $ limit of the Reissner-N\"{o}rdstrom-de Sitter solution (Eq.~(\ref{s25i})), in each of which the surface gravities of the black hole and the cosmological horizons become equal.       

The semiclassical tunneling method \cite{Kraus:1994}-\cite{Paddy3:2002}
is an alternative approach to model particle creation by black holes using relativistic single particle quantum mechanics in the WKB approximation scheme. The goal of this method is to compute the imaginary part of the `particle' action which gives the emission or absorption probability from the 
event horizon. From the expression of these probabilities one identifies the 
temperature of the radiation. The earliest work in this context can be found in \cite{Kraus:1994}. Following these works an approach called the null geodesic method was developed in \cite{Wilczek:2000, Parikh2:2004}. There is also another way to model black hole evaporation via tunneling called the complex path analysis \cite{Paddy1:1999, Paddy2:2001, Paddy3:2002} which we will discuss in this Chapter. This method involves writing down in the semiclassical limit $\hbar \to 0$ a Hamilton-Jacobi equation from the matter equations of motion, treating the horizon as a singularity in the complex plane and then complex integrating the equation across that singularity to obtain an imaginary contribution for the particle action. 
Both these two alternative approaches have received great attention during last few years. Since both of these methods deal only with the near horizon geometry,
they can be useful alternatives particularly when the spacetime has no
well defined asymptotic structure or infinities.

So far as we neglect the backreaction of the matter fields, the 
temperature of the radiation or the Hawking temperature should not depend upon the parameters, e.g. mass, spin, and charge, of the particle species.  
The Smarr formula for black hole mechanics predicts that this temperature is proportional to the surface gravity
of the event horizon for a stationary black hole with a Killing horizon \cite{Bardeen}. This is known as the universality of the Hawking radiation.
The complex path analysis approach has been successfully applied to scalar emissions as well as to spinor emissions separately for a wide class of stationary black holes giving the expected expressions of Hawking temperatures in terms of the horizons' surface gravities. To tackle Dirac equation in this approach the usual method has been employed, i.e. finding a proper representation of the general $\gamma$ matrices in terms of the Minkowskian $\gamma$'s and the metric functions and then making the variable separation. For an exhaustive review and list of references on this we refer our reader to \cite{rb}.
Thus the universality of the Hawking temperature has been proved case by case for a wide variety of black holes via the complex path method. Can we prove this universality from a more general point of view?

In particular, in this Chapter we will show that for the Dirac spinors we do not need 
to work with any particular representation of the $\gamma$ matrices in the semiclassical framework. We will demonstrate in a coordinate independent way that for an arbitrary spacetime with any number of dimensions, the equations of motion for a Dirac spinor, a vector, spin-$2$ meson and spin-${\displaystyle\frac{3}{2}}$ fields reduce to the Klein-Gordon equations in the semiclassical limit $\hbar \to 0$ for the usual WKB ansatz. The equations for a charged Dirac spinor reduce to that of a charged scalar. This clearly shows that at the semiclassical level all those different equations of motion of various particle species are equivalent and it is
sufficient to deal with the scalar equation only. We will also present for a stationary spacetime with some reasonable geometrical properties and a Killing horizon, a general coordinate independent expression for the emission probability and the temperature of 
radiation. We will see that this temperature is independent of any parameter concerning the particle species. Having proven the universality of particle emission from an arbitrary
Killing horizon, we will discuss Hawking radiation in the Schwarzschild-de Sitter spacetime explicitly. But before we go into that, we will present below an alternative derivation of the Smarr formula (\ref{s52}) using the mass function derived in \cite{Abbott:82}.
To compare our results with the literature, we will set $G=1$ throughout this Chapter. 
%%%%%%%%%%%%%%%%%%%%
\section{The Smarr formula}
A mass function for the Schwarzschild-de Sitter spacetime is given by \cite{Abbott:82}
\begin{eqnarray}
U=M.
\label{var1}
\end{eqnarray}
We will perform the variation of this mass function subject to the change of the black hole mass parameter $M$, assuming $\Lambda$ to be a universal constant. 

As we have seen in Chapter 1, the Schwarzschild-de Sitter spacetime 
 has three horizons $(r_{\rm{H}},~r_{\rm{C}},~r_{\rm{U}})$ when 
$3M\sqrt{\Lambda}<1$. The black hole $(r_{\rm{H}})$ and the cosmological horizon $(r_{\rm{C}})$ are given by
\begin{eqnarray}
r_{\rm{H}}=\frac{2}{\sqrt{\Lambda}}\cos\left[\frac{1}{3}
\cos^{-1}\left(3M\sqrt{\Lambda}\right)+\frac{\pi}{3}\right],~
r_{\rm{C}}=\frac{2}{\sqrt{\Lambda}}\cos\left[\frac{1}{3}
\cos^{-1}\left(3M\sqrt{\Lambda}\right)-\frac{\pi}{3}\right].\nonumber\\
\label{var2}
\end{eqnarray}
Let us first consider the black hole horizon $(r_{\rm{H}})$. Since the spacetime is spherically symmetric, we define the area of the horizon to be 
\begin{eqnarray}
A_{\rm{H}}=4 \pi r_{\rm{H}}^2.
\label{var3}
\end{eqnarray}
Squaring the first of Eq.s (\ref{var2}) and substituting Eq.~(\ref{var3}) into it, we find
\begin{eqnarray}
A_{\rm{H}}=\frac{16 \pi}{\Lambda}\cos^2\left(\frac{\theta}{3}+\frac{\pi}{3}\right)
&\Rightarrow& \cos\left(\frac{\theta}{3}+\frac{\pi}{3}\right)=\sqrt{  \frac{\Lambda A_{\rm{H}} }{16 \pi}  } \nonumber\\
&\Rightarrow&\frac{\theta}{3}=\cos^{-1}\sqrt{\frac{\Lambda A_{\rm{H}} }{16 \pi}  }-\frac{\pi}{3}, 
\label{var3'}
\end{eqnarray}
where $\cos\theta=3M\sqrt{\Lambda}$, so that,
\begin{eqnarray}
\cos\theta =3M\sqrt{\Lambda}= \cos \left(3\cos^{-1}\sqrt{\frac{\Lambda A_{\rm{H}} }{16 \pi}  }-\pi\right)&=&-\cos\left(3 \cos^{-1}\sqrt{\frac{\Lambda A_{\rm{H}} }{16 \pi}  } \right)\nonumber\\
&=&-4\left(\frac{\Lambda A_{\rm H} }{16\pi}\right)^{\frac32}+3\left(\frac{\Lambda A_{\rm H} }{16\pi}\right)^{\frac12},\nonumber\\
\label{var3'1}
\end{eqnarray}
where the identity $\cos\theta=4\cos^3\frac{\theta}{3}-3\cos\frac{\theta}{3}$ has been used.
Eq.s (\ref{var3'1}) thus give
\begin{eqnarray}
M(A_{\rm{H}})=-\frac{4 \Lambda}{3}
\left(\frac{A_{\rm{H}}}{16 \pi}
\right)^{\frac{3}{2}}+\left(\frac{A_{\rm{H}}}{16 \pi}
\right)^{\frac{1}{2}}.
\label{var4}
\end{eqnarray}
Now we rewrite the mass function $U$ in Eq.~(\ref{var1}) in
terms of the new variable, i.e. the black hole horizon area $A_{\rm{H}}$,  
\begin{eqnarray}
U(A_{\rm{H}})=-\frac{4 \Lambda}{3}\left
(\frac{A_{\rm{H}}}{16 \pi}
\right)^{\frac{3}{2}}+\left(\frac{A_{\rm{H}}}{16 \pi}
\right)^{\frac{1}{2}}.
\label{var5}
\end{eqnarray}
We take the variation of Eq.~(\ref{var5}) to get
\begin{eqnarray}
\delta U(A_{\rm{H}})=\left[-\frac{2 \Lambda} {\left(16 
\pi\right)
^{\frac{3}{2}}} \left(A_{\rm{H}}\right)^{\frac{1}{2}}
+\frac{1}{2 \left(16 \pi A_{\rm{H}}\right)^{\frac{1}{2}}}\right]\delta 
A_{\rm{H}}.
\label{var5'}
\end{eqnarray}
Let $\kappa_{\rm{H}}$ be the surface gravity of the black hole event horizon. It is given by the derivative of the norm of the timelike Killing field at the black hole horizon \cite{Wald:1984rg},
\begin{eqnarray}
\kappa_{\rm{H}}=\frac{1}{2}\partial_{r}\left( 1-\frac{2M}{r}
-\frac{\Lambda r^2}{3}\right)_{r=r_{\rm{H}}}=
\left(\frac{M}{{r_{\rm{H}}}^2}
-\frac{\Lambda r_{\rm{H}}}{3}\right).
\label{var6}
\end{eqnarray}
Substituting Eq.s (\ref{var3}), (\ref{var4}) into it we find
\begin{eqnarray}
\kappa_{\rm{H}}=-\frac{\Lambda}{2}\sqrt{ \frac{A_{\rm H} }{4\pi}  }+\sqrt{\frac{\pi}{A_{\rm H}}}.
\label{var6'}
\end{eqnarray}
Combining this with Eq.~(\ref{var5'}) we obtain
\begin{eqnarray}
\delta U(A_{\rm{H}})=\frac{\kappa_{\rm{H}}}
{8\pi}\delta 
A_{\rm{H}}.
\label{var7}
\end{eqnarray}
Similarly we have
\begin{eqnarray}
\delta U(A_{\rm{C}})=-\frac{\kappa_{\rm{C}}}
{8\pi}\delta A_{\rm{C}},
\label{var8}
\end{eqnarray}
where $A_{\rm{C}}$ and $\kappa_{\rm{C}}$ are respectively the area and the surface gravity of the cosmological horizon,
\begin{eqnarray} 
A_{\rm{C}}=4\pi r_{\rm C}^2,\quad
\kappa_{\rm C}=-\frac{1}{2}\partial_{r}\left( 1-\frac{2M}{r}
-\frac{\Lambda r^2}{3}\right)_{r=r_{\rm{C}}}=\frac{\Lambda}{2}\sqrt{ \frac{A_{\rm C} }{4\pi}}-\sqrt{\frac{\pi}{A_{\rm C}}}.%\nonumber\\
\label{var8}
\end{eqnarray}
Combining Eq.s (\ref{var7}) and (\ref{var8}) we obtain 
\begin{eqnarray}
\kappa_{\rm{H}}\delta A_{\rm{H}}+\kappa_{\rm{C}}\delta A_{\rm{C}}=0,
\label{var9}
\end{eqnarray}
which is the Smarr formula first derived in \cite{Gibbons:1977mu} using a different mass function.

We note also that  Eq.~(\ref{var7}) or Eq.~(\ref{var8}) are formally similar to Eq.~(\ref{s51}). This indicates that both black hole and the cosmological horizons of the Schwarzschild-de Sitter spacetime should have similar individual thermodynamical properties and there may be thermal radiation coming from both them at temperatures ${\displaystyle\frac{\kappa_{\rm H}}{2\pi}}$ and ${\displaystyle\frac{\kappa_{\rm C}}{2\pi}}$ respectively. To see this is really the case, we will now go into the study of particle creation via semiclassical complex path analysis. To exhibit the quantum nature of particle emission, we will retain $\hbar$ in the following.

%%%%%%%%%%%%%%%%%%%%%%%%%%%%%%%%%%%%%%%%%%%%%%%%%%%%%%%%%%%%%
\section{Particle creation via complex path}
\subsection{Reduction of the semiclassical Dirac equation into scalar equations}

Let us start by considering a spacetime of dimension $n$, and a metric $g_{ab}$ defined on it. We consider the Dirac equation
\begin{eqnarray}
i\gamma^a\nabla_a\Psi 
=-\frac{m}{\hbar}\Psi.
\label{s1}
\end{eqnarray} 
$\nabla_a$ is the spin covariant derivative defined by $\nabla_a\Psi:=
\left(\partial_a+\Gamma_a\right)\Psi$, where $\Gamma_a$ are the spin connection. The matrices $\gamma^a(x)$ are the curved space generalization of the Minkowskian $\gamma^{(\mu)}$. We expand $\gamma^a$ in an orthonormal basis $e_{(\mu)}^{a}$, 
$\gamma^a=\gamma^{(\mu)}e_{(\mu)}^{a} :\mu=0,~1,~2,\dots,~(n-1)$, where the Greek
indices within bracket denote the local Lorentz indices. In terms of $\gamma^{(\mu)}$ and $e_{(\mu)}^{a}$, the spin connection matrices $\Gamma_a$ take the form (see e.g. \cite{Witten:81}),
\begin{eqnarray}
\Gamma_a=\frac18\left[\gamma^{(\mu)},~\gamma^{(\nu)}\right]_{-}e_{(\mu)}^{b}\nabla_a e_{(\nu)b}.
\label{s1ad}
\end{eqnarray} 
We also have by definition
$g^{ab}e^{(\mu)}_{a}e^{(\nu)}_b=\eta^{(\mu)(\nu)}$ where
$\eta^{(\mu)(\nu)}$ is the inverse metric for
the $n$-dimensional Minkowski spacetime. The $\gamma^{(\mu)}$ satisfy the well known anti-commutation relation: $\left[\gamma^{(\mu)},~\gamma^{(\nu)}\right]_+=2 \eta^{(\mu)(\nu)} \bf{I}$, where $\bf{I}$ denotes the $n\times n$ identity matrix.  
 
The expansion of $\gamma^a$ in terms of the orthonormal basis $\{e_{(\mu)}^{a}\}$, and the anti-commutation relation for $\gamma^{(\mu)}$'s give 
\begin{eqnarray}
\left[\gamma^a,~\gamma^b\right]_+=2g^{ab}\bf{I}.
\label{s2}
\end{eqnarray} 
Now we square Eq.~(\ref{s1}) by acting with $i\gamma^b\nabla_b$ on both sides from left, producing
\begin{eqnarray}
\frac{1}{2}\left[\gamma^b,~\gamma^a\right]_+
\nabla_b\nabla_a\Psi+
\frac{1}{4}\left[\gamma^b,~\gamma^a\right]_-
\left[\nabla_b,~\nabla_a\right]\Psi+
\left(\gamma^b\nabla_b \gamma^a\right)\nabla_a \Psi=
\frac{m^2}{\hbar^2}\Psi.\nonumber\\
\label{s3}
\end{eqnarray} 
Using $\nabla_a\Psi=\partial_a \Psi+\Gamma_a\Psi$, the commutativity of the partial derivatives and the anti-commutation relation for $\gamma^a$ in Eq.~(\ref{s2}), Eq.~(\ref{s3}) becomes 
\begin{eqnarray}
\nabla_a\nabla^a\Psi+
\frac{1}{4}\left[\gamma^a,~\gamma^b\right]_-
\left[\partial_{[a}\Gamma_{b]}+\Gamma_{[a}\Gamma_{b]} \right]\Psi+
\left(\gamma^b\nabla_b \gamma^a\right)\nabla_a \Psi=
\frac{m^2}{\hbar^2}\Psi.
\label{s4}
\end{eqnarray} 
We will look at Eq.~(\ref{s4}) semiclassically. We choose the usual WKB ansatz for the 4-component wave function 
\begin{eqnarray}
\Psi  
&=&\left( 
\begin{array}{c}
f_1(x)e^{\frac{i I_1(x)}{\hbar}} \\ 
f_2(x)e^{\frac{i I_2(x)}{\hbar}} \\ 
f_3(x) e^{\frac{i I_3(x)}{\hbar}}\\ 
f_4(x)e^{\frac{i I_4(x)}{\hbar}}\\
\end{array}
\right),
\label{s5}
\end{eqnarray}
where $f_i(x)$ and $I_i(x)$ are independent of $\hbar$.
We substitute this into Eq.~(\ref{s4}). Since we are neglecting backreaction of the matter field, the metric functions do not depend upon $\hbar$. Thus $\Gamma_a$ given in Eq.~(\ref{s1ad}) are
independent of $\hbar$. Then it is clear that in the semiclassical limit $\hbar \to 0$, on the left hand side of Eq.~(\ref{s4}) only the first term survives because only this one contains some second derivatives of $\Psi$, which are of ${\cal {O}}\left(\hbar^{-2}\right)$. The single derivative terms coming from the Laplacian will certainly not survive in the semiclassical limit, but we will formally keep the Laplacian $\nabla_a\nabla^a$ intact till later when we will discuss its expansion explicitly. Thus in the semiclassical limit, the WKB ansatz (\ref{s5}) implies that Eq.~(\ref{s4}) can formally be represented by four Klein-Gordon equations
\begin{eqnarray}
\nabla_a\nabla^a\Psi
-\frac{m^2}{\hbar^2}\Psi=0.
\label{s6}
\end{eqnarray} 
If we consider a Dirac particle with a charge $e$ coupled to a classical gauge field $A_a$, the spin covariant derivative $\nabla_a$ in Eq.~(\ref{s1}) is replaced
by the gauge covariant derivative $\widetilde{\nabla}_a \equiv \nabla_a-\frac{ie}{\hbar} A_a$, so that the equation of motion becomes
\begin{eqnarray}
i\gamma^a\nabla_a\Psi +\frac{e}{\hbar}\gamma^a A_a\Psi
=-\frac{m}{\hbar}\Psi.
\label{s7}
\end{eqnarray} 
We now apply from the left $\left(i\gamma^b\nabla_b+\frac{e}{\hbar}\gamma^b A_b\right)$ on both sides of this equation. Using Eq.s~(\ref{s2}), (\ref{s3}) and (\ref{s4}) we obtain
\newpage
\begin{eqnarray}
\nabla_a\nabla^a\Psi+
\frac{1}{4}\left[\gamma^a,~\gamma^b\right]_-
\left[\partial_{[a}\Gamma_{b]}+\Gamma_{[a}\Gamma_{b]}\right]\Psi+ 
\left(\gamma^b\nabla_b \gamma^a\right)\nabla_a \Psi 
-\frac{e^2}{\hbar^2}\left(A_bA^b\right)\Psi \nonumber \\+\frac{2ie}{\hbar}A^a
\nabla_a\Psi
-\frac{ie}{\hbar}\left[\left(\gamma^b\nabla_b \gamma^a\right)
A_a +\frac{1}{4}\left[\gamma^a,~\gamma^b\right]_-F_{ab}
 +\left(\nabla_a A^a\right)
\right]\Psi=
\frac{m^2}{\hbar^2}\Psi,%%\nonumber\\
\label{s8}
\end{eqnarray} 
where $F_{ab}=\nabla_{[a} A_{b]}$. We now substitute the general ansatz of Eq.~(\ref{s5}) into Eq.~(\ref{s8}) and take the semiclassical limit $\hbar \to 0$. Since $A^b$ is a classical gauge field, both $A_b$ and $F_{ab}$ are independent of $\hbar$. We keep only the terms of ${\cal{O}}(\hbar^{-2})$ to
see that in this limit Eq.~(\ref{s8}) can formally be represented by four equations
\begin{eqnarray}
\nabla_a\nabla^a\Psi 
-\frac{e^2}{\hbar^2}\left(A_bA^b\right)\Psi
+\frac{2ie}{\hbar}A^a\nabla_a\Psi-
\frac{m^2}{\hbar^2}\Psi=0,
\label{s9}
\end{eqnarray} 
each of which has the form of the equation of motion of a scalar particle with charge $e$ and mass $m$.

What have we seen so far? We have dealt with neutral and charged Dirac spinors and have explicitly shown in a coordinate independent way that, for the semiclassical WKB ansatz all those equations of motion are equivalent
to that of scalars in any arbitrary spacetime. 
We will show explicitly in Section 5.3 that similar conclusions hold also for the Proca field, massive spin-$2$ and spin-${\displaystyle\frac{3}{2}}$ fields.
 But before that we wish to discuss the explicit expansions and the near horizon limits of Eq.s~(\ref{s6}), (\ref{s9}) in a stationary spacetime containing a Killing horizon. We will address only the charged Dirac spinor or equivalently the charged scalar, since the other case is equivalent to setting $e=0$.

%%%%%%%%%%%%%%%%%%%%%%%%%%%%%%%%%%%%%%%%%%%%%%%%%%%%%%%%%%%%%%%%%%%%%
\subsection{Particle emission from a Killing horizon}
\subsubsection{Derivation of the general formula}

We wish to present in the following a general coordinate independent expression
for the emission or absorption probability from a Killing horizon in a stationary spacetime. Let us first construct the geometrical setup using definitions and assumptions we make. 

We consider an $n$-dimensional stationary spacetime endowed with Killing fields $(\xi_a,~\{\phi^{i}_a\})$, where $i=1,2,\dots,m$. $\xi_a$ is the timelike Killing field which generates stationarity and $\{\phi^{i}_a\}$ are the spacelike Killing fields generating other isometries 
of the spacetime, for example, spherical or axisymmetry etc. However we do not need to
 specify these spacelike isometries explicitly. The assumption of stationarity will let us provide a meaningful notion of the `particle' energy. We assume that the Killing fields commute with each other,
\begin{eqnarray}
[\xi,~\phi^i]^a=\pounds_{\xi}\phi^{ia}=0,\quad [\phi^i,~\phi^j]^a=\pounds_{\phi^i}\phi^{ja}=0,
\label{lm0}
\end{eqnarray} 
for all $i,j=1,2,\dots, m$.
We assume that the spacetime can be foliated into a family of spacelike hypersurfaces $\Sigma$ of dimension $(n-1)$, orthogonal to a timelike vector field $\chi^a$ with norm $-\beta^2$.
We further assume that the hypersurface orthogonal vector field $\chi^a$, orthogonal to $\{\phi^{i}_a\}$ or
any spacelike field, can be written as a linear combination of all the Killing fields
\begin{eqnarray}
\chi_a=\xi_a+\alpha^i(x)\phi^i_a,\quad \chi^a\chi_a=-\beta^2, 
\label{lm1}
\end{eqnarray} 
where $\{\alpha^i(x)\}\vert_{i=1}^{m}$ are smooth spacetime functions. If we set $\{\alpha^i(x)\}\vert_{i=1}^{m}=0$, we recover an $n$-dimensional static spacetime.
Since $\chi_a$ is orthogonal to all $\left\{\phi^i_a\right\}$, the functions $\alpha^i(x)$ can be determined by solving $m$ algebraic equations constructed from contracting Eq.~(\ref{lm1}) by $\phi^{ja}$,
\begin{eqnarray}
\xi_a\phi^{ja}+\alpha^i(x)\left(\phi^i_a\phi^{ja}\right)=0,~j=1,2,\dots, m.
\label{lm1'}
\end{eqnarray} 
Thus $\alpha^i(x)$ are functions of the inner products $(\xi\cdot\phi^i,~\phi^i\cdot\phi^j)$. Then Eq.~(\ref{lm0}) implies
\begin{eqnarray}
&&\pounds_{\xi}\alpha^i(x)=0=\pounds_{\phi^j}\alpha^{i}(x),\nonumber\\
&&\pounds_{\chi}\alpha^{i}(x)=\pounds_{\xi}\alpha^i(x)+\alpha^j\pounds_{\phi^j}\alpha^i(x)=0,~
{\rm for~all}~i,j=1,2,\dots,m.
\label{lm1''}
\end{eqnarray} 
Then following exactly the same procedure described in Chapter 2 we can show that  over any
$\beta^2=0$ surface ${\cal{H}}$, the functions $\alpha^i(x)$ become constants and hence the 
vector field $\chi^a$ is Killing over ${\cal{H}}$,
\begin{eqnarray}
\chi^a\chi_a\vert_{{\cal{H}}}=-\beta^2\vert_{{\cal{H}}}=0,~\chi^a\vert_{{\cal{H}}}=\chi_{\rm{H}}^a:~\nabla_{(a}\chi_{{\rm{H}}b)}=0.
\label{ad1}
\end{eqnarray} 
 This means that the null surface ${\cal{H}}$ is the true or Killing horizon of the spacetime.
We note that
$\chi^a$ is not necessarily a Killing field everywhere because $\alpha^i(x)$ are in general
neither zero
nor constants but it is Killing at least over ${\cal{H}}$ by our construction. 
 
Let us now write the spacetime metric $g_{ab}$ as
\begin{eqnarray}
g_{ab}=-\beta^{-2}\chi_a\chi_b+\lambda^{-2}R_aR_b +\gamma_{ab},
\label{e1}
\end{eqnarray} 
where $R^a$ is a spacelike vector field orthogonal to $\chi^a$, and $\lambda^2$ is the norm of $R_a$. $\gamma_{ab}$ represents the $(n-2)$-dimensional spacelike portion of the metric well behaved on or in an infinitesimal neighborhood of ${\cal{H}}$, orthogonal to both $\chi_a$ and $R_a$. 

Using Killing's equation we have $\nabla_{(a}\chi_{b)}=\phi^i_{(a}\nabla_{b)}\alpha^i(x)$, so that
\begin{eqnarray}
\chi^a\chi^b\nabla_a\chi_b=-\frac{1}{2}\chi^a\nabla_a\beta^2
=\frac12\chi^a\chi^b\phi^i_{(a}\nabla_{b)}\alpha^i(x)=0,
\label{lm2}
\end{eqnarray} 
where we have used the orthogonality $\chi^a\phi_a^i=0 $.
Eq.~(\ref{lm2}) shows that $\nabla_a\beta^2$ is everywhere orthogonal to $\chi^a$ and hence it is spacelike when $\chi^a$ is timelike, so we may choose $R_a=\nabla_a\beta^2$ in Eq.~(\ref{e1}).

To look at the behavior of $\nabla_a\beta^2$ over ${\cal{H}}$ we follow the same procedure
described in Chapter 2. We write $\chi_a=\rho\nabla_a u$ to have  over ${\cal{H}}$ 
\begin{eqnarray}
\nabla_a\beta^2=-2\kappa\chi_{{\rm{H}}a},
\label{g1}
\end{eqnarray} 
where $\kappa$ is a function over ${\cal{H}}$. 
Eq.~(\ref{g1}) shows that due to the torsion-free condition, $\chi_{{\rm{H}}[a}\nabla_{b} \chi_{{\rm{H}}c]}=0$ which means ${\cal{H}}$ is a null hypersurface.
Eq.~(\ref{g1}) also shows that $\nabla_a \beta^2$ is null over ${\cal{H}}$ since $\chi_{\rm H}^a$ is null over ${\cal{H}}$, vanishing both as ${\cal{O}}(\beta^2)$.  
We note that the choice $R_a=\nabla_a\beta^2$ is not unique, we could have multiplied $\nabla_a\beta^2$ by some function non-diverging over ${\cal{H}}$. But we will retain this choice for our convenience.  
  
An expression for $\kappa$ can easily be found from Eq.~(\ref{g1}) and the Frobenius condition \cite{Wald:1984rg},
\begin{eqnarray}
4\kappa^2=\frac{\left(\nabla^a\beta^2\right)\left(\nabla_a\beta^2\right)}{\beta^2}\Bigg \vert_{{\cal{H}}}. 
\label{g2'}
\end{eqnarray} 
Then it turns out that $\kappa$ is a constant over the horizon. We call $\kappa$ to be the Killing horizon's surface gravity. 

Let $R$ be the parameter along  $R^a=\nabla^{a}\beta^2$. Then we have  
\begin{eqnarray}
R^aR_a=\lambda^2=\left(\nabla_a\beta^2\right)\left(\nabla^a\beta^2\right)=R^a\nabla_a\beta^2=\frac{d\beta^2}{dR},
\label{g2}
\end{eqnarray} 
which along with Eq.~(\ref{g2'}) means over ${\cal{H}}$ we have
\begin{eqnarray}
\frac{1}{\beta^2}\frac{d\beta^2}{dR}=4\kappa^2.
\label{g3}
\end{eqnarray} 
With the choice of $R^a$ we have made, it is clear that the metric (\ref{e1}) becomes doubly null over ${{\cal{H}}}$. We note that Eq.~(\ref{e1}) can readily be realized in its doubly null form for a static spherically symmetric spacetime by employing the usual $(t,~r_{\star})$ coordinates, where $r_{\star}$ is the tortoise coordinate, as we have seen in Chapter 4 for the Schwarzschild-de Sitter spacetime.

For $n > 4$, the uniqueness and other general properties of spacetimes
are not very well understood and there may exist more general stationary spacetimes than mentioned above. However we will see that for known stationary exact solutions
 the above construction will be sufficient.

Let us now expand explicitly Eq.~(\ref{s9}) with the ansatz of Eq.~(\ref{s5}). We find for $\hbar \to 0$,
\begin{eqnarray}
-g^{ab}\partial_a I\partial_b I-e^2g_{ab}A^a A^b-2e A^b\partial_b I -m^2=0,
\label{g4'}
\end{eqnarray} 
where we have suppressed the index of $I$ since each of them satisfy the same equation.
Substituting the expression of $g_{ab}$ from  Eq.~(\ref{e1}) into it we find 
\begin{eqnarray}
\lambda^2\left(\chi^a\partial_a I-ef\right)^2
-\beta^2\left(R^a\partial_a I+
eg\right)^2-\left(\beta\lambda\right)^2\left[
\gamma_{ab}\partial^aI\partial^b I
+e^2\gamma_{ab}A^aA^b\right.\nonumber\\\left.-
2e\gamma_{ab}A^a\partial^b I
+m^2\right]=0, 
\label{g4}
\end{eqnarray} 
where $f=-\chi^aA_a$, and $g=R_aA^a$. Now we will look at Eq.~(\ref{g4}) in the near horizon limit. By our
assumption the metric functions $\gamma_{ab}$ are well behaved over the horizon ${\cal{H}}$. So $\gamma_{ab}A^aA^b$ is non divergent over ${{\cal{H}}}$. Also examples with
$g\neq 0$ seem to be unknown in the literature. So we will set $g=0$ in Eq.~(\ref{g4}) and write Eq.~(\ref{g4}) in the near horizon limit as
\begin{eqnarray}
\lambda^2\left(\chi^a\partial_a I-ef\right)^2
-\beta^2\left(R^a\partial_a I\right)^2-\left(\beta\lambda\right)^2\left[
\gamma_{ab}\partial^aI\partial^b I-
2e\gamma_{ab}A^a\partial^b I
\right]=0. 
\label{g5}
\end{eqnarray} 
To further simplify Eq.~(\ref{g5}), let us choose an orthogonal
basis $\left\{m^a_{i}\right\}_{i=1}^{n-2}$ for $\gamma_{ab}$ and let
$\theta_i$ be the parameter along each $m^a_{i}$. Let us consider the first term within the square brackets. This is a sum of the 
squares of $(n-2)$ Lie derivatives:
\begin{eqnarray}
\gamma_{ab}\partial^aI\partial^b I= \frac{1}{m_1^2}(\pounds_{m_1}I)^2+
\frac{1}{m_2^2}(\pounds_{m_2}I)^2+\dots, 
\label{g5f}
\end{eqnarray} 
where $m_i^2$ is the norm of each $m^a_{i}$, by our definition which are non-zero finite over ${\cal{H}}$. Since $I$ is a scalar those Lie derivatives are partial derivatives along the respective parameters : 
\begin{eqnarray}
\pounds_{m_i}I=m_i^a\partial_a I=\partial_{\theta_i}I,\quad {\rm for~all }~i=1,~2,\dots,n-2. 
\label{g5ff}
\end{eqnarray} 
We will now check whether the terms within the square bracket in Eq.~(\ref{g5}) are divergent on ${\cal{H}}$. Let us suppose that infinitesimally close to ${{\cal{H}}}$ the following divergence occur
\begin{eqnarray}
\gamma_{ab}\partial^aI\partial^bI=\frac{D(x)}{\beta^2},
\label{g6}
\end{eqnarray} 
where $D(x)$ is bounded on or in an infinitesimal vicinity of ${\cal{H}}$ and independent of
$\beta$ at leading order. Then Eq.~(\ref{g3}) implies that $D(x)$ is also independent of $R$ over ${\cal{H}}$
\begin{eqnarray}
\partial_{R}D(x)=\left(\partial_{\beta^2}D(x)\right)\frac{d\beta^2}{dR}=4\kappa^2\beta^2\left(\partial_{\beta^2}D(x)\right)\to 0.
\label{g7}
\end{eqnarray} 
%
%Also by our choice $R_a=\nabla_a\beta^2$, whose norm is $\lambda^2$, vanishes over ${\cal{H}}$ as ${\cal{O}}(\beta^2)$ by Eq.~(\ref{g1}). So the function $D(x)$ is also independent of $\lambda$ in the leading order
%over ${\cal{H}}$. 
Since the metric functions $\gamma_{ab}$ are well
behaved over ${\cal{H}}$ the divergence of $\gamma_{ab}\partial^aI\partial^bI$ arises from the Lie derivatives $(\partial_{\theta_i}I)^2$.
For simplicity we will suppose that the divergence comes from a single Lie derivative which is the $i$-th one. We can easily generalize our calculations for more than one diverging term. Let us take near the horizon
\begin{eqnarray}
\partial_{\theta_i}I=\pm \frac{C_{i}(x)}{\beta},
\label{g8}
\end{eqnarray} 
where $C_{i}(x)$ is a non-diverging function independent of $\beta$ in the leading order on or infinitesimally close to ${\cal{H}}$, and hence by Eq.~(\ref{g7}) is independent of $R$ over ${\cal{H}}$. 

Thus by our construction
the divergence of the second term within the square bracket in Eq.~(\ref{g5}) comes from $(\partial_{\theta_i}I)$ which, by Eq.~(\ref{g8}) is ${\cal{O}}(\beta^{-1})$. So this
term can be neglected with respect to the quadratic term $(\partial_{\theta_i}I)^2$, which
is divergent over ${\cal{H}}$ as ${\cal{O}}(\beta^{-2})$. Hence
comparing Eq.s~(\ref{g6}), (\ref{g8}) we have 
\begin{eqnarray}
D(x)=\frac{C_{i}^2(x)}{m_i^2}.
\label{g9f}
\end{eqnarray} 
 Using Eq.~(\ref{g3}) we obtain from Eq.~(\ref{g8}) the following divergence on or infinitesimally close to ${\cal{H}}$,
\begin{eqnarray}
\frac{\partial^2I}{\partial R\partial{\theta_i}}=\mp\frac{2\kappa^2
C_{i}(x)}{\beta}.
\label{g9}
\end{eqnarray} 
On the other hand we can write Eq.~(\ref{g5}) near ${\cal{H}}$ in the leading order now as
\begin{eqnarray}
\left(\partial_R I\right)=\pm\frac{\lambda}{\beta}\left[\left(\chi^a\partial_a I-ef\right)^2-D(x)\right]^{\frac12}.
\label{e6}
\end{eqnarray} 
We will take the partial derivative of Eq.~(\ref{e6}) with respect to $\theta_i$ over ${\cal{H}}$. 
By Eq.s (\ref{g2}), (\ref{g3}) we have $\left(\frac{\lambda}{\beta}\right)^2\Bigg\vert_{{\cal{H}}}=4\kappa^2$ which is a constant over ${\cal{H}}$. This means that $\partial_{\theta_i}\kappa=0$ over ${\cal{H}}$. Since the vector field $\chi_{\rm{H}}^a$ is Killing  
over ${\cal{H}}$, the term $\left(\chi_{\rm{H}}^a\partial_a I-ef\right)$ is a conserved
quantity, i.e. a constant. We interpret this term to be the conserved
effective energy $E$ of a `particle' with charge $e$, with $-ef$ as the electrostatic potential energy on the horizon.
 So using Eq.~(\ref{g9}) and the commutativity of the partial derivatives we find that the partial derivative of Eq.~(\ref{e6}) with respect to $\theta_i$ gives the following ${\cal{O}}(\beta^{-1})$ divergence over ${\cal{H}}$
\begin{eqnarray}
\frac{\partial^2I}{\partial \theta_i \partial R}=&\mp&  
\frac{\kappa \partial_{\theta_i}D(x)}{\left[
E^2-D(x)\right]^{\frac{1}{2}} }=\mp \frac{2\kappa^2
C_{i}(x)}{\beta}=\mp \frac{2\kappa^2
[m_i^2(x)D(x)]^{\frac12}}{\beta}\nonumber\\
&\Rightarrow&\partial_{\theta_i}D(x)=\frac{2\kappa
[m_i^2(x)D(x)\left(E^2-D(x)\right)]^{\frac12}}{\beta},
\label{e8}
\end{eqnarray} 
using Eq. (\ref{g9f}). Taking partial derivative with respect to $\beta$ and using the commutativity of the partial derivatives, we find ${\displaystyle\d_{\theta_i}(\d_{\beta}D(x))}$ to be divergent as ${\cal{O}}(\beta^{-2})$ on ${\cal{H}}$. Since $\beta$ is a constant $(=0)$ tangent to ${\cal{H}}$, we have $\d_{\theta_i}\beta=0$ on ${\cal{H}}$, and 
Eq.~(\ref{e8}) thus contradicts the fact that $D(x)$ is bounded and independent of $\beta$ in the leading order on ${\cal{H}}$. So Eq.~(\ref{g6}) cannot be true. Similarly we can show that the term $\gamma_{ab}\partial^aI\partial^bI$
 cannot be divergent as ${\cal{O}}(\beta^{-n})$ for any $n>2$. Thus $\beta^2\gamma_{ab}\partial^aI\partial^bI=0$ on the horizon.

With all these, we now integrate Eq.~(\ref{g5}) across the horizon
\begin{eqnarray}
I_{\pm} 
=\pm\int_{{\cal{H}}}
\frac{\lambda}{\beta}\left(\chi_{\rm{H}}^a\partial_a I-ef\right)dR=\pm\int_{{\cal{H}}}\frac{
\left(\chi_{\rm{H}}^a\partial_a I-ef\right)}{2\kappa}\frac{d\beta^2}{\beta^2},
\label{e}
\end{eqnarray} 
where in the last step we have used Eq.~(\ref{g3}). Since $\beta^2=0$
on $\cal{H}$, the above integration cannot be performed in real space.
So we have to complexify the path and lift the singularity in the complex plane. 

We will now integrate Eq.~(\ref{e}) across ${\cal{H}} $ along an appropriate complex path or contour
containing the singularity $\beta^2=0$ following the prescription of \cite{Kraus:1994}-\cite{Paddy3:2002}. Since both the quantities
$\left(\chi_{\rm{H}}^a\partial_a I-ef\right)$ and $\kappa$ are constants on ${\cal{H}}$, we can take them out from the integration. The multiple sign comes from the fact that 
 there will be modes which are incoming as well as which are outgoing. For $+(-)$ sign in Eq.~(\ref{e})
we choose anti-clockwise contours in the upper-half (lower-half) complex planes yielding
\begin{eqnarray}
I_{+}=\frac{i\pi\left(\chi_{\rm{H}}^a\partial_a I-ef\right)}{2\kappa},\quad I_{-}=-\frac{i\pi\left(\chi_{\rm{H}}^a\partial_a I-ef\right)}{2\kappa}.
\label{e'}
\end{eqnarray} 
On the other hand, if we take for $+(-)$ sign clockwise contours in the lower-half (upper-half)
complex planes we find instead
\begin{eqnarray}
I_{+}=-\frac{i\pi\left(\chi_{\rm{H}}^a\partial_a I-ef\right)}{2\kappa},\quad I_{-}=\frac{i\pi\left(\chi_{\rm{H}}^a\partial_a I-ef\right)}{2\kappa}.
\label{e''}
\end{eqnarray} 
From the ansatz (\ref{s5}) we see that the probability densities $\left(\sim \left|e^{i \frac{ I}{\hbar} }\right|^2\right)$ associated with solutions (\ref{e'}) are
\begin{eqnarray}
P_{+}\sim \left|e^{iI_+}\right|^2=\exp\left(\frac{-\pi\left(\chi_{\rm{H}}^a\partial_a I-ef\right)}{\hbar\kappa}\right),\quad P_{-}\sim \left|e^{iI_-}\right|^2=\exp\left(\frac{\pi\left(\chi_{\rm{H}}^a\partial_a I-ef\right)}{\hbar\kappa}\right),\nonumber\\
\label{pe'}
\end{eqnarray} 
whereas the probability densities corresponding to (\ref{e''}) are
\begin{eqnarray}
P_{+}\sim \exp\left(\frac{\pi\left(\chi_{\rm{H}}^a\partial_a I-ef\right)}{\hbar\kappa}\right),\quad P_{-}\sim \exp\left(\frac{-\pi\left(\chi_{\rm{H}}^a\partial_a I-ef\right)}{\hbar\kappa}
\right).
\label{pe''}
\end{eqnarray} 
Now we have to identify the emission and absorption probabilities. To do this we recall that classically there could be no emission from a Killing horizon. Taking $\hbar\to 0$ limit we find $P_+\to 0 $ in Eq.~(\ref{pe'}) and $P_{-}\to 0$ in Eq.~(\ref{pe''}) whereas the others diverge. So we identify $P_{+}
(P_{-})$ as emission probability $P_{\rm E}$ in Eq.~(\ref{pe'}) (Eq.~(\ref{pe''})) and the others as the absorption probabilities $P_{\rm A}$.
In any case taking the ratio of the single particle emission to absorption probability we find
\begin{eqnarray}
\frac{P_{\rm E}}{ P_{\rm A}}\sim \exp\left(\frac{-\left(\chi_{\rm{H}}^a\partial_a I-ef\right)}{\frac{\hbar\kappa}{2\pi}}
\right).
\label{temp}
\end{eqnarray} 
We have interpreted earlier the term $\left(\chi_{\rm{H}}^a\partial_a I-ef\right)$ as the conserved energy of a particle.
Then Eq. (\ref{temp}) shows that the emission from a Killing horizon is thermal and the emitted 
particles have a temperature proportional to the Killing horizon's surface gravity, 
\begin{eqnarray}
T_{\cal{H}}=\frac{\hbar\kappa}{2\pi},
\label{tempf}
\end{eqnarray} 
which one expects from the predictions of the black hole thermodynamics, also we have shown that this is true for any Killing horizon as well. 

In the next Section we will demonstrate that the known non-trivial stationary solutions satisfy our assumptions. Then we shall go into discussing the case
of the Schwarzschild-de Sitter spacetime. Precisely, by taking one particular solution it
will be sufficient to show that a vector field $\chi^a$ exists,
which can be written as a linear combination of commuting Killing fields
in the form of Eq.~(\ref{lm1}), that $\chi^a$ is orthogonal to a family of spacelike hypersurfaces $\Sigma$, and $\chi^a$ becomes null and Killing over a surface $\cal{H}$ defining the Killing horizon. We have seen that all the other things follow from this.

 %%%%%%%%%%%%%%%%%%%%%%
\subsubsection{Some explicit examples}
%%%%%%%%%%%%%%%%%%%%%%
Let us start with the simplest case of a Killing horizon in the flat spacetime, namely
the Rindler spacetime 
\begin{eqnarray}
ds^2=-a^2x^2dt^2+dx^2+dy^2+dz^2,
\label{rind}
\end{eqnarray}
where $a$ is a constant having the dimension of inverse length. For many interesting geometrical properties of the Rindler spacetime we refer our reader to e.g. \cite{Wald:1984rg, Birrell}.  

We first note that this spacetime has a timelike Killing field $(\partial_t)^a $ with norm 
$-a^2x^2$, which becomes null at $x=0$. We mentioned earlier that the necessary and sufficient condition for a subspace to form a hypersurface is the existence of a Lie algebra among the vectors spanning that subspace. Since the coordinate vector fields $(\partial_x)^a$, $(\partial_y)^a$, $(\partial_z)^a$ commute with each other, the spacelike 3-surfaces 
spanned by these vector fields form a family of spacelike hypersurfaces, $\Sigma$. Thus the Rindler spacetime trivially satisfies our assumptions with $\chi^a=(\partial_t)^a$, and $x=0$ is the Killing horizon, called the Rindler horizon. The surface gravity $\kappa$ of the Rindler horizon can be computed from Eq.~(\ref{g2'})
\begin{eqnarray}
\kappa=a,
\label{rind2}
\end{eqnarray}
and thus the temperature of emission from the Rindler horizon is $ {\displaystyle T_{\cal{H}}=\frac{\hbar a}{2\pi}} $, which matches with the Unruh temperature \cite{Birrell}.    

 Next we consider the charged Kerr black hole
\begin{eqnarray}
ds^2=-\frac{\Delta-a^2\sin^2\theta}{\Sigma}dt^2
-\frac{2a\sin^2\theta\left(r^2+a^2-\Delta\right)}{\Sigma}
dtd\phi\nonumber \\+ \frac{\left(r^2+a^2\right)^2-
\Delta a^2 \sin^2\theta}{\Sigma}\sin^2\theta d\phi^2 
+\frac{\Sigma}{\Delta}dr^2+\Sigma d\theta^2,
\label{e10}
\end{eqnarray} 
where
\begin{eqnarray}
\Sigma=r^2+a^2\cos^2\theta,\quad \Delta(r)=r^2+a^2+Q^2-2Mr\geq 0.
\label{e10'}
\end{eqnarray} 
 $a$ and $Q$ are the parameters specifying rotation and charge respectively. The gauge field of this solution is ${\displaystyle A_a=-\frac{Qr}{\Sigma}\left[(dt)_a-a\sin^2\theta(d\phi)_a\right]}$.

We first define ${\displaystyle\chi^a:=(\partial_{t})^a-\frac{g_{t\phi}}{g_{\phi\phi}}
(\partial_{\phi})^a}$, such that $\chi_a(\partial_{\phi})^a=0$ everywhere.
The coordinate Killing fields $(\partial_t)^a$ and $(\partial_{\phi})^a$ commute.  
For $\Delta \to 0$ we have $\chi_a\chi^a=-\beta^2\approx-\frac{\Delta \Sigma}{\left(r^2+a^2
\right)^2-\Delta a^2 \sin^2\theta}\leq 0$. So $\chi^a$ is timelike 
for $\Delta>0$ and becomes null at
\begin{eqnarray}
\Delta(r)=0\Rightarrow r_{\rm H}=M\pm\sqrt{M^2-a^2-Q^2}.
\label{e10'}
\end{eqnarray} 
The subspace spanned by the coordinate vector fields $(\partial_r)^a$,
$(\partial_{\theta})^a$, $(\partial_{\phi})^a$ commute with each other, thereby forming a family of spacelike hypersurfaces. At $r=r_{\rm H}$,
we find that,
\begin{eqnarray}
\chi^a\vert_{{\cal{H}}}=
\chi_{\rm{H}}^a=(\partial_{t})^a-\frac{g_{t\phi}}{g_{\phi\phi}}(r_{\rm{H}})(\partial_{\phi})^a=(\partial_{t})^a+\frac{a}{r_{\rm{H}}^2+a^2}(\partial_{\phi})^a, 
\label{e10'f}
\end{eqnarray} 
which is Killing and null. Thus we have specified the required hypersurface orthogonal vector field $\chi^a$ which becomes null and Killing over the horizon. It is the larger root of Eq.~(\ref{e10'}) which denotes the black hole event horizon and concerns us.

Thus we see that the charged Kerr-black hole spacetime satisfies our assumptions.
The emission probability is given by Eq.~(\ref{temp}), with  
the surface gravity of the black hole horizon is computed to be   
\begin{eqnarray}
\kappa=\frac{\left(M^2-a^2-Q^2 \right)^{\frac12} } {2M\left[ M+(M^2-a^2-Q^2)^{\frac12} \right]-Q^2 },
\label{e11}
\end{eqnarray} 
and ${\displaystyle f=-A_{a}\chi_{\rm{H}}^a=-\frac{Q r_{\rm{H}}}{r_{\rm{H}}^2+a^2}}$. The temperature of emission or the Hawking temperature is given by Eq.~(\ref{tempf}), which was earlier obtained 
in \cite{Kerner:2008qv, Li:2008zra} by explicitly solving the semiclassical Dirac equation by the method of separation of variables.

We will consider next some examples from higher dimensions. First we consider non-extremal rotating charged black hole solution of five dimensional minimal supergravity with two different rotation parameters $(a,~b)$ written in the Boyer-Lindquist coordinates \cite{Chong:2005hr},
\begin{eqnarray}
ds^2 = &-&\left[\frac{\Delta_{\theta}\left(1+g^2r^2\right)}{\Sigma_a\Sigma_b}-
\frac{\Delta_{\theta}^2\left(2m\rho^2-q^2+2abqg^2\rho^2
\right)}{\rho^4\Sigma_a^2\Sigma_b^2}
\right]dt^2+\frac{\rho^2}{\Delta_r}dr^2+\frac{\rho^2}{\Delta_\theta}
d\theta^2 \nonumber\\
&+& \left[\frac{\left(r^2+a^2\right)\sin^2\theta}{\Sigma_a}+
\frac{a^2 \left(2m\rho^2-q^2\right)\sin^4\theta +2abq\rho^2\sin^4\theta }{\rho^4 \Sigma_a^2}\right] d\phi^2 \nonumber\\
&+&\left[\frac{\left(r^2+b^2\right)\cos^2\theta}{\Sigma_b}+
\frac{b^2 \left(2m\rho^2-q^2\right)\cos^4\theta+2abq\rho^2\cos^4\theta }
{\rho^4 \Sigma_b^2}\right] d\psi^2\nonumber\\
&-&\frac{2\Delta_{\theta}\sin^2\theta\left[a\left(2m\rho^2-q^2\right)
+ bq\rho^2\left(1+a^2g^2\right) \right]}
{\rho^4\Sigma_a^2\Sigma_b}dtd\phi \nonumber\\
&-&\frac{2\Delta_{\theta}\cos^2\theta\left[b\left(2m\rho^2-q^2\right)
+ aq\rho^2\left(1+b^2g^2\right) \right]}
{\rho^4\Sigma_a\Sigma_b^2}
dtd\psi\nonumber\\
&+&\frac{2\sin^2\theta\cos^2\theta\left[ab\left(2m\rho^2-q^2\right)
+ q\rho^2\left(a^2+b^2\right) \right]}
{\rho^4\Sigma_a\Sigma_b}
d\phi d\psi,%\nonumber\\
\label{e12}
\end{eqnarray} 
where $\rho^2=\left(r^2+a^2\cos^2\theta+b^2\sin^2\theta\right)$, $\Delta_{\theta}=\left(1-a^2g^2\cos^2\theta-b^2g^2\sin^2\theta\right)$, $\Sigma_a=(1-a^2g^2)$, $\Sigma_b=(1-b^2g^2)$ and $\Delta_r=\left[\frac{(r^2+a^2)(r^2+b^2)(1+g^2r^2)+q^2+2abq}{r^2}-2M\right]\geq0$. The parameters $M,~a,~b,~q$ specify respectively 
the mass, angular momenta and the charge of the black hole and $g$ is a real positive constant. The gauge field corresponding to the charge $q$ is given by $ A_a=\frac{\sqrt{3}q}{\rho^2}\left(\frac{\Delta_{\theta}}{\Sigma_a\Sigma_b}(dt)_a-\frac{a\sin^2\theta}{\Sigma_a}(d\phi)_a 
-\frac{b\cos^2\theta}{\Sigma_b}(d\psi)_a\right)$. 

We first note that the solution (\ref{e12}) has three commuting coordinate Killing vector fields $(\partial_t)^a$, $(\partial_{\phi})^a$ and $(\partial_{\psi})^a$. Also, the 
spacelike coordinate basis vector fields $(\partial_{\phi})^a $, $(\partial_{\psi})^a $, $(\partial_{\theta})^a $, $(\partial_{r})^a $ commute with each other, so that the spacelike 4-surfaces spanned by them are hypersurfaces. Let us next construct a vector field $\chi^a$,
\begin{eqnarray}
\chi^a 
:=(\partial_{t})^a-\frac{\left(g_{t\phi}g_{\psi\psi}-g_{t\psi}g_{\phi\psi}\right)}
{\left(g_{\phi\phi}g_{\psi\psi}-(g_{\psi\phi})^2\right)}
(\partial_{\phi})^a
-\frac{\left(g_{t\psi}g_{\phi\phi}-g_{t\phi}g_{\phi\psi}\right)}
{\left(g_{\phi\phi}g_{\psi\psi}-(g_{\psi\phi})^2\right)}(\partial_{\psi})^a,
\label{5d1}
\end{eqnarray} 
so that $\chi_a(\partial_{\phi})^a=0=\chi_a(\partial_{\psi})^a$ everywhere. Thus the vector field $\chi^a$ is orthogonal to the family of spacelike hypersurfaces $\Sigma$, spanned by
 $(\partial_{\phi})^a $, $(\partial_{\psi})^a $, $(\partial_{\theta})^a $, $(\partial_{r})^a $.
Also as $\Delta_r\to 0$, the norm of $\chi^a$ is
$\chi^a\chi_a
=-\beta^2=-\frac{\rho^2r^4\Delta_r} {\left[(r^2+a^2)(r^2+b^2)+abq
\right]^2}+{\cal{O}}({\Delta_r^2})\leq 0$. Thus $\chi^a$ becomes null over
the surface $\Delta_r=0$ and timelike outside it. Let $r=r_{\rm H}$ be the largest root
of $\Delta_r=0$. Then at $r= r_{\rm H}$, the vector field $\chi^a$ becomes
\begin{eqnarray}
\chi_{\rm{H}}^a 
=(\partial_{t})^a+\Omega_{\phi}(\partial_{\phi})^a
+\Omega_{\psi}(\partial_{\psi})^a,
\label{e14}
\end{eqnarray} 
where
\begin{eqnarray}
\Omega_{\phi}=- \frac{\left(g_{t\phi}g_{\psi\psi}-g_{t\psi}g_{\phi\psi}\right)}
{\left(g_{\phi\phi}g_{\psi\psi}-(g_{\psi\phi})^2\right)}
\Bigg\vert_{r=r_{\rm{H}}}
=\frac{a(r_{\rm{H}}^2+b^2)(1+g^2r_{\rm{H}}^2)+bq}
{(r_{\rm{H}}^2+a^2)(r_{\rm{H}}^2+b^2)+abq}, \nonumber\\
\Omega_{\psi}=-\frac{\left(g_{t\psi}g_{\phi\phi}-g_{t\phi}g_{\phi\psi}\right)}
{\left(g_{\phi\phi}g_{\psi\psi}-(g_{\psi\phi})^2\right)}
\Bigg\vert_{r=r_{\rm{H}}}=
\frac{b(r_{\rm{H}}^2+a^2)(1+g^2r_{\rm{H}}^2)+aq}
{(r_{\rm{H}}^2+a^2)(r_{\rm{H}}^2+b^2)+abq}.
\label{e13}
\end{eqnarray} 
Thus we have constructed the timelike vector field $\chi^a$ orthogonal to a family of spacelike
hypersurfaces $\Sigma$, and which becomes null and Killing on the surface $r=r_{\rm H}$. Thus $r=r_{\rm H} $ is the Killing or black hole horizon ${\cal{H}}$ of the spacetime (\ref{e12}). The ratio of the emission to absorption probabilities and the Hawking temperature of this Killing horizon are given by Eq.s (\ref{temp}), (\ref{tempf}), with  
\begin{eqnarray}
\kappa=\frac{r_{\rm H}^4\left[1+g^2(2r_{\rm H}^2+a^2+b^2 )\right] -(ab+q)^2 }{r_{\rm H}\left[ (r_{\rm H}^2+a^2) (r_{\rm H}^2+b^2)+abq\right] },
\label{e15}
\end{eqnarray} 
and $f=-A_a\chi_{\rm{H}}^a=-\frac{\sqrt{3}q r_{\rm{H}}}{(r^2+a^2)(r^2+b^2)+abq}$. This matches with the prediction from the Smarr formula of (\ref{e12}) derived in \cite{Chong:2005hr}, as well as the result of \cite{Li:2010zzd} obtained by explicit solution of the semiclassical Dirac equation by method of separation of variables.

 It can be easily verified using the same methods as above that Eq.s (\ref{temp}), (\ref{tempf}) hold and recover the desired results for the (4+1)-dimensional stationary solutions with Killing horizons, such as squashed Kaluza-Klein black hole \cite{Kurita:2008mj, Ishihara:2005dp}, a black string \cite{Kurita:2008mj, Horowitz:2002ym}, black hole solutions of $z = 4$ Horava-Lifshitz gravity \cite{Park:2009zra} and toroidal black hole solutions of \cite{Rinaldi:2002tc}. We shall not go into demonstrating them here.

Our scheme also applies very easily to an $n$-dimensional Myres-Perry black hole with a single rotation parameter $a$ \cite{Myers:1986un}, \newpage
\begin{eqnarray}
ds^2 =-dt^2+(r^2+a^2)\sin^2\theta d\phi^2+\frac{\mu}{
r^{n-5}\Sigma}\left(dt-a\sin^2\theta d\phi\right)^2\nonumber\\+
\frac{r^{n-5} \Sigma} {r^{n-5}(r^2+a^2)-\mu}dr^2
+\Sigma d\theta^2+r^2\cos^2\theta d\Omega^{n-4},
\label{e16}
\end{eqnarray} 
where the parameters $\mu,~a$ represent respectively the mass and angular momentum of the black hole, $\Sigma=r^2+a^2\cos^2\theta$ and $d\Omega^{n-4}$ represents the metric over an $(n-4)$-sphere.
It is easy to check that the required vector field $\chi^a$ is given by $(\partial_t)^a-\frac{g_{t\phi}}{g_{\phi\phi}}(\partial_{\phi})^a$. 

After this necessary digression for checking the validity of our assumptions for different cases,
finally let us discuss the scenario for the Schwarzschild-de Sitter spacetime (\ref{s22i}). 
As we have seen earlier, this spacetime has a timelike Killing field $\chi^a=(\partial_t)^a$
orthogonal to the family of spacelike hypersurfaces spanned by $(\partial_r)^a$, $(\partial_{\theta})^a$ and $(\partial_{\phi})^a$. For $3M\sqrt{\Lambda}\leq1$, the norm of the timelike
Killing field vanishes at two points $r_{\rm H}\leq r_{\rm C}$. 
Thus $r_{\rm H}$ and $r_{\rm C}$ are the Killing horizons of the spacetime namely, the black hole and the cosmological horizon. Thus Eq.s (\ref{temp}), (\ref{tempf}) hold good with $f=0$ for this case. 
The surface gravities $\kappa_{\rm H}$ and $\kappa_{\rm C} $ of the two horizons are given by Eq.s (\ref{var6}), (\ref{var8}). Then Eq.s (\ref{temp}), (\ref{tempf}) say that there will be thermal emissions from both the Killing horizons and the temperatures of emission will be ${\displaystyle\frac{\kappa_{\rm H} \hbar }{2\pi}}$ and $\displaystyle{ \frac{\kappa_{\rm C} \hbar }{2\pi}}$ respectively.
 Similar results hold also for other stationary de Sitter black hole spacetimes, such as the Reissner-N\"{o}rdstrom-de Sitter or the Kerr-Newman-de Sitter spacetimes. 
 
%%%%%%%%%%%%%%%%%%%%%%%%%%%%%%%%%%%%%%%%%%%%%%%%%%%%%%%%%%%%%%%%%%%%%%%%%%%%%%%%%%%%%%%%%%%%
\section{Vector, spin-$2$ and spin-${\displaystyle\frac{3}{2}}$ fields}
We have seen in Section 5.2.1 that the equation of motion for a Dirac spinor reduces to scalar equations in the semiclassical WKB framework.
We will show below that the equations of motion for Proca, massive spin-$2$ and spin-${\displaystyle\frac{3}{2}}$ fields also reduce to the scalar equations in the semiclassical framework. Let us first consider the equation of motion for a Proca field $A^b$, 
\begin{eqnarray}
\nabla_{a}F^{ab} = \frac{m^{2}}{\hbar^2} A^{b},
\label{v1}
\end{eqnarray} 
where $F_{ab}=\nabla_{[a} A_{b]}$. Eq.~(\ref{v1}) can be written as 
\begin{eqnarray}
\nabla_a\nabla^a A_b -R_{b}{}^{a}A_a-
\nabla_b\left(\nabla_a A^a\right)=\frac{m^{2}}{\hbar^2} A_b,
\label{v2}
\end{eqnarray} 
where $R_{ab}$ is the Ricci scalar.
But Eq.~(\ref{v1}) implies that $\nabla_a A^a=0$ identically.
Now let us choose a set of orthonormal basis $\left\{e_{a}^{(\mu)}\right\}$. We expand the vector field $A_a$ in this basis, $A_b=e_{b}^{(\mu)}A_{(\mu)}$. With this
expansion and the fact that $\nabla_a A^a=0$, Eq.~(\ref{v2}) becomes
\begin{eqnarray}
e_{b}^{(\mu)}\nabla_a\nabla^a A_{(\mu)}+ A_{(\mu)}\nabla_a\nabla^a 
e_{b}^{(\mu)}+2\nabla_a A_{(\mu)}\nabla^a e_{b}^{(\mu)} 
-R_{b}{}^{(\mu)}A_{(\mu)}
=\frac{m^{2}}{\hbar^2}A_{(\mu)}e_{b}^{(\mu)},\nonumber\\
\label{v3}
\end{eqnarray} 
which after contracting both sides by $e^{b}_{(\nu)}$ reduces to   
\begin{eqnarray}
\nabla_a\nabla^a A_{(\nu)}+ A_{(\mu)}e^{b}_{(\nu)}
\nabla_a\nabla^a 
e_{b}^{(\mu)}+2e^{b}_{(\nu)}\nabla_a A_{(\mu)}\nabla^a 
e_{b}^{(\mu)} 
-R_{(\nu)}{}^{(\mu)}A_{(\mu)}
=\frac{m^{2}}{\hbar^2} A_{(\nu)}.\nonumber\\
\label{v4}
\end{eqnarray} 
 We choose the usual WKB ansatz for each $A_{(\nu)}$ :$A_{(\nu)}=f_{\nu}(x) e^{\frac{i I_{\nu}(x)}{\hbar}}$, where the repeated indices are not summed here and the functions $f$ and $I$ are independent of $\hbar$. 
Substituting this into Eq.~(\ref{v4}), we take the semiclassical limit $\hbar\to 0$. 
It immediately turns out that in the semiclassical limit Eq.~(\ref{v4}) can be formally represented by Klein-Gordon equations for the $n$ scalars $A_{(\nu)}$,
\begin{eqnarray}
\nabla_a\nabla^a A_{(\nu)}
-\frac{m^{2}}{\hbar^2} A_{(\nu)}=0,
\label{v5}
\end{eqnarray} 
with $\nu=0,~1,~2,\dots,~(n-1)$. When each of the Eq.s~(\ref{v5}) is
explicitly expanded and the near horizon limit is taken we get Eq.~(\ref{e}) with $e=0$. Thus Eq.s (\ref{temp}) and (\ref{tempf}) hold for this case also.

Next we consider the massive spin-$2$ field $\pi_{ab}$ satisfying the Fierz-Pauli equation \cite{pauli}
\begin{eqnarray}
\nabla_c\nabla^c \pi_{ab}-\frac{m^2}{\hbar^2}\pi_{ab}=0,
\label{h1}
\end{eqnarray} 
where $\pi_{ab}$ are symmetric tensor fields. As before we expand $\pi_{ab}$ in orthonormal basis, $\pi_{ab}=e^{(\mu)}_a e^{(\nu)}_b\pi_{(\mu)(\nu)}$. In the
semiclassical limit and for the WKB ansatz, Eq.~(\ref{h1}) can effectively be represented by $\frac{n(n+1)}{2}$ Klein-Gordon equations for the scalars $\pi_{(\mu)(\nu)}$ 
\begin{eqnarray}
\nabla_c\nabla^c \pi_{(\mu)(\nu)}-
\frac{m^2}{\hbar^2}\pi_{(\mu)(\nu)}=0,
\label{h2}
\end{eqnarray} 
and thus similar conclusions hold for this case also.

             %%%          SPIN 3/2

Finally we will address the spin-$\frac{3}{2}$ fields satisfying the Rarita-Schwinger equation \cite{rarita}. The tunneling phenomenon for this field was addressed in \cite{Yale:2008kx}
for the Kerr black hole by explicitly solving the equations of motion in the near horizon limit.

 The Rarita-Schwinger equation in a curved spacetime reads
\begin{eqnarray}
i\gamma^a\nabla_a\Psi_b =-\frac{m}{\hbar}\Psi_b,
\label{h3}
\end{eqnarray} 
where $\Psi_b$ is a spinor. The $\gamma$'s are matrices satisfying the anti-commutation relation similar to the
Dirac $\gamma$'s: $\left[\gamma^a,~\gamma^b \right]_+=2g^{ab}\bf{I}$.
 The spin-covariant derivative $\nabla$ is defined as $\nabla_a \Psi_b:=\left(\partial_a + \Gamma_a\right)\Psi_b$, where 
$\Gamma_a$ are the spin connection matrices. 

Due to the similarity of the spin-${\displaystyle\frac{3}{2}}$ fields with the Dirac spinors discussed in Section 5.2.1, we will apply the same method here to show
that $\Psi_b$ satisfies the Klein-Gordon equation in the semiclassical WKB
framework. We square Eq.~(\ref{h3}) by applying $i\gamma^c\nabla_c$ from
left. A little computation using the definition of the spin-covariant derivative $\nabla_a$, the anti-commutation relation satisfied by the $\gamma$'s, and also the commutativity of the partial derivatives yields as before,
\begin{eqnarray}
\nabla_a\nabla^a\Psi_b+
\frac{1}{4}\left[\gamma^a,~\gamma^c\right]
 \left(\partial_{[a}\Gamma_{c]}+\Gamma_{[a}\Gamma_{c]}\right)\Psi_b+
\left(\gamma^c\nabla_c \gamma^a\right)\nabla_a \Psi_b=
\frac{m^2}{\hbar^2}\Psi_b.
\label{h4}
\end{eqnarray} 
So as in the previous cases it immediately follows then for the usual ansatz
\begin{eqnarray}
\Psi_a  
&=&\left[ 
\begin{array}{c}
A_a(x)e^{\frac{i I_1(x)}{\hbar}}\\ 
B_a(x)e^{\frac{i I_2(x)}{\hbar}} \\
C_a(x)e^{\frac{i I_3(x)}{\hbar}} \\
D_a(x)e^{\frac{i I_4(x)}{\hbar}}\\
\end{array}
\right], 
\label{h5}
\end{eqnarray}
Eq.~(\ref{h4}) reduce to Klein-Gordon equations in the semiclassical limit. We can easily generalize this result for a charged spin-$\displaystyle{\frac{3}{2}}$ particle coupled to a gauge field by replacing the spin covariant derivative by the gauge spin covariant derivative. This gives 
charged Klein-Gordon equations.   

%%%%%%%%%%%%%%%%%%%%%%%%%%%%%%%%%%%%%%%%%%%%%%%%%%%%%%%%%%%%%%%%%%%%%%%

Let us now summarize our results. In this Chapter our main goal was to address thermodynamics, and particle creation in the Schwarzschild-de Sitter spacetime by complex path method. 
 In doing so, we have put the complex path approach for stationary spacetimes in a general framework. 
 We have dealt with some well known physical matter fields and shown for any arbitrary spacetime in a coordinate independent way that in the semiclassical WKB
 framework all those field equations of motion are equivalent to the scalar equations. We have done this without choosing any particular 
basis of the vector fields or the $\gamma$ matrices. We needed to assume only that a metric $g_{ab}$ can be defined
on the spacetime which guarantees the existence of the orthonormal basis $\left\{e^{(\mu)}_a\right\}$. So it is clear that as far as the semiclassical level is concerned it is sufficient to work only with scalars for any arbitrary spacetime. 

 We further presented a general coordinate independent expression for the emission probability from an arbitrary stationary Killing horizon with some reasonable geometrical properties. It was shown that for such spacetimes the emission is always thermal and the temperature is given in terms of the Killing horizon's surface gravity as ${\displaystyle \frac{\kappa\hbar}{2\pi}}$, thereby proving the universality of particle emissions from Killing horizons through a very general approach. 

This helped us to discuss particle creation in stationary de Sitter black spacetimes. For such spacetimes there are two kind of Killing horizons -- one is the black hole and the other is the cosmological horizon. We have demonstrated that the semiclassical complex path method let us treat particle emissions from both the horizons in an equal footing. We addressed explicitly the case for the Schwarzschild-de Sitter spacetime. Although we note that our calculations clearly show that for any arbitrary stationary de Sitter black hole spacetime, Eq.s (\ref{temp}) and hence (\ref{tempf}) hold, and thus the two horizons always radiate thermally, and the temperature of emission from those horizons will always be proportional to their respective surface gravities.

%%%%%%%%%%%%%%%%%%%%%%%%%%%%%%%%%%%%%%%%%%%%%%%%%%%%%%%%%%%%%%%%%%%%%%%%%%%%%%%%%%%%%%%%%%%%%%%%%%%%%%%%%%%%%%%%%%%%%%%%%%%%%%%%%%%%%%%%%%%%%%%%%%%%%%%%%%%%%%%%%%%%%%%%%%%%%%%%%%%%%%%%%%%%%%%%%%%%%%%%%%%%%%%%%%%%%%%%%%%%%%%%%%%%%%%%%%%%%%%%%%%%%%%%%%%%%%%%%%%%%%%%%%%%%%%%%%%%%%%%%%%%%%%%%%%%%%%%%%%%%%%%%%%%%%%%%%%%%%%%%%%%%%%%%%%%%%%%%%%%%%%%%%%%

%%%%%%%%%%%%%%%%%%%%%%%%%%%%%%%%%%%%%%%%%%%%%%%%%%%%%%%%%%%%%%%%%%%%%%%%%%%%%%%%%%%%%%%%%%%
\chapter{Summary}
In this thesis we have studied some properties of black hole spacetimes endowed with a positive cosmological constant $\Lambda$. We know from exact solutions that the inclusion of a positive $\Lambda$ into the Einstein equations gives rise to an outer null hypersurface
under some reasonable conditions. This outer null hypersurface acts as an outer boundary of the spacetime and is known as the cosmological event horizon. In all stationary exact and known solutions with $\Lambda>0$, this boundary is a Killing horizon. Due to this boundary an observer located inside the cosmological horizon cannot refer to the region behind that and thus any precise notion of asymptotic is lost. 
Our main goal in this thesis was to investigate the role or effect of $\Lambda$ and this outer boundary of the spacetime in gravity. The motivation of this study comes from the recent observations which indicate that there is a strong possibility that our universe is indeed endowed with a small but positive $\Lambda$ \cite{Riess:1998cb,Perlmutter:1998np}.

 In Chapter 1 we reviewed briefly the history of $\Lambda$ and
 elaborated our motivation to study gravity with this. We considered some exact stationary solutions with positive $\Lambda$ and discussed the properties of the cosmological event horizon. We reviewed black hole no hair theorems, geodesic motion in cosmic string spacetimes and thermodynamics and Hawking radiation, which are addressed in the remaining part of the thesis.     
In Chapter 2 we established a general criterion for the existence of the cosmological event horizons
in static and stationary axisymmetric spacetimes. We found that the energy-momentum tensor must violate the strong energy condition, at least over some portion of a spacelike hypersurface in our region of interest. In Chapter 3 we discussed various classical no hair theorems for black hole spacetimes endowed with a positive $\Lambda$, i.e. endowed with a cosmological horizon.
We considered both static and stationary axisymmetric spacetimes.
We found for static spacetimes a clear exception of the no hair theorem for the Abelian Higgs model---we found a spherically symmetric electrically charged solution sitting in the false vacuum of the Higgs field. This has no $\Lambda=0$ analogue. This comes from the non-trivial boundary effect at the cosmological horizon. In particular, this indicates that the existence of the cosmological horizon may change the local physics considerably. In Chapter 4 we constructed static cosmic Nielsen-Olesen string spacetimes with $\Lambda>0$. We considered both free, infinitely long string and a string piercing the horizons of the Schwarzschild-de Sitter spacetime. The conical singularity terms were estimated also. For a free cosmic string, we discussed the geodesic motion and demonstrated the repulsive effect of positive $\Lambda$. In Chapter 5 we discussed thermodynamics of the Schwarzschild-de Sitter spacetime and Hawking or Hawking like radiation
via the semiclassical complex path method. We proved the universality of particle emission from any Killing horizon of a stationary spacetime by deriving a general formula. This helped us to discuss the particle creation by the black hole and the cosmological horizon in an equal footing. We also note that since the general formula for Hawking radiation in this Chapter was derived on the basis of some geometrical properties of the spacetime in a coordinate independent way, the result also applies well to any arbitrary stationary black hole spacetime with a Killing horizon in a de Sitter universe.  

We have mentioned in each of the Chapters the possible extensions or generalizations of the problems we discussed. Here we emphasize separately one of the most interesting open problems 
in the de Sitter or de Sitter black hole spacetimes. Precisely, this is the construction of a quantum field theoretic description of the particle creation or Hawking radiation in such spacetimes. Unlike the flat spacetime, there exists no preferred coordinate system in a curved spacetime and so the concept of particles or vacuum states in curved spacetimes are observer dependent. It has been shown for the Schwarzschild spacetime that there exist a certain class of observers or vacuum states which can register thermal radiation (see e.g. \cite{Birrell} and references therein), the temperature of the radiation being given uniquely by the surface gravity of the Killing horizon. In the Schwarzschild-de Sitter spacetime, on the other hand, there are two Killing horizons which radiate thermally at temperatures proportional to their respective surface gravities. So, what will be the vacuum states for observers receiving radiations from both the horizons? Or, what will be the response function for a particle detector? 

Also, we recall that Hawking's original calculations give a clear mechanism of particle creation by black holes by considering an object undergoing gravitational collapse to form a black hole \cite{Hawk} at late times. Can we construct an analogous description for de Sitter black holes also? The main obstacle to this is, unlike the asymptotically flat spacetimes, we cannot set our boundary conditions at future and past null infinities for this case due to the existence of the cosmological horizon.

\appendix
\chapter{ Derivation of Eq.~(\ref{s42i})}

We consider a test particle moving along a timelike or null geodesic $u^a$ in the de Sitter spacetime (\ref{dsin3}). The norm $k$ of $u^a$ is
\begin{eqnarray}
k=g_{ab}u^bu^b=-\left(1-\frac{\Lambda r^2}{3}\right)\dot{t}^2+
\left(1-\frac{\Lambda r^2}{3}\right)^{-1}\dot{r}^2 +r^2\dot{\theta}^2
+r^2\sin^2\theta \dot{\phi}^2,
\label{ad50}
\end{eqnarray}
where $k=-1~(0)$ if $u^a$ is timelike (null), and the `dot' denotes differentiation with
respect to some parameter $\tau$ along the geodesic. We can reduce this motion to an effective one dimensional central force problem in the following way. If $\zeta^a$ is any Killing field, the quantity $u^a\zeta_a$ is conserved along any geodesic $u^a$,
\begin{eqnarray}
u^a\nabla_a\left(u^b\zeta_b\right)=\frac12u^au^b\nabla_{(a}\zeta_{b)}+\zeta_b\left(u^a\nabla_{a}u^b\right)=0.
\label{energy'}
\end{eqnarray}
 We consider the four Killing fields of the de Sitter spacetime,
\begin{eqnarray}
\zeta_0^a&=& (\partial_t)^a,\quad
\zeta_1^a=-\sin\phi (\partial_\theta)^a-\cot\theta\cos\phi  (\partial_\phi)^a, \nonumber\\
\zeta_2^a&=&\cos\phi (\partial_\theta)^a-\cot\theta\sin\phi  (\partial_\phi)^a,
\quad\zeta_3^a=(\partial_\phi)^a.
\label{ad51}
\end{eqnarray}
The first one is the timelike Killing field whereas the remaining three are spacelike and generate rotations over a 2-sphere. The conserved quantities associated with them are
\begin{eqnarray}
%%\begin{subequation}
E&=&-g_{ab}u^a\zeta_0^b=\left(1-\frac{\Lambda r^2}{3}\right)\dot{t}, \nonumber\\
L_1&=&g_{ab}u^a\zeta_1^b=-r^2\left(\sin\phi \cdot\dot{\theta}+\sin\theta\cos\theta\cos\phi
\cdot\dot{\phi}  \right), \nonumber\\
L_2&=&g_{ab}u^a\zeta_2^b=r^2\left(\cos\phi \cdot\dot{\theta}-\sin\theta\cos\theta\sin\phi\cdot\dot{\phi} \right),
\nonumber\\
L_3&=&g_{ab}u^a\zeta_3^b=r^2\sin^2\theta\cdot\dot{\phi},
\label{ad52}
%%\end{subequation}
\end{eqnarray}
where the first one can be regarded as the conserved energy and the remaining can be regarded as the conserved orbital angular momenta along the geodesic. From Eq.s~(\ref{ad52}) we have
\begin{eqnarray}
L_1^2+L_2^2+L_3^2=L^2=r^4\left(\dot{\theta}^2+\dot{\phi}^2\sin^2\theta\right).
\label{ad53}
\end{eqnarray}
Using the first of Eq.s~(\ref{ad52}) and  Eq.~(\ref{ad53}), we eliminate $\dot{t},
 ~\dot{\theta}$ and $\dot{\phi}$ from Eq.~(\ref{ad50}) to have
\begin{eqnarray}
k=-\frac{E^2}{\left(1-\frac{\Lambda r^2}{3}\right)}+\left(1-\frac{\Lambda r^2}{3}\right)^{-1}\dot{r}^2+\frac{L^2}{r^2},
\label{ad54}
\end{eqnarray}
which can be rewritten as  
\begin{eqnarray}
\frac{1}{2}\dot{r}^2+\psi (r,~L)=\frac{1}{2}E^2,
\label{geo1}
\end{eqnarray}
where the effective potential $\psi (r,~L)$ is given by
\begin{eqnarray}
\psi (r,~L)=\frac{1}{2}
\left(1-\frac{\Lambda r^2}{3}\right)
\left(\frac{L^2}{r^2}-k\right).
\label{geo2}
\end{eqnarray}
Thus Eq.~(\ref{geo1}) represents an effective non-relativistic central force motion of a unit mass test particle of energy ${\displaystyle\frac12 E^2}$. 
%

%%%%%%%%%%%%%%%%%%%%%%%%%%%%%%%%%%%%%%%%%%%%%%%%%%%%%%%%%%%%%%%%%%%%%%%%%%%%%%%%%%%%%%%%%%%%%%%

%%%%%%%%%%%%%%%%%%%%%%%%%%%%%%%%%%%%%%%%%%%%%%%%%%%
%\begin{thebibliography}{0}
%%% \frenchspacing
%\baselineskip=23pt

%\cite{Mandelstam:1974pi}

\pagebreak
%%\chapter*{Publications and in preparation }
\addcontentsline{toc}{chapter}{Publications and in preparation}
\chapter*{Publications and in preparation }

\hskip .37cm$1)$
   `Black-hole no hair theorems for a positive cosmological constant',\\
Sourav Bhattacharya, Amitabha Lahiri; Phys.Rev.Lett.{\bf 99}:201101 (2007), 
[arXiv:gr-qc/0702006]$^{^\star}$.

$2)$
   `Effect of a positive cosmological constant on cosmic strings',\\
Sourav Bhattacharya, Amitabha Lahiri; Phys.Rev.D{\bf 78}:065028 (2008),
arXiv:0807.0543[gr-qc]$^{^\star}$.

\hskip .04cm$3)$ `G\"{o}del black hole, closed timelike horizon and the study of particle emissions',\\
Sourav Bhattacharya, Anirban Saha; Gen.Rel.Grav.{\bf 42}:1809 (2010), arXiv:0904.3441[gr-qc].

$4)$
   `A Note on Hawking radiation via complex path analysis',\\
Sourav Bhattacharya; Class.Quant.Grav.{\bf 27}:205013 (2010), arXiv:0911.4574[gr-qc]$^{^\star}$.

$5)$
 `On the existence of cosmological event horizons',\\
Sourav Bhattacharya, Amitabha Lahiri; Class.Quant.Grav.{\bf 27}:165015 (2010), arXiv:1001.1162[gr-qc]$^{^\star}$.

$6)$
 `Cosmic strings with positive $\Lambda$',\\
Sourav Bhattacharya, Amitabha Lahiri; prepared for MG12 proceedings, arXiv:1003.0295[gr-qc]$^{^\star}$.

$7)$
     `No hair theorems for stationary axisymmetric black holes',\\
Sourav Bhattacharya, Amitabha Lahiri; Phys. Rev. D {\bf 83}, 124017 (2011),
arXiv:1102.0053[gr-qc]$^{^\star}$.

$8)$
`Mass function for the Schwarzschild-de Sitter black hole and thermodynamics',
Sourav Bhattacharya, Amitabha Lahiri (in preparation).

$\star$~Included in this thesis.
%\begin{center}
%%%%%%%%%%%%%%%%%%%%%%%%%%%%%%%%%%%% 
%%%%%%%%%%%%%% FIG   %%%%%%%%%%%%%%%%%%%%%%%%%%%%%%%%%%%
%\begin{figure}[h]
%\centering%keepaspectratio
%\rotatebox{0}{
%\includegraphics[height=10cm,width=8cm]{sou_pre.pdf}}
%%\includegraphics[height=10.0cm,keepaspectratio]{sou_pre.ps}}
%\includegraphics[height=2.5cm,width=10.0cm]{bhatta1.eps}}
%\caption{The Kruskal diagram for the de Sitter spacetime (\ref{anex9'}) with each point understood over a 2-sphere.}
%\end{figure}
%%%%%%%%%%%%%%%%%%%%%%%
%\end{center} 
%\vskip 1cm

%%%%%%%%%%%%%%%%%%%%%%%%%%%%%%
%%%%%%%%%%%%%%%%%%%%%%%%%%%%%%%%%%%%%%%%% 

\addcontentsline{toc}{chapter}{Bibliography}
\begin{thebibliography}{99}
\label{bb}
%A
%B
%C
%D
%E
%%%%%%%%%%%%%%%%%%%%%%%%%%%%%%%%%%%%%%%%%%%%%%%%%%%%%%%%%%%%%%%%%%%%%%%%%%%%
%\cite{Wald:1984rg}
\bibitem{Wald:1984rg}
  R.~M.~Wald,
  ``General Relativity,''
%\href{http://www.slac.stanford.edu/spires/find/hep/www?irn=1334239}{SPIRES
%entry} 
{\it  Chicago, Usa: Univ. Pr. ( 1984)}.
\vskip .2cm

%%\cite{weinbergbook}
\bibitem{weinbergbook}
  S.~Weinberg,
  ``Gravitation and Cosmology,''
%\href{http://www.slac.stanford.edu/spires/find/hep/www?irn=7886489}{SPIRES entry}
{\it  John Wiley and Sons, New York (1972)}.
\vskip .2cm

%\cite{Weinberg:2008zzc}
\bibitem{Weinberg:2008zzc}
  S.~Weinberg,
  ``Cosmology,''
%\href{http://www.slac.stanford.edu/spires/find/hep/www?irn=7886489}{SPIRES entry}
{\it  Oxford, UK: Oxford Univ. Pr. (2008)}.


\vskip .2cm
%\cite{Hawking:1973uf}
\bibitem{Hawking:1973uf}
  S.~W.~Hawking and G.~F.~R.~Ellis,
  ``The Large scale structure of spacetime,''
%\href{http://www.slac.stanford.edu/spires/find/hep/www?irn=6991262}{SPIRES entry}
{\it  Cambridge University Press, Cambridge, (1973)}.

\vskip .2cm
\bibitem{Hubble}
  E.~Hubble,
  %``Observational Evidence from Supernovae for an Accelerating Universe and a
  %Cosmological Constant,''
  Proc.\ N.\ A.\ S.\  {\bf 15}, 168 (1929).


\vskip .2cm
%\cite{Riess:1998cb}Perlmutter:1998np
\bibitem{Riess:1998cb}
  A.~G.~Riess {\it et al.}  [Supernova Search Team Collaboration],
  %``Observational Evidence from Supernovae for an Accelerating Universe and a
  %Cosmological Constant,''
  Astron.\ J.\  {\bf 116}, 1009 (1998).
  %% [arXiv:astro-ph/9805201].
  %%CITATION = ASTRO-PH 9805201;%%

\vskip .2cm
%\cite{Perlmutter:1998np}
\bibitem{Perlmutter:1998np}
  S.~Perlmutter {\it et al.}  [Supernova Cosmology Project Collaboration],
  %``Measurements of Omega and Lambda from 42 High-Redshift Supernovae,''
  Astrophys.\ J.\  {\bf 517}, 565 (1999).
  %% [arXiv:astro-ph/9812133].
  %%CITATION = ASTRO-PH 9812133;%%

\vskip .2cm
\bibitem{pilar}
   P.~Ruiz-Lapuente,
  ``Dark Energy,''
%\href{http://www.slac.stanford.edu/spires/find/hep/www?irn=6991262}{SPIRES entry}
{\it  Cambridge University Press, Cambridge, (2010)}.

\vskip .2cm
\bibitem{copeland}
E.~J.~Copeland, M.~Sami and S.~Tsujikawa, 
Int. \ J. \ Mod. \ Phys. \  D {\bf 15}, 1753 (2006).
%%[arXiv:hep-th/0603057].

\pagebreak
%\cite{Kastor:1992nn}
\bibitem{Kastor:1992nn}
  D.~Kastor and J.~H.~Traschen,
  %``Cosmological multi - black hole solutions,''
  Phys.\ Rev.\  D {\bf 47}, 5370 (1993).
  %%[arXiv:hep-th/9212035].
  %%CITATION = PHRVA,D47,5370;%%

\vskip .2cm
%\cite{Carter:1968ks}
\bibitem{Carter:1968ks}
  B.~Carter,
  %``Hamilton-Jacobi and Schrodinger separable solutions of Einstein's
  %equations,''
  Commun.\ Math.\ Phys.\  {\bf 10}, 280 (1968).
  %%CITATION = CMPHA,10,280;%%

\vskip .2cm
%\cite{Wald:1984rgnew1}
\bibitem{Wald:1984rgnew1}
  R.~M.~Wald,
  %%``Cosmological multi - black hole solutions,''
  Phys.\ Rev.\  D {\bf 28}, 2118 (1983).
  %%[arXiv:hep-th/9212035].
  %%CITATION = PHRVA,D47,5370;%%


\vskip .2cm
%\cite{Gibbons:1977mu}
\bibitem{Gibbons:1977mu}
  G.~W.~Gibbons and S.~W.~Hawking,
  %``Cosmological Event Horizons, Thermodynamics, And Particle Creation,''
  Phys.\ Rev.\  D {\bf 15}, 2738 (1977).
  %%CITATION = PHRVA,D15,2738;%%

\vskip .2cm
\bibitem{Birrell}
  N.~D.~Birrell and P.~C.~W.~Davies,
  ``Quantum fields in curved space,''
%\href{http://www.slac.stanford.edu/spires/find/hep/www?irn=1845780}{SPIRES entry}
{\it  Cambridge University Press, Cambridge (1982).}


\vskip .2cm
%\cite{Gourgoulhon:2005ng}
\bibitem{Gourgoulhon:2005ng}
  E.~Gourgoulhon and J.~L.~Jaramillo,
  %``A 3+1 perspective on null hypersurfaces and isolated horizons,''
  Phys.\ Rept.\  {\bf 423}, 159 (2006).
%  [arXiv:gr-qc/0503113].
  %%CITATION = PRPLC,423,159;%%


\vskip .2cm
%\cite{Chandrasekhar:1985kt}
\bibitem{Chandrasekhar:1985kt}
  S.~Chandrasekhar,
  ``The mathematical theory of black holes,''
%\href{http://www.slac.stanford.edu/spires/find/hep/www?irn=1845780}{SPIRES entry}
{\it  Oxford, UK: Clarendon (1992).}


\vskip .2cm
%\cite{Chrusciel:1994sn}
\bibitem{Chrusciel:1994sn}
  P.~T.~Chrusciel,
  %``'No hair' theorems: Folklore, conjectures, results,''
  Contemp.\ Math.\  {\bf 170}, 23 (1994).
  %% [arXiv:gr-qc/9402032].
  %%CITATION = GR-QC 9402032;%%

\vskip .2cm
%\cite{Heusler:1998ua}
\bibitem{Heusler:1998ua}
  M.~Heusler,
  %``Stationary Black Holes: Uniqueness And Beyond,''
  Living Rev.\ Rel.\  {\bf 1}, 6 (1998).
  %%CITATION = 00222,1,6;%%

\vskip .2cm
%\cite{Bekenstein:1998aw}
\bibitem{Bekenstein:1998aw}
  J.~D.~Bekenstein,
 %{\it Black holes: Classical properties, thermodynamics, and heuristic
 %quantization,} 
Cosmology and Gravitation, M. Novello, ed. (Atlantisciences, France 2000), pp. 1-85,
(arXiv:gr-qc/9808028).
  %%CITATION = GR-QC 9808028;%%


\vskip .2cm
%\cite{Bekenstein:1971hc}
\bibitem{Bekenstein:1971hc}
  J.~D.~Bekenstein,
  %``Nonexistence of baryon number for static black holes,''
  Phys.\ Rev.\ D {\bf 5}, 1239 (1972).
  %%CITATION = PHRVA,D5,1239;%%


\vskip .2cm
%\cite{Adler:1978dp}
\bibitem{Adler:1978dp}
  S.~L.~Adler and R.~B.~Pearson,
  %``'NO HAIR' THEOREMS FOR THE ABELIAN HIGGS AND GOLDSTONE MODELS,''
  Phys.\ Rev.\ D {\bf 18}, 2798 (1978).
  %%CITATION = PHRVA,D18,2798;%%


\vskip .2cm
%\cite{Lahiri:1993vg}
\bibitem{Lahiri:1993vg}
A.~Lahiri,
%``The no hair theorem for the Abelian Higgs model,''
Mod.\ Phys.\ Lett.\ A {\bf 8}, 1549 (1993).
%% [gr-qc/9207008].
%%CITATION = GR-QC 9207008;%%


\vskip .2cm
%\cite{Price:1971gc}
\bibitem{Price:1971gc}
  R.~H.~Price,
  %``Nonspherical perturbations of relativistic gravitational
  %  collapse. II. Integer spin and zero rest-mass fields,''
  Phys.\ Rev.\ D {\bf 5}, 2439 (1972).
  %%CITATION = PHRVA,D5,2419;%%


\vskip .2cm
%\cite{Chambers:1994sz}
\bibitem{Chambers:1994sz}
  C.~M.~Chambers and I.~G.~Moss,
  %``A Cosmological No Hair Theorem,''
  Phys.\ Rev.\ Lett.\  {\bf 73}, 617 (1994).
  %% [arXiv:gr-qc/9406036].
  %%CITATION = GR-QC 9406036;%%


\vskip .2cm
%\cite{Torii:1998ir}
\bibitem{Torii:1998ir}
  T.~Torii, K.~Maeda and M.~Narita,
  %``Toward the No-scalar hair conjecture in asymptotic de Sitter
  %spacetime,'' 
  Phys.\ Rev.\ D {\bf 59}, 064027 (1999).
  %% [arXiv:gr-qc/9809036].
  %%CITATION = GR-QC 9809036;%%

\pagebreak

%\cite{Martinez:2002ru}Izquierdo:2005ku
\bibitem{Martinez:2002ru}
  C.~Martinez, R.~Troncoso and J.~Zanelli,
  %``De Sitter black hole with a conformally coupled scalar field in  four
  %dimensions,''
  Phys.\ Rev.\ D {\bf 67}, 024008 (2003).
  %% [arXiv:hep-th/0205319].
  %%CITATION = HEP-TH 0205319;%%
\vskip .2cm

%\cite{Bekenstein:1972ky}
\bibitem{Bekenstein:1972ky}
  J.~D.~Bekenstein,
  %``Nonexistence of baryon number for black holes. ii,''
  Phys.\ Rev.\  D {\bf 5}, 2403 (1972).
  %%CITATION = PHRVA,D5,2403;%
\vskip .2cm

%\cite{Skakala:2009ss}
\bibitem{Skakala:2009ss}
  J.~Skakala and M.~Visser,
  %``Birkhoff-like theorem for rotating stars in (2+1) dimensions,''
  arXiv:0903.2128 [gr-qc].
  %%CITATION = ARXIV:0903.2128;%%
\vskip .2cm


%\cite{Sen:1998bj}
\bibitem{Sen:1998bj}
  S.~Sen and N.~Banerjee,
  %``No scalar hair theorem for a charged axially symmetric stationary black
  %hole,''
Pramana \ {\bf 56}, 487 (2001).
% arXiv:gr-qc/9809064.
  %%CITATION = GR-QC/9809064;%%
\vskip .2cm

%\cite{Mazur:2000pn}
\bibitem{Mazur:2000pn}
  P.~O.~Mazur, arXiv:hep-th/0101012,
 %% ``Black hole uniqueness theorems,'' 
an earlier version published in Proceedings of the 11th International Conference on General Relativity and Gravitation, ed. M. A. H. MacCallum, Cambridge University Press, Cambridge 1987, pp. 130-157. 
\vskip .2cm

\bibitem{Robinson}
D.~C.~Robinson, ``Four decades of black hole uniqueness
theorems'', in {\it The Kerr Spacetime:
Rotating Black Holes in General Relativity}, eds. D L Wiltshire,
M Visser \& S M Scott, {\it Cambridge University Press, 2009.}
\vskip .2cm


%\cite{Boucher:1983cv}
\bibitem{Boucher:1983cv}
  W.~Boucher, G.~W.~Gibbons and G.~T.~Horowitz,
  %``A Uniqueness Theorem For Anti-De Sitter spacetime,''
  Phys.\ Rev.\  D {\bf 30}, 2447 (1984).
  %%CITATION = PHRVA,D30,2447;%%
\vskip .2cm

%\cite{AyonBeato:2004if}
\bibitem{AyonBeato:2004if}
  E.~Ayon-Beato, C.~Martinez, J.~Zanelli,
  %``Birkhoff's theorem for three-dimensional AdS gravity,''
  Phys.\ Rev.\  {\bf D70}, 044027 (2004).
%%  [hep-th/0403227].
\vskip .2cm

%\cite{Suneeta:2003bj}
\bibitem{Suneeta:2003bj}
  V.~Suneeta,
  %``Quasinormal modes for the SdS black hole: An analytical approximation
  %scheme,''
  Phys.\ Rev.\  D {\bf 68}, 024020 (2003).
%  [arXiv:gr-qc/0303114].
  %%CITATION = PHRVA,D68,024020;%%
\vskip .2cm

%\cite{Chambers:1994ap}
\bibitem{Chambers:1994ap}
 C.~M.~Chambers and I.~G.~Moss,
  %``Stability of the Cauchy horizon in Kerr-de Sitter spacetimes,''
 Class.\ Quant.\ Grav.\  {\bf 11}, 1035 (1994).
%  [arXiv:gr-qc/9404015].
  %%CITATION = CQGRD,11,1035;%%
\vskip .2cm


%\cite{Rindler:2007zz}
\bibitem{Rindler:2007zz}
  W.~Rindler and M.~Ishak,
  %``The Contribution of the Cosmological Constant to the Relativistic Bending
  %of Light Revisited,''
  Phys.\ Rev.\  D {\bf 76}, 043006 (2007).
 % [arXiv:0709.2948 [astro-ph]].
  %%CITATION = PHRVA,D76,043006;%%
\vskip .2cm


%\cite{Ishak:2008ex}
\bibitem{Ishak:2008ex}
M.~Ishak,
        %``Light Deflection, Lensing, and Time Delays from Gravitational Potentials
        %and Fermat's Principle in the Presence of a Cosmological Constant,''
Phys.\ Rev.\  D {\bf 78}, 103006 (2008).
%[arXiv:0801.3514 [astro-ph]].
        %%CITATION = PHRVA,D78,103006;%%
\vskip .2cm

%\cite{Ishak:2008zc}
\bibitem{Ishak:2008zc}
  M.~Ishak, W.~Rindler and J.~Dossett,
  %``More on Lensing by a Cosmological Constant,''
  Mon.\ Not.\ Roy.\ Astron.\ Soc.\  {\bf 403}, 2152 (2010).
 % [arXiv:0810.4956 [astro-ph]].
  %%CITATION = MNRAA,403,2152;%%
\vskip .2cm

%\cite{Ishak:2010zh} 
\bibitem{Ishak:2010zh}
  M.~Ishak and W.~Rindler,
  %``The Relevance of the Cosmological Constant for Lensing,''
  Gen.\ Rel.\ Grav.\  {\bf 42}, 2247 (2010).
 % [arXiv:1006.0014 [astro-ph.CO]].
  %%CITATION = GRGVA,42,2247;%%
\pagebreak

\bibitem{Schucker:2007}
  T.~Schucker, 
Gen.\ Rel. \ Grav.\ {\bf 41}, 67 (2009).
%% arXiv:0712.1559[astro-ph]. 
\vskip .2cm


\bibitem{Garfinkle:1985}
  D.~Garfinkle,  
  Physical\ Review \ D  {\bf 32}, 1323 (1985).
  \vskip .2cm


\bibitem{Vilenkin1:1981}
  L.~H.~Ford and A.~Vilenkin,
  J.\ Phys.\ A:\ Math. Gen.\  {\bf 14}, 2353 (1981).
 \vskip .2cm


%%\bibitem{Brihaye2:2008}
%%Y.~Brihaye and B.~Hartmann, 
%%arXiv: 0806.0536[hep-th].


\bibitem{Vilenkin3:1981}
  A.~Vilenkin,
  Phys. \ Rev. \ D {\bf 23}, 852 (1981).
\vskip .2cm



\bibitem{Vilenkin4:1981}
 M.~Aryal, L.~H.~Ford and A.~Vilenkin,
  Phys. \ Rev. \ D {\bf 34}, 2263 (1986).
\vskip .2cm



\bibitem{Hiscock:1985}
  W.~Hiscock,
  Phys. \ Rev. \ D  {\bf 31}, 3288 (1985).
 \vskip .2cm



\bibitem{Gott:1985}
  J.~R.~Gott III,
  Astrophys. \ J.  {\bf 288}, 422  (1985).
\vskip .2cm


\bibitem{Gregory:1995}
R.~Gregory, A.~Achucarro and K.~Kuijken,
Phys. \ Rev. \ D  {\bf 52}, 5729 (1995).
\vskip .2cm 


\bibitem{Nielsen:1973cs}
  H.~B.~Nielsen and P.~Olesen,
  Nucl. \ Phys. \ B  {\bf 61}, 45 (1973).
\vskip .2cm



\bibitem{Vilenkin2:2000}
  A.~Vilenkin and E.~P.~S.~Shellard,
  ``Cosmic Strings and Other Topological Defects'', {\it Cambridge University Press (2000)}.
 \vskip .2cm

%\cite{misner1}
\bibitem{misner1}
R.~ Arnowitt, S.~Deser and C.~W.~Misner, 
Phys. \ Rev. \ {\bf 117}, 1595 (1960). 
\vskip .2cm


%\cite{misner2}
\bibitem{misner2}
R.~ Arnowitt, S.~Deser and C.~W.~Misner, 
Phys. \ Rev. \ {\bf 118}, 1100 (1960). 
\vskip .2cm


%\cite{misner3}
\bibitem{misner3}
R.~ Arnowitt, S.~Deser and C.~W.~Misner, 
Phys. \ Rev. \ {\bf 122}, 997 (1961). 
\vskip .2cm


%\cite{geroch}
\bibitem{geroch}
R.~Geroch and G.~T.~Horowitz,
Ann. \ Phys. \  {\bf 117}, 1 (1979).
\vskip .2cm


%\cite{penrose}
\bibitem{penrose}
R.~Penrose, R.~D.~Sorkin, E.~Woolgar,
arXiv:gr-qc/9301015.
\vskip .2cm



%\cite{shon1}
\bibitem{shon1}
R.~Schon and S.~T.~Yau, 
Commun.\ Math. \ Phys. \ {\bf 65}, 45 (1979).
\vskip .2cm


%\cite{shon2}
\bibitem{shon2}
R.~Schon and S.~T.~Yau, 
Commun.\ Math. \ Phys. \ {\bf 79}, 231 (1981).
\vskip .2cm


%\cite{Witten:81}
\bibitem{Witten:81}
  E.~Witten,
  %``A New Proof of the Positive Energy Theorem,''
 Commun.\ Math.\ Phys.\ {\bf 80}, 381 (1981).
\vskip .2cm

\pagebreak

%\cite{Hawking:83}
\bibitem{Hawking:83}
 S.~W.~Hawking {\it et al},
  %``Positive Mass Theorems for Black Holes,''
  Commun.\ Math.\ Phys.\ {\bf 88}, 295 (1983).
 \vskip .2cm


%\cite{Abbott:82}
\bibitem{Abbott:82}
  L.~F.~Abbott and S.~Deser,
  %``Stability of gravity with a cosmological constant,''
  Nucl.\ Phys.\ B \ {\bf 195}, 76 (1982).
 \vskip .2cm


%\cite{Shiromizu:94}
\bibitem{Shiromizu:94}
T.~Shiromizu, 
%% ``Positivity of gravitational mass in asymptotically de 
%Sitter spacetimes"
Phys. \ Rev. \ D {\bf 49}, 5026 (1994).
\vskip .2cm


%\cite{kastor}
\bibitem{kastor}
D.~Kastor and J.~Traschen,
Class. \ Quant. \ Grav. {\bf 13}, 2753 (1996).
\vskip .2cm

%\cite{Smarr:1972kt}
\bibitem{Smarr:1972kt}
  L.~Smarr,
  %``Mass Formula For Kerr Black Holes,''
  Phys.\ Rev.\ Lett.\  {\bf 30}, 71 (1973)
  [Erratum-ibid.\  {\bf 30}, 521 (1973)].
  %%CITATION = PRLTA,30,71;%%
\vskip .2cm

%\cite{Hawking:72}
\bibitem{Hawking:72}
 S.~W.~Hawking, 
 Commun.\ Math.\ Phys. \ {\bf 25}, 152 (1972).
\vskip .2cm


\bibitem{Roman}
T.~A.~Roman,
Gen.\ Rel.\ Grav.\ {\bf 20}, 359 (1988).
\vskip .2cm



%\cite{Bekenstein:1973ur}
\bibitem{Bekenstein:1973ur}
  J.~D.~Bekenstein,
  %``Black holes and entropy,''
  Phys.\ Rev.\  D {\bf 7}, 2333 (1973).
  %%CITATION = PHRVA,D7,2333;%%
\vskip .2cm


\bibitem{Bardeen}
J.~M.~Bardeen, B.~Carter and S.~W.~Hawking,
Commun.\ Math.\ Phys.\ {\bf 31}, 161 (1973). 
\vskip .2cm


%\cite{Padmanabhan:2003gd}
\bibitem{Padmanabhan:2003gd}
  T.~Padmanabhan,
  %``Gravity and the thermodynamics of horizons,''
  Phys.\ Rept.\  {\bf 406}, 49 (2005).
%  [arXiv:gr-qc/0311036].
  %%CITATION = PRPLC,406,49;%%
\vskip .2cm


%\cite{Hawking1}
\bibitem{Hawk}
S.~W.~Hawking,
 Commun.\ Math.\ Phys. \ {\bf 43}, 199 (1975).
 \vskip .2cm

%\cite{Hartle:1976tp}
\bibitem{Hartle:1976tp}
  J.~B.~Hartle and S.~W.~Hawking,
  %``Path Integral Derivation Of Black Hole Radiance,''
  Phys.\ Rev.\  D {\bf 13}, 2188 (1976).
  %%CITATION = PHRVA,D13,2188;%%
\vskip .2cm

%%%%%%%%%%%%%%%%%%%%%%%%%%%%%%%%%%%%%%%%%%%%%%%%%%%%%

%%%%%%%%%%%%%%%%%%%%%%%%%%%%%%%%%%%%%%%%%%%%%%%%%%%%%%%%%%%%%%%%%%%%%%%%%%%%%%%%%%%%%%%%%%%%%%%%%%%%%%%%%%%%%%%%%%%%%%%%%%%%%%%%%%%
%\cite{Kraus:1994}
\bibitem{Kraus:1994}
 P.~Kraus and F.~Wilczek,
%[arxiv:gr-qc/9406042], 
Nucl. \ Phys.\ B \ {\bf 437}, 231 (1995).
\vskip .2cm


%\cite{Kraus:1996}                                                     
%%\bibitem{Kraus:1996}
 %%P.~Kraus and E.~Keski-Vakkuri,
 %%Nucl. \ Phys.\ B \ {\bf 491}, 249 (1997).  
 %%[arXiv:hep-th/9610045]



%\cite{Wilczek:2000}
\bibitem{Wilczek:2000}
 M.~K.~Parikh and F.~Wilczek,
 %``Hawking Radiation as Tunneling,''
 Phys. \ Rev. \ Lett. {\bf 85}, 5042 (2000).
 %%[arXiv:hep-th/9907001].  
\vskip .2cm


%\cite{Parikh2:2004}
\bibitem{Parikh2:2004}  
 M.~K.~Parikh,
 %``A Secret Tunnel Through The Horizon,''
 Int.\ J. \ Mod. \ Phys.\  {\bf D13}, 2351 (2004).
 %%[arXiv:hep-th/0405160].  
\vskip .2cm


%\cite{Paddy1:1999}
\bibitem{Paddy1:1999}
 K.~Srinivasan and T.~Padmanabhan,
 Phys. \ Rev. \ D {\bf 60}, 24007 (1999).

\vskip .2cm

\bibitem{Paddy2:2001}
S.~Shankaranarayanan, K.~Srinivasan and T.~Padmanabhan, 
 %``Method of complex paths and general covariance of Hawking radiation,''
 Mod. \ Phys. \ Lett. {\bf A 16}, 571 (2001).
 %%[arXiv:gr-qc/0007022v2].  
 
\pagebreak

%\cite{Paddy3:2002}
\bibitem{Paddy3:2002}
S.~Shankaranarayanan, T.~Padmanabhan and K.~Srinivasan,  
 %``Hawking radiation in different coordinate settings: Complex paths approach,''
 Class. \ Quant. \ Grav. {\bf 19}, 2671 (2002).
 %%[arXiv:gr-qc/0010042v4].  



\bibitem{rb} 
R.~Banerjee and B.~R.~Majhi,
 JHEP \ {\bf 0806}, 095 (2008). 
\vskip .2cm



%\cite{Zhu:2008hn}
%%\bibitem{Zhu:2008hn}
 %% T.~Zhu and J.~R.~Ren,
  %``Corrections to Hawking-like Radiation for a Friedmann-Robertson-Walker
  %Universe,''
 %% Eur.\ Phys.\ J.\  C {\bf 62}, 413 (2009).
 % [arXiv:0811.4074 [hep-th]].
  %%CITATION = EPHJA,C62,413;%%


%\cite{Zhu:2009wa}
\bibitem{Zhu:2009wa}
  T.~Zhu, J.~R.~Ren and D.~Singleton,
  %``Hawking-like radiation as tunneling from the apparent horizon in a FRW
  %Universe,''
  Int.\ J.\ Mod.\ Phys.\  D {\bf 19}, 159 (2010).
 % [arXiv:0902.2542 [hep-th]].
  %%CITATION = IMPAE,D19,159;%%
\vskip .2cm

  %%CITATION = ARXIV:0910.3934;%%
%Pizzi arXiv:0904.4572v2 and Belinski arXiv:gr-qc/0607137v1 , arXiv:0910.3934v1)





%\cite{Kerner:2008qv}
\bibitem{Kerner:2008qv}
  R.~Kerner and R.~B.~Mann,
  %``Charged Fermions Tunnelling from Kerr-Newman Black Holes,''
  Phys.\ Lett.\  B {\bf 665}, 277 (2008).
  %[arXiv:0803.2246 [hep-th].
  %%CITATION = PHLTA,B665,277;%%
\vskip .2cm

%\cite{Li:2008zra}
\bibitem{Li:2008zra}
  R.~Li and J.~R.~Ren,
  %``Hawking radiation of Dirac particles via tunneling from Kerr black hole,''
  Class.\ Quant.\ Grav.\  {\bf 25}, 125016 (2008).
 % [arXiv:0803.1410 [gr-qc]].
  %%CITATION = CQGRD,25,125016;%%
\vskip .2cm


%\cite{Li:2010zzd}
\bibitem{Li:2010zzd}
  H.~L.~Li,
  %``Charged Dirac particles' Hawking radiation via tunneling from the general
  %non-extremal rotating charged black hole of D=5 minimal gauged
  %supergravity,''
  Eur.\ Phys.\ J.\  C {\bf 65}, 547 (2010).
  %%CITATION = EPHJA,C65,547;%%
\vskip .2cm
 
  %\cite{Hashimoto:2003}
%\bibitem{Hashimoto:2003}
 %E.~Gimon and A.~Hashimoto,
 %``Black Holes in Go¨del Universes and pp Waves,''
 %Phys.\ Rev.\ Lett. \ {\bf 91}, 021601 (2003).
 %%[arXiv:hep-th/0304181]


 %\cite{Kerner:2007jk}
%\bibitem{Kerner:2007jk}
%  R.~Kerner and R.~B.~Mann,
  %``Tunnelling from Goedel black holes,''
 % Phys.\ Rev.\  D {\bf 75}, 084022 (2007)
  %[arXiv:hep-th/0701107].
  %%CITATION = PHRVA,D75,084022;%%



 %\cite{Chen:2008ys}
%\bibitem{Chen:2008ys}
 % S.~B.~Chen, B.~Wang and J.~l.~Jing,
  %``Scalar emission in a rotating G\'{o}del black hole,''
 % Phys.\ Rev.\  D {\bf 78}, 064030 (2008).
  %[arXiv:0806.2177 [gr-qc]].
  %%CITATION = PHRVA,D78,064030;%%



%\cite{Hawking:72}
%\bibitem{Hawking:72}
% S.~W.~Hawking, 
% Commun.\ Math.\ Phys. \ {\bf 25}, 152 (1972).






%\cite{Chen:2009bja}
\bibitem{Chen:2009bja}
  D.~Y.~Chen, H.~Yang and X.~T.~Zu,
  %``Hawking radiation of black holes in the $z = 4$ Horava-Lifshitz gravity,''
  Phys.\ Lett.\  B {\bf 681}, 463 (2009).
  %[arXiv:0910.4821 [gr-qc]].
  %%CITATION = PHLTA,B681,463;%%
\vskip .2cm



 %\cite{Yale:2008kx}
\bibitem{Yale:2008kx}
  A.~Yale and R.~B.~Mann,
  %``Gravitinos Tunneling from Black Holes,''
  Phys.\ Lett.\  B {\bf 673}, 168 (2009).
  %[arXiv:0808.2820 [gr-qc]].
  %%CITATION = PHLTA,B673,168;%%
\vskip .2cm
%%%%%%%%%%%%%%%%%%%%%%%%%%%%%%%%%%%%%%%%%%%%%%%%%%%%%%%%%%%%%%%%%%%%%%%%%%%%%%%%%%%%%%%%%%%%%%%%%%%%%%%%%%%%%%%%%%%%%%%%%%%%%%%%%%%%%%%%%%
%\cite{Goheer:2002vf}
\bibitem{Goheer:2002vf}
  N.~Goheer, M.~Kleban and L.~Susskind,
  %``The trouble with de Sitter space,''
  JHEP {\bf 0307}, 056 (2003).
%  [arXiv:hep-th/0212209].
  %%CITATION = JHEPA,0307,056;%%

\vskip .2cm
%\cite{Saida:2009ss}
\bibitem{Saida:2009ss}
  H.~Saida,
  %``To what extent is the entropy-area law universal ? -- Multi-horizon and
  %multi-temperature spacetime may break the entropy-area law --,''
  Prog.\ Theor.\ Phys.\  {\bf 122}, 1515 (2010).
 % [arXiv:0910.2510 [gr-qc]].
  %%CITATION = PTPKA,122,1515;%%

\vskip .2cm
%\cite{Saida:2009up}
\bibitem{Saida:2009up}
  H.~Saida,
  %``de Sitter thermodynamics in the canonical ensemble,''
  Prog.\ Theor.\ Phys.\  {\bf 122}, 1239 (2010).
%  [arXiv:0908.3041 [gr-qc]].
  %%CITATION = PTPKA,122,1239;%%
\vskip .2cm

%\cite{Urano:2009xn}
\bibitem{Urano:2009xn}
  M.~Urano, A.~Tomimatsu and H.~Saida,
  %``Mechanical First Law of Black Hole Spacetimes with Cosmological Constant
  %and Its Application to Schwarzschild-de Sitter Spacetime,''
  Class.\ Quant.\ Grav.\  {\bf 26}, 105010 (2009).
 % [arXiv:0903.4230 [gr-qc]].
  %%CITATION = CQGRD,26,105010;%%
\vskip .2cm


%\cite{Carter:69}
\bibitem{Carter:69}
B.~Carter, 
 J. \ Math. \ Phys. \ {\bf 10}, 70 (1969).
\vskip .2cm


\bibitem{Izquierdo:2005ku}
  G.~Izquierdo and D.~Pavon,
  %``Dark Energy And The Generalized Second Law,''
  Phys.\ Lett.\  B {\bf 633}, 420 (2006).
%%  [arXiv:astro-ph/0505601].
  %%CITATION = PHLTA,B633,420;%%
\vskip .2cm



\bibitem{Allen:90}
T.~ J.~Allen, M.~J.~Bowick, and A.~Lahiri,
 Phys. \ Lett. \ B \ {\bf 237},
47 (1990).
\vskip .2cm

%\cite{Bowick:1988xh}
\bibitem{Bowick:1988xh}
  M.~J.~Bowick et al,
  %``Axionic Black Holes And A Bohm-Aharonov Effect For Strings,''
  Phys.\ Rev.\ Lett.\  {\bf 61}, 2823 (1988).
  %%CITATION = PRLTA,61,2823;%%
\vskip .2cm

%\cite{Achucarro:1995nu}
%%\bibitem{Achucarro:1995nu}
 %% A.~Achucarro, R.~Gregory and K.~Kuijken,
  %``Abelian Higgs hair for black holes,''
 %% Phys.\ Rev.\ D {\bf 52}, 5729 (1995).
  %% [arXiv:gr-qc/9505039].
  %%CITATION = GR-QC 9505039;%%

%\cite{Volkov:1998cc}
\bibitem{Volkov:1998cc}
For a comprehensive review including original references, see
  M.~S.~Volkov and D.~V.~Gal'tsov,
  %``Gravitating non-Abelian solitons and black holes with
  %Yang-Mills  fields,'' 
  Phys.\ Rept.\  {\bf 319}, 1 (1999).
  %% [arXiv:hep-th/9810070].
  %%CITATION = HEP-TH 9810070;%%
\pagebreak

\bibitem{Brihaye:2006kn}
  Y.~Brihaye {\it et al.}
  %% , B.~Hartmann, E.~Radu and C.~Stelea,
  %``Cosmological monopoles and non-abelian black holes,''
  Nucl.\ Phys.\  B {\bf 763}, 115 (2007).
  %% [arXiv:gr-qc/0607078].
  %%CITATION = NUPHA,B763,115;%%
\vskip .2cm



%\cite{Coleman:1991ku}
\bibitem{Coleman:1991ku}
  S.~R.~Coleman, J.~Preskill and F.~Wilczek,
  %``Quantum hair on black holes,''
  Nucl.\ Phys.\ B {\bf 378}, 175 (1992).
  %% [arXiv:hep-th/9201059].
  %%CITATION = HEP-TH 9201059;%%
\vskip .2cm


%\cite{Lahiri:1992yz}
\bibitem{Lahiri:1992yz}
A.~Lahiri,
%``An Alternative scenario for nonAbelian quantum hair,''
Phys.\ Lett.\ B {\bf 297}, 248 (1992).
%% [hep-th/9202045].
%%CITATION = HEP-TH 9202045;%%
\vskip .2cm


%\cite{Dvali:2006az}
\bibitem{Dvali:2006az}
  G.~Dvali,
  %``Black holes with quantum massive spin-2 hair,''
  Phys.\ Rev.\ D {\bf 74}, 044013 (2006).
  %% [arXiv:hep-th/0605295].
  %%CITATION = HEP-TH 0605295;%%
\vskip .2cm


%\cite{Cruz:1999gd, Schleich:2009uj, Bernabeu:2009ug,Schleich:2009ix }
\bibitem{Cruz:1999gd}
  J.~Cruz {\it et al},
  %``Integrable models and degenerate horizons in two-dimensional gravity,''
  Phys.\ Rev.\  D {\bf 61}, 024011 (2000).
 %% [arXiv:hep-th/9906187].
  %%CITATION = PHRVA,D61,024011;%%
\vskip .2cm


%%Carter:1968ks
%\cite{Schleich:2009uj}
\bibitem{Schleich:2009uj}
  K.~Schleich and D.~M.~Witt,
  %``A simple proof of Birkhoff's theorem for cosmological constant,''
J. \ Math. \ Phys. \ {\bf 51}, 112502 (2010).
 %% arXiv:0908.4110 [gr-qc].
  %%CITATION = ARXIV:0908.4110;%%
\vskip .2cm

%\cite{Bernabeu:2009ug}
\bibitem{Bernabeu:2009ug}
  J.~Bernabeu, C.~Espinoza and N.~E.~Mavromatos,
  %``Cosmological Constant and Local Gravity,''
  Phys.\ Rev.\  D {\bf 81}, 084002 (2010).
 %% [arXiv:0910.3637 [gr-qc]].
  %%CITATION = PHRVA,D81,084002;%%
\vskip .2cm

%\cite{Schleich:2009ix}
\bibitem{Schleich:2009ix}
  K.~Schleich and D.~M.~Witt,
  %``What does Birkhoff's theorem really tell us?,''
  arXiv:0910.5194 [gr-qc].
  %%CITATION = ARXIV:0910.5194;%%
\vskip .2cm




%%%%%%%%%%%%%%%%% NO HAIR UP %%%%%%%%%%%%%%%%%%

%\cite{Perivolaropoulos:2005wa}
\bibitem{Perivolaropoulos:2005wa}
 L.~Perivolaropoulos,
 %``The rise and fall of the cosmic string theory for cosmological
 %perturbations,''
 Nucl.\ Phys.\ Proc.\ Suppl.\  {\bf 148}, 128 (2005).
%%  [arXiv:astro-ph/0501590].
 %%CITATION = NUPHZ,148,128;%%
\vskip .2cm

%\cite{Vilenkin:1981zs}
%%\bibitem{Vilenkin:1981zs}
%% A.~Vilenkin,
 %``Gravitational Field Of Vacuum Domain Walls And Strings,''
%% Phys.\ Rev.\  D {\bf 23}, 852 (1981).
 %%CITATION = PHRVA,D23,852;%%
\vskip .2cm

\bibitem{Tian:1986}
 Q.~Tian,
  Phys. \ Rev. \ D  {\bf 33}, 3549 (1986).
 %%CITATION = PHRVA,D33,3549;%%
\vskip .2cm


%\cite{Linet:1986sr}
\bibitem{Linet:1986sr}
  B.~Linet,
  %``The static, cylindrically symmetric strings in general relativity with
  %cosmological constant,''
  J.\ Math.\ Phys.\  {\bf 27}, 1817 (1986).
  %%CITATION = JMAPA,27,1817;%%
\vskip .2cm

%\cite{BezerradeMello:2003ei}
\bibitem{BezerradeMello:2003ei}
  E.~R.~Bezerra de Mello, Y.~Brihaye and B.~Hartmann,
  %``Strings in de Sitter space,''
  Phys.\ Rev.\  D {\bf 67}, 124008 (2003).
  %%[arXiv:hep-th/0302212].
  %%CITATION = PHRVA,D67,124008;%%
\vskip .2cm



%\cite{Ghezelbash:2002cc}
\bibitem{Ghezelbash:2002cc}
  A.~M.~Ghezelbash and R.~B.~Mann,
  %``Vortices in de Sitter spacetimes,''
  Phys.\ Lett.\  B {\bf 537}, 329 (2002).
 %% [arXiv:hep-th/0203003].
  %%CITATION = PHLTA,B537,329;%%
\vskip .2cm

%\cite{Deser:1983dr}
\bibitem{Deser:1983dr}
  S.~Deser and R.~Jackiw,
  %``Three-Dimensional Cosmological Gravity: Dynamics Of Constant Curvature,''
  Annals Phys.\  {\bf 153}, 405 (1984).
  %%CITATION = APNYA,153,405;%%
\vskip .2cm


%\cite{Brihaye:2008uy}
\bibitem{Brihaye:2008uy}
  Y.~Brihaye and B.~Hartmann,
  %``Cosmic strings in a spacetime with positive cosmological constant,''
Phys.\ Lett. \ B  {\bf 669}, 119 (2008). 
%% arXiv:0806.0536[hep-th].
  %%CITATION = ARXIV:0806.0536;%%
\vskip .2cm
%\cite{Garfinkle:1985hr}
%\bibitem{Garfinkle:1985hr}
 % D.~Garfinkle,
  %``General Relativistic Strings,''
  %Phys.\ Rev.\  D {\bf 32}, 1323 (1985).
  %%CITATION = PHRVA,D32,1323;%%

%\cite{Aryal:1986sz}
%\bibitem{Aryal:1986sz}
%  M.~Aryal, L.~H.~Ford and A.~Vilenkin,
  %``COSMIC STRINGS AND BLACK HOLES,''
%  Phys.\ Rev.\  D {\bf 34}, 2263 (1986).
  %%CITATION = PHRVA,D34,2263;%%


%\cite{J.L.Synge:1960zz}
\bibitem{J.L.Synge:1960zz}
  J.~L.~Synge,
   ``Relativity: The General Theory,''
%\href{http://www.slac.stanford.edu/spires/find/hep/www?irn=7684894}{SPIRES entry}
{ \it North-Holland, Amsterdam, (1960)}.
\pagebreak
%\cite{Traschen:1999zr}
\bibitem{Traschen:1999zr}
  J.~H.~Traschen,
  %``An Introduction to black hole evaporation,''
  arXiv:gr-qc/0010055.
  %%CITATION = GR-QC/0010055;%%
\vskip .2cm


%\cite{Chong:2005hr}
\bibitem{Chong:2005hr}
  Z.~W.~S.~Chong {\it et al},
 %M.~Cvetic, H.~Lu and C.~N.~Pope,
  %``General non-extremal rotating black holes in minimal five-dimensional
  %gauged supergravity,''
  Phys.\ Rev.\ Lett.\  {\bf 95}, 161301 (2005).
 % [arXiv:hep-th/0506029].
  %%CITATION = PRLTA,95,161301;%%
\vskip .2cm

%\cite{Kurita:2008mj}
\bibitem{Kurita:2008mj}
  Y.~Kurita and H.~Ishihara,
  %``Thermodynamics of Squashed Kaluza-Klein Black Holes and Black Strings -- A
  %Comparison of Reference Backgrounds --,''
  Class.\ Quant.\ Grav.\  {\bf 25}, 085006 (2008).
 % [arXiv:0801.2842 [hep-th]].
  %%CITATION = CQGRD,25,085006;%%
    \vskip .2cm

%\cite{Ishihara:2005dp}
\bibitem{Ishihara:2005dp}
  H.~Ishihara and K.~Matsuno,
  %``Kaluza-Klein black holes with squashed horizons,''
  Prog.\ Theor.\ Phys.\  {\bf 116}, 417 (2006).
  %[arXiv:hep-th/0510094].
  %%CITATION = PTPKA,116,417;%%
\vskip .2cm

%\cite{Tomizawa:2008hw}
%\bibitem{Tomizawa:2008hw}
 % S.~Tomizawa {\it et al}, 
  %``Squashed Kerr-Godel Black Holes - Kaluza-Klein Black Holes with Rotations
  %of Black Hole and Universe -,''
 % Prog.\ Theor.\ Phys.\  {\bf 121}, 823 (2009).
  %[arXiv:0803.3873 [hep-th]].
  %%CITATION = PTPKA,121,823;%%


%\cite{Horowitz:2002ym}
\bibitem{Horowitz:2002ym}
  G.~T.~Horowitz and K.~Maeda,
  %``Inhomogeneous near-extremal black branes,''
  Phys.\ Rev.\  D {\bf 65}, 104028 (2002).
  %[arXiv:hep-th/0201241].
  %%CITATION = PHRVA,D65,104028;%%
\vskip .2cm


 %\cite{Park:2009zra}
\bibitem{Park:2009zra}
  M.~I.~Park,
  %``The Black Hole and Cosmological Solutions in IR modified Horava Gravity,''
  JHEP {\bf 0909}, 123 (2009).
  %[arXiv:0905.4480 [hep-th]].
  %%CITATION = JHEPA,0909,123;%%
\vskip .2cm


%\cite{Rinaldi:2002tc}
\bibitem{Rinaldi:2002tc}
  M.~Rinaldi,
  %``Toroidal black holes and T-duality,''
  Phys.\ Lett.\  B {\bf 547}, 95 (2002).
  %[arXiv:hep-th/0208026].
  %%CITATION = PHLTA,B547,95;%%
\vskip .2cm

%\cite{Myers:1986un}
\bibitem{Myers:1986un}
  R.~C.~Myers and M.~J.~Perry,
  %``Black Holes In Higher Dimensional Space-Times,''
  Annals Phys.\  {\bf 172}, 304 (1986).
  %%CITATION = APNYA,172,304;%%

\vskip .2cm

\bibitem{pauli} 
M.~Fierz and W.~Pauli,
Proc. \ Roy. \ Soc. \ (Lond.) \ {\bf A 173}, 211 (1939). 
\vskip .2cm


%\bibitem{Fang}
%G.~L.~Fang,
%Astrophys.\ Space \ Sci.\  {\bf 325}, 1 (2010).
%%\vskip 2cm


\bibitem{rarita}
W.~Rarita and J.~Schwinger,
 %“On a Theory of Particles with Half-Integral Spin”,
 Phys. \ Rev. \ {\bf 60}, 61 (1941).






%%%%%%%%%%%%%%%%%%%%%%%%%%%%%%%%%%%%%%%%%%%%%%%%%%%%%%%%%%%%%%%%%%%%%%%%%%
%\cite{Gross:1973ju}
%\bibitem{Gross:1973ju}
 % D.~J.~Gross and F.~Wilczek,
  %``Asymptotically Free Gauge Theories. 1,''
  %Phys.\ Rev.\  D {\bf 8}, 3633 (1973).
  %%CITATION = PHRVA,D8,3633;%%
%\cite{Politzer:1973fx}
%\bibitem{Politzer:1973fx}
 % H.~D.~Politzer,
  %``RELIABLE PERTURBATIVE RESULTS FOR STRONG INTERACTIONS?,''
  %Phys.\ Rev.\ Lett.\  {\bf 30}, 1346 (1973).
  %%CITATION = PRLTA,30,1346;%%

%\bibitem{Bali:2000gf}
 % G.~S.~Bali,
  %``QCD forces and heavy quark bound states,''
  %Phys.\ Rept.\  {\bf 343}, 1 (2001)
  %[arXiv:hep-ph/0001312].
  %%CITATION = PRPLC,343,1;%%
  
  
%\bibitem{hooftstringnotes}
% G.~ 't Hooft,
%INTRODUCTION TO STRING THEORY
%version 14-05-04



%\bibitem{Mandelstam:1974pi}
%  S.~Mandelstam,
  %``Vortices And Quark Confinement In Nonabelian Gauge Theories,''
 % Phys.\ Rept.\  {\bf 23} (1976) 245.
  %%CITATION = PRPLC,23,245;%%
%Mandelstam S 1976 Phys. Rep. 23 245
%\cite{Nambu:1975ba}
  
    
\end{thebibliography}
\end{document}